\documentclass[twoside,chapterprefix,bibtotoc,pagesize,letterpaper]{scrreprt}

\usepackage{ucs}
\usepackage[utf8x]{inputenc}
\usepackage[english]{babel}
\usepackage[intlimits]{amsmath}
\usepackage{amsfonts,amssymb}
\usepackage{bbm}
\usepackage{graphicx}
\usepackage[hang,small,bf]{caption}
\usepackage{subfig}
\usepackage[vflt]{floatflt}
\usepackage{tensor}
\usepackage{fancyhdr}
\usepackage[arrow,matrix]{xy}
\usepackage{xspace}
\usepackage{ifpdf}

  \usepackage{url}
  \usepackage[thref,thmmarks,amsmath]{ntheorem}
  \newcommand{\hypertarget}[1]{}
  
  \newcommand{\texorpdfstring}[2]{#1}
  \newcommand{\Defref}[1]{\thref{#1}}
  \newcommand{\Theoremref}[1]{\thref{#1}}
  \newcommand{\Lemref}[1]{\thref{#1}}
  \newcommand{\Corref}[1]{\thref{#1}}
  \newcommand{\Propref}[1]{\thref{#1}}
  \newcommand{\phantomsection}{}
  \newcommand{\hypersetup}[1]{}

\areaset[0.75cm]{15.5cm}{21.5cm}

\fancypagestyle{plain}{
\fancyhf{}
\fancyfoot[C]{\thepage}

}

\theoremstyle{plain}
\theorembodyfont{\normalfont}
\theoremsymbol{\ensuremath{_\Box}}
\newtheorem{Def}{Definition}[chapter]

\newtheorem{Theorem}[Def]{Theorem}
\newtheorem{Cor}[Def]{Corollary}
\newtheorem{Lem}[Def]{Lemma}
\newtheorem{Prop}[Def]{Proposition}
\newtheorem{Conj}[Def]{Conjecture}

\theoremheaderfont{\sc}\theorembodyfont{\small\upshape}
\theoremstyle{nonumberplain}
\theoremseparator{:}
\theoremsymbol{\rule{1ex}{1ex}}
\newtheorem{Proof}{\underline{Proof}}

\theoremstyle{break}
\theoremheaderfont{\normalfont\small}
\theorembodyfont{\it\small\flushleft}
\theoremsymbol{}

\newcommand{\N}{\mathbb N}
\newcommand{\Z}{\mathbb Z}
\newcommand{\R}{\mathbb R}
\newcommand{\C}{\mathbb C}
\newcommand{\Stwo}{\mathbb S^2}
\newcommand{\SoXSt}{\ensuremath{\mathbb S^1\times\mathbb S^2}\xspace}
\renewcommand{\S}{\ensuremath{\mathbb S^3}\xspace}
\newcommand{\tildeS}{\tilde{\mathbb S}^3}
\newcommand{\T}{\ensuremath{\mathbb T^3}\xspace}
\newcommand{\U}{\mathrm{U(1)}}
\newcommand{\SU}{\mathrm{SU(2)}}

\newcommand{\scalarpr}[2]{\left<#1,#2\right>}
\newcommand{\normadapt}{\mathrm{Norm}^{(\textit{adapt})}}
\newcommand{\normelec}{\mathrm{Norm}^{(\textit{elec})}}
\newcommand{\normdiffexact}{\mathrm{Norm}^{(\textit{diffexact})}}
\newcommand{\normeinstein}{\mathrm{Norm}^{(\textit{einstein})}}
\newcommand{\normweyl}{\mathrm{Norm}^{(\textit{weyl})}}
\newcommand{\normkilling}{\mathrm{Norm}^{(\textit{killing})}}
\newcommand{\lieder}[2]{\mathcal L_{#1}{#2}}
\newcommand{\Connection}[3]{\Gamma\indices{_{#1}^{#2}_{#3}}}
\newcommand{\Connectionh}[3]{\gamma\indices{_{#1}^{#2}_{#3}}}
\newcommand{\Connectiond}[3]{\Gamma\indices{_{#1}_{#2}_{#3}}}
\newcommand{\ConnS}[4]{S(#1)\indices{_{#2}^{#3}_{#4}}}
\newcommand{\ConnectionWeyl}[3]{\hat\Gamma\indices{_{#1}^{#2}_{#3}}}
\newcommand{\Commutator}[3]{C\indices{^{#1}_{#2}_{#3}}}

\newcommand{\tr}{\mathrm{tr}\,}
\newcommand{\scri}{\ensuremath{\mathcal J}\xspace}
\newcommand{\scrim}{\ensuremath{\mathcal J^-}\xspace}
\newcommand{\scrip}{\ensuremath{\mathcal J^+}\xspace}
\newcommand{\scripm}{\ensuremath{\mathcal J^\pm}\xspace}
\newcommand{\idmatrix}{\mathbbm{1}}
\newcommand{\Eqref}[1]{Eq.~\eqref{#1}}
\newcommand{\Eqsref}[1]{Eqs.~\eqref{#1}}
\newcommand{\Sectionref}[1]{Section~\ref{#1}}
\newcommand{\Chapterref}[1]{Chapter~\ref{#1}}
\newcommand{\Partref}[1]{Part~\ref{#1}}
\newcommand{\Figref}[1]{Fig.~\ref{#1}}
\newcommand{\Fignref}[1]{Fig.~\ref{#1}}
\newcommand{\initially}[1]{\left.{#1}\right|_{*}}
\newcommand{\Mathematica}{\textit{Mathematica}\xspace}
\newcommand{\term}[1]{\textbf{#1}}

\graphicspath{{figures/}}

\begin{document}

\ifpdf
\hypersetup{pdfauthor=Florian Beyer}
\hypersetup{pdftitle=Asymptotics and Singularities in Cosmological Models with positive Cosmological Constant}
\hypersetup{pdfsubject=Mathematical Cosmology} 
\hypersetup{pdfkeywords=PhD Thesis}
\fi

\renewcommand{\figurename}{Fig.}

\bibliographystyle{hplain}

\begin{titlepage}
  \begin{centering} 
    {\Huge\bf
      Asymptotics and singularities\\      
      in cosmological models\\
      with positive cosmological constant\\
    }

    \vspace{2cm}
    {\Large\textbf{Dissertation}}\\
    zur Erlangung des akademischen Grades\\
    ``doctor rerum naturalium''\\
    (Dr.\ rer.\ nat.)\\
    in der Wissenschaftsdisziplin ``Theoretische Physik''
    \vspace{1cm}\\
    {\large von}\\
    \vspace{1cm}
    {\Large {\bf Florian Beyer}} \\
    \vspace{2cm}
    {\large eingereicht bei der} \\
    \vspace{5mm}
    {\Large{Mathematisch-Naturwissenschaftlichen Fakult\"at \\
        \vspace{1mm}
        der Universit\"at Potsdam}}  \\
    \vspace{1cm}
    {\large durchgef\"uhrt in Golm am} \\
    \vspace{5mm}
    {\Large{Max Planck Institut f\"ur Gravitationsphysik}}\\
    \vspace{1cm}
    {\large unter der Betreuung von}  \\
    \vspace{5mm}
    {\Large{Prof.~Dr.~Helmut Friedrich}}\\
    \vfill
    {\Large Potsdam, im Mai 2007} \\    
    
  \end{centering}
\end{titlepage}

\pagestyle{empty}
\chapter*{Anmerkung\markboth{Anmerkung}{Anmerkung}}
\vspace{1ex}
\vfill
\begin{flushleft}
\Large
In der vorliegenden Version wurden einige Druckfehler und inhaltliche
Ungenauigkeiten  beseitigt, so dass 
sie sich geringf\"ugig von der offiziellen Version unterscheidet, die beim
Pr\"ufungssekretariat eingereicht wurde.

\vspace{1cm}

This version differs from the original version of this dissertation
handed in at the university of Potsdam by some corrections and
clarifications. 
\end{flushleft}
\vfill

\newpage
\,\quad

\pagestyle{fancy}

\renewcommand{\contentsname}{Table of Contents}  
\tableofcontents


\pagestyle{empty}
\phantomsection
\addcontentsline{toc}{chapter}{Abstract} 
\section*{Abstract}
{\small
These are exciting times for cosmologists. From the observational
point of view, many highly accurate data sets are available nowadays,
and other, even more sophisticated measurement techniques are
currently being developed. The data and their analyses, coming from an
increasing set of observational sources, keep drawing a more and more
consistent picture. In particular, models in the class of homogeneous
and isotropic solutions of Einstein's field equations can be fitted
successfully. It is, however, surprising that these models correctly
represent the data from so many different sources: on the one hand
because of their simplicity, and on the other hand due to the fact
that a new matter component, the so called dark energy, which amounts
about $70\%$ of the content of the universe, is required. It is
particularly astonishing that its gravitational interaction must be
repulsive, driving accelerated cosmic expansion. In any case, despite
certain theoretical arguments against this, we are justified, due to
excellent agreement with all observations so far, to consider
cosmological models with a non-vanishing cosmological constant - at
least until the forthcoming high-precision observational data become
available.

These are also exciting times for cosmologists from the fundamental
point of view. Both rigorous mathematical, but also numerical and
computational techniques are developing rapidly. Recently they were
successfully used in solving certain outstanding issues of fundamental
importance to general relativity in the cosmological setting. At least
this was possible in certain important cases; the general case,
however, is still open. These issues, in particular the cosmic
censorship conjecture, the BKL-conjecture and the cosmic no-hair
paradigm, are motivated both by our fundamental perspectives on any
quantitative physical theory as general relativity, but also by the
observational facts. Understanding these issues would at first provide
information on how far general relativity can be considered as a
well-defined physical theory. Next, it would yield a characterization
of phenomena which can occur within Einstein's theory and hence also
have to be taken into account in the models of our universe. Thus it
would enable us to decide, if the assumptions made for the interpretations
of the observations are justified.

For the study of these outstanding issues, I consider in this thesis
spacetimes which show accelerated expansion in the future driven by a
non-vanishing cosmological constant as suggested by the
observations. Therefore I describe the development of a new numerical
code concentrating on spacetimes with spatial topologies which are
non-trivial from the numerical point of view. I start by discussing
the underlying ideas and expectations for advantages and disadvantages
of my approach based on spectral methods with explicit regularizations
at the coordinate singularities in comparison to other methods. Then I
analyze my code particularly in non-trivial situations with
cosmological singularities. I am able to obtain the first numerical
results for spacetimes with Gowdy symmetry and spatial 3-sphere
topology. Beside findings about properties of certain gauge conditions
in such situations, I discover interesting evidence: first, on the
non-linear stability property of this class of spacetimes within a
more general class, and second, on some properties of certain Cauchy
horizons. Although more studies are necessary to draw reliable
conclusions about these latter issues, these investigations are able
to lay the foundation for many possibilities of future research, where
my methods can be applied. But the application of my code is not
restricted to questions concerning the properties of Einstein's theory
at this fundamental level. One can also think of applications more
related to observational problems. However, they are not yet
considered in this thesis.
}

\pagestyle{fancy}

\chapter{Preface}
\label{ch:introduction}
\markright{Preface}

When, for the first time in 1998, the authors of
\cite{Perlmutter98,Riess98} reported on their analyses 
of observational data collected from supernovae explosions of type Ia, the 
standard picture  physicists had about our universe was changed
drastically.
On the basis of various assumptions whose
justifications are partly still under investigation, their results 
not only implied that our
cosmos is currently expanding -- this was 
known since the times of Hubble \cite{Hubble29} -- they rather
claimed that
this expansion is \textit{accelerated}. 
There are strong hints that the matter types, that can be studied within the
scope of modern 
laboratory experiments, are not able to drive such an acceleration. An
exotic, so 
far not directly observed matter component must be postulated to
explain these findings. It was christened dark energy, a
name chosen to reflect our current lack of
understanding.  From observations starting in the early
1960s and with 
increasing accuracy in the following decades, one already
knew that the universe can be considered as homogeneous
and isotropic on sufficiently large length scales. The
measurements of the cosmic
microwave background (CMB) confirmed this picture in particular; the first
measurements were done by Penzias and Wilson in 1965, and the first satellite
mission to explore the temperature distribution was the COBE mission
in 1992; see \cite{Partridge95} for a review. In 2004, the WMAP
satellite was launched and the three year data \cite{Spergel06} increased
the accuracy much further. 
It was shown that the fluctuations around the temperature
$2.725$ K are only of the order $10^{-5}$. Indeed, at the time of
recombination, the universe was much more homogeneous than today, and
apparently these tiny inhomogeneities were just right to explain the
formation of the present structure of the universe.

Modern observations, taking into account not only supernovae and the CMB but
also various other sources, keep drawing a more and more
consistent picture which confirms the earlier results from above.
The dimensionless present energy density quantity for
dark energy is roughly $\Omega_{\Lambda}\sim 0.7$ for most of these
measurements, and for the residual matter one has $\Omega_{m}\sim
0.3$. Hence dark energy strongly 
dominates our present universe. 
Note that only a few percent of this ``residual matter'' is
of known type; another deep lack of understanding referred to as the
``dark matter problem''.
The latest state of knowledge of the
cosmological parameters is 
summarized in \cite{Sanchez06}, where the authors focus on results
obtained from measurements of the
CMB. Further data recently obtained
from
supernovae observations including a review of the underlying
analysis techniques can be found in \cite{Perivolaropoulos06}. For a very
comprehensive review of the current observational status, analysis
techniques and the large field of theoretical physics concerning
dark energy 
models, see
\cite{Copeland06}. The last review points to one of the particular
problems in our current picture of the universe: although 
the observations suggest consistently that the equation of state
of dark energy 
is $p=w\rho$ with $w$ around $-1$, we still do not know what dark energy is
actually
made of. According to the opinion of many physicists, there is so far
no completely convincing 
explanation for it, apart from ad hoc models, which would
be in accordance with all fundamental principles of theoretical physics. In
particular, the cosmological constant, which can be considered as a
matter field with equation of state parameter $w=-1$, is considered as
problematic. Nevertheless, it is the simplest
``candidate'' for dark energy and is presently in excellent
agreement with the observations.
In this thesis, such problems are not be elaborated on, and it will be
just assumed that dark energy is ``made'' of a 
non-vanishing cosmological constant. Indeed, already this
non-dynamical model for dark energy 
is able to generate dynamics of high complexity whose
analysis is non-trivial, and it is the purpose of this thesis to shed
further light on such interesting and fundamental phenomena. 
The understanding of spatially inhomogeneous models with
their enormously
complicated character is particularly important for the interpretation
of the 
observed present structure 
of our cosmos, including the fluctuations in the
CMB temperature distribution.

The first published attempt to explain the apparent isotropy and
homogeneity of our universe without appealing to special initial
conditions at the big bang was due to Hoyle et al.\ \cite{Hoyle63} and
was motivated 
by the knowledge about linear\footnote{Nowadays we know that the
  de-Sitter spacetime
is even non-linearly stable, see \Sectionref{sec:non-linear stability of
  de-Sitter}.} stability properties of the de-Sitter 
spacetime. 
The idea of {inflation} was introduced in \cite{Guth81}, and since
then it has become very prominent. 
In this model the cosmos expands exponentially fast
during ``inflationary epochs''. At least one such
epoch took place in the early universe, and apparently we happen to be
in another such expanding phase presently. The term
``inflation'' usually refers to the first of these two epochs; it is
sometimes also called ``super-expansion'' since it is believed to have
involved about $50-80$ e-folds, i.e.\ the scale factor of the universe
was at least $e^{50}$-times smaller before than after inflation. One can expect
the matter distribution and hence the physical processes  driving the
acceleration of the expansion to be very different today than 
in the early 
universe, and so the physical models for both phases must take this
into account. We are not going into this here. Inflation is
supposed to be a solution of a couple of problems for the understanding of our
universe. 
An important one is, that one can 
hope that by such a rapid blow-up process of physical
scales  inhomogeneities are smoothed out and maybe even isotropy
is attained, which would explain why our universe is so homogeneous
and isotropic.
However, at the first sight it seems that this would
strongly depend on the conditions at the beginning of the inflationary
epoch. Nevertheless, there is the 
so called cosmic no-hair conjecture (\Sectionref{sec:cosmicnohair})
which claims that there is a large class of inflationary 
solutions of Einstein's field equations attracted by the de-Sitter
solution. If this held, then inflation would be a very
natural explanation for the apparent homogeneity and isotropy of our
universe. Another important 
problem addressed by inflation is the 
``horizon problem''. The underlying question is: how can it be explained
that the universe is homogeneous on scales corresponding to regions
which have never been in
causal contact?
``Super-expansion'' in the early universe can indeed serve as a
solution to this.  

Within the family of FLRW-models, i.e.\ the spatially homogeneous and
isotropic solutions of Einstein's field equations, the currently
accepted values of the 
cosmological parameters imply that the curvature density $\Omega_k$ of
the spatial slices is very small. Note that this does not imply that
the Ricci tensor of the spatial slices and hence the curvature
is small; nevertheless let us assume for a moment that this is the
case.  Many cosmologists conclude from this that the
spatial slices have the topology of $\R^3$, leading to the
``concordance model'' of our universe. However, such a conclusion
cannot be drawn from local measurements considering only
FLRW-models. Namely, even within this family, 
the geometry 
of the spatial slices can still be any factor 
space of $\R^3$ with vanishing curvature, or of $\S$ with small
positive curvature, or of $\mathbb H^3$ with small negative
curvature. In any case, our universe is \textit{not} spatially
homogeneous, and 
by measurements taking the inhomogeneities into account there is a
chance to deduce the topology or at least to exclude certain
topologies.
Namely, although the concordance model for our
universe is able to reproduce large parts of the power spectrum
of the CMB fluctuations, 
there seem to be deviations in the 
lowest multipole moments. It is maybe possible to explain these by a
different topology of 
the spatial slices than $\R^3$, see \cite{Aurich06,Caillerie07} and
references  therein. With the currently available data, however, it
seems not yet to be
possible to conclude about this issue.
In our investigations we will restrict to compact
spatial slices. The main reason for this 
assumption, which is consistent with all observations so far, is
simplicity from the mathematical, and 
naturalness from the physical point of view in situations, where one is
not willing to accept that there is an infinite 
amount of matter in the universe. 
Later in this thesis I will particularly focus on
spatial $\S$-topology, and the discussion in \cite{Caillerie07} shows that this
topology is indeed a candidate in order to explain the low multipole
deviations in the CMB data.
However, let us not make this restriction yet and allow any compact
spatial topology.

The research presented in this thesis focuses
on the investigation of fundamental outstanding issues of 
cosmological solutions of Einstein's field equations\footnote{EFE are
  the only field equations for gravity which will be assumed in this thesis.} 
$G_{\mu\nu}+\lambda g_{\mu\nu}=T_{\mu\nu}$ (EFE); in
particular of inhomogeneous ones. Here $G_{\mu\nu}$ is the Einstein
tensor, $\lambda$ is the cosmological constant with $\lambda>0$ and
$T_{\mu\nu}$ is the 
stress energy tensor of the matter; the units have been chosen such
that Newton's constant $G$ takes the value $1/8\pi$, more details
are given later. 
In our discussions, the term
``cosmological solution'' refers to globally hyperbolic solutions
of EFE with compact Cauchy surfaces. 
Global
hyperbolicity reflects the fundamental point of view that a physical
theory must  
be deterministic, i.e.\ we must be able to deduce the
evolution of a physical
system from it, for instance our universe, when its state at a given
time is known. This is deeply connected to the causality principle.
The assumption of global hyperbolicity takes care of these
issues as we will discuss later. 
Fortunately, global hyperbolicity turns out to be a natural
requirement in Einstein's theory, because 
there is a well-defined notion
of maximal globally hyperbolic developments
in the Cauchy problem of
Einstein's equations with
its fundamentally physical motivation.
However, there
are also examples of solutions of EFE which are not globally
hyperbolic. 
The $\lambda$-Taub-NUT spacetimes (\Sectionref{sec:TaubNUT}) for instance, 
which will play a role in this thesis work, can be extended in
non-unique ways through Cauchy horizons and
possess closed causal curves. Now, the idea behind strong cosmic censorship
(\Sectionref{sec:SCC}), which has been confirmed only in special
situations so far, 
is that spacetimes, which are extendible in such a way,
should not occur as solutions of the Cauchy problem of Einstein's
field equations in a generic manner. 

I will only consider the vacuum case in this
thesis because it turns out that
vacuum spacetimes already show many complicated phenomena. Due to the
presence of the cosmological constant in my 
investigations, the considered solutions thus represent spacetimes
that are dominated by dark energy. In the time direction of expansion,
this is actually a good approximation to our real universe since, as was
said above, dark energy dominates over other matter fields
presently. In the 
collapsing time direction the results of investigations in vacuum clearly
have only limited validity for modeling our universe; nevertheless, their
study is important in order to identify generic features and to shed
light on many outstanding issues  in general relativity. 

Another outstanding fundamental
issue concerning cosmological solutions which I want to address in
this thesis, besides the strong cosmic
censorship and the cosmic no-hair conjectures already mentioned
above, is the so called
BKL-conjecture
(\Sectionref{sec:bklconjecture}), which claims to describe the
properties of generic gravitational singularities. 
Understanding these issues would, at first, provide
information on how far general relativity can be considered as a
well-defined physical theory. Next, it would yield a characterization
of phenomena which can occur within Einstein's theory and hence also
have to be taken into account in the models of our universe. Thus it
would enable us to decide, whether the assumptions made in order to interpret
the observations are justified. 
For their investigation, I consider the class of future
asymptotically de-Sitter (FAdS) spacetimes 
(\Sectionref{sec:fads}) since these show accelerated future expansion
consistent with the cosmic no-hair
picture in a natural way. By means of his 
conformal field 
equations (\Sectionref{sec:conformal_geom}), Friedrich has
worked out a Cauchy problem for this
class of spacetimes and proved
its well-posedness (\Sectionref{sec:initial_data}). For this, the future
conformal boundary 
$\scrip$, being spacelike in this case, 
is considered as 
the ``initial'' hypersurface where ``initial'' data, subject to certain
constraint conditions, can be prescribed. Then the conformal field   
equations are used to evolve these data into the past.
This allows us
to construct FAdS spacetimes with prescribed future asymptotics, all of
them in
agreement with a generalized cosmic no-hair picture. Although their
future 
behavior is well 
understood, this is not so for their past behavior. 
We know from Friedrich's results
that each data set of \scrip determines a unique corresponding FAdS
solution; however, there is not a complete
understanding of the question which data to choose in order to obtain a certain
past behavior. A particular result is Friedrich's stability of the
de-Sitter spacetime (\Sectionref{sec:non-linear
  stability of de-Sitter}), stating that all
solutions ``close'' to the de-Sitter spacetime at $\scrip$
develop a smooth past conformal boundary and hence obey the
generalized cosmic no-hair picture in the past, too. In general,
however,  it is not 
even clear which types of past behaviors can occur
at all. For example, from the cosmic censorship
point of view one would 
like to exclude that Cauchy horizons form in the past generically, but only
special subclasses of solutions have been studied successfully
regarding this so far.
One can draw a rough incomplete picture
of these issues (\Sectionref{sec:situation_FADS})
based on singularity 
theorems by Andersson and Galloway
(\Sectionref{sec:singularitytheoremGA}), the non-linear stability
result of the de-Sitter spacetime  and the famous Yamabe theorem. The
presence of this latter theorem in the discussion already indicates
that the topology of $\scrip$ plays a subtle role. 
In this thesis I will restrict to the cases when $\scrip$
has either $\T$- or $\S$-topology.
In any case,
there are many outstanding problems,  and we can expect that all
fundamental issues raised above will become 
important. 

Motivated by the fruitful interplay of rigorous and numerical
analysis in the field of relativistic cosmology, I decided to 
study the problems above numerically. Numerical
investigations often have the power to discover and describe phenomenology
without relying on ad hoc approximations. Such information
can sometimes be used to construct analytic rigorous
descriptions. 
Indeed, in our research field this has for instance happened for
the ``chaotic'' mixmaster
singularities and the spiky 
features in Gowdy spacetimes
(\Sectionref{sec:gowdyphenom}). 
Certainly, a numerical investigation
of 
the solution space of EFE can only give us hints about the possible
occurrence of certain
phenomena. The cases, that can be computed during the
life time of a numerical analyst, can in particular only make up a
subset of ``measure zero'' within the set of all cases, and hence it
is part of the art of numerical relativity to choose the considered
cases wisely in order to be able to draw reliable conclusions.

For my studies, I need to implement
Friedrich's Cauchy problem including his conformal 
field equations numerically.  My particular interest in
spacetimes with $\S$-topology complicates the numerical
treatment,  since $\S$ cannot be
covered by a single regular coordinate patch. In contrast,
$\T$, the other manifold of interest here, can be treated with a
single patch ``closed'' by means 
of periodic boundary conditions. In the
numerical relativity community there are mainly two distinct
approaches to situations with such ``non-trivial'' topologies
(\Sectionref{sec:nontrivialtopologies}): 
single patch methods which try to deal 
with associated coordinate singularities, and multipatch methods. The
latter ones have become quite reliable recently; however I decided
to avoid the technical problems involved and to try an alternative single patch
approach based on spectral methods
(\Sectionref{sec:pseudospsectr_background}) for this thesis. 
By making use of the Lie group properties
of $\SU$, I establish that all smooth functions on $\S$ are represented by
Fourier series of special type, when expressed with respect to my choice
of coordinates. This knowledge is exploited in order 
to explicitly regularize the formally singular terms in
Einstein's field equations at the
coordinate singularities (\Chapterref{ch:treatmentofS3}). I find that
this approach makes it possible to treat the 
cases of $\S$- and of $\T$-topology (and even a subcase of
$\Stwo\times\mathbb S^1$) with one common numerical
spectral infrastructure, as will be demonstrated in this thesis.

In the applications I restrict to Gowdy
symmetry (\Sectionref{sec:relevantsymmclass}), although
the code is implemented under less
restrictions requiring so far only $\U$-symmetry. 
There are several
reasons for restricting to the Gowdy class. One reason is the
reduction of
the problem to a simpler case, which is still highly non-trivial,
though, as a first  
step. Furthermore, in the Gowdy class there exists quite a 
large number of rigorous and numerical results
(\Sectionref{sec:gowdyphenomon_real}) and hence many ideas
also for the outstanding problems. One of these
is the discussion of $\S$-Gowdy solutions,
both numerically and rigorously, apart from partial rigorous
results. In any case, note that most of these results about the Gowdy
class are restricted to $\lambda=0$. On
the basis of the ``matter does not matter'' argument
(\Sectionref{sec:bklconjecture}), it is expected that 
the ``presence'' of $\lambda$ does not effect the qualitative features
of singularities. However, one cannot exclude that the cosmological
constant can 
``prevent'' gravitational collapse due to the repulsive forces
associated with it. This can happen either everywhere or 
only in certain regions, and so it is maybe possible to discover
new interesting phenomena.

In
\Partref{part:preliminaries} of this thesis, I summarize the necessary
underlying background material, 
fix the notation and give an overview of relevant existing results and
techniques.  
\Partref{part:treatment} is started with a discussion of the choice of
method based on the actual application problems. Then I describe my
implementation of the code, taking care in particular that the
``non-trivial'' spatial $\S$-topology case can be treated together
with $\T$-topology within a common 
numerical infrastructure. 
Since my applications will be restricted to Gowdy symmetry, I also
discuss certain issues related to this particularly on $\S$. 
The last topic of \Partref{part:treatment} is the
construction of those 
initial data sets which will be used in the numerical experiments
later. In
\Partref{part:analysis} of the thesis,
I apply my numerical method; I analyze the behavior of the code, the
numerical performance, errors  
and convergence, in particular also in singular
situations. 
Further, I discuss differences of my method and other existing 
methods in the literature to conclude about advantages and
disadvantages, focusing on the aspects relevant for my applications.
Besides these tests of the method, I also obtain preliminary results about
the non-linear stability properties of $\S$-Gowdy spacetimes within classes of
spacetimes of lower symmetry. I also start to investigate the
stability of Cauchy horizons in the $\lambda$-Taub-NUT family under
non-linear perturbations. 

In
this thesis I had to give the development and evaluation of 
the method slightly higher priority than the study of actual
fundamental problems of general relativity; 
many of those latter investigations, which have motivated this thesis
project, are left as future research possibilities. I summarize and
discuss them in the light of my results together with
expectations for necessary modifications of my code at
the end of 
\Partref{part:analysis}.
Among those outstanding studies are investigations of the 
properties of singularities in $\S$- and $\T$-Gowdy spacetimes, in
particular also 
of the influence of the cosmological constant. 
A far future aim is the study of
``generic singularities'' in cosmological spacetimes and hence
the BKL-conjecture. I hope that further analyses of the Cauchy
horizon issue lead to interesting insights
concerning the strong cosmic censorship conjecture. Other interesting
research projects are related to the study of the topology of the
stability region of the de-Sitter spacetime and the properties of the
corresponding solutions. For example, near the boundary of this set one
can expect to 
find solutions which one can maybe interpret as cosmological black
hole spacetimes. 
But the application
of my code is not restricted to questions concerning the properties of
Einstein's theory only at this fundamental level. One can also think of
applications related to observational problems, e.g. predictions about the
distribution of the cosmic microwave background fluctuations or the
search for primordial gravitational waves with the planned LISA space
telescope \cite{LISA}. However, they are not yet considered in this thesis.

\part{Preliminaries --- Listing the underlying facts}
\label{part:preliminaries}

\chapter{Mathematical preliminaries}
\section{Elements of causal theory}
\label{ch:LorentzGeometry}

We assume that the very basic notions, like Lorentz manifold,
timelike, null and spacelike vector fields and corresponding integral
curves, time orientation etc.\ are familiar to the reader.  Here we
list only those 
notions and facts which are of particular relevance for this thesis
following the presentations in \cite{oneil,LorentzGeometry}. Further
details can be found in these references.

Let $(M,g)$ be a smooth connected oriented
time-oriented Lorentz manifold of dimension $n$
with signature $(-,+,\ldots,+)$. In most cases we will consider
manifolds of $4$ dimensions. However, there will be situations,
hopefully clear from the 
context, when the general $n$-dimensional case is
discussed. Sometimes, when we want to emphasize the 
Lorentzian structure, we write instead $N+1$ dimensions such that $N$ is
the dimension of ``space''. A
synonym for such Lorentz manifolds is the term \term{spacetimes}.
If a spacetime has a smooth conformal
compactification in the sense of \Sectionref{sec:conformal_geom}, and
if, in that 
given context, we are 
interested in the conformal properties of the spacetime we
will often write $(\tilde M,\tilde g)$ (instead of $(M,g)$) for the
original spacetime  
and $(M,g,\Omega)$ for the conformal spacetime. We will try to make such a
change of notation as transparent for the reader as possible. Further
notation that is used in the context of conformal spaces is
introduced in \Sectionref{sec:conformal_geom}.

We use the
notation $p\ll q$ (with respect to $M$) for two points $p,q\in M$ if $q$
is in the chronological future of $p$, i.e.\ there is a timelike curve
in $M$ from $p$ to $q$, and $p<q$ (with respect to $M$) if $q$ is in
the causal future of $p$. Further, we employ the standard notation for
the causal sets $I_\pm(p)$ and $J_\pm(p)$.
\begin{Def}
  \label{def:causalityconditions}
  We say that $(M,g)$ satisfies the \term{chronology condition} if
  there are no closed timelike curves, the \term{causality condition}
  if the are no closed causal curves and the \term{strong causality
    condition} if for each $p\in M$ and any neighborhood $U\subset M$
  of $p$ there is a neighborhood $V\subset U$ of $p$ such that each
  causal curve which starts and ends in $V$ is completely in $U$.
\end{Def}
The strong causality condition implies the causality condition and the
causality condition implies the chronology condition. The inverse
implications are not true. For example, let $U$ be an open subset of
$M$ with $p\in U$ and $\{q_n\}\subset U$ a sequence of points with
$\lim_{n\rightarrow \infty} q_n=p$ but $q_n\not=p$ for all
$n$. Suppose that for each $n\in\N$ there is a causal curve $\gamma_n$
from $p$ to $q_n$ that leaves $U$ for some parameter time. Now, if $M$
was strongly causal we would be able to find a neighborhood $V$ of $p$ such that
all causal curves starting and ending in $V$ stay in $U$ which is not
the case. Even though $M$ is not strongly causal, it can still satisfy the
causality condition.

\begin{Def}
  Let $c:[a,b]\rightarrow M$ be a curve in $M$. We define its
  \term{length} by
  \[L[c]:=\int_a^b \sqrt{|g_{c(t)}(\dot c(t),\dot c(t))|}\, dt.\]
  For $p,q\in M$, the \term{Lorentzian distance} is
  \[\tau(p,q):=
  \begin{cases}
    \sup\left\{L[c]\,\bigl|\,c:[a,b]\rightarrow M, c(a)=p, c(b)=q\right\} 
    & \text{for } p<q,\\
    0 & \text{for } p\not<q.
  \end{cases}
  \]
  The distance of a point $p\in M$ and a set $A\subset M$ is
  \[\tau(A,p):=\sup_{q\in A} \tau(q,p).\]
\end{Def}

\begin{Def}
  A subset $A\subset M$ is called \term{achronal} (resp.\ \term{acausal}) if
  there is no pair of points $p,q\in A$ with $p\ll q$ (resp.\ $p<q$)
  with respect to $M$. 
\end{Def}
Any achronal spacelike hypersurface in a smooth Lorentz manifold is
acausal. 

\begin{Def}
  We call a subset $A\subset M$ 
  \term{Cauchy surface} if it is hit by each
  inextendible timelike curve in $M$ exactly once. 
\end{Def}
If
$A\subset M$ is a Cauchy surface, then it is an achronal closed
topological hypersurface. Even every inextendible
null curve must hit $A$ but not necessarily only once.
If $A\subset M$ is a Cauchy surface we
find the disjoint decomposition 
\[M=I_-(A)\,\dot\cup\, A\,\dot\cup\, I_+(A)\]
where $\dot\cup$ is the disjoint union.
Furthermore, any two Cauchy surfaces in $M$ are
homeomorphic.

\begin{Def}[Cauchy development]
  \label{def:CauchyDev}
  Let $A\subset M$ be achronal. Then $D_+(A)$, the {future Cauchy
    development} of $A$, is defined as the set of all points $p\in M$ such that
  all past inextendible causal curves in $M$ through $p$ hit
  $A$. Analogously define $D_-(A)$, the {past Cauchy
    development} of $A$. The {Cauchy
    development} of $A$ is $D(A):=D_+(A)\cup D_-(A)$.
\end{Def}
If $\Sigma\subset M$ is a Cauchy surface of $M$ then $D(\Sigma)=M$. If
$\Sigma$ is an achronal subset
and $D(\Sigma)=M$, then $\Sigma$ is a Cauchy surface of $M$.
If $A\subset M$ is a closed achronal subset then $\overline{D^+(A)}$
coincides with the set of all points $p\in M$ such that
all past inextendible \textit{timelike} curves through $p$ hit
$A$; analogously for the past case.

\begin{Def}
A subset
$A\subset M$ is called \term{globally hyperbolic} if 
$A$ satisfies the strong causality condition and for all $p,q\in A$
the set $J_+(p)\cap J_-(q)$ is compact and in $A$.
\end{Def}
An important fact is the following. If $A\subset M$ is achronal, then
$\mathring{D}(A)$ is globally hyperbolic. This is not necessarily true for
$D(A)$, see the remark after Theorem~2.7.3 in
\cite{LorentzGeometry}. However, from this it follows that $M$ is globally
hyperbolic if it has a Cauchy surface because any Cauchy surface $\Sigma$ is
achronal and $M=D(\Sigma)$, being open, equals its interior.
One can show that $D(A)$ is globally hyperbolic (and open) if we
assume that $A$ is actually an acausal topological 
hypersurface in $M$. 

\begin{Theorem}
  \label{th:existence_of_maximal_geodesic}
  Let $\Sigma\subset M$ be a closed achronal spacelike (hence acausal)
  hypersurface and 
  pick $p\in D(\Sigma)$. Then there is a geodesic from $\Sigma$ to $p$
  with length $\tau(\Sigma,p)$ 
  which is orthogonal to $\Sigma$, has no focal point before $p$ and is
  timelike as long as $p\not\in \Sigma$.
\end{Theorem}

\begin{Def}[Cauchy horizon]
  Let $A\subset M$ be achronal. Then
  \[H_+(A):=\overline{D_+(A)}\backslash I_-(D_+(A))=
  \left\{p\in \overline{D_+(A)}\, |\, I_+(p)\cap D_+(A)=\emptyset\right\}\]
  is called {future Cauchy horizon} of $A$. Analogously define the
  {past Cauchy horizon} $H_-(A)$ of $A$. The {Cauchy
    horizon} of $A$ is $H(A):=H_+(A)\cup H_-(A)$.
\end{Def}
For $A$ achronal, $H_{\pm}(A)$ is a closed achronal subset
of $M$. If $A$ is additionally closed then 
$\partial D_{\pm}(A)=H_{\pm}(A)\cup A$.
Now, let $\Sigma$ be a closed acausal topological hypersurface. Then
$H_+(\Sigma)$ is a closed achronal topological hypersurface given by
$H_+(\Sigma)=\overline{D_+(\Sigma)}\backslash D_+(\Sigma)$.
Any
point of $H_+(\Sigma)$ is the starting point of a past directed
inextendible null 
geodesic without conjugate points lying completely in
$H_+(\Sigma)$. In particular, such null curves cannot hit
$\Sigma$. Further $\Sigma$ is a Cauchy surface of $M$ if and only if
$H(\Sigma)=\emptyset$. Hence, under these conditions $\Sigma$ is a
Cauchy surface in particular if all inextendible null geodesics in $M$
hit
$\Sigma$. If $H(\Sigma)\not=\emptyset$, we will often say that $\Sigma$
\textit{has} a Cauchy horizon, while if $H(\Sigma)=\emptyset$, we say
that $\Sigma$ \textit{has no} Cauchy horizon.

\begin{Theorem}
  \label{th:mainsplittheorem}
  \renewcommand{\labelenumi}{(\roman{enumi})}
  Let $(M,g)$ be a connected time-oriented Lorentz manifold. Then
  these statements are equivalent:
  \begin{enumerate}
  \item $(M,g)$ is globally hyperbolic.
  \item $(M,g)$ has a (topological) Cauchy surface.
  \item $(M,g)$ has a smooth spacelike Cauchy surface.
  \item There is a smooth $3$-surface $S$ such that $(M,g)$ is
    isometric to $(\R\times S,\bar g)$ with  
    \[\bar g=-\beta d\tau^2+h_\tau\]
    where $\beta:\R\times S\rightarrow \R^{>0}$ is smooth and $h_\tau$
    is a family of smooth Riemannian metrics on $S$. Each surface
    $\{t\}\times S$ corresponds to a smooth spacelike Cauchy surface
    of $(M,g)$.
  \end{enumerate}
\end{Theorem}

In the proofs of the singularity theorems, this causal theory is applied.
We will say more about some of these theorems in
\Sectionref{sec:singularitytheoremGA}.

\section{Geometry of \texorpdfstring{$\S$}{S3}}
\label{sec:geometryS3}
In this thesis we are particularly interested in the manifold
$\S$. Here we discuss some necessary background material and introduce
coordinates which are well adapted to the symmetries which we will
deal with mostly. The background
material can be found for instance in 
the books by Berger \cite{berger1,berger2}; however, we adapt the
language and notation to our purposes here.

\subsection{Coordinates}
\label{sec:coordinatesS3}
We consider $\S$ as the embedded submanifold of $\R^4$ given by
\[\S=\left\{(x_1,x_2,x_3,x_4)\in\R^4,\;x_1^2+x_2^2+x_3^2+x_4^2=1\right\}.\]
Since $x_i$ are four smooth functions on $\S$ any choice of three of
them forms local coordinates on a suitable subset of $\S$. 
However, we will make use of other local coordinates, namely the \term{Euler
parametrization} of $\S$,
\begin{gather}
  \label{eq:euler_param}
  \begin{split}
    x_1&=\cos\chi\cos(\rho_1+\rho_2),\quad x_2=\cos\chi\sin(\rho_1+\rho_2),\\
    x_3&=\sin\chi\cos(\rho_1-\rho_2),\quad x_4=\sin\chi\sin(\rho_1-\rho_2)
  \end{split}
\end{gather}
in terms of the coordinate functions,
\[(\chi,\rho_1,\rho_2) \in \left]0,\pi/2\right[\times 
\left[0,2\pi\right[\times\left[0,2\pi\right[.\] 
These cover smoothly the dense subset of $\S$ given by
\begin{gather}
\label{eq:definition_tildeS}
\begin{split}
\tildeS:=
\Biggl\{\Bigl.\Bigl(x_1(\chi,\rho_1,\rho_2),x_2(\chi,\rho_1,\rho_2),
      &x_3(\chi,\rho_1,\rho_2),x_4(\chi,\rho_1,\rho_2)\Bigr)\,\Bigr|\\
&(\chi,\rho_1,\rho_2) \in \left]0,\pi/2\right[\times 
\left[0,2\pi\right[\times\left[0,2\pi\right[\Biggr\}.
\end{split}
\end{gather}
The points on $\S$ given
by the limits $\chi\rightarrow 0,\pi/2$ are not smoothly covered and
constitute coordinate singularities.

\begin{figure}[t]
  \centering
  \includegraphics[width=0.49\linewidth]{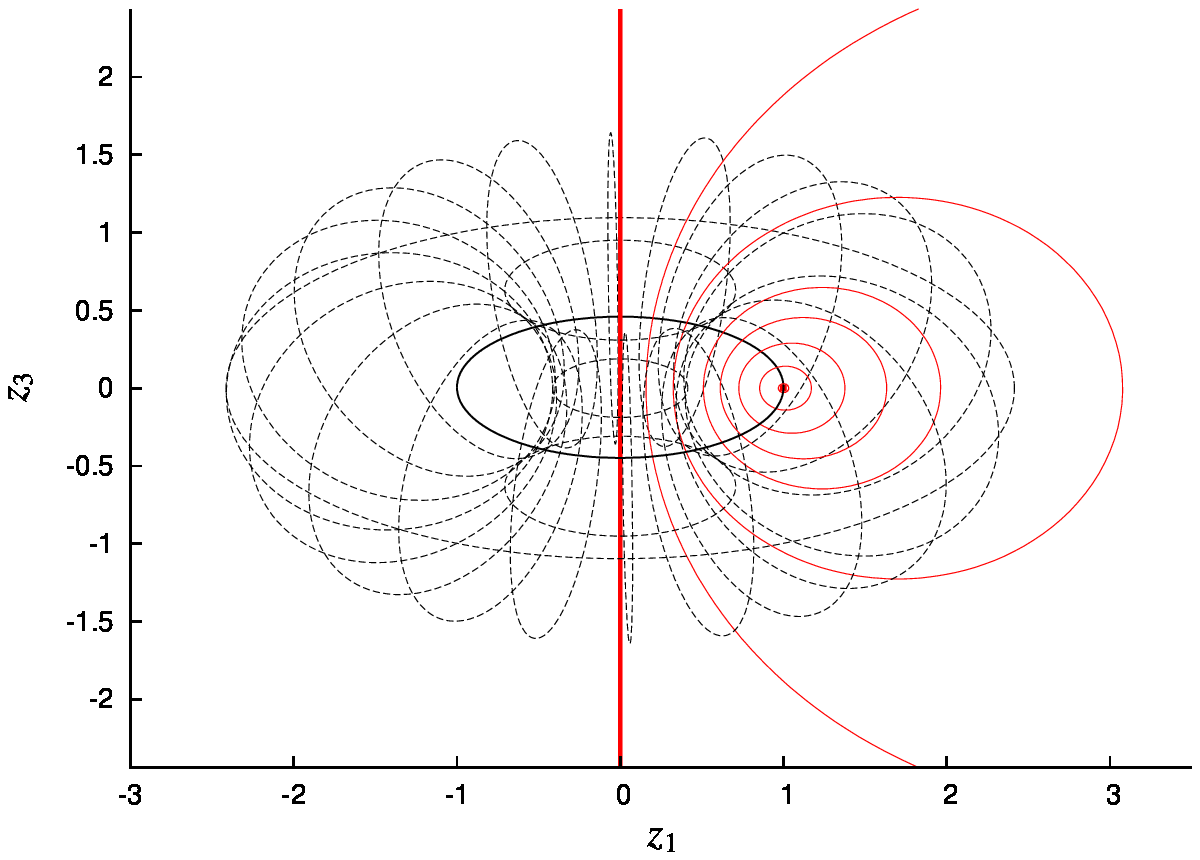}
  \includegraphics[width=0.49\linewidth]{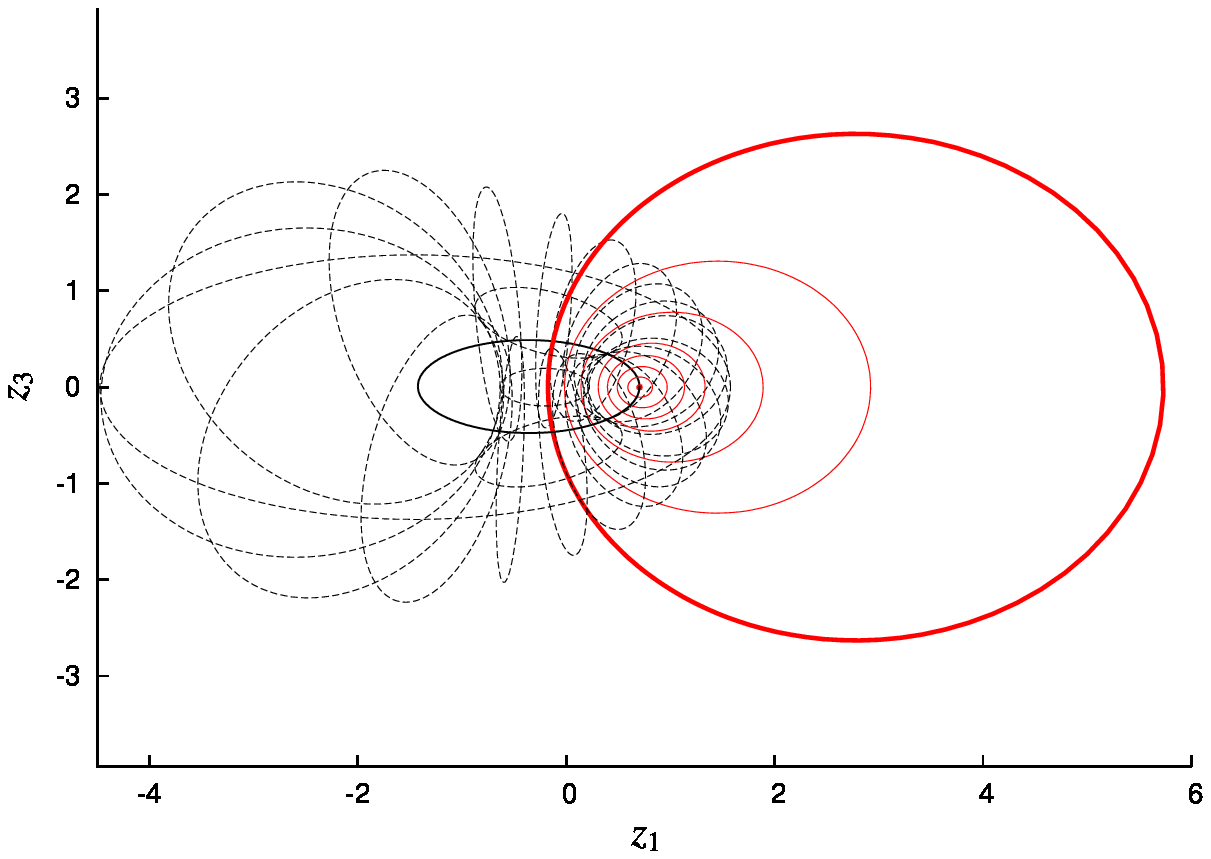}
  \caption{Visualization of the Euler Parametrization by stereographic
    projections}  
  \label{eq:viz_euler_param}
\end{figure}
\Figref{eq:viz_euler_param} visualizes
the geometry of the Euler 
Parametrization. The left picture shows the stereographic projection
with respect to the $x_4$-north pole
\[z_1=\frac{x_1}{1-x_4},\quad z_2=\frac{x_2}{1-x_4},\quad
  z_3=\frac{x_3}{1-x_4}.\] 
Set
\begin{equation}
  \label{eq:lambda12}
  \lambda_1:=\rho_1+\rho_2,\quad \lambda_2:=\rho_1-\rho_2.
\end{equation}
The red ellipses are integral curves of $\partial_{\lambda_2}$ for
$\lambda_1=0$ and for various values of $\chi$; note that for $\chi=0$
this curve is a point while for $\chi=\pi/2$ it corresponds to the
$z_3$-axis. For any $\chi\in]0,\pi/2[$, the integral
curves of $\partial_{\lambda_1}$ and $\partial_{\lambda_2}$ 
generate tori; one of that is drawn by the dashed black curves. For
$\chi=0$ this torus degenerates to a circle in the $z_1$-$z_2$-plane
which is drawn as a solid black curve, and for $\chi=\pi/2$ the
degeneracy is such that the torus becomes the $z_3$-axis. The fact
that this curve is a circle of infinite radius is caused by the choice
of the $x_4$-northpole as the reference point for the stereographic
projection and hence is not related to the geometry of the Euler
parametrization per se. To make this clear, the same curves are drawn
in the right 
picture, but this 
time another point in the intersection of the $x_1$-$x_4$-plane in
$\R^4$ and $\S$ is used as the reference point for the stereographic
projections. Here it becomes more obvious that the two degenerate
tori, i.e.\ that submanifolds of $\S$ that are \textit{not} smoothly
covered by the Euler Parametrization,
can be considered as ``two linked circles'' \cite{chrusciel1990}. The
relation to the \term{Clifford parallelism} is discussed in the books
by Berger \cite{berger1,berger2}.

\subsection{Identification with \texorpdfstring{$\SU$}{SU(2)} 
  and smooth global frames}
\label{sec:identification_su2s3}
The group $\SU$ is the set of complex unitary $2\times 2$-matrices 
with unit determinant together with matrix multiplication. Considered
as a subset of $\R^4$ it obtains a natural smooth manifold
structure. It is a well known fact that there is diffeomorphism
between $\S$ and $\SU$
\begin{equation}
\label{eq:identification_S3SU2}
\Psi:\S\rightarrow\SU,\quad (x_1,x_2,x_3,x_4)\mapsto 
\begin{pmatrix}
  x_1+i x_2 & -x_3+i x_4\\
  x_3+i x_4 & x_1-i x_2
\end{pmatrix}
\end{equation}
which can be used to transport the group 
structure of $\SU$ to $\S$. Hence, both $\SU$ and $\S$ can be
considered as identical Lie groups via the map $\Psi$. In the
following we will not distinguish between $\S$ and $\SU$ anymore
having always the identification map $\Psi$ in mind.

On any group, hence in particular on $\S$,  we can define left and right
translation maps 
\[L, R: \S\times\S\rightarrow\S,\quad (u,v)\mapsto L_u(v):=uv, \quad
  (u,v)\mapsto R_u(v):=vu.\]
On Lie groups, the maps $L_u$ and $R_u$ are diffeomorphisms from the
group to itself for each element $u$.
Those maps can be 
employed to construct smooth global frames: Choose a basis of the
tangent space at the unit element of the group and use the push
forward of $L$ or $R$ to transport this basis smoothly
to any other point of the group. More specifically on $\SU$, we choose
the Pauli matrices\footnote{Note that our normalization misses
  the standard factor $1/2$ and is chosen such that the
  frame $\{Y_a\}$ is orthonormal on the standard unit sphere.}
\begin{equation}
  \label{eq:Def_TildeYa}
  \tilde Y_1=\begin{pmatrix}0&i\\i&0\end{pmatrix},\quad
  \tilde Y_2=\begin{pmatrix}0&-1\\1&0\end{pmatrix},\quad
  \tilde Y_3=\begin{pmatrix}i&0\\0&-i\end{pmatrix}
\end{equation}
as elements of $T_e(\SU)$. When we write this we consider $\SU$
as a Lie subgroup of $\mathrm{GL}(2,\C)$ such that the Lie algebra of
$\SU$ is a subalgebra of $\mathfrak{gl}(2,\C)$.
Now let us define for any $u\in\SU$
\begin{equation}
\label{def:YZ}
(Y_a)_u:=(L_u)_*(\tilde Y_a),\quad (Z_a)_u:=(R_u)_*(\tilde Y_a).
\end{equation}
Clearly, the frame $\{Y_a\}$ is left invariant while the frame
$\{Z_a\}$ is right invariant; both are smooth global frames on
$\S$. In terms of the standard coordinates 
$(x_1,\ldots,x_4)$ on $\R^4$ the coordinate components of those fields,
considering $\SU$ as a subset of $\R^4$ according to
\Eqref{eq:identification_S3SU2}, yields the following expressions
\nobreak
\begin{subequations}
\label{eq:standard_frames_x}
\begin{align}
  Y_1&=-x_4\partial_{x_1}-x_3\partial_{x_2}+x_2\partial_{x_3}+x_1\partial_{x_4}\\
  Y_2&=-x_3\partial_{x_1}+x_4\partial_{x_2}+x_1\partial_{x_3}-x_2\partial_{x_4}\\
  Y_3&=-x_2\partial_{x_1}+x_1\partial_{x_2}-x_4\partial_{x_3}+x_3\partial_{x_4}\\
  Z_1&=-x_4\partial_{x_1}+x_3\partial_{x_2}-x_2\partial_{x_3}+x_1\partial_{x_4}\\
  Z_2&=-x_3\partial_{x_1}-x_4\partial_{x_2}+x_1\partial_{x_3}+x_2\partial_{x_4}\\
  Z_3&=-x_2\partial_{x_1}+x_1\partial_{x_2}+x_4\partial_{x_3}-x_3\partial_{x_4}.
\end{align}
\end{subequations}
With respect to the Euler parametrization they have the representation
\nobreak
\begin{subequations}
\label{eq:coordinate_repr_standard_frame}
\begin{align}
  Y_1&=\sin 2\rho_1\,\partial_\chi
  -\frac 12\cos 2\rho_1\,
  \left[(\tan\chi-\cot\chi)\partial_{\rho_1}
    +(\tan\chi+\cot\chi)\partial_{\rho_2}\right]\\
  Y_2&=\cos 2\rho_1\,\partial_\chi
  +\frac 12\sin 2\rho_1\,
  \left[(\tan\chi-\cot\chi)\partial_{\rho_1}
    +(\tan\chi+\cot\chi)\partial_{\rho_2}\right]\\
  Y_3&=\partial_{\rho_1}\\
  Z_1&=-\sin 2\rho_2\,\partial_\chi
  +\frac 12\cos 2\rho_2\,
  \left[(\tan\chi-\cot\chi)\partial_{\rho_1}
    +(\tan\chi+\cot\chi)\partial_{\rho_2}\right]\\
  Z_2&=\cos 2\rho_2\,\partial_\chi
  +\frac 12\sin 2\rho_2\,
  \left[(\tan\chi-\cot\chi)\partial_{\rho_1}
    +(\tan\chi+\cot\chi)\partial_{\rho_2}\right]\\
  \label{eq:Z3Euler}
  Z_3&=\partial_{\rho_2}.
\end{align}
\end{subequations}
In these expressions, the coordinate singularities at 
$\chi\rightarrow 0,\pi/2$
are explicit.
Moreover, the following relations turn out to be useful
\nobreak
\begin{subequations}
\label{eq:expressZbyY}
\begin{align}
  Z_1=&(\cos2\rho_1\cos2\rho_2\cos2\chi-\sin2\rho_1\sin2\rho_2)Y_1\\
  +&(-\sin2\rho_1\cos2\rho_2\cos2\chi-\cos2\rho_1\sin2\rho_2)Y_2\notag\\
  +&(\cos2\rho_2\sin2\chi)Y_3,\notag\\
  Z_2=&(\cos2\rho_1\sin2\rho_2\cos2\chi+\sin2\rho_1\cos2\rho_2)Y_1\\
  +&(-\sin2\rho_1\sin2\rho_2\cos2\chi+\cos2\rho_1\cos2\rho_2)Y_2\notag\\
  +&(\sin2\rho_2\sin2\chi)Y_3,\notag\\
  \label{eq:expressZ3byY}
  Z_3=&-\cos2\rho_1\sin 2\chi Y_1+\sin2\rho_1\sin 2\chi Y_2+\cos2\chi Y_3.
\end{align}
\end{subequations}

Of great importance will be the commutator relations
\nobreak
\begin{subequations}
\label{eq:commutator_rel_Y}
\begin{align}
  \label{eq:commutator_rel_Ya}
  {}[Y_a,Y_b]&=2\sum_{c=1}^3\epsilon\indices{_a_b_c}Y_c,\\
  [Y_a,Z_b]&=0
\end{align}
\end{subequations}
where $\epsilon\indices{_a_b_c}$ is the totally antisymmetric symbol
  with $\epsilon\indices{_1_2_3}=1$.
Note that the non-standard factor $2$ in \Eqref{eq:commutator_rel_Ya}
is due to our non-standard normalization of the 
Pauli matrices in
\Eqsref{eq:Def_TildeYa}.

The integral curves of the vector fields $Y_a$ and $Z_a$ are circles on
$\S\subset\R^4$. Later on, we will deal with functions $f$ on $\S$
that are  
constant along the circles generated by $Z_3$, i.e.\ $Z_3(f)=0$. I will
call those functions 
{$\U$-symmetric}.

\subsection{Hopf fibration}
\label{sec:Hopf_fibration}
Consider the following map $\Pi:\S\rightarrow\R^3$
\[(y_1,y_2,y_3)=\left(2(x_1x_3+x_2x_4),\,2(x_2x_3-x_1x_4),\,
x_1^2+x_2^2-x_3^2-x_4^2\right).\]
One checks that 
$\text{Im }\Pi=\Stwo\subset\R^3$ and, indeed, we can consider $\Pi$ as a
surjective map $\S\rightarrow\Stwo$ which we will always do in the following.
One checks straight forwardly that $\Pi$ is smooth.
Further set for any $p\in\Stwo$, $(\hat Y_3)_p:=\Pi_*(Y_3)_q$ for a
$q\in\Pi^{-1}(\{p\})$. Independent of the choice of $p$ and $q$ we have that
$(\hat Y_3)_p=0$.  
Further, for any $p\in\Stwo$ there is a neighborhood $U\subset\Stwo$ and
a diffeomorphism $\Phi_{U}:\Pi^{-1}(U)\rightarrow U\times\mathbb S^1$ that
is compatible, i.e.\ $\pi_1\circ\Phi_U=\Pi$.  Here $\pi_1$ is the
projection on the first factor. This is local
triviality. Hence $\Pi$ can be considered as the
projection map of a smooth fiber bundle $\S\rightarrow\Stwo$ with
structure group $\U$ generated by the fibers tangential to $Y_3$. In
particular, the group $\U$ generated by $Y_3$
acts on $\S$ such that the 
quotient manifold obtains a natural smooth structure and is
diffeomorphic to $\Stwo$.
This bundle is called
\term{Hopf bundle} and $\Pi$ is referred to as \term{Hopf fibration}.

In fact, it turns out that in terms of the Euler parametrization of
$\S$, the Hopf fibration takes the form
\[(y_1,y_2,y_3)=(\sin 2\chi\cos 2\rho_2, \sin 2\chi\sin 2\rho_2,
\cos 2\chi)\]
which means that the Euler parametrization is related in this simple
way to the standard coordinates on $\Stwo$. With this, the local
trivializations of the Hopf bundle can be written (in a sloppy fashion) as 
$\S\rightarrow\Stwo\times\mathbb S^1$, $(\chi,\rho_1,\rho_2)\mapsto
((2\chi,2\rho_2),\rho_1)$.

Now choose a smooth local section in the bundle and define
$(\hat Y_a)_{\Pi(q)}:=\Pi_*(Y_a)_q$ for all $q$ in the image of the
section. If the domain of the section is sufficiently small, it yields
a smooth local frame $\{\hat Y_1,\hat Y_2\}$ on  
$\Stwo$, since we have already found that $\hat Y_3=0$. With the
Euclidean scalar 
product inherited from $\R^3$ we get
\[\scalarpr{\hat Y_1}{\hat Y_1}=\scalarpr{\hat Y_2}{\hat Y_2}=4, \quad
\scalarpr{\hat Y_1}{\hat Y_2}=0.\]
Let $e_{\chi}:=\partial_{2\chi}$ and
$e_{\rho_2}:=\partial_{2\rho_2}/\sin 2\chi$ be the standard
orthonormal frame on $\Stwo$ with respect to the standard Euclidean
metric of $\R^3$. Then it turns out that
\[\hat Y_1=2\left(\sin2\rho_1 e_\chi-\cos2\rho_1 e_{\rho_2}\right),\quad
\hat Y_2=2\left(\cos2\rho_1 e_\chi+\sin2\rho_1 e_{\rho_2}\right),
\]
thus they are, apart from the factor $2$, just the standard frame
$(e_\chi,e_{\rho_2})$ rotated by 
the angle $2\rho_1$.

For the coordinate transformation
\[\bar x_1=x_1,\,\bar x_2=x_2,\,\bar x_3=x_4,\,\bar x_4=x_3\]
the corresponding left and right invariant frame fields transform
\[\bar Y_1=Z_2,\,\bar Y_2=Z_1,\,\bar Y_3=Z_3,
\,\bar Z_1=Y_2,\,\bar Z_2=Y_1,\,\bar Z_3=Y_3\]
according to the relations \Eqsref{eq:standard_frames_x}. Thus, in the
new coordinates the fibers of the Hopf fibration are tangent to
$\bar Z_3$. Thus, it turns out that in the expression that relate the Euler
parametrization of $\S$ to the standard coordinates on $\Stwo$ the
coordinate functions $\rho_1$ and $\rho_2$ just change their roles;
the same happens for the frame expressions.

\subsection{Generalized Fourier series on
  \texorpdfstring{$\SU$}{SU(2)}}
\label{sec:S3GenFourier}

\subsubsection{Explicit representations}
\label{sec:fourier_series_su2}
In this thesis, we will use a spectral method to do numerical
calculations on spacetimes with spatial \S-topology. For this it is
crucial to study generalized Fourier series.
Reference \cite{sugiura} gives the basic elements of
harmonic analysis and the Peter-Weyl Theorem. This theory is now
applied to the 
$\SU$-case to construct a basis for $L^2(\SU)$. This procedure leads
to the well known spin-spherical harmonics and is indicated, although
without details, in \cite{sugiura}. 

Here, the space $L^2(\SU)$ is defined with respect to the standard Haar
measure induced by the standard metric on the unit sphere.
Let\footnote{Convention for the whole thesis: $0\in\N$.} $n\in\N$ and
$V^n$ be the space of 
complex homogeneous polynomials in two complex variables of degree
$n$ with the basis
\begin{equation*} 
  \left\{\varphi^n_i(z_1,z_2):=z_1^i\,
    z_2^{n-i},\,i=\{0,\ldots,n\}\right\}.
\end{equation*}
Define the map
\[T: \C^2\times\SU\rightarrow\C^2,\quad ((z_1,z_2),u)\mapsto
T_u(z_1,z_2):=(z_1,z_2)\cdot u\]
where the dot denotes matrix multiplication.
The action
\[U^n: \SU\times V^n\rightarrow V^n,\quad (u,f)\mapsto U^n_u(f):=f\circ T_u\]
is a unitary representation\footnote{Continuous with respect to the
  canonical topology on $V^n$.} of $\SU$ for each $n\in\N$ if we
choose the scalar product\footnote{Convention for scalar products:
  linear in the first 
  argument, antilinear in the second argument.} on $V^n$ determined by  
\begin{equation}
\label{eq:scalarpr_representation}
\left<\varphi^n_i,\varphi^n_k\right>=k! (n-k)!\delta_{ik}.
\end{equation}
One can prove that for each $n$, this representation is
irreducible. Furthermore, it is an important fact that any irreducible
unitary representation of $\SU$ is equivalent to $U^n$ for one
$n\in\N$. We define the unitary matrix elements of these representations as
($n\in\N$, $i,k\in\{0,\ldots,n\}$)
\[w^n_{jk}: \SU\rightarrow\C,\quad
u\mapsto\frac 1{\sqrt{j!(n-j)!k!(n-k)!}}
\left<\varphi^n_j,\varphi^n_k\circ T_u\right>.\]
These functions are smooth and according to the Peter-Weyl theorem
constitute a  complete orthonormal basis for $L^2(\SU)$.
Using the representation (cf.\ \Eqref{eq:identification_S3SU2})
\[u=\begin{pmatrix}
  g_1 & -\bar{g_2}\\
  g_2 & \bar{g_1}
\end{pmatrix}\]
we obtain by means of \Eqref{eq:scalarpr_representation} after some algebra
\[w^n_{jk}(u)=\sqrt{\frac {j!(n-j)!}{k!(n-k)!}}
\sum_{\substack{l\in\{0,\ldots,k\}\\\cap\{j+k-n,\ldots,j\}}}
(-1)^{j-l}
\begin{pmatrix}
  k\\l
\end{pmatrix}
\begin{pmatrix}
  n-k\\j-l
\end{pmatrix}
\bar{g_1}^l g_1^{n-k-j+l}\bar{g_2}^{k-l}g_2^{j-l}.
\]
Taking into account \Eqref{eq:identification_S3SU2} and
\eqref{eq:euler_param} we find
\[g_1=\cos\chi e^{i(\rho_1+\rho_2)},
\quad g_2=\sin\chi e^{i(\rho_1-\rho_2)},\]
thus
\begin{gather}
\label{eq:wnik}
\begin{split}
  w^n_{jk}(\chi,\rho_1,\rho_2)
  =&\sqrt{\frac {j!(n-j)!}{k!(n-k)!}}\,\cdot
  e^{i(n-2k)\rho_1}e^{i(n-2j)\rho_2}\times\\
  &\times \sum_{\substack{l\in\{0,\ldots,k\}\\\cap\{j+k-n,\ldots,j\}}}
  \left[(-1)^{j-l}
    \begin{pmatrix}
      k\\l
    \end{pmatrix}
    \begin{pmatrix}
      n-k\\j-l
    \end{pmatrix}
    \cos^{n-k-j+2l}\chi\sin^{k+j-2l}\chi\right].
\end{split}
\end{gather}
These are the \term{spin-spherical harmonics}.

From the results\footnote{Or by
directly checking with \Eqref{eq:wnik} and
\Eqsref{eq:coordinate_repr_standard_frame}.} in \cite{sugiura}, we can easily
derive 
\nobreak
\begin{subequations}
  \label{eq:frame_on_basis}
  \begin{align}
    Y_1(w^n_{jk})&=-i\left(
      \sqrt{k(n-k+1)}\,w^n_{j,k-1}+\sqrt{(k+1)(n-k)}\,w^n_{j,k+1}
    \right)\\
    Y_2(w^n_{jk})&=
    \sqrt{k(n-k+1)}\,w^n_{j,k-1}-\sqrt{(k+1)(n-k)}\,w^n_{j,k+1}\\
    Y_3(w^n_{jk})&=-i(2k-n)\,w^n_{jk}.
  \end{align}
\end{subequations}

\subsubsection{Expansions of smooth functions on \texorpdfstring{$\SU$}{SU(2)}}
In \cite{sugiura}, we find the following fundamental results. 
\begin{Theorem}
  Let $f:\SU\rightarrow\C$ be a $C^2$-function. The series
  \[\sum_{n=0}^\infty(n+1)\sum_{j,k=0}^n\left(f,w^n_{jk}\right)w^n_{jk}\]
  converges absolutely and uniformly to $f$.
  Here
  $(\cdot,\cdot)$ denotes the standard $L^2$-scalar product on $\SU$.
\end{Theorem}
\begin{Theorem}
  \label{th:smooth_fourier}
   Let $f:\SU\rightarrow\C$ be continuous. The function $f$ is
   $C^\infty$ if and only if its Fourier coefficients 
   \[a^n_{ik}:=(n+1)\left(f,w^n_{jk}\right)\]
   are rapidly decreasing in $n$. 
\end{Theorem}
Hence, under these conditions 
\begin{equation}
  \label{eq:standard_repr_smooth_fct_s3}
  f=\sum_{n=0}^\infty\sum_{j,k=0}^n a^n_{ik} w^n_{ik}
\end{equation}
converges pointwise absolutely and uniformly.

We need some uniform estimates for the functions
$w^n_{ik}$. The following Lemma is not formulated explicitly in
\cite{sugiura} but can be deduced easily from the arguments there.
\begin{Lem}
  \label{lem:estimates_basis}
  For all $n\in\N$, $i,k\in\{0,\ldots,n\}$ we have the following
  estimates (supremum norm on $\S$)
  \[\left\|w^n_{ik}\right\|\le 1,\quad
  \left\|Y_1(w^n_{ik})\right\|,\left\|Y_2(w^n_{ik})\right\|\le n+1,\quad 
  \left\|Y_3(w^n_{ik})\right\|\le n.\]
  \begin{Proof}
    For each $n$ and at each point $u\in\SU$, the values of
    $w^n_{ik}$ form a unitary matrix in $i$ and $k$; hence
    \[\sum_{k=0}^n w^n_{ik}\bar w^n_{lk}=\delta_{il}.\]
    In particular, for a given $i$ one finds
    \[1=\sum_{k=0}^n w^n_{ik}\bar w^n_{ik}
    =\sum_{k=0}^n \left|w^n_{ik}\right|^2\]
    which implies the first claim. With this estimate, the other
    claims can easily be 
    obtained by means of the relations \Eqsref{eq:frame_on_basis} and
    the simple fact that
    \[k(n-k+1)\le\frac {(n+1)^2}4,\quad\forall 0\le k\le n+1.\]
  \end{Proof}
\end{Lem}

We apply this Lemma to prove the following simple but important Proposition.
\begin{Prop}
  \label{prop:swap_sum_derivative}
  Let $f\in C^\infty(S^3)$ be given by the representation
  \[f=\sum_{n\in\N}\sum_{j,k=0}^n a^n_{jk} w^n_{jk}\]
  with $a^n_{jk}\in\C$.
  Then for all $a=1,2,3$,
  \[Y_a(f)=\sum_{n\in\N}\sum_{j,k=0}^n a^n_{jk} Y_a(w^n_{jk}).\]
  \begin{Proof}
    According to \Theoremref{th:smooth_fourier} the coefficients
    $a^n_{jk}$ are rapidly decreasing in $n$ and the convergence of
    the representation formula is pointwise absolute and
    uniform. As soon as we have proven that
    $\sum_{n\in\N}\sum_{j,k=0}^n a^n_{jk} Y_a(w^n_{jk})$ also
    converges uniformly on $\S$ we can choose local coordinates in a
    neighborhood of any point of $\S$ such that $Y_a$ (for a fixed
    $a=1,2,3$) corresponds locally to one coordinate derivative. Then,
    we can apply 
    the result from one-dimensional calculus to prove the claim. Why
    do we have uniform convergence of the latter series? Since the
    coefficients are rapidly 
    decreasing and the $Y_a(w^n_{ik})$ can be estimated uniformly by means of
    \Lemref{lem:estimates_basis} as polynomials in $n$, the terms
    $a^n_{jk} Y_a(w^n_{jk})$ are rapidly decreasing in each point and
    hence can be uniformly estimated by a converging series constant
    on $\S$ (majorant). This establishes
    uniform convergence on $\S$ and hence the claim.
  \end{Proof}
\end{Prop}

\subsubsection{\texorpdfstring{$\U$}{U(1)}-symmetric functions on 
\texorpdfstring{$\SU$}{SU(2)}}
Later, we will restrict to functions with the following symmetry.
\begin{Def}
  \label{def:U1symm}
  A smooth function $f$ on $\S$ is called \term{$\U$-symmetric} if
  $Z_3(f)=0$ everywhere.
\end{Def}
The name is motivated by the fact that $Z_3$ generates a smooth
effective action of the $\U$-group on $\S$. The quotient manifold of
$\S$ and the action of 
the group generated by $Z_3$ is $\Stwo$, which follows by means of the
Hopf fibration (\Sectionref{sec:Hopf_fibration}).

Now, let us fix  the subbasis
\[w_{np}:=w^n_{n/2,n/2+p}\]
for $n\in 2\N$ and $p\in\{-n/2,\ldots,n/2\}$,
cf.\ \Eqref{eq:wnik}.
If $p\ge 0$ we get from \Eqref{eq:wnik}
\nobreak
\begin{subequations}
\label{eq:wnp_representation_explicit}
\begin{gather}
\begin{split}
  w_{np}=&(-1)^{n/2}\,\left(\frac n2\right)!\,
  \frac {1}{\sqrt{(\frac n2+p)!(\frac n2-p)!}}\,\times\\
  &\times\sum_{l=p}^{n/2}
  \left[(-1)^{l}
    \begin{pmatrix}
      \frac n2+p\\l
    \end{pmatrix}
    \begin{pmatrix}
      \frac n2-p\\\frac n2-l
    \end{pmatrix}
    \cos^{2(l-p)}\chi\sin^{n-2l}\chi\right]
  \left(\frac 12\sin 2\chi\, e^{-2i\rho_1}\right)^p.
\end{split}
\end{gather}
One can check furthermore that
\begin{equation}
  \label{eq:wnp_representation_pnegative}
  w_{n,p}=(-1)^p\,\overline{w_{n,-p}}
\end{equation}
\end{subequations}
so that the expression for $p<0$ can be derived from these two results.

\begin{Cor}
  \label{cor:expansion_U1symmetric}
  A smooth function $f:\S\rightarrow\C$ is $\U$-symmetric
  if and
  only if there is a representation 
  \begin{equation}
    \label{eq:wnp_representation}
    f=\sum_{n\in 2\N}\sum_{p=-n/2}^{n/2} a_{n,p}w_{np}
  \end{equation}
  with $a_{n,p}\in\C$ rapidly decreasing. The series representation is
  then converging absolutely and uniformly to $f$. If $f$ is a real valued
  function then
  \begin{equation}
    \label{eq:reality_wnp}
    a_{n,0}\in\R,\quad a_{n,p}=(-1)^p\bar a_{n,-p}\quad
    \forall n\in 2\N,\,n/2\ge p\ge 1.
  \end{equation}
  \begin{Proof}
    This follows easily because the subbasis $w_{np}$ of $w^n_{ik}$
    consists exactly of those functions which do \textit{not} depend
    on $\rho_2$; recall $Z_3=\partial_{\rho_2}$. The reality condition
    is implied by \Eqref{eq:wnp_representation_pnegative}.
  \end{Proof}
\end{Cor}

\chapter{Einstein's field equations}

\section{Conventions and notation}
\label{sec:convnotation}

Let us now introduce the notation and conventions which will be used 
to write tensors, the $3+1$-split and the field equations.
See \Chapterref{ch:LorentzGeometry} for further conventions
on the manifolds.

\paragraph{Coordinates, frames, tensors and indices}
We will mostly deal with globally hyperbolic Lorentz manifolds $(M,g)$
of dimension   
$3+1$. Our notation for
{$3+1$-splits}, see 
\Sectionref{sec:maximaldevelopments}, for which 
one chooses a foliation of $M$  
in terms of spacelike Cauchy surfaces, is as follows. Let
$\{x^\mu\}$ be local coordinates on 
$M$ where Greek indices $\mu,\nu,\ldots=0,1,2,3$ denote spacetime
coordinate indices. For the coordinates in such a $3+1$-split, we will write
$\{t,x^\alpha\}$ with $t=x^0$,
where $\{x^\alpha\}$ (Greek letters
$\alpha,\beta,\ldots=1,2,3$ represent spatial coordinate indices)
represent local coordinates 
on each leaf $\Sigma_{t_0}:=\{p\in M,\,t(p)=t_0\}$ of the foliation.
We assume that those local ``spatial'' coordinates are dragged along
the congruence determined by $\partial_t$ in a smooth manner.

Associated with any smooth semi-Riemannian manifold is the bundle of
orthonormal frames. Any local section in the bundle, i.e.\ any local
orthonormal frame, will be written as $\{e_i\}$ where Latin letters
$i,j,\ldots=0,1,2,3$ indicate spacetime frame indices. Note, that any
globally hyperbolic $4$-dim.\ Lorentz manifold with 
compact orientable spacelike Cauchy surfaces is
parallelizable\footnote{In fact, due to Geroch, all orientable
  globally hyperbolic 
  4-manifolds are parallelizable. A modern reference is
  \cite{parker} which also summarizes the older results.} since, 
due to Stiefel's Theorem \cite{Stiefel36} (see also \cite{Milnor74}
for the background),
such Cauchy surfaces are parallelizable and
$M$ is isometric to $\Sigma_{t_0}\times I$ according to
\Theoremref{th:mainsplittheorem}. However, there are often reasons not
to choose such a global frame, but to work rather with some
collection of locally defined ones. In any case, having
chosen an orthonormal frame $\{e_i\}$, globally or not, we make the
convention that 
$e_0$ is timelike (future or past directed) and call the residual
frame fields $\{e_a\}$ ``spatial'' (Latin letters
$a,b,\ldots=1,2,3$ denote spatial frame indices). But note that this
is not supposed 
to suggest 
that the subframe $\{e_a\}$ is tangent to hypersurfaces
orthogonal to $e_0$. Indeed, it is not (yet) required that
$e_0$ is hypersurface forming.
 The dual frame of
$\{e_i\}$ will mostly be denoted by $\{\sigma^i\}$. The pairing of
any vector space and its dual space will be written as
$\scalarpr{\cdot}{\cdot}$.

Assume that coordinates and an orthonormal frame on $M$ are given as above. For
tensor fields on $M$ we use the following notations. Let for 
example $T:T_pM\times T^*_pM\rightarrow\R$ be a tensor at $p$,
$V\in T_pM$ a tangent vector and $\omega\in T^*_p M$ a covector. We
will use all of the following three ways of writing the same
aspect, namely $T(V,\omega)$, $T\indices{_\mu^\nu}V^\mu\omega_\nu$
and $T\indices{_i^j}V^i\omega_j$. The index notation is based on the
relations
\[T\indices{_\mu^\nu}=T(\partial_\mu,dx^\nu), \quad
T\indices{_i^j}=T(e_i,\sigma^j)\]
together with Einstein's summation convention.
Similar expressions are used for general tensors. The ``abstract''
tensor, i.e.\ the 
multilinear map itself, will be denoted equivalently by $T$,
$T\indices{_\mu^\nu}$ and $T\indices{_i^j}$ and it should be clear
from the context in which situation, for example,
$T\indices{_\mu^\nu}$ means the ``abstract'' tensor and in which it means
the number $T(\partial_\mu,dx^\nu)$. This is the so called
\term{abstract index notation}. 

Indices are shifted in the usual way from up
to down by the metric $g$ and from down to up with the inverse of
$g$. In situations where more than one metric is present it is made
clear  
which metric is used for index manipulations. Finally, we
use standard 
notations for symmetrization and antisymmetrization of tensors, e.g.\
$T_{(\mu\nu)}$ and $T_{[\mu\nu]}$ respectively such that
\[T_{\mu\nu}=T_{(\mu\nu)}+T_{[\mu\nu]}.\]

\paragraph{Connection and curvature}
Assume that a connection\footnote{Note that we will abuse the terminology
  and will often not distinguish between
``covariant derivatives'' and ``connections''. This can be done since
each implies the other, cf.\ \cite{kobayashi}.}
on $M$ is given.  Mostly, we will deal 
with Levi-Civita connections denoted by $\nabla$ of a metric $g$ on
$M$; in \Sectionref{sec:conf_weyl} also Weyl connections
$\hat\nabla$ are introduced and
the corresponding notation is fixed there.
The \term{Christoffelsymbols} of $\nabla$ are defined by
\begin{equation}
\label{eq:christoffel}
\Connection\mu\nu\sigma
:=\scalarpr{dx^\nu}{\nabla_{\partial_\mu}\partial_\sigma}
\end{equation}
and similarly the \term{orthonormal frame connection coefficients},
also called 
\term{Ricci rotation coefficients}, fulfill
\[\Connection ijk
:=\scalarpr{\sigma^j}{\nabla_{e_i}e_k},\quad 
\Connectiond ijk:=\Connection ilk g_{jl}.\]
Note, that the Christoffelsymbols and the frame connection
coefficients have different algebraic properties.

The \term{curvature tensor} (or
\term{Riemann tensor}) of a Levi-Civita connection $\nabla$ (but in
principle for any connection) is 
\begin{equation}
  \label{eq:RiemannDef}
  R(X,Y)Z=\nabla_X\nabla_Y
  Z-\nabla_Y\nabla_X Z-\nabla_{[X,Y]} Z
\end{equation}
for arbitrary $C^2$-vector fields $X,Y,Z$. 
In a coordinate basis (and similar with respect to an orthonormal
frame) we write 
\begin{equation}
\label{eq:indexRiemann}
R\indices{^{\mu}_\nu_\lambda_\rho}
=\scalarpr{dx^\mu}{R(\partial_\lambda,\partial_\rho)\partial_\nu}.
\end{equation}
By contraction of the first and third index one
constructs the \term{Ricci tensor} $R_{\mu\nu}$
and the \term{Ricci scalar} $R$ as the trace of
$R\indices{^\mu_\nu}$. 

\paragraph{Second fundamental form}
Let $n$ be a smooth unit timelike vector field. Its pointwise orthogonal
complements form a smooth distribution\footnote{See \cite{lee} for an
  introduction into the theory of smooth distributions and the
  Frobenius theorem. For instance if $V\in D$, then $V$ is a smooth
  vector field such that $V_p\perp n$ for each $p\in M$.} $D\subset
TM$. We define the following bilinear 
map
\nobreak
\begin{subequations}
\label{eq:2ndFundamentalForm}
\begin{equation}
  \hat\chi:D\times D\rightarrow\R:\quad (V,W)\mapsto g(\nabla_V n,W).
\end{equation}
Let $P$ be the operator that projects any vector field in $TM$
pointwise into $D$ orthogonally. Then we define
\begin{equation}
  \chi:TM\times TM\rightarrow\R:\quad\chi(V,W):=\hat\chi(PV,W).
\end{equation}
In
particular, $\chi$ is a smooth rank $2$ covariant tensor field on $M$.
It has the following property (choose without loss of generality
$V,W\in D$)
\begin{align*}
  \chi(V,W)&=g(\nabla_V n,W)=-g(n,\nabla_V W)
  =-g(n,\nabla_W V)-g(n,[V,W])\\
  &=g(\nabla_W n,V)-g(n,[V,W])=\chi(W,V)-g(n,[V,W]).
\end{align*}
Hence the map $\chi$ is symmetric if and only if the distribution
$D$ is involutive, or in other words the field $n$ is hypersurface
orthogonal. 
Usually, only if $n$ is hypersurface orthogonal, $\chi$ is
called \term{$2$nd fundamental form} of the hypersurfaces orthogonal
to $n$. However, for the sake of simplicity, we will always refer to
$\chi$ as the $2$nd fundamental form (of $n$).
Let an orthonormal frame $\{e_i\}$ be given with $e_0=n$, hence $\{e_a\}$
is spatial. Then one obtains
\begin{equation}
  \chi\indices{_a_b}=\Connection 0ca g_{cb}=\Connection a0b,\quad
  \chi_{0a}=\chi_{a0}=0.
\end{equation}
\end{subequations}

\paragraph{Commutator quantities}
Now, we
introduce the {commutator quantities} $\Commutator ijk$ by  
\begin{equation}
  \label{eq:commutator_fcts}
  \Commutator kij:=\scalarpr{\sigma^k}{[e_i,e_j]}.
\end{equation}
The mutual dependence of the connection
coefficients $\Connection ijk$ and the
commutation functions $\Commutator ijk$ of the orthonormal frame
$\{e_i\}$ is expressed on the one hand as 
\[g(\nabla_{e_i}e_j,e_l)=\frac 12\left\{
  g([e_i,e_j],e_l)-g([e_j,e_l],e_i)+g([e_l,e_i],e_j)
\right\}\]
or equivalently in index notation
\begin{equation}
  \label{eq:commutator2connection}
  \Connectiond ilj=\frac 12\left(
  \Commutator mij g_{ml}-\Commutator mjl g_{mi}+\Commutator mli g_{mj}
\right).
\end{equation}
On the other hand, the torsion freeness of the connection implies
\[\Commutator kij=\Connection ikj-\Connection jki.\]
Lie brackets satisfy the \term{Jacobi identities}
\[[[e_i,e_j],e_k]+[[e_j,e_k],e_i]+[[e_k,e_i],e_j]=0,\]
which lead to the $16$
independent equations for the commutator quantities
\begin{equation}
  \label{eq:jacobiid}
  e\tensor[_{[k}]{(C}{^l_i_{j]}})
  +C\tensor[^m_{[i}_j]{C}{^l_{k]}_m}=0.
\end{equation}

\paragraph{Volume form}
Choose an orientation of $M$
such that the 
frame $\{\partial_0, \partial_1, \partial_2, \partial_3\}$ has positive
orientation. Then we may introduce
the form 
\begin{equation}
  \label{eq:spacetimevolumeform}
  \eta:=-\sqrt{\det g}\, dx^0\wedge dx^1\wedge dx^2\wedge dx^3,
\end{equation}
which is the volume form of $(M,g)$ up to sign\footnote{We employ here
the convention of \cite{vanElst96}.}.
Here $\det g$ is the determinant of the matrix $(g_{\mu\nu})$. In
particular, if $\{e_i\}$ is an oriented orthonormal 
frame, then
\[\eta_{0123}=\eta(e_0,e_1,e_2,e_3)=-1.\]
The tensor $\eta_{ijkl}$ is completely antisymmetric. The
corresponding tensor with upper indices $\eta^{ijkl}$ is also
completely antisymmetric and fulfills $\eta^{0123}=1$. For the volume
form on the orthogonal complement of a timelike
unit vector field $n$ we use the convention
\[\epsilon=-\eta(n,\cdot,\cdot,\cdot).\]
In particular, if $\{e_i\}$ is as above, 
then for $n=e_0$,
\[\epsilon_{123}=\epsilon(e_1,e_2,e_3)=1,\quad \epsilon^{123}=1.\]

\paragraph{Einstein's field equations}
For the whole thesis we choose units such that the speed of light is
$c=1$ and Newton's constant is $G=1/8\pi$. Then \term{Einstein's field
equations} take the form
\nobreak
\begin{subequations}
\begin{equation}
  \label{eq:EFE}
  G_{\mu\nu}+\lambda g_{\mu\nu}=T_{\mu\nu}\quad\text{(EFE)}.
\end{equation}
The \term{Einstein tensor} is
\[G_{\mu\nu}:=R_{\mu\nu}-\frac 12 Rg_{\mu\nu},\]
the constant $\lambda$ is the \term{cosmological constant} and
$T_{\mu\nu}$ is the \term{energy-momentum tensor} of the matter.
In vacuum, given by $T_{\mu\nu}=0$, \Eqref{eq:EFE}
is
\begin{equation}
\label{eq:EFE_vacuum}
G_{\mu\nu}+\lambda g_{\mu\nu}=0\quad\Leftrightarrow\quad
R_{\mu\nu}=\lambda g_{\mu\nu}.
\end{equation}
\end{subequations}
We should point out, that EFE are the only field equations for gravity
in this thesis; in particular we do not consider
modified theories etc.


\section{The Cauchy problem of Einstein's field equations}
\label{sec:maximaldevelopments}

Let $(M,g)$ be a $4$-dim.\ Lorentz manifold which
solves Einstein's field equations in vacuum \Eqref{eq:EFE_vacuum}.
We assume that $M$ is globally hyperbolic so that there is a smooth global
time function $t$ on $M$ and a foliation of spacelike Cauchy surfaces
$\Sigma_t$ which are the $t=const$ surfaces, cf.\
\Theoremref{th:mainsplittheorem}. As before, let us assume 
that we can choose local coordinates 
$\{x^\alpha\}$ on each  $\Sigma_t$ such that $(t,x^\alpha)$ are local
coordinates on $M$ and such that the vector field $\partial_t$ is
smooth. Denote the future pointing normal field of the foliation by $n$.
Then we have the orthogonal decomposition
\begin{equation}
  \label{eq:lapseshift}
\partial_t=N n+\beta,\quad g(\beta,n)=0,
\end{equation}
where $N$ is the \term{lapse function} and $\beta$ the \term{shift
  vector field}.  Let us further choose an orthonormal frame $\{e_i\}$
with $e_0=n$.
Since $(M,g)$ is a solution of EFE, the induced metric $h$ and the $2$nd
fundamental form $\chi$ on a given $\Sigma_t$ satisfy the following
\term{constraint equations}
\nobreak
\begin{subequations}
  \label{eq:ADMconstraints}
  \begin{align}
    \label{eq:Hamconstraint}
    &r-\chi_{ab}\chi^{ab}+(\tr \chi)^2=2\lambda\\
    \label{eq:Momconstraint}
    &D_a (\tr\chi)-D_b\chi\indices{_a^b}=0.
  \end{align}
\end{subequations}
These equations correspond, up to factors, to the 
$(0,0)$- and $(0,a)$-components of the first version of
\Eqref{eq:EFE_vacuum}.
Here $D$ and $r$ are the Levi-Civita covariant derivative and the
scalar curvature, respectively, of $h$. 
If the triple $(\Sigma_t,h,\chi)$ with a Riemannian metric $h$ and a
symmetric covariant $2$-tensor field $\chi$ satisfies the constraint equations,
then it is 
called \term{(vacuum) initial data set} and the pair
$(h,\chi)$ the \term{(induced) data} on $\Sigma_t$. Note that
constraint equations only involve internal derivatives of the given
spatial hypersurface. In contrast to that, the
$(a,b)$-components of \Eqref{eq:EFE_vacuum} constitute
\term{evolution equations}, 
since they include also derivatives in $n$-direction. 

The
\term{Cauchy problem} of EFE is the problem of finding the solution of
EFE corresponding to a prescribed initial data set.  We
are not going to give a discussion about all these issues in full
generality here; such 
can be found in \cite{Friedrich96}. It is however important to point
out that there is a 
large freedom in the way evolution and constraint equations can be
extracted from EFE. First one has to choose the variables by which one
wants to express the equations. Moreover, there are different
possibilities of combining evolution and constraint equation. Further,
one has the choice if one wants to use first or second order systems
in time and in space. This leads to various \term{formulations} of
EFE, some of them are listed in \cite{Friedrich96}. Moreover, one has
to make choices for the gauge.  In particular, Friedrich's
notion of gauge source functions which are further discussed in
\Sectionref{sec:cfe} plays an important role. 

Well-posedness of a given formulation of the Cauchy problem of EFE
involves the following aspects. First, the evolution equations
including the gauge prescription,
possibly after
having restricted to a special class of gauges, must be hyperbolic since
this allows us to deduce short time
existence, uniqueness and stability\footnote{For the theory of
  hyperbolic partial 
  differential equations, in particular symmetric hyperbolic ones, we
  refer the reader to \cite{john82,Majda84,RendalNHS}.}. The evolution
equations are 
then called \term{(hyperbolically) reduced EFE}. However, is the solution
of these reduced equations corresponding to a given initial data set also
a solution of the (full) EFE? Yes, clearly, if and only if the
solution additionally satisfies the constraint equations on each leaf
of the foliation.
To check if this is the case,
one derives the evolution equations for the constraint quantities, the
so called \term{subsidiary system}. Having brought all terms to one
side of the constraint equation system, one can schematically
write the constraints as $C=0$; for instance in
\Eqsref{eq:ADMconstraints} bring $2\lambda$ to the
left hand side. Then $C$ is called
\term{constraint 
  quantity} (or constraint violation quantity). Differentiating the
constraint system with respect to time and after a few
manipulations, typically involving the contracted Bianchi identities and
the evolution equations, one obtains the subsidiary system. Some examples
for these kind of computations can be found in
\cite{Friedrich96}. Now, if it is possible to show that the subsidiary
system implies that if $C=0$ 
initially, then $C\equiv 0$ for all times, one has
proven that the solution of the evolution equations corresponding to an
arbitrary initial data set is a 
solution of Einstein's field equations and hence the given formulation
of the Cauchy problem is well-posed.
In many important cases the subsidiary system turns out to be a
symmetric hyperbolic 
homogeneous system.

In this thesis, we are only going to use two formulations of Einstein's field
equations, namely the conformal field equations and the commutator
field equations which we present in the following two sections. To be
precise, the conformal field equations are not just a special
formulation of EFE but actually an extension thereof; see
\Sectionref{sec:conformal_geom}. 

To make the previous discussion more precise and to
formulate what 
one means by ``cosmic censorship'' eventually, we introduce a bit more
nomenclature. 
A \term{(vacuum) development} of a vacuum initial data set
$(\Sigma,h,\chi)$ is a 
Lorentz manifold $(M,g)$, that satisfies the vacuum EFE
\eqref{eq:EFE_vacuum}, together with an embedding 
$\mathbf{i}:\Sigma\rightarrow M$ such that the pullback of the induced data on
$\mathbf{i}(\Sigma)$ agrees with $h$ and $\chi$. Such a development is
called \term{Cauchy development} if the embedding $\mathbf{i}$ can be
chosen such that $\mathbf{i}(\Sigma)$ is a Cauchy surface of $M$. Let
$(M,g)$ and $(M',g')$ be two developments of a given initial data set
$(\Sigma,h,\chi)$ with maps $\mathbf{i}$ and $\mathbf{i'}$ as
above. If there is a map $\Psi:M\rightarrow M'$ which is an isometry
onto its image such
that $\mathbf{i'}(\Sigma)=\Psi\circ\mathbf{i}(\Sigma)$, then $M'$ is
called an \textbf{extension} of 
$M$. A Cauchy development $(M,g)$ of a given initial data set is called
\term{maximal Cauchy development} if there is no further Cauchy
development of the same initial data set which is an extension of $(M,g)$.
An
important contribution by Choquet-Bruhat and Geroch \cite{choquet69}
building on earlier work by Choquet-Bruhat \cite{choquet52} is the
following theorem.
\begin{Theorem}
  \label{th:MCD}
  Let $(\Sigma,h,\chi)$ be a vacuum initial data set. Then there exists a
  unique (up to isometry) maximal Cauchy development of $(\Sigma,h,\chi)$.
\end{Theorem}
Hence, for a given fixed initial data set, the maximal Cauchy development is
an extension of \textit{all} corresponding Cauchy developments.
In fact, the authors proved this theorem for the case $\lambda=0$, but it is
straight forward to generalize their arguments.  This result shows that the
requirements for a formulation of the Cauchy problem of EFE to be
well-posed discussed above can indeed
be met. The proof of the authors of \cite{choquet52,choquet69} used
the harmonic gauge.  

Note that the notion of maximal Cauchy 
developments presented here with the underlying notion of the Cauchy
problem of EFE is related naturally to the Lorentz
geometrical notion of Cauchy developments in
\Defref{def:CauchyDev}. Let $M$ be the maximal Cauchy development in
the sense here of the initial data set $(\Sigma,h,\chi)$ and $M'$
another development of the same data which is an extension of $M$. For
simplicity, identify $M$ with its image under the corresponding
isometry so that 
$M\subset M'$,
and identify $\Sigma$ with $\mathbf{i}(\Sigma)$, where $\mathbf{i}$ is an
embedding as above such that $\mathbf{i}(\Sigma)$ is a Cauchy surface
of $M$. It follows from the discussion in
\Chapterref{ch:LorentzGeometry} 
that $M=D(\Sigma)\subset M'$.
If $M\not=M'$,
then the Cauchy horizon $H(\Sigma)$ (with respect to 
$M'$) is
non-empty. While $M$ is globally hyperbolic this is then not necessarily
the case for $M'$. Hence, it is important to realize that a maximal
Cauchy development of a given initial data set might not be maximal
among all extensions. Furthermore, if 
$M\not=M'$ then
one can expect generically that there are further
extensions of $M$ which are neither extensions of $M'$ nor
extended by $M'$ because the uniqueness result of \Theoremref{th:MCD}
holds only for maximal Cauchy developments. See also
\cite{Chrusciel94,Chrusciel06} for further discussions on this
aspect. An example is the family of Taub-NUT spacetimes (or their
modification to the case $\lambda>0$ presented in \Sectionref{sec:TaubNUT})
which allow several non-globally hyperbolic
extensions which not even satisfy the
chronology condition (\Defref{def:causalityconditions}).
In this
case, the Cauchy problem of EFE cannot 
determine a unique maximal spacetime for a given initial data
set, and in this sense Einstein's theory
loses its deterministic power. One hopes that such pathological
solutions are  in some sense
non-generic which is the underlying idea of cosmic censorship
elaborated in \Sectionref{sec:cosmiccensorship}. 

For the whole discussion above we assumed that it is actually
possible to find solutions of the constraint equations and hence to
construct initial data sets. For a general extensive discussion on
this issue, see \cite{bartnik04}. In this thesis, we restrict the
construction of initial data to a
special situation; cf.\ \Sectionref{sec:initial_data}.

It should be pointed out that the concept of
well-posedness of an initial (boundary) value problem restricts its
attention to only certain
properties of the solutions like
short-time existence, uniqueness and continuous dependence on
the 
initial data. Hence, this concept can be considered as being
necessary, otherwise one cannot expect that
reasonable conclusions about the solutions of the problem can be
drawn, for instance by numerical 
means. Often, from the analytical point of view, proving
well-posedness is ``all one can hope for'' because further results, 
for instance about long-time existence, blow-up conditions and sharper
energy estimates, are out of reach of the available techniques at hand.
However, from the numerical point of view such information
beyond well-posedness would often be as valuable as the knowledge about
well-posedness itself. For instance, with sharper estimates about the
deviations 
of the solutions under perturbations than usually accessible in
particular in the non-linear case, the errors in the computations could
be estimated more reliably. Motivated by this issue, one
often encounters the phrase
``well-posedness is not enough'' in the numerical literature.

\section{Conformal field equations}
\label{sec:conformal_geom}

\subsection{Conformal geometry}
\label{sec:conf_weyl}

More details on the following discussions can be found in
\cite{Friedrich2002} and references therein.

\paragraph{Conformal spaces and Weyl connections}
Let $(M,g)$ be a smooth pseudo-Rie\-mann\-ian manifold of dimension
$n>3$; not
necessarily a solution of Einstein's field equations. Here we 
discuss a modification of the notion of pseudo-Riemannian manifolds to 
manifolds which carry the following structure.
\begin{Def}
  A (pseudo-Riemannian) \textbf{conformal structure} $\mathcal C$ on a manifold
  $M$ is a set of all 
  locally defined (pseudo-Riemannian) metrics whose domains of
  definition exhaust $M$ 
  and who are related by conformal rescalings in the
  intersections of their domains of definitions. The pair
  $(M,\mathcal C)$ is
  called a \textbf{conformal space}. 
\end{Def}
If not stated otherwise, we will
always assume that the metrics and conformal factors involved
are smooth. A globally defined smooth metric $g$ on $M$ determines a
unique conformal structure denoted as $\mathcal C_g$. In the later
applications, $M$ will actually be a manifold with boundary; however,
the notion of conformal spaces can be extended to that
situation. 
If  there is a
metric $\tilde g\in\mathcal C$ on an open subset of
$M$ that is a solution of Einstein's
field equations in vacuum with $\lambda>0$, then we call $\tilde g$ a
\term{physical   metric}. 

The Levi-Civita connection of $g\in\mathcal C$
is the unique
connection determined by the covariant derivative $\nabla$ that is
metric with respect to $g$ and torsionfree. In an analogous way,
\term{Weyl connections} form the natural class of connections with
respect to
a conformal structure $\mathcal C$. 
The analogy of Levi-Civita connections on pseudo-Riemannian manifolds
and Weyl connections on conformal spaces is best formulated in the
language of connections on orthonormal and conformal frame bundles,
respectively; we do not go into this, but see e.g.\ \cite{ConfGeodesics}.
Weyl connections
are determined by covariant
derivatives
$\hat\nabla$ which are torsionfree and \mbox{compatible} with the conformal
structure $\mathcal C$ in the following manner. Choose $g\in\mathcal C$;
then there is a $1$-form $f$ such that
\begin{equation}
\label{eq:Weylconnection}
\hat\nabla_\mu g_{\nu\rho}=-2 f_\mu g_{\nu\rho}.
\end{equation}
Given any other metric $\tilde g=\Omega^{-2} g\in\mathcal C$, the same
Weyl connection is determined by the $1$-form
\begin{equation*}
  \tilde f=f+\Omega^{-1}d\Omega.
\end{equation*}
Hence, a Weyl connection is determined uniquely by the family of
$1$-forms of the form 
$\{f+\Omega^{-1}d\Omega\,\bigl|\,\text{$\Omega>0$ smooth}\}$ in this
sense. Any  
metric in $\mathcal C$ can be chosen as a representative for
determining a Weyl connection by a $1$-form $f$, any other metric in
$\mathcal C$ requires the transformed $1$-form above to represent the
same Weyl connection. 
 Note
that Levi-Civita connections are just special cases of Weyl
connections. For instance, choosing $f=0$ with respect to $g\in\mathcal
C$, then $\hat\nabla$ is the
Levi-Civita connection of $g$. Accordingly, choosing
$f=-\Omega^{-1}d\Omega$, then $\hat\nabla$ is the Levi-Civita
connection of the metric $\tilde g=\Omega^{-2} g$. In general, if $f$
is an exact (closed)
$1$-form, then $\hat\nabla$ is (locally) the Levi-Civita connection of
a metric in  $\mathcal C$. 

The Christoffelsymbols of a Weyl connection $\hat\nabla$ are
written similarly as \Eqref{eq:christoffel},
\[\ConnectionWeyl \mu\nu\rho
:=\scalarpr{dx^\nu}{\hat\nabla_\mu\partial_\rho},\]
and
are related to the Levi-Civita connection coefficients of the metric
$g$ by
\begin{equation}
  \label{eq:levi_civita2Weyl1}
  \ConnectionWeyl\mu\rho\nu=\Connection\mu\rho\nu+\ConnS{f}\mu\rho\nu
\end{equation}
with
\begin{equation*}
  \ConnS{f}\mu\rho\nu:=\delta\indices{^\rho_\mu} f_\nu
  +\delta\indices{^\rho_\nu}
  f_\mu-g_{\mu\nu}g^{\rho\lambda}f_\lambda.
\end{equation*}
Here $f$ is the $1$-form that determines $\hat\nabla$ with respect to
$g$. Note that the object $\ConnS{f}\mu\rho\nu$ is a tensor field, i.e.\ it
has the same form in particular with respect to an orthonormal frame
of $g$. Further one should be aware of the fact that Weyl connection
coefficients obey a different algebra than their Levi-Civita counterparts.

\paragraph{Weyl curvature quantities}
The curvature tensor of a Weyl connection $\hat\nabla$ (in fact of
any connection) is
\[\hat R(X,Y)Z=\hat\nabla_X\hat\nabla_YZ
-\hat\nabla_Y\hat\nabla_X Z-\hat\nabla_{[X,Y]} Z\]
in accordance with \Eqref{eq:RiemannDef}.
With the same index convention as in \Eqref{eq:indexRiemann},
contracting the Riemann tensor yields the Ricci tensor
\[\hat R_{\nu\rho}=\hat R\indices{^{\mu}_\nu_\mu_\rho}.\]
We define the \term{Schouten tensor} as
\begin{equation}
\label{eq:defSchouton}
\hat L_{\mu\nu}=\frac1{n-2}\left(\hat
  R_{(\mu\nu)}-\frac{n-2}n\hat R_{[\mu\nu]}
  -\frac 1{2(n-1)}g_{\mu\nu}g^{\rho\sigma}\hat R_{\rho\sigma}\right),
\end{equation}
with $g\in\mathcal C$ arbitrary, i.e.\ this definition does not depend
on the choice of $g\in\mathcal C$. The Schouten tensor plays the role
of the tracefree part of the Ricci tensor roughly. Now, one can write
\begin{equation}
\label{eq:def_weyl}
C\indices{^{\mu}_\nu_\lambda_\rho}
:=\hat R\indices{^{\mu}_\nu_\lambda_\rho}
-2\left(g\indices{^{\mu}_{\,[\lambda}} \hat L\indices{_{\rho]}_\nu}
  -g_{\nu[\lambda} \hat L_{\rho]}^{\,\,\mu}
  -g\indices{^\mu_\nu}\hat L_{[\lambda\rho]}\right),
\end{equation}
where again $g\in\mathcal C$ arbitrary and
where $C\indices{^{\mu}_\nu_\lambda_\rho}$ is the \term{conformal
Weyl tensor}. This important object is independent of the choice of the
Weyl connection. In particular, this definition of the Weyl tensor
corresponds to usual one given in many textbooks
when $\hat\nabla$ is
a Levi-Civita connection since in this case the
Schouten tensor is symmetric and the definition simplifies accordingly.

Let $(M,\mathcal C)$ be a conformal space, $g\in\mathcal C$ and
$\hat\nabla$ the Weyl 
connection determined by the 1-form $f$ with respect to $g$. Let us
define 
\[\hat R:=\hat R_{\mu\nu}g^{\mu\nu}\]
which is the first object in this section which depends on the choice of
$g\in\mathcal C$.
The following relations can be derived
\begin{gather}
  \hat R_{\nu\rho}
  =R_{\nu\rho}-(n-1)\nabla_\rho f_\nu
  +\nabla_\nu f_\rho
  +(n-2)f_\nu f_\rho
  -g_{\nu\rho}g^{\lambda\sigma}
  \left(\nabla_\lambda f_\sigma+(n-2)f_\lambda f_\sigma\right),\\
  \hat R=R-(n-1)g^{\lambda\sigma}
  \left(2\nabla_\lambda f_\sigma+(n-2)f_\lambda f_\sigma\right),\\
  \label{eq:relate_Weyl_L_and_LeviCivita_L}
  L_{\mu\nu}-\hat L_{\mu\nu}=\nabla_\mu f_\nu-f_\mu f_\nu+\frac
  12g_{\mu\nu}g^{\lambda\sigma}f_\lambda f_\sigma,\\
  \label{eq:shouton_weyl}
  \hat\nabla_\rho\hat L_{\mu\nu}-\hat\nabla_\mu\hat
  L_{\rho\nu}=\nabla_\rho L_{\mu\nu}-\nabla_\mu
  L_{\rho\nu}+f_\lambda C\indices{^\lambda_\nu_\rho_\mu}.
\end{gather}
For the special case when $\hat\nabla$ is the Levi-Civita
connection of a metric $\tilde g\in\mathcal C$ with $g=\Omega^2\tilde
g$ one obtains
\begin{align}
  \label{eq:ConformalRicci}
  R_{\nu\rho}&=\tilde
  R_{\nu\rho}
  -\frac{n-2}\Omega\nabla_\nu\nabla_\rho\Omega
  -g_{\nu\rho}g^{\lambda\delta}
  \left(\frac 1\Omega\nabla_\lambda\nabla_\delta\Omega
    -\frac{n-1}{\Omega^2}\nabla_\lambda\Omega\nabla_\delta\Omega\right)\\
  \label{eq:scaling_R}
  \tilde R&=\Omega^{2}R
  +2(n-1)\Omega\,\nabla^\lambda\nabla_\lambda\Omega
  -n(n-1)\nabla^\lambda\Omega\nabla_\lambda\Omega.
\end{align}
Here we use the notation that all curvature quantities related to
$\tilde g$ are denoted with a tilde while those corresponding to $g$
come without tildes. In particular, here 
$\tilde R:=\tilde R_{\mu\nu}\tilde g^{\mu\nu}$.

Of fundamental interest are the \term{structure equations}.
  For every torsion free connection, in particular for a Weyl
  connection $\hat\nabla$, one has for a (not necessarily orthonormal) frame
  $\{e_i\}$ ($1$st structure equation of Cartan)
  \nobreak
  \begin{subequations}
    \label{eq:cartan}
  \begin{equation}
    \label{eq:cartan1}
    {}[e_i,e_k]
    =(\ConnectionWeyl ijk-\ConnectionWeyl kji)e_j.
  \end{equation}
  The $2$nd structure equation states for any connection, in
  particular for $\hat\nabla$,
  \begin{equation}
    \label{eq:cartan2}
    \hat R\indices{^i_j_k_l}
    =e_k(\ConnectionWeyl lij)
    -e_l(\ConnectionWeyl kij)
    +\ConnectionWeyl lmj\ConnectionWeyl kim	
    -\ConnectionWeyl kmj\ConnectionWeyl lim
    -\left(\ConnectionWeyl kml-\ConnectionWeyl lmk\right)\ConnectionWeyl mij.
  \end{equation}
\end{subequations}

\paragraph{Conformal geodesics}
\begin{Def}[Conformal geodesics]
  \label{def:conformal_geodesic}
  Let $(M,\mathcal C)$ be a conformal space. A conformal geodesic is a
  curve $x:]-\epsilon,\epsilon[\rightarrow M$ with
  parameter $\tau$ 
  determined together with a Weyl connection $\hat\nabla$ along the
  curve, written   as the pair 
  $(x(\tau),\hat\nabla)$, that fulfill 
  \begin{equation}
      \label{eq:conformal_geodesic}
    \hat\nabla_{\dot x}\dot x=0,\quad \hat L_{\mu\nu}\dot x^\mu=0.
  \end{equation}
\end{Def}
Let $g$ be an arbitrary metric in the conformal structure and $\nabla$
its Levi-Civita connection. Then the conformal geodesic equations 
\Eqref{eq:conformal_geodesic} are equivalent to
\nobreak
\begin{subequations}
\label{eq:conformal_geodesic2}
\begin{align}
  \bigl(\nabla_{\dot x}\dot x\bigr)^\mu
  +\ConnS{ f}\lambda\mu\rho\,\,\dot x^\lambda\dot x^\rho&=0\\
  \bigl(\nabla_{\dot x} f\bigr)_\rho
  -\frac 12 f_\mu\ConnS{ f}\lambda\mu\rho\,\,\dot
  x^\lambda&= L_{\lambda\rho}\dot x^\lambda,
\end{align}
\end{subequations}
where $f$ is the $1$-form that determines $\hat\nabla$ with respect
$g$. These equations are conformally invariant, i.e.\ choosing another
metric in the conformal structure, the canonical transformation of
\Eqsref{eq:conformal_geodesic2} yields an equivalent system. In
particular, the solution corresponding to
appropriately transformed initial data is the same pair
$(x(\tau),\hat\nabla)$. 
 The theory of ordinary
differential equations easily implies that for given initial data,
$\initially{\dot x}$ and $\initially f$ at $p\in M$, and for an
arbitrary metric 
$g\in\mathcal C$ there is a unique conformal
geodesic locally in the parameter starting in $p$ with these data.
Conformal geodesics have the same meaning for conformal spaces with
Weyl connections as (metric) geodesics for pseudo-Riemannian
manifolds with Levi-Civita connections: they are projections of 
horizontal curves in the relevant principal bundles, cf.\ again
\cite{ConfGeodesics} for more details. Other topics related to
conformal geodesics are discussed and further related literature is
given in
\cite{ConfGeodesics,Friedrich2002,Friedrich03}.

\paragraph{Conformal Weyl tensor decomposition  and the Bianchi system}
The conformal Weyl tensor inherits all the symmetries of the Riemann
tensor but additionally fulfills
$C\indices{^\mu_\nu_\mu_\rho}=0$.
Any tensor with these algebraic properties can decomposed as
follows. Let $n$ be a smooth future pointing timelike congruence which
is not necessarily hypersurface forming. Introduce an orthonormal
frame $\{e_i\}$ with respect to a metric $g\in\mathcal C$ such that
$e_0=n$. Let \mbox{$h=g+\sigma^0\otimes\sigma^0$} be the induced metric on
the orthonormal complement of $n$. Then we define
\begin{equation}
  \label{eq:electricmagnetic}
  E_{ij}:=h\indices{_i^k}h\indices{_j^l}C_{kmlp}n^mn^p,\quad
  B_{ij}:=h\indices{_i^k}h\indices{_j^l}C^*_{kmlp}n^mn^p
\end{equation}
where $E$ is the \term{electric part} and $B$ is the \term{magnetic
  part} of $C$. The dual $C^*$ is defined as
\[C^*_{ijkl}=-\frac 12 C_{ijmp}\eta\indices{^m^p_k_l};\]
the minus sign is due to our convention for $\eta$ in
\Eqref{eq:spacetimevolumeform}. The tensor $E$ has the
properties
\[E_{ij}n^j=0,\quad E_{[ij]}=0,\quad E\indices{^i_i}=0,\]
the same for $B$. With the tensors $E$ and $B$ we can write the
conformal Weyl tensor as
\begin{equation}
  \label{eq:Weyl_algebraicsplit}
  C_{ijkl}=2\left(-l\tensor[_{j[k}]{E}{_{l]i}}+l\tensor[_{i[k}]{E}{_{l]j}}
    -n\tensor[_{[k}]{B}{_{l]m}}\epsilon\indices{^m_i_j}
    -n\tensor[_{[i}]{B}{_{j]m}}\epsilon\indices{^m_k_l}
  \right),
\end{equation}
here
$l:=g+2\sigma^0\otimes\sigma^0$. This formula is from
\cite{FriedrichNagy}, but note the different choice of signature.

Let two metrics $g,\tilde g\in\mathcal C$ be given such that
$g=\Omega^2\tilde g$ with $\Omega>0$ smooth.
For the conformal Weyl tensor one can 
derive the following invariance 
relation
\begin{equation}
  \label{eq:conformal_covariance}
  \nabla_\mu(\Omega^{3-n}C\indices{^\mu_\nu_\lambda_\rho})=
  \Omega^{3-n}\tilde \nabla_\mu C\indices{^\mu_\nu_\lambda_\rho}.
\end{equation}
Now, let $\tilde g\in\mathcal C$ be a solution of Einstein's field
equation in vacuum with an arbitrary cosmological constant
$\lambda$. Then  the \textbf{rescaled
  Weyl tensor} defined by
\begin{equation}
  \label{eq:rescaled_weyl}
  W\indices{^\mu_\nu_\lambda_\rho}
  :=\Omega^{3-n}C\indices{^\mu_\nu_\lambda_\rho}
\end{equation}
fulfills the \textbf{(vacuum) Bianchi system}
\begin{equation}
  \label{eq:bianchi_system}
  \nabla_\mu W\indices{^\mu_\nu_\lambda_\rho}=0
\end{equation}
where $\nabla$ is the Levi-Civita connection associated with $g$. This
follows from the conformal invariance relation
\Eqref{eq:conformal_covariance}, the Bianchi identities and the vacuum
field equations \Eqref{eq:EFE_vacuum}. The Bianchi system is
conformally invariant. Let $\bar g=\Theta^2 g$ so that
$\bar\Omega=\Omega/\Theta$ is the conformal factor that relates 
$\bar g$ and $\tilde g$. For the rescaled Weyl tensor we have
\[\bar W=\bar\Omega^{3-n} C=\bar\Omega^{3-n}\Omega^{n-3} W=\Theta^{n-3}W.\]
Then from the invariance relation \Eqref{eq:conformal_covariance} we
get that
\[\bar\nabla_\mu \bar W\indices{^\mu_\nu_\lambda_\rho}=0.\]
But note that the rescaled Weyl tensor itself is \textit{not} 
conformally invariant. 

The
Bianchi system forms the heart of the conformal field
equations, as we discuss in the next section. Because the rescaled
Weyl tensor has the same algebraic 
properties as the conformal Weyl tensor, the same algebraic split as
in \Eqref{eq:Weyl_algebraicsplit} can be done. When we speak about the
tensors $E$ and $B$ in the remainder of this thesis we will always
mean, with respect to a $g\in\mathcal C$ and a timelike unit vector
field $n$, the electric and magnetic
part of the rescaled Weyl tensor.

\subsection{Conformal field equations}
\label{sec:cfe}
Penrose \cite{penrose1963,penrose1979} suggested that a useful
description of the
properties of ``infinity'' of general relativistic spacetimes is to consider
\term{conformal infinity}. 
In this picture, a spacetime 
$(\tilde M,\tilde g)$ with its  
associated conformal structure $\mathcal C$ is ``extended''
conformally to a 
manifold
with boundary $M$ such that the conformal factor associated with any
metric in the conformal 
class vanishes at this boundary. More details on this extension
process are 
discussed in \Sectionref{sec:fads_def}. 
Penrose concluded that such point of view is in agreement with the
field equations because he had analyzed certain classes of explicit
solutions of 
EFE, for example also the de-Sitter spacetime discussed in
\Sectionref{sec:deSitter}. 
Penrose's original motivation was to study isolated systems, and for
those his idea was particularly promising because it yielded a natural
geometric description for the
analysis by Bondi et al.\ done before with respect to special
asymptotic coordinate systems.
However, it remained unclear and indeed the source of controversial
debates, if 
the concept of conformal infinity is compatible with solutions of
Einstein's field equations for a large class of physically interesting
cases. See for instance the review
\cite{Friedrich2002} for further information.

In this section we elaborate
on the problem to formulate EFE such that they make sense also at
conformal infinity. Recalling the relation \Eqref{eq:ConformalRicci}
shows that this cannot be done in a naive manner since formally
singular terms spoil the analysis.
Friedrich found a regular formulation of the equations called
\term{conformal field equations} which we will discuss in the
following. For more details, a useful review reference is
\cite{Friedrich2002}. 
The discussion in this section applies in vacuum for arbitrary
$\lambda$. Friedrich has also worked on some cases with
matter, for instance \cite{friedrich91}. Later, but not yet now, we will
restrict to the vacuum case with $\lambda>0$ for the whole rest of the thesis. 
Note
that Friedrich's formulation of the field equations is restricted to
the $3+1$-case which is the relevant case for this thesis. Generalization
of his formalism to higher dimensions can be found in
\cite{Anderson04,Anderson05b}.

From the equations that were
listed in the last section one can collect the 
following system for the derived quantities of the conformal metric
$g$ in $3+1$ with conformal factor 
$\Omega$ such that $g=\Omega^2\tilde g$ with $\tilde g$ a physical
metric (\Sectionref{sec:conf_weyl}),
\nobreak
\begin{subequations}
\label{eq:regular_conformal_field_equations}
\begin{align}
  \label{eq:regular_conformal_field_equations_torsionfree}
  e^\mu_{k,\nu}e^\nu_j-e^\mu_{j,\nu}e^\nu_k
  &=\left(\Connection jik-\Connection kij\right)e_i^\mu\\
  e_k(\Connection lij) &-e_l(\Connection kij) +\Connection
  lmj\Connection kim -\Connection kmj\Connection lim
  -\left(\Connection kml-\Connection lmk\right)\Connection mij\\
  &=2\left(g\indices{^{i}_{\,[k}}L\indices{_{l]}_j}
    -g_{j[k}L_{l]}^{\,\,i}\right)
  +\Omega W\indices{^{i}_j_k_l}\\
  \label{eq:regular_conformal_field_equations_DefSigma}
  \nabla_i\Omega&=\Sigma_i\\
  \label{eq:regular_conformal_field_equations_SecondOrderOmega}
  \nabla_i\Sigma_j&=-\Omega L_{ij}+sg_{ij}\\
  \label{eq:regular_conformal_field_equations_s}
  \nabla_i s&=-\Sigma^j L_{ji}\\
  \label{eq:regular_conformal_field_equations_L}
  \nabla_k L_{ij}-\nabla_i L_{k j}
  &=\Sigma_l\,W\indices{^{l}_j_k_i}\\
  \label{eq:regular_conformal_field_equations_Bianchi}
  0&=\nabla_i W\indices{^i_j_k_l}\\
  \label{eq:regular_conformal_field_equations_lambda}
  \lambda&=3(2\Omega s-\Sigma^i\Sigma_i).
\end{align}
\end{subequations}
These are the \term{conformal field equations} (CFE).
The first two equations are obtained from the structure equations
\Eqsref{eq:cartan} in the special case $f=0$ (with respect to the
conformal metric $g$) with the decomposition \Eqref{eq:def_weyl} and
the definition of the rescaled Weyl tensor for $n=4$. The third equation can be
considered as the
definition of the quantity
$\Sigma_i$ to be the differential of $\Omega$. 
\Eqref{eq:EFE_vacuum} for $\tilde g$, and \Eqsref{eq:ConformalRicci} and
\eqref{eq:scaling_R} yield a second order system of equations for the conformal
factor $\Omega$ which is written in $1$st-order form
\Eqsref{eq:regular_conformal_field_equations_DefSigma} and 
\eqref{eq:regular_conformal_field_equations_SecondOrderOmega}.
Here, $s$ is defined by
\begin{equation}
  \label{eq:Defs}
  s:=\frac{1}{24}\Omega R
  +\frac {1}4\nabla_i\Sigma^i.
\end{equation}
The system \Eqsref{eq:regular_conformal_field_equations_DefSigma} and 
\eqref{eq:regular_conformal_field_equations_SecondOrderOmega} for
$\Omega$ is clearly overdetermined, however, 
\Eqref{eq:regular_conformal_field_equations_s} is the corresponding
integrability 
condition. Moreover, note that
\Eqref{eq:Defs} is part of 
\Eqsref{eq:regular_conformal_field_equations_SecondOrderOmega}. 
Next, \Eqref{eq:regular_conformal_field_equations_L} is derived from
\Eqref{eq:shouton_weyl} for $f=-\Omega^{-1}d\Omega$ together with
Einstein's vacuum field equations for $\tilde g$ and the definition of the
rescaled Weyl tensor. The last two equations are the Bianchi system
\Eqref{eq:bianchi_system} and an equation, so far the
only one involving the cosmological constant $\lambda$,
obtained from the trace of the vacuum field equations and the
transformation behavior of the Ricci scalar
\Eqref{eq:scaling_R}.

If a conformal spacetime $(M,g,\Omega)$ (with its derived
quantities) satisfies
\Eqsref{eq:regular_conformal_field_equations}, then the
conformally rescaled spacetime $(M,\Theta^{-2}g,\Theta\Omega)$ with
arbitrary $\Theta>0$ fulfills the
same equations. This is referred to as
\term{conformal gauge freedom}. 
The way the system above is constructed we also have that any
conformal rescaling of a 
solution of EFE $(\tilde M,\tilde g)$ in vacuum with arbitrary
$\lambda$ and with arbitrary conformal factor $\Omega>0$ satisfies 
\Eqsref{eq:regular_conformal_field_equations}. In particular consider
the physical 
quantities themselves given by $\Omega\equiv 1$. 
So for $\Omega\equiv 1$, these equations are just a special
representation of 
EFE. Of further relevance is that we have found 
that for any solution $(M,\Omega,g)$ of 
\Eqsref{eq:regular_conformal_field_equations} 
with a given cosmological
constant $\lambda$, the spacetime 
$(\tilde M,\tilde g=\Omega^{-2}g)$ is a solution of EFE in vacuum with
that $\lambda$, where $\tilde M$ is that subset of $M$ determined by
$\Omega>0$. As we will see later, that part of $M$ where $\Omega$
vanishes can be considered as the conformal boundary of $\tilde M$. In
this sense the conformal field equations are an extension of Einstein
field equations since in principle
they can be 
used to compute spacetimes including
their conformal infinity. 

In any case, before one can really make such
statements one has to see if a well-posed Cauchy problem can
be formulated for these equations. We make only a few remarks about this
here and report on the rigorous results on this issue in the case
$\lambda>0$ in \Sectionref{sec:initial_data}.
Friedrich
\cite{Friedrich85} succeeded in performing a hyperbolic
reduction of the system above with the full coordinate, frame and
conformal gauge freedom incorporating even the case when the conformal
factor vanishes
somewhere. Since we will not use the full gauge freedom
in the following, we only make a few statements
about this, but see \cite{Friedrich85} for the complete argument.
The crucial idea is that of 
\term{gauge source functions}. These are
functions, that on the one hand can be prescribed arbitrarily, i.e.\ the
hyperbolicity 
properties of the evolution equations are independent of them. On the
other hand they
determine a gauge, and, vice versa, given
an arbitrary gauge, one is able to determine the corresponding
gauge source functions. Friedrich showed that a hyperbolic reduction
of the
system \Eqsref{eq:regular_conformal_field_equations} with
gauge source functions for
the coordinate, the frame and the conformal gauge can be found
such that the full gauge freedom is preserved.

Having fixed the coordinate and frame gauge somehow,
\Eqsref{eq:regular_conformal_field_equations_torsionfree} can  
be split into evolution and constraint equations for the frame
component functions with respect to the coordinates; we skip the
discussion of gauge source functions 
corresponding to coordinate and frame gauge here. The 
$2$nd structure equation implies evolution equations for
the connection coefficients. The definition of $\Sigma_i$,
\Eqref{eq:regular_conformal_field_equations_DefSigma}, can be seen as
yielding an evolution equation for the conformal factor; the following
one an evolution equation  
for $\Sigma_i$ and \Eqref{eq:regular_conformal_field_equations_s} for $s$.
Now, \Eqref{eq:regular_conformal_field_equations_s} is the first
equation with a slight complication because it yields evolution
equations for all components of the Schouten tensor
except for $L_{00}$.  From
the definition of the Schouten tensor
(\Eqref{eq:defSchouton} with $f=0$), one finds that its trace
must fulfills $L=R/6$. Hence
$L_{00}=L_{11}+L_{22}+L_{33}-R/6$.
Now, Friedrich showed that $R$ can be considered as a conformal gauge source
function in the sense above. Having chosen $R$ arbitrarily, $L_{00}$
can be
computed as soon as the other components of the Schouten tensors are
determined from their evolution equations derived from
\Eqref{eq:regular_conformal_field_equations_s}. The Ricci scalar $R$
is indeed a gauge source
function for the conformal gauge 
since the choice of $R$ influences the evolution of $\Omega$ via
\Eqref{eq:Defs} which is part of
\Eqref{eq:regular_conformal_field_equations_SecondOrderOmega}. 
We do not write down in full generality how the
Bianchi
system \Eqref{eq:regular_conformal_field_equations_Bianchi} can be
reduced to a system of evolution and constraint equations for the
quantities of the rescaled Weyl tensor; this will be done only in a
special gauge in \Sectionref{sec:LCCGG}. The general case can be
found in \cite{Friedrich85} and, in a non-spinorial fashion, in
\cite{FriedrichNagy}.
Surprisingly, it turns out the Bianchi system implies
exactly enough evolution equations for the rescaled Weyl tensor
components if and only if the dimension of 
spacetime is $n=4$. Hence, a system of equations which includes the Bianchi
system can only lead to a well-posed formulation if
$n=4$ which we always assume. It turns out that Friedrich's
system of evolution equations is symmetric hyperbolic. 

In the reduction procedure which we have just sketched, one yields
a quite large number of constraint equations and it is a tedious but
eventually successful task to prove that they propagate
\cite{Friedrich85}.

Thus, in
principle, it is shown that one can find well-posed formulations of
the Cauchy problem of the conformal field equations. However, in many
important special cases there are still problems. For example, in the
asymptotically flat case with $\lambda=0$, one would like to formulate
the Cauchy problem such that the initial hypersurface is a Cauchy
surface reaching spatial infinity. But generically, the conformal
structure is not smooth at this point and hence well-posedness in the
sense above is not sufficient. This issue is still under
investigations; see for instance \cite{Friedrich2002}. Another example
is the case of 
spacetimes of Anti-de-Sitter type with $\lambda<0$. It turns out that
there are no Cauchy surfaces at all and one has to formulate an
initial boundary value problem for the conformal field equations
\cite{AntiDeSitter}. 
A well-posed Cauchy problem for $\lambda>0$ incorporating the
conformal field equations will be discussed in
\Sectionref{sec:initial_data}.

Friedrich's implementation of $R$ as a gauge source function seems
necessary to obtain both hyperbolic evolution equations and the full
conformal gauge freedom. But, from the geometric point of view
this approach is not optimal because, for example, the prescription of
$R$ yields 
no a priori information on the position of a $\Omega=0$-surface, if it
exists.
If one is willing to give up the full gauge freedom, then
one can do the following. We construct a
special geometrically inspired gauge for the coordinates, the frame
and the conformal factor. 
The hope is that this gauge is general enough at least for some of
the problems in mind and yields a priori information on the position
of a possible $\Omega=0$-surface. It is the \term{conformal Gauß
  gauge} introduced in 
\cite{AntiDeSitter}; Friedrich calls the conformal field equations in
this gauge \term{general conformal field equations} (GCFE) despite of the
fact that these equations do \textit{not} involve the full 
gauge freedom. The name is motivated by the fact that the full freedom
related to the conformal structure, i.e.\ both the choice of conformal
metric and of the Weyl
connection, is used to derive these equations.
It is clear that we can express
\Eqsref{eq:regular_conformal_field_equations} also by means of
an arbitrary Weyl connection compatible with the conformal structure of the 
conformal metric $g$. However, it has currently not been studied which 
the most general conditions for $f$ are such that symmetric hyperbolic
reductions can be obtained; only the following special case has been
considered so far. 

The idea for the conformal Gauß gauge is as follows; note the
similarity with the construction of standard Gauß gauges. Let $\Sigma$ be a
smooth spacelike hypersurface in a Lorentzian conformal space
$(M,\mathcal C)$. Choose a smooth conformal frame $\{c_i\}$ at $\Sigma$ and a
metric $g\in\mathcal C$ such that $\{c_i\}$ is orthonormal with
respect to $g$ on $\Sigma$. Further require that $c_0$ is orthogonal to 
$\Sigma$ (i.e.\ timelike).  Further, prescribe a
smooth ($4$-dimensional) $1$-form $\omega$ at $\Sigma$. Now consider the
conformal geodesic equations \Eqsref{eq:conformal_geodesic2} written
with respect to the metric $g$. Starting in an arbitrary point
$p$ of $\Sigma$ with initial data $\initially{\dot x}=c_0(p)$ and
$\initially{f}=\omega(p)$ we obtain a unique conformal geodesic
$(x(t),\hat \nabla)$ starting in $p$ where $\hat\nabla$ is determined
along that curve by $f$ with respect to $g$. This can be
done for any point in $\Sigma$ and we get a congruence of conformal
geodesics covering a neighborhood $U$ of $\Sigma$. We can choose $U$
so small that there is exactly one of these curves passing through
each point of it. Then, $f$ and $\dot x$ become smooth (co)tangent
vector fields in $U$. Hence $f$ determines a Weyl connection
$\hat\nabla$ on $U$ as above. Now, let us construct a
frame $\{e_i\}$ in $U$. Set
\nobreak
\begin{subequations}
  \label{eq:CGG_frame}
  \begin{equation}
    e_0=\dot x
  \end{equation}
  and determine the spatial part of the frame by
  \begin{equation}
    \hat\nabla_{\dot x} e_a=0\text{ with } \initially{e_a}=c_a.
  \end{equation}
\end{subequations}
From the underlying theory one knows that this frame exists on a open
subset of $U$ which, after having shrunken $U$ accordingly, equals
$U$. Further it is conformal with respect to $g$, i.e.\ 
\[g(e_i,e_j)=\Theta^2\eta_{ij}\]
with $\eta_{ij}$ the standard Minkowski metric. One can find that
\[\Theta(t)=e^{-\int_0^t f_0(t')dt'}\]
along a given conformal geodesic where $t=0$ corresponds to the
starting point. In the same way as above, $\Theta$ can be considered
as a smooth positive function on $U$. After a conformal rescaling of
$g$ with this 
conformal factor $\Theta$ we can assume without loss of generality
that the frame $\{e_i\}$ is orthonormal with respect to $g$. 
With respect to this $g$ we then have on $U$,
\begin{equation}
  \label{eq:f0vanish}
  f_0=0.
\end{equation}
In fact, this particular
metric $g$ could have been used from the beginning in this
construction since the whole construction is invariant under
rescalings with smooth conformal factors which are identically one on $\Sigma$.
Note that by the
definition of conformal geodesics we have
\[\hat L_{0i}=0\]
on $U$. Moreover,
the Weyl connection coefficients obey 
\[\ConnectionWeyl 0ij=0\]
due to \Eqsref{eq:CGG_frame} on $U$. As a side remark, the fact
that the frame is parallel transported along $e_0$ with respect to
$\hat\nabla$ implies that it is Fermi transported along $e_0$ with
respect to $\nabla$; see \Sectionref{sec:comm_geometry} for the
definition of Fermi transport.

Now we can fix coordinates $\{t,x^\alpha\}$ on $U$ as follows.
Prescribe local coordinates
$\{x^\alpha\}$ on $\Sigma$ and set $e_0=\partial_t$ with $t=0$ on
$\Sigma$. This means that the spatial coordinates of $\Sigma$ are dragged along
the constructed
congruence of conformal geodesics. This also means in view of
\Eqref{eq:lapseshift} that, with respect to the conformal metric $g$, the
lapse function is identically one and the shift vector
vanishes. 

So, for a given hypersurface $\Sigma$ with local coordinates
$\{x^\alpha\}$, frame $\{c_i\}$ and $1$-form 
$\omega$ as above, we have constructed coordinates $\{t,x^\alpha\}$,
a distinguished metric $g\in\mathcal C$, an orthonormal frame
$\{e_i\}$ with respect to $g$ and a Weyl connection $\hat\nabla$
determined by $f$ 
with respect to $g$ in a neighborhood $U$
of $\Sigma$. Writing the conformal field equations with this choice of 
gauge leaves no further gauge freedom.
In \cite{AntiDeSitter}, Friedrich proves that in this conformal Gauß
gauge 
the conformal factor $\Omega$ is a $2$nd-order polynomial in time 
\begin{equation}
\label{eq:Omega2ndOrderPolyn}
\Omega(t,x^\alpha)=\Omega_0(x^\alpha)+\Omega_1(x^\alpha)t
+\Omega_2(x^\alpha)t^2
\end{equation}
if $\tilde g$ is a physical metric in vacuum with an arbitrary
cosmological constant, and $\tilde g=\Omega^{-2} g$ with $g$ as just
constructed. Moreover, he 
showed that the $1$-form
\begin{equation}
  \label{eq:def_d}
  d:=\Omega f+d\Omega=\Omega\tilde f
\end{equation}
fulfills
\begin{equation}
  \label{eq:CGG_d}
  d_0=e_0(\Omega),\quad d_a=d^{*}_a(x^\alpha).
\end{equation}
Note, that $d_0=e_0(\Omega)$ is equivalent to
\Eqref{eq:f0vanish}. It turns out that the functions $d^{*}_a$,
$\Omega_0(x^\alpha)$, $\Omega_1(x^\alpha)$ and $\Omega_2(x^\alpha)$
are not completely
free in the special case discussed in \Sectionref{sec:initial_data} when $\Sigma$ corresponds to $\scri$.

Now, in conformal Gauß gauge, a reduction of the conformal field
equations, i.e.\ the general conformal field equation (GCFE), is given
by \cite{Friedrich2002}
\nobreak
\begin{subequations}
\label{eq:gcfe}
\begin{align}
  \partial_0 e_k^\mu&=-\ConnectionWeyl kj0e_j^\mu\\
  \label{eq:Connection_in_gcfe}
  \partial_0\ConnectionWeyl lij
  &=-\ConnectionWeyl lm0\ConnectionWeyl mij
  +\Omega W\indices{^{{i}}_{j}_{0}_{l}}
  +g\indices{^{{i}}_{{0}}} \hat L\indices{_{{l}}_{j}}
  -g_{{j}{0}} \hat L_{{l}}^{\,\,{i}}+g\indices{^{i}_{j}}\hat
  L_{{l}{0}}\\
  \label{eq:Schouton_in_gcfe}
  \partial_0\hat L_{ij}
  &=d_l W\indices{^l_j_0_i}-\ConnectionWeyl il0\hat L_{lj}\\
  \label{eq:bianchi_system_in_gcfe}
  \nabla_iW\indices{^i_j_k_l}&=0\\
  \Omega(t)&=\Omega_0+t\Omega_1+t^2\Omega_2\\
  d_0&=\dot\Omega,\quad d_a=d^{*}_a,
\end{align}
for the unknowns
\begin{equation}
  \label{eq:GCFEunknowns}
  u=\left(e_a^\mu, \ConnectionWeyl ijk, \hat L_{ab}, \hat L_{a0},
    W\indices{^i_j_k_l}, d_i, \Omega\right). 
\end{equation}
\end{subequations}
For brevity we have still not yet written down a reduction of the
Bianchi system here. See \Sectionref{sec:LCCGG} for the case of
relevance for this thesis. Given a symmetric hyperbolic reduction of
the Bianchi system, it follows directly
that the general conformal field equations \Eqsref{eq:gcfe} are
symmetric hyperbolic, since all other equations are either algebraic
or ODEs. Further, one can check that the constraints propagate.

Once more note that the system \Eqsref{eq:gcfe} does not have the full
gauge freedom anymore; a conformally rescaled solution of this
system is not a solution of this system anymore in general. However,
considering such a solution as a solution of the original system
\Eqsref{eq:regular_conformal_field_equations} (to be more precise, its
generalization written in terms of Weyl connections)
with corresponding
gauge source functions, the rescaled solution is also a
solution of the original system, but with 
correspondingly different gauge source functions.



\section{Commutator field equations} 
\label{sec:commutatorfieldequations}
\subsection{Introduction}
In this section we introduce that formulation of Einstein's
field equations which we refer to as \term{commutator field
  equations}\footnote{This is terminology is not standard, and maybe
  not even well-chosen. However, this is how the following set of
  equations will be 
referred to in this thesis.} and which will be used in
\Sectionref{sec:RunsCosmFE} to compute numerical solutions
alternatively to the conformal field equations. 
It  was developed to study issues in
cosmological spacetimes by the authors
of the three main references \cite{vanElst96,vanElst01,Andersson03}.
The main variables of this system are the geometric
commutator quantities introduced in \Sectionref{sec:comm_geometry} of
a distinguished timelike congruence. In
\Sectionref{sec:comm_field_eqs} we
summarize the steps in \cite{vanElst01} to 
formulate a consistent symmetric hyperbolic reduction for the
case of Gowdy spacetimes with $\T$ spatial slices
and arbitrary  
cosmological constant in timelike area gauge. Further
all relevant constraint equations are derived.

\subsection{Geometry of timelike congruences}
\label{sec:comm_geometry}
Let $(M,g)$ be a smooth $4$-dim.\ Lorentzian manifold as above and $u$ a
smooth, 
future directed, timelike, not necessarily hypersurface orthogonal, 
vector field of unit length\footnote{Before, we called such a vector fields
  $n$. However, here
we want to stay consistent with the notation in \cite{vanElst96}.}. Let
$h$ be the induced metric  
on the orthogonal complement of $u$. We can 
write (notation similar to \cite{vanElst96})
\[(\nabla_\mu u)_\nu=-u_\mu \dot u_\nu+\chi_{\mu\nu}\]
with the \term{acceleration} of $u$,
\begin{equation}
  \label{eq:dot_u}
  \dot u^\mu:=h\indices{^\mu_\nu}(\nabla_u u)^\nu=(\nabla_u u)^\mu,
\end{equation}
and $\chi_{\mu\nu}$ the $2$nd fundamental form of $u$ introduced
in \Eqsref{eq:2ndFundamentalForm}.
In particular, $\dot u$ and $\chi$ are spatial with
respect to $u$, i.e.\
\[\dot u_\nu u^\nu=\chi_{\mu\nu}u^\nu=\chi_{\mu\nu}u^\mu=0.\] 
We make the following further split
\nobreak%
\begin{subequations}
\label{eq:split2ndFF}
\begin{equation}
\chi_{\mu\nu}=\sigma_{\mu\nu}+H h_{\mu\nu}-\omega_{\mu\nu}.
\end{equation}
Here $\sigma$ is the symmetric part of $\chi$ called \term{shear tensor} of
u, the \term{Hubble scalar} is
\begin{equation}
H=\frac 13 \tr \chi,
\end{equation}
\end{subequations}
and $\omega$ is (up to sign) the antisymmetric part of $\chi$ called
\term{twist tensor} 
of $u$. Since $\omega$ is spatial with respect to $u$, its dual,
called \term{twist vector},
\[\omega^\mu:=\frac 12\eta^{\mu\nu\rho\sigma}\omega_{\nu\rho}u_\sigma\]
contains the same information as the twist tensor since
\[\omega_{\mu\nu}=\eta_{\mu\nu\rho\sigma}\omega^\rho u^\sigma.\]
Recall the conventions for the volume form $\eta$ in
\Sectionref{sec:convnotation}. 

Now, let $\{e_i\}$ be a smooth orthonormal frame with $e_0=u$ and the
commutator functions as in \Eqref{eq:commutator_fcts}.
Let us\footnote{This is similar to the decomposition by Bianchi in his
  classification of all real 
$3$-dim.\ Lie algebras (\Sectionref{sec:relevantsymmclass}).}
write 
\[\Commutator abc=2 a\tensor[_{[b}]{\delta}{^a_{c]}}
+\epsilon_{bcd} n^{da}\]
where $(a_c)$ is an $\R^3$-valued function and $(n^{ab})$ is symmetric
$3\times 3$-matrix valued function. Such a decomposition is always possible.
Furthermore, introduce the \term{frame angular velocity} 
\[\Omega^a:=\frac 12 \epsilon\indices{^a^b^c} 
g(\nabla_{e_0} e_c,e_b).\]
The frame $\{e_i\}$ is called \term{Fermi transport}ed along
$e_0$ if $\Omega^a=0$; it is called \term{parallel
  transport}ed along $e_0$ if it is Fermi transported and 
$\dot u=0$.
After straight forward algebra one obtains \cite{vanElst96}
\nobreak
\begin{subequations}
\label{eq:frame_commutator}
  \begin{align}
    \label{eq:frame_commutator1}
    {}[e_0,e_a]&=\dot u_a e_0-\left(H \delta\indices{_a^b}+\sigma\indices{_a^b}
      -\epsilon\indices{_c_a^b}(\omega^c-\Omega^c)\right)e_b\\
    [e_a,e_b]&=-2\epsilon_{abc}\omega^c e_0
    +\left(2 a\tensor[_{[a}]{\delta}{^c_{b]}}+\epsilon_{abd} n^{dc}\right)e_c.
  \end{align}
\end{subequations}
In these expressions one sees that $e_0$ is
hypersurface orthogonal if and only if $\omega^i=0$. On the other hand
for a fixed $a=1,2,3$, the vector field $e_a$ is hypersurface
orthogonal if and only if the three relations
\begin{equation}
  \label{eq:easurfaceorthogonal}
  n^{aa}=0,\quad
  \sigma\indices{_d^a}-\epsilon\indices{_e_d^a}(\omega^e-\Omega^e)=0,\quad
   d=1,2,3\not=a
\end{equation}
hold.

\subsection{Field equations}
\label{sec:comm_field_eqs}
Here we outline the derivation of the commutator field equations
\cite{vanElst96,vanElst01} based on the 
quantities just introduced for the $\T$-Gowdy case. Since there are
lengthy expressions and 
calculations involved we only write down the results; more details can
be found in these references. I have checked all
computations with Mathematica and compared all resulting expressions
with those of these references. More details on the generation of
numerical code
are given in \Sectionref{sec:implequations}. Note that it is an
outstanding problem to formulate similar equations 
for Gowdy spacetimes with spatial $\S$-topology which would be of
particular interest for our research.

Gowdy spacetimes, see
\Sectionref{sec:relevantsymmclass}, have two commuting spatial Killing
vector fields whose integral curves are closed curves and which can be
chosen as two spatial coordinate fields. So 
let us determine coordinates 
$(t,x,y_1,y_2)$ such that $\partial_{y_1}$ and $\partial_{y_2}$ are
the two Gowdy Killing vector fields. Further we can always assume that
the shift
vector vanishes, see \Eqref{eq:lapseshift}, i.e.\ $\partial_t$ is
orthogonal to the $t=const$-hypersurfaces everywhere. Let us suppose
that we can construct an orthonormal frame $\{e_i\}$ of the form
\begin{equation}
  \label{eq:GowdyONF}
  e_0=N^{-1}\partial_t,\quad e_1=e\indices{_1^1}\partial_x,\quad
  e_A=e\indices{_A^2}\partial_{y_1}+e\indices{_A^3}\partial_{y_2}
\end{equation}
such that $e_0$ is hypersurface orthogonal, the spatial frame
$\{e_a\}$ is Gowdy invariant on each $t=const$-hypersurface and the
vector field $e_1$ is 
hypersurface orthogonal. Here $N$ is the so far unspecified lapse
function and $A,B=2,3$.
Due to Gowdy
symmetry all functions 
that are introduced here have vanishing $e_2$ and $e_3$ derivatives.
For the reader who already knows the properties of Gowdy spacetimes,
one should point out that these assumptions above exclude the case of
non-vanishing twist constants.

This choice of gauge has the following impact on the
commutator quantities obtained with the identification $e_0=u$. Since
$e_0$ is hypersurface orthogonal we have 
$\omega^i=0$. The surface forming property of $e_1$ implies, according
to \Eqsref{eq:easurfaceorthogonal}, that $n^{11}=0$ and that
$\sigma_{12}=\Omega^3$, $\sigma_{13}=-\Omega^2$.

Now projecting \Eqsref{eq:frame_commutator1} onto the coordinate basis
and taking the coordinate and frame choice above into
account provides us with the conditions
$\sigma_{12}=\sigma_{13}=0$, $\Omega^3=\Omega^2=0$, $\dot u_A=0$,
$a_A=0$, $n^{B1}=0$. Moreover, one obtains evolution and constraint
equations for the 
variables $e\indices{_a^\beta}$, which we do not write down yet at this
point, and the following gauge constraint 
\begin{equation}
  \label{eq:gaugefixing}
  e_1(N)-N\dot u_1=0.
\end{equation}

For the symmetric tracefree $3$-tensor $\sigma_{ab}$ with
$\sigma_{12}=\sigma_{13}=0$ we can introduce
the quantities $\sigma_+$, $\sigma_-$ and $\sigma_\times$ such that
\[(\sigma_{ab})=
\begin{pmatrix}
  -2\sigma_+ & 0 & 0\\
  0 & \sigma_++\sqrt 3\sigma_- &\sqrt 3\sigma_\times\\
  0 & \sqrt 3\sigma_\times & \sigma_+-\sqrt 3\sigma_-
\end{pmatrix};\]
similarly
\[(n^{ab})=
\begin{pmatrix}
  0 & 0 & 0\\
  0 & n_++\sqrt 3n_- &\sqrt 3n_\times\\
  0 & \sqrt 3n_\times & n_+-\sqrt 3n_-
\end{pmatrix}.\]

Now, let us express the frame components of the Ricci tensor $R_{ij}$ in
terms of the connection coefficients, a formula derived from
\Eqref{eq:cartan2} by contraction choosing $\hat\nabla$ to be the Levi-Civita
connection of $g$. Then, let us express the connection coefficients in
terms of the $\Commutator ijk$ quantities by means of
\Eqref{eq:commutator2connection} and then use the relations
\Eqsref{eq:frame_commutator} so that finally the Ricci tensor is expressed
completely in terms of the geometric quantities of the congruence
$u=e_0$ introduced in the previous section. Note
that the Ricci tensor is symmetric if the Jacobi
identities \Eqref{eq:jacobiid} are satisfied. Indeed, the Riemann
tensor has all desired symmetries if and only if this is the
case. We can express now, say, the second version of EFE
in vacuum 
with arbitrary cosmological constant \Eqref{eq:EFE_vacuum}
in terms of these relations. In particular, as
we have just stated, the antisymmetric part of them are equivalent to
the Jacobi identities. Since we will discuss the equations implied by
the Jacobi identities in a moment, it is sufficient to ignore the
antisymmetric part of EFE at this point.
The
$(0,0)$-component of 
EFE yields the well known 
\term{Raychaudhuri equation}
which is an 
evolution equation for the Hubble scalar $H$. The combination of the components
$(0,0)+(1,1)+(2,2)+(3,3)$ yields the \term{generalized Friedmann
  equation} which has its name from its importance for the study of
homogeneous and isotropic cosmologies. The Friedmann equation is just
the Hamiltonian constraint \Eqref{eq:Hamconstraint} which is also
called \term{Gauß constraint}. The equations obtained from the
$(0,1)$, $(0,2)$ and $(0,3)$ components correspond to the momentum
constraints \Eqref{eq:Momconstraint}. By adding the right components
of the Jacobi 
identities to these three equations one can get rid of all time
derivatives such 
that they can be interpreted as divergence
constraints for the shear tensor. It turns out
that, with the choices above, only one of them is non-trivial which we
refer to as the 
\term{Codazzi constraint}. 
The tracefree part of the spatial components of EFE yields evolution
equations for $\sigma_{ab}$.
With this, all of the
$10$ independent 
components of (the symmetric part of) EFE have been considered. 

Now, let us consider the Jacobi identities.
By choosing the appropriate combinations of the $16$ equations implied
by \Eqref{eq:jacobiid},
one is provided with
$6$ evolution equations for the $6$ quantities of the symmetric matrix
$n^{ab}$, while the three evolution equations for $n^{11}$, $n^{12}$
and $n^{13}$ are identically
satisfied by the conditions above. Similarly we obtain $3$ evolution
equations for the quantities $(a_a)$ while, analogously, only that for $a_1$
is non-trivial. The $3$ evolution equations for $\omega^a$ are also
consistent with the condition $\omega^a=0$ above. The $4$ residual
equations implied by the Jacobi identities turn out to be identically
satisfied as well.

Since we assume vacuum with arbitrary cosmological constants,
there are no further equations. We write down the total
implied system in a moment.

Let us introduce the quantities
\[\alpha:=H-2\sigma_+,\quad\beta:=H+\sigma_+\]
and substitute $H$ and $\sigma_+$ by these quantities. From the
Raychaudhuri equation and the evolution equation for $\sigma_+$, we
can derive evolution equations for $\alpha$ and $\beta$. It is easy to
check that
\[\beta=\frac 12(\chi_{22}+\chi_{33})\]
and hence it is the mean curvature of the Gowdy group orbits with
respect to the normal $e_1$ within a
given $t=const$-slice. So $\beta$ has a similar meaning as the Hubble scalar
has for spatially homogeneous spacetimes; both being the mean
curvature of the symmetry orbits. Thus,
similarly as the Hubble scalar is useful to construct scale invariant
quantities in the spatially homogeneous case where the group orbits
are the full $3$-slices, the quantity $\beta$ can be used to obtain
scale invariant quantities in the Gowdy case here; this is done
next. 
Define
\begin{gather*}
  (\mathcal N^{-1}, E\indices{_1^1}, E\indices{_2^2}, E\indices{_2^3}, 
  E\indices{_3^2}, E\indices{_3^3})
  :=(N^{-1},e\indices{_1^1}, e\indices{_2^2}, e\indices{_2^3}, 
  e\indices{_3^2}, e\indices{_3^3})/\beta\\
  (\dot U,A,1-3\Sigma_+,\Sigma_-,N_\times,\Sigma_\times,N_-,N_+,R):=
  (\dot u_1,a_1,\alpha,\sigma_-,n_\times,\sigma_\times,n_-,n_+,
  \Omega_1)/\beta\\
  \Omega_{\Lambda}:=\lambda/(3\beta^2).
\end{gather*}
In accordance to that we define $E_0:=e_0/\beta$ and $E_1:=e_1/\beta$,
hence
\[E_0=\mathcal N^{-1}\,\partial_{t}, \quad E_1=E\indices{_1^1}\,\partial_{x}.\]
Furthermore set
\begin{equation}
\label{eq:defqr}
E_0(\beta)=:-(q+1)\beta,\quad 
E_1(\beta)=:-r\,\beta.
\end{equation}
From all the equations, which we mentioned above, it is straight
forward to derive the corresponding system for the $\beta$-rescaled
quantities\footnote{I
  point out that many of the following equations in my \LaTeX\ code are taken
  from the freely available \TeX\ code of \cite{vanElst01} which I
  modified
  for my purposes. The reason for this was that I had problems to
  convert the 
  equations from my \Mathematica file with
  reasonable effort to \LaTeX\ myself. However, the equations in my
  \Mathematica file are exactly the same as those here, except for the
  specialization to vacuum here.}.
The evolution equations for the main variables are
\nobreak
\begin{subequations}
\label{eq:evolcomm}
\begin{align}
E_0(E\indices{_1^1})
=&  (q+3\Sigma_+)\,E\indices{_1^1} \\
\label{eq:evolSigmaPlus}
3\,E_0(\Sigma_+)
=&  -3\,(q+3\Sigma_+)\,(1-\Sigma_+)
+ 6\,(\Sigma_++\Sigma_-^{2}+\Sigma_\times^{2})- 3\,\Omega_\Lambda  \\ 
&  -(E_1-r+\dot U-2A)(\dot U)\notag \\
\label{eq:evolA}
E_0(A)
=&  (q+3\Sigma_+)\,A + r-\dot U \\
\label{eq:evolNplus}
E_0(N_+)
=&  (q+3\Sigma_+)\,N_+ + 6\,(\Sigma_-\,N_-+\Sigma_\times\,N_\times)
- (E_1-r+\dot U)(R) \\
%
E_0(\Omega_\Lambda)
=&  2\,(q+1)\,\Omega_\Lambda \\
%
E_0(\Sigma_-) + E_1(N_\times)
=&  (q+3\Sigma_+-2)\,\Sigma_- - 2\,N_+\,N_-
+ (r-\dot U+2A)\,N_\times - 2\,R\,\Sigma_\times \\
%
E_0(N_\times) + E_1(\Sigma_-)
=& (q+3\Sigma_+)\,N_\times + 2\,\Sigma_\times\,N_+
+ (r-\dot U)\,\Sigma_- + 2\,R\,N_- \\
%
E_0(\Sigma_\times) - E_1(N_-)
=&  (q+3\Sigma_+-2)\,\Sigma_\times - 2\,N_+\,N_\times
- (r-\dot U+2A)\,N_- + 2\,R\,\Sigma_- \\
\label{eq:evolNminus}
E_0(N_-) - E_1(\Sigma_\times)
=& (q+3\Sigma_+)\,N_- + 2\,\Sigma_-\,N_+
- (r-\dot U)\,\Sigma_\times - 2\,R\,N_\times.
\end{align}
\end{subequations}
From the evolution equation for $\beta$ and the Codazzi constraint we
get the following algebraic equations for $q$ and $r$
\begin{align}
\label{eq:qalb}
q & = \frac{1}{2} + \frac{1}{2}\,(2\dot U-A)\,A
+ \frac{3}{2}\,(\Sigma_-^{2}+N_\times^{2}+\Sigma_\times^{2}+N_-^{2})
- \frac{3}{2}\,\Omega_\Lambda \\
\label{eq:ralb}
r & = -\,3\,A\,\Sigma_+ - 3\,(N_\times\,\Sigma_--N_-\,\Sigma_\times).
\end{align}
The constraint equations following from the Gauß constraint and the
definition of $\Omega_\Lambda$ are
\begin{align}
\label{eq:constrGauss}
0 &= {\mathcal C}_\text{Gauß} \ = \ -\,\frac{2}{3}\,(E_1-r)(A)
+ A^{2} + N_\times^{2} + N_-^{2} - 1 + 2\Sigma_+
+ \Sigma_-^{2} + \Sigma_\times^{2} + \Omega_\Lambda \\
\label{eq:constrLambda}
0 &= {\mathcal C}_{\Lambda} \ = \ (E_1-2r)(\Omega_\Lambda).
\end{align}
The constraint that follows from the gauge fixing constraint
\Eqref{eq:gaugefixing} takes the form 
\begin{equation}
\label{eq:gaugeconstr}
0 = {\mathcal C}_{\text{gauge}}
:= E_1(\mathcal N) + (r-\dot U)\mathcal N.
\end{equation}
The decoupled evolution equations for the other frame
variables are 
\nobreak
\begin{subequations}
\label{eq:evolEresidual}
\begin{align}
  E_0(E\indices{_2^2})&=(q-\sqrt{3}\Sigma_-)E\indices{_2^2}
  -(R+\sqrt{3}\Sigma_\times)E\indices{_3^2}\\
  \label{eq:evolE23}
  E_0(E\indices{_2^3})&=(q-\sqrt{3}\Sigma_-)E\indices{_2^3}
  -(R+\sqrt{3}\Sigma_\times)E\indices{_3^3}\\
  E_0(E\indices{_3^2})&=(q+\sqrt{3}\Sigma_-)E\indices{_3^2}
  +(R-\sqrt{3}\Sigma_\times)E\indices{_2^2}\\
  E_0(E\indices{_3^3})&=(q+\sqrt{3}\Sigma_-)E\indices{_3^3}
  +(R-\sqrt{3}\Sigma_\times)E\indices{_2^3},
\end{align}
\end{subequations}
and their constraints take the form
\nobreak
\begin{subequations}
\label{eq:constrEresidual}
\begin{align}
  \label{eq:constrE22}
  0&=(E_1-A-\sqrt{3} N_\times-r)(E\indices{_2^2})
  +(\sqrt{3} N_--N_+)E\indices{_3^2}\\
  \label{eq:constrE23}
  0&=(E_1-A-\sqrt{3} N_\times-r)(E\indices{_2^3})
  +(\sqrt{3} N_--N_+)E\indices{_3^3}\\
  \label{eq:constrE32}
  0&=(E_1-A+\sqrt{3} N_\times-r)(E\indices{_3^2})
  +(\sqrt{3} N_-+N_+)E\indices{_2^2}\\
  \label{eq:constrE33}
  0&=(E_1-A+\sqrt{3} N_\times-r)(E\indices{_3^3})
  +(\sqrt{3} N_-+N_+)E\indices{_2^3}.
\end{align}
\end{subequations}
Finally, the integrability condition for $\beta$
to be determined from \Eqsref{eq:defqr} is
\begin{equation}
  \label{eq:integrabbeta}
  E_0(r)-E_1(q)=r(q+3\Sigma_+)-(q+1)(r-\dot U)
  +(q+1)\frac{{\mathcal C}_{\text{gauge}}}{\mathcal N}.
\end{equation}
One checks straight forwardly that all constraints propagate for
solutions of the evolution system. Further, the integrability condition
is implied by the evolution system and the constraints. Also note that
there are no evolution 
equations for the gauge quantities $\mathcal N$, $\dot U$ and $R$.

The next step is to fix the residual gauge freedom. 
Let us first prescribe
\[E\indices{_2^3}=0.\] 
Then its evolution equation \Eqref{eq:evolE23}
and its constraint \Eqref{eq:constrE23} are only solved if
\[R=-\sqrt{3}\Sigma_\times\quad\text{and}\quad N_+=\sqrt{3} N_-.\] 
Fortunately, the evolution equations of $N_-$, \Eqref{eq:evolNplus}, and
$N_+$, \Eqref{eq:evolNminus}, are consistent
with these choices, namely they become equal up to a factor. 
For the
\term{separable area gauge} 
we make the following further choice. For the gauge quantity $\dot U$
we can set 
\[\dot U=r.\] 
In this case the gauge constraint
\Eqref{eq:gaugeconstr} requires that $\mathcal N$ is a function of $t$
only and we use the freedom of choosing the time coordinate to make this
constant
\begin{equation}
\label{eq:choicelapse}
\mathcal N=\mathcal N_0=const.
\end{equation}
Such a time coordinate is in close analogy with that in inverse mean curvature
flows. Indeed, one should be concerned at this point if it is possible to find
hyperbolic evolution equations in this gauge at all. In fact, in the main
evolution equations \Eqsref{eq:evolcomm} there are problematic spatial
derivative terms. But it turns out that one can find an important
special case in which the system becomes symmetric hyperbolic. 
Namely, let us further assume that $A=0$ which is consistent with
\Eqref{eq:evolA} in separable area gauge.
Then the Gauß constraint
\Eqref{eq:constrGauss} becomes algebraic and can be used to determine
$\Sigma_+$. One can check that the evolution equation of
$\Sigma_+$ \Eqref{eq:evolSigmaPlus} is identically fulfilled if we do
so. Skipping this evolution equation from the system yields a
symmetric hyperbolic system where the quantities 
$(E\indices{_1^1},N_+,\Omega_\Lambda,\Sigma_-,N_\times,\Sigma_\times,N_-)$
are determined by the residual symmetric hyperbolic evolution
subsystem of \Eqsref{eq:evolcomm} and the quantities 
$(q, r, \Sigma_+)$ are determined algebraically by \Eqsref{eq:qalb},
\eqref{eq:ralb} and \eqref{eq:constrGauss} respectively. In
particular, the Gauß constraint is explicitly satisfied. The only
constraint left for the main system is \Eqref{eq:constrLambda}. The
decoupled system for the other frame quantities is constrained by the
residual three 
of \Eqsref{eq:constrEresidual}. Under all these conditions the gauge is
called \term{timelike area gauge} and we will say a few more words
about its interpretation in a moment. It is particularly interesting to note
that for a vanishing cosmological constant $\Omega_\Lambda=0$ the main
evolution system becomes completely unconstrained. However, this
is not the case for $\lambda\not=0$. 

We want to write down the final form of the equations in timelike area
gauge now. It is
important to realize that, with the choice of equations above,
$\Sigma_+$ only enters via the combination
$q+3\Sigma_+$ which equals $2-3\Omega_{\Lambda}$ using the Gauß constraint.
Thus, the residual evolution equations of the main system simplify to
\nobreak
\begin{subequations}
\label{eq:commFE_timelikeareagauge_Evolution}
\begin{align}
\label{eq:E11dot}
E_0(E\indices{_1^1})
=&  (2-3\Omega_{\Lambda})\,E\indices{_1^1} \\
\label{eq:OmegaLambdaDot}
E_0(\Omega_\Lambda)
=&  2\,(q+1)\,\Omega_\Lambda \\
%
E_0(\Sigma_-) + E_1(N_\times)
=&  -3\Omega_{\Lambda}\,\Sigma_- - 2\sqrt{3}N_-^2
+ 2\sqrt{3}\Sigma_\times^2\\
%
E_0(N_\times) + E_1(\Sigma_-)
=& (2-3\Omega_{\Lambda})\,N_\times + 2\sqrt{3}\,\Sigma_\times\,N_-
- 2\sqrt{3}\,\Sigma_\times\,N_- \\
%
E_0(\Sigma_\times) - E_1(N_-)
=& -3\Omega_{\Lambda}\,\Sigma_\times - 2\sqrt{3}\,N_-\,N_\times
- 2\sqrt{3}\,\Sigma_\times\,\Sigma_- \\
E_0(N_-) - E_1(\Sigma_\times)
=& (2-3\Omega_{\Lambda})\,N_- + 2\sqrt{3},\Sigma_-\,N_-
+ 2\sqrt{3}\,\Sigma_\times\,N_\times,
\end{align}
with 
\begin{equation}
\label{eq:qgauge}
q = \frac{1}{2} 
+ \frac{3}{2}\,(\Sigma_-^{2}+N_\times^{2}+\Sigma_\times^{2}+N_-^{2})
- \frac{3}{2}\,\Omega_\Lambda,
\end{equation}
\end{subequations}
which is clearly a symmetric hyperbolic system.

The reason for the name ``timelike area gauge'' is the following. The
area density\footnote{The symbol $\mathcal A$ for the area density
  should 
  \textit{not} be confused with the symbol for the quantity $A$.}
$\mathcal A$ of the symmetry orbits is 
\begin{equation}
  \label{eq:orbitareadensity}
  \mathcal A^{-1}
  =e\indices{_2^2}e\indices{_3^3}-e\indices{_2^3}e\indices{_3^2}.
\end{equation}
One can easily check using the evolution equations
\Eqsref{eq:evolEresidual} and the constraints
\Eqsref{eq:constrEresidual} that
\[E_0(\mathcal A)=2\mathcal A,\quad
E_1(\mathcal A)=-2 A\mathcal A.\]
Thus, under the assumption $A=0$ made above, the area density is
constant on each $t=const$-slice. In fact we have
\[\mathcal A=l_0^2 e^{2\mathcal N_0 t}\]
where $l_0$ represents some spatial scale constant. The authors in
\cite{Andersson03} choose $\mathcal N_0=-1/2$ which implies in
particular, as for all negative values of $\mathcal N_0$, that
the symmetry orbits are shrinking
with increasing time. 

It is of interest to note that the polarized Gowdy case, which is
given by the condition that the Gowdy Killing vector fields can be
chosen to be orthogonal everywhere, is the invariant subset
of the state space given by 
\[\Sigma_\times=N_-=0.\] 

Finally, we want to remark that it seems to be difficult, maybe even
impossible, to regularize the system on $\scri$ without major modifications.

\chapter{Cosmological spacetimes}
\label{ch:cosmolog_spacetimes}

\section{Introduction}
\label{sec:cosm_intro}
In this thesis we want to consider cosmological spacetimes.
A standard definition in mathematical cosmology is the following.
\begin{Def}
  A \term{cosmological spacetime} is a spacetime with compact Cauchy
  surfaces. If a
cosmological spacetime is additionally a solution of EFE it will be called
\term{cosmological solution}.
\end{Def}
This definition has already been motivated in \Chapterref{ch:introduction}.
For many issues below, compactness of the spatial slices is not
necessary; however, we restrict to that case.  

The study of spacetimes with isometries plays a particular role in the
cosmological setting.
Although isometries are
clearly not restricted to this setting,
we start this chapter by discussing a few relevant topics
related to symmetry in \Sectionref{sec:isometries}.  After that,
fundamental issues for cosmological spacetimes like
strong cosmic censorship, the BKL-conjecture etc.\ are discussed and 
important known results in the literature are listed
(\Sectionref{sec:cosmiccensorship}).  
Finally in \Sectionref{sec:fads}, we turn to the
class of future asymptotically de-Sitter spacetimes, which will be the
class of interest for this thesis. 

\section{Isometries}
\label{sec:isometries}

\subsection{Preliminaries}
In the following we only make a few comments on topics of
particular relevance 
for this thesis.
For an introduction into the underlying theory, the reader
is referred to \cite{MacCallum79,Wainwright,exactsol}.

Let $(M,g)$ be a smooth Lorentzian manifold. For a global 
\term{group of transformations}
one assumes that there is a Lie group $G$ which acts 
smoothly on $M$. We write such an action as 
$G\times M\rightarrow M$, $(u,p)\mapsto up$. For any fixed $u\in G$, the
map $\Theta_u:M\rightarrow M$, $\Theta_u(p)=up$ is a
diffeomorphism. The action is called \term{effective} if $e\in G$ is the only
element $u\in G$ such that $\Theta_u=\mathrm{id}_M$. A group of
transformation is 
an \term{isometry group} (\term{group of motions}) if additionally 
$\Theta_u^* g=g$ for all $u\in G$. Then, each element $\xi$ of the 
Lie algebra of {Killing vector fields} (KVF), which are those
vector fields generated by the action of the isometry group, fulfills
the Killing equation
\begin{equation}
  \label{eq:killingeq}
  \lieder{\xi} g=0.
\end{equation}
Also, the Lie derivatives of the derived
curvature quantities of $g$ vanish along $\xi$. 
Effectiveness of the action implies that this Lie algebra is
isomorphic to the Lie algebra of the isometry 
group. Indeed, we will assume that the actions of all transformation
groups in the rest of this thesis are effective. There is also the
notion of local groups of 
transformations. This plays a particular role when we have a Lie
algebra of 
vector fields given and have to decide if this algebra is generated by
a group of transformations. Such issues are discussed in
\cite{Hall87}, summarizing work by Palais \cite{Palais57}, but we will
not give further details. In this thesis, all transformation groups
are global if not stated otherwise.

Assume now that an orthonormal frame $\{e_i\}$ is given on $M$. Let $\xi$
be a Killing vector field. The Killing equation \Eqref{eq:killingeq}
is equivalent to  
\begin{equation}
  \label{eq:killing_eq_onf}
  g([\xi,e_{(i}],e_{j)})=0.
\end{equation}
Moreover, consider the following important notion.
\begin{Def}
  An orthonormal frame $\{e_i\}$ is called \term{$\xi$-invariant} (or
  \term{group invariant}) if $[\xi,e_i]=0$ on $M$.
\end{Def}
One sees that the existence of an orthonormal frame which is
$\xi$-invariant implies that $\xi$ is a Killing vector
field. But on the other hand, orthonormal frames do not need to be
$\xi$-invariant in general if $\xi$ is a Killing vector field.

The physically relevant isometry classes for cosmological spacetimes have
spacelike Killing  
vector fields and some of those are introduced in
\Sectionref{sec:relevantsymmclass}. Before that in
\Sectionref{sec:isometry_transport}, we give a discussion motivated by the
following question of interest. 
Assume that an
initial data set $(\Sigma,h,\chi)$
for Einstein's field equations in the sense of
\Sectionref{sec:maximaldevelopments} is given so that there is an
effective group of motions
$p\mapsto up$ on $\Sigma$ such that
also the second fundamental form $\chi$ is invariant (i.e.\
its Lie derivatives along all KVFs of $\Sigma$ vanish). 
Now consider $\Sigma$ as embedded into the corresponding solution $(M,g)$ of
EFE. Then, is there
always an extension of the KVFs of $\Sigma$ to spacelike KVFs of $M$
on a
neighborhood of $\Sigma$ in $M$?
The answer is yes. For instance, Chru\'{s}ciel \cite{chrusciel1990} argues
that if the coordinate gauge is harmonic with $t=0$ corresponding to
the embedding of $\Sigma$ then one sees directly that
$(t,p)\mapsto (t,up)$ determines a group of motion acting effectively
on $M$ with spacelike KVFs for
$t$ small enough. In any case, in our work we will employ
other gauges. Founding on this knowledge, however, we can
investigate 
how a spacelike
(spacetime) Killing vector field behaves with respect to a given timelike
congruence which defines a foliation. In particular we try to find
a condition on the foliation under which $(t,p)\mapsto (t,up)$ is a
spacetime isometry 
if $p\mapsto up$ is an isometry on the spacelike $t=0$-surface.

\subsection{Transport of Killing vector fields along timelike
  congruences} 
\label{sec:isometry_transport}

In this section  we try to identify a simple condition under which
a spacelike KVF is adapted to the foliation generated by a timelike
congruence in the sense of the previous discussion.

In the following assume that $(M,g)$ is a smooth Lorentzian
manifold with local coordinates $\{x^\mu\}$ and that $\{e_i\}$ is a
smooth orthonormal frame with $e_0$ 
timelike, but not (yet) necessarily hypersurface orthogonal. We consider
the timelike congruence 
generated by $e_0$. Let us start with 
a simple observation.
\begin{Lem}
  \label{lem:cor1}
  Let $\xi$ be a smooth Killing vector field and $[\xi,e_0]=0$ along
  $e_0$,
  for instance $\{e_i\}$ a $\xi$-invariant
  frame. Then $e_0(\xi^\mu)\equiv 0$, i.e.\ the coordinate components of
  $\xi$ are constant along the integral curves of $e_0$, if and only if
  $\xi(e\indices{_0^\mu})\equiv 0$ along the integral curves of $e_0$.
  \begin{Proof}
    This follows directly, writing $[\xi,e_0]=0$ in the coordinate basis.
  \end{Proof}
\end{Lem}
The main result of this section is the following simple Proposition.
\begin{Prop}
  \label{prop:xiconst}
  Let $\xi$ be a smooth Killing vector field such that $g(e_0,\xi)=0$
  at $p\in M$. Suppose the frame is given such that for the vector
  field $\dot u$ 
  (\Eqref{eq:dot_u} for $u=e_0$) one has $g(\dot u,\xi)= 0$ along the
  integral curve of $e_0$ starting at $p$. Then $g(e_0,\xi)= 0$
  along the integral curve of $e_0$ starting in $p$. If we further
  assume that $e_0$ is a hypersurface orthogonal vector field, then
  $[\xi,e_0]= 0$ along the integral curve of $e_0$ starting in
  $p$. Hence, if  
  $\xi(e\indices{_0^\mu})=0$ along the curve, 
  then according to \Lemref{lem:cor1}, $\xi^\mu=const$ along the
  curve.
  \begin{Proof}
    We have
    \[\nabla_{e_0}(g(\xi,e_0))=g(\nabla_{e_0}\xi,e_0)+g(\xi,\nabla_{e_0}e_0)
    =g([e_0,\xi],e_0)+g(\nabla_{\xi}e_0,e_0)+g(\xi,\dot u).\]
    The first term on the right hand side vanishes due to
    \Eqref{eq:killing_eq_onf}; the 
    second is zero  
    because $e_0$ is a unit vector field. For the third one we note that
    $\dot u=\nabla_{e_0}e_0$, cf.\
    \Eqref{eq:dot_u}. This can be considered as an ODE for the
    function $g(e_0,\xi)$ along the integral curve of $e_0$. With the
    initial data $g(e_0,\xi)=0$ at $p$, the unique solution is
    $g(e_0,\xi)\equiv 0$. 

    For the second claim, we have $g([\xi,e_0],e_0)\equiv 0$ due to
    \Eqref{eq:killing_eq_onf}. Moreover, due to the same equation, 
    $g([\xi,e_0],e_a)=-g([\xi,e_a],e_0)\equiv 0$, because $\xi$ is
    orthogonal to $e_0$, i.e.\ spatial, and because $e_0$ is
    hypersurface orthogonal, i.e.\ the integral surfaces of the
    orthogonal vector fields are involutive.
  \end{Proof}
\end{Prop}
Note, that in particular in Gauß gauge, $\dot u=0$ and $e_0$ is
hypersurface forming and hence \Propref{prop:xiconst} can be
applied. Now let the coordinates and the frame be given such that
$e_0=\partial_t$, i.e.\ lapse and shift in \Eqref{eq:lapseshift} are
trivial, and $e_0$ is hypersurface orthogonal. Then under the further
conditions of \Propref{prop:xiconst}, the spacetime isometry
generated by $\xi$ can be written as $(t,p)\mapsto (t,up)$
where $p\mapsto up$ is the isometry generated by $\xi\bigr|_{t=0}$
considered as a vector field tangent to the $t=0$-Cauchy surface
orthogonal to $e_0$.

A couple of further results of this kind, but which are not so relevant for
this thesis, can be found in \cite{andersson04a}.

\subsection{Some relevant symmetry classes}
\label{sec:relevantsymmclass}
Let us  start with $3+1$-dimensional spatially homogeneous spacetimes
for which the 
orbits coincide with $3$-dimensional spacelike Cauchy surfaces; i.e.\
the isometry groups act transitively on the Cauchy surfaces.
 From the  
general 
theory \cite{MacCallum79} one knows that under this condition
the maximal dimension of the isometry 
groups is $6$. In this maximally symmetric case, the isotropy subgroup
is $3$-dimensional and so the orbits must have constant
curvature. The corresponding spatially homogeneous 
and isotropic solutions of EFE are the
\term{Friedmann-Lemaitre-Robertson-Walker spacetimes}
(FLRW). Further 
details on this important class are in \cite{Wainwright}, including a
qualitative analysis based 
on dimensionless quantities. The symmetry requires that the
matter fields are of perfect fluid type with stress energy tensors of
the form
\begin{equation}
\label{eq:perfectfluid}
  T_{\mu\nu}=\rho u_\mu u_\nu+p(g_{\mu\nu}+u_\mu u_\nu).
\end{equation}
Here, $u^\mu$ is the unit $4$-velocity vector field of the fluid,
$\rho$ the matter density and $p$ 
the (isotropic) pressure. Note, that a cosmological constant can be
considered as a perfect fluid with equation of state $p=-\rho$ so that
$\rho$ equals the value of the cosmological constant $\lambda$.
The importance of the FLRW models, at least for
suitable matter fields, stems from the fact that they are the simplest
parametrized models which have been successfully fitted to
observational data; see some discussion in 
\Chapterref{ch:introduction}. Further note, that the $6$-dimensional
isometry groups with spacelike orbits can be subgroups of larger
isometry groups in this class of spacetimes. 
For example, for the $3+1$-dimensional
de-Sitter spacetime (\Sectionref{sec:deSitter}) the total isometry
group is $10$-dimensional.

One can show \cite{MacCallum79} that there can be no
$5$-dimensional isometry group acting on $3$-dimensional
hypersurfaces 
transitively. Let us hence continue with the $4$-dimensional case;
again the following cosmological spacetimes are not required to be
solutions of EFE. It
turns out that there are two possibilities. The first one 
arises when the isometry group has a $3$-dimensional subgroup that
acts simply-transitively on the spacelike Cauchy surfaces. It is
called \term{LRS-Bianchi} case (Locally Rotationally Symmetric),
since then the isometry group consists of a $3$-dimensional Bianchi
group, see below, and a $1$-dimensional additional symmetry.
Of
particular interest for our studies will be the family of
$\lambda$-Taub-NUT spacetimes, see \Sectionref{sec:TaubNUT}, which is
of type LRS-Bianchi IX. The
second case is realized when the $4$-dimensional isometry group does
not have such a $3$-dimensional subgroup.
Then it turns out that the only allowed topology of the
spacelike Cauchy surfaces is 
$\Stwo\times\mathbb S^1$, and this case is called
\term{Kantowski-Sachs} \cite{KantowskiSachs,Collins77,MacCallum79}.  

If the isometry group is $3$-dimensional and the action is transitive
on $3$-dimensional spacelike slices,
then it acts
simply-transitively on the spacelike Cauchy surfaces. This is the
\term{Bianchi} case, since it was Bianchi who first classified
$3$-dimensional real Lie 
algebras; see \cite{Wainwright} for an introduction to the Bianchi
classification and the definition of the Bianchi types I to IX and the
classes Bianchi-A and Bianchi-B. 
The relation of the allowed spatial topologies for (local) Bianchi
isometry groups and the
Thurston geometries is given in \cite{andersson04a}. 
The theory of solutions of EFE of Bianchi type is well elaborated in 
\cite{Wainwright} using dynamical system techniques. Of particular
importance for such qualitative analyses are
scale invariant quantities. This leads to the equations by Wainwright
and Hsu \cite{Wainwright89} which are analogous to the
commutator field equations described in
\Sectionref{sec:commutatorfieldequations}. We discuss some results for
this class of
solutions in \Sectionref{sec:gowdyphenom}. But note, that the analysis
of Bianchi solutions in
\cite{Wainwright} is obsolete in so far as the Bianchi
IX case is concerned, see below. 

Let us now reduce the dimension of the isometry group further to
$2$. Clearly, such spacetimes cannot be spatially homogeneous
anymore. However, we still assume that the orbits are subsets of
spacelike Cauchy surfaces, in particular the Killing vector fields are
spacelike. 
Further let us restrict to
global actions of the group $\U\times\U$. Comments on the motivation
for this restriction and further details can be looked up in
\cite{andersson04a,Wainwright} and the references below. Spacetimes with
such an isometry group were discussed first by 
Gowdy \cite{Gowdy73}. Some implicit assumptions made by Gowdy were
removed in \cite{chrusciel1990}; further his arguments and results
were clarified and 
extended. The assumption of a global smooth effective isometric 
$U(1)\times U(1)$-action 
on smooth connected orientable $3$-manifold has the following
implications (see references in \cite{chrusciel1990}). First, 
the associated
Killing vector fields commute because the group is Abelian. Secondly,
the action is
unique up to equivalence.
Next,
the only admissible topologies of such 
$3$-manifolds are the $3$-torus $\T$, the $3$-sphere $\S$ (or lens
spaces which are always 
included implicitly in the following discussions) and the $3$-handle
$\mathbb S^1\times\Stwo$; 
if $\U\times\U$ is a local isometry group then further
topologies are possible, see \cite{Tanimoto01}, but this cannot be
elaborated here. Further, the twist
quantities, which are defined e.g.\ in \cite{chrusciel1990} but which
we will not discuss here further, turn out to be 
constant for any Gowdy spacetime which satisfies EFE in vacuum with an
arbitrary cosmological constant.  
One finds that
non-vanishing twist constants can only occur when the topology
is $\T$.
If
the twist constants vanish, the solutions are
called \term{Gowdy spacetimes}. However, even if they do not vanish, I
will sometimes denote the group $\U\times\U$ as \term{Gowdy
  group} in this thesis. Gowdy spacetimes, where the Killing vector
fields of the Gowdy isometry group can be chosen to be
orthogonal everywhere, are referred to as 
\term{polarized Gowdy spacetimes}. I will list more details, in particular on
results concerning cosmic censorship in
\Sectionref{sec:gowdyphenom}. 

Let us decrease the dimension of the isometry group even
further to the $1$-dimensional case. Most investigations on solutions
of EFE of this kind were done so
far for the case when the isometry group is $\U$; a few more details
on this are listed in \Sectionref{sec:gowdyphenom}. The most general
class of 
spacetimes is given when there are no symmetries. Since all
spacetimes with symmetries have to be considered as non-generic
in the space of all solutions of EFE, real
statements about the character of generic solutions cannot be
made before this general class can be controlled. The situation is
that the
techniques both on the rigorous analytical as on the numerical
side are not sufficient yet for such general studies.
 However relevant techniques are progressing enormously
so that there is hope for deeper understanding in the near future. In
the meantime, careful investigations of special classes of 
spacetimes are important to develop the tools and to obtain first
ideas, which kinds phenomena are characteristic for solutions of EFE.


\section{Fundamental issues for cosmological solutions}
\label{sec:cosmiccensorship}

In this section we summarize some important outstanding problems and
issues for cosmological solutions which we want to investigate in this
thesis. In fact, many of these problems are not
restricted to this class of spacetimes, however, I list them here to
keep the presentation as compact as possible.

\subsection{Incompleteness, extendibility  and cosmic
censorship}
\label{sec:SCC}
Singularity theorems in general relativity,
some of them are quoted in \Sectionref{sec:singularitytheoremGA}, 
give conditions under which some or all causal geodesics in a globally
hyperbolic Lorentz manifold must cease to exist after a
finite affine parameter time. The way these theorems
make use of the field equations usually does not allow to make
statements about 
the reason for incompleteness. One possibility is that the geodesics run
into some sort of curvature singularities; but it might also be that
the globally hyperbolic spacetime is extendible to a non-globally
hyperbolic one and that the geodesics just
leave the globally 
hyperbolic region.  In the second case,
if the extension into the non-globally hyperbolic region is regular in
an appropriate sense, not necessarily smooth as considered in
\Sectionref{sec:maximaldevelopments}, 
there are well-defined Cauchy horizons. These separate the
globally hyperbolic ``predictable'' region from the non-globally
hyperbolic rest with the possibility of many kinds of pathologies. For
example, closed 
causal curves cannot be excluded. In addition, one can expect
generically that there are 
various non-equivalent extensions through the horizon. An example for
this is the $\lambda$-Taub-NUT
family (\Sectionref{sec:TaubNUT}). More examples with this
kind of behavior are
known for instance in the class of polarized Gowdy solutions, see
\cite{chrusciel91a,Chrusciel06}. 

If these pathological properties
were generic among solutions of general relativity, Einstein's theory
would disagree with our fundamental belief and, to some degree,
experience about causality and deterministic laws of nature. One has to 
accept that there are some solutions of EFE with such ``bad''
behavior, but one would like to find that EFE somehow excludes it
generically. This is the issue of \term{strong cosmic censorship}
(SCC) formulated as follows
in the class of cosmological solutions in
vacuum with arbitrary cosmological constants.
\begin{Conj}
  Let $\Sigma$ be a compact manifold of dimension $3$. Then for a generic
  vacuum initial data set $(\Sigma,h,\chi)$, the corresponding maximal
  Cauchy development is maximal among all developments of this
  data set.
\end{Conj}
Two aspects are left open in this formulation of the strong cosmic
censorship conjecture, namely, what is meant by ``generic'' and the
choice of
the class of extensions and developments. We say more about this in a moment.
Roughly speaking, as formulated by Clarke et al.\ \cite{clarke80}, we hope that
generically, ``whole'' solutions of EFE are globally hyperbolic.
In the presence of black holes one might weaken this conjecture and
state that violations of global hyperbolicity are only allowed when
they are hidden inside an event horizon. This is the notion of
\term{weak cosmic censorship}. Weak cosmic censorship is usually
considered in the case of asymptotically flat solutions; one of the first
rigorous results was obtained by
Christodoulo \cite{christodoulou99} who proves weak cosmic censorship
for spherically symmetric Einstein-scalar field equations. In any
case, weak cosmic censorship will not be considered further in this
thesis.

The first point left open in the conjecture is the type of
extensions; usually one considers $C^2$-extensions within a fixed
class of solutions of EFE of interest. Eventually of course, one would
like to be able
to study the class of all solutions of EFE, but for the time
being with the tools currently available it is almost always ambitious enough
to restrict to simpler feasible settings. The reason why 
extensions of $C^2$-type  are considered is not discussed
here; see comments in the relevant references listed below. The
standard way to show that a  
solution is $C^2$-inextendible is to prove that the
\term{Kretschmann scalar}
\begin{equation}
  \label{eq:KretschmannDef}
  \kappa:=R_{\mu\nu\rho\sigma}R^{\mu\nu\rho\sigma}
\end{equation}
blows up along all causal geodesics approaching the relevant points. 
In our considerations, assuming a physical metric $\tilde g$ in vacuum
with arbitrary cosmological constant, an
important formula in the notation of 
\Sectionref{sec:conformal_geom} is
\begin{equation}
  \label{eq:physKretschConfQuant}
  \tilde \kappa=24 \frac{\lambda^2}9 +8\Omega^6(|E|^2-|B|^2).
\end{equation}
Here $\tilde\kappa$ is the Kretschmann scalar of $\tilde g$, the
conformal metric $g$ is 
given by $g=\Omega^2\tilde g$ 
with conformal factor $\Omega>0$, and $E$ and $B$ are the electric and
magnetic parts of the 
rescaled Weyl tensor (\Eqref{eq:rescaled_weyl})
with respect to the timelike congruence
determined by $e_0$ of an orthonormal frame $\{e_i\}$ with respect
to the conformal metric.
Here, 
\[|E|^2:=\sum_{a,b=1}^3 |E_{ab}|^2,\]
similar for $|B|^2$.

The second aspect which is left open in the formulation of the SCC
conjecture above is the notion of ``genericity''. For this, one
assumes that a reasonable topology on
the set of all initial data sets corresponding to the class of
solutions under consideration can be introduced. Then
``generic'' subclasses are required to be dense. Currently, one has no idea if
such a topology exists in general and how it looks.
However, in the
class of Bianchi and Gowdy spacetimes (under further restrictions) for
instance,
this topology has been found and SCC has been confirmed rigorously, see
below. 

In \cite{clarke80}, the authors give an overview of the discussions about
extensions and SCC up to the
beginning of the $1980$s, and name the relevant references. Some newer
results are given in the following sections. 
Note that the cosmic censorship conjecture was essentially formulated by
Penrose \cite{Penrose69,penrose1979}.

One more comment is in place here. It might seem that SCC is a purely
academic problem since we know that GR must be substituted by some
sort of quantum gravity in strong field regimes anyway. Although this
might be true, the example of Taub-NUT spacetimes shows us that problems
related to the question of extendibility are not restricted to strong
field regimes. Indeed, an observer who would cross the corresponding
Cauchy horizon in the Taub-NUT case would not feel particularly strong
gravitational 
forces. Hence, there is the potential danger that a quantum theory of
gravity built on some of the same principles as general relativity
would suffer from the same problems. Thus, it is crucial to check if
general relativity obeys strong cosmic censorship and so is
a self-consistent theory
in agreement with our fundamental view about nature. If general
relativity turned out to violate strong cosmic censorship, it would be
important 
for reasonable formulations of quantum theories of gravity to identify
the responsible underlying principles and avoid or modify them.

\subsection{Cauchy horizons in cosmological solutions}
\label{sec:CHcosm}
Related to the issues of extendibility and SCC is the fact that there
are classes of solutions with smooth Cauchy horizons, some of them
were mentioned before. Here, we present some general results about
solutions with
Cauchy horizons.

A first step in the direction to prove that solutions with smooth
Cauchy horizons 
are non-generic is the following result \cite{moncrief83}. Let an
analytic vacuum (or electrovacuum) spacetime with $\lambda=0$ be given
which has a compact orientable null hypersurface ruled by closed null
generators 
(in the sense of a fiber bundle). Then the spacetime has a non-trivial
Killing symmetry with Killing fields normal to the horizon.
If the surface is ``non-degenerate'' it must be a Cauchy horizon and
the Killing field changes from being spacelike in the globally
hyperbolic region to timelike in the residual part of spacetime.
Later, Isenberg and Moncrief extended this result by removing the
bundle condition in \cite{isenberg85}; the key ingredient in this
analysis was the use of Seifert fibrations. The analyticity requirement is
relaxed to smoothness in \cite{Friedrich98}. Hence, in this situation,
the existence 
of compact smooth Cauchy
horizons ruled by closed null generators implies a symmetry and hence
the spacetimes are 
non-generic. Isenberg and Moncrief \cite{isenberg85} also comment
about the case of Cauchy horizons ruled by generators with non-closed
orbits but were not able to treat that case. 

Moncrief \cite{moncrief84}
constructs all analytic 
solutions of the vacuum field equations ($\lambda=0$) with compact
Cauchy horizons of $\S$-topology ruled by closed null generators which
fiber the horizon in the sense of the Hopf fibration
(\Sectionref{sec:Hopf_fibration}). This result relies on earlier work
\cite{Moncrief82} where Moncrief treats the same question for horizons
of $\T$-topology with a global product bundle fibration.
The idea is to study a singular initial value problem
where the Cauchy horizon is the initial hypersurface; by this the
field equations 
can be reduced to Fuchsian form. One can show that there is an infinite
dimensional family of solutions of this type which is, however, due
to the symmetry properties above non-generic.

Recently Chru\'{s}ciel \cite{Chrusciel06}
has worked out the uniqueness and maximality theory of general
conformal boundary extensions, including extension through null hypersurfaces
like Cauchy horizons. His results cannot be discussed here despite of
their fundamental importance.

\subsection{BKL-conjecture}
\label{sec:bklconjecture}
Consider a $C^2$-inextendible geodesically incomplete
cosmological solution of EFE. The \term{BKL-conjecture} is an attempt
to describe the properties of gravitational singularities in generic such
cases. First investigations in this direction
were done by Khalatnikov and Lifshitz (KL) \cite{lifshitz63} which were then
improved together with Belinskii in \cite{belinskii70,belinskii82}
(BKL). A  review 
with a summary of newer results can be found in \cite{berger98} and
most recent numerical studies are in
\cite{garfinkle04a,Curtis05}. Investigations into the direction of a
precise formulation of the conjecture can be found in
\cite{Uggla03,Heinzle07}. We report on rigorous results in special
classes of solutions in \Sectionref{sec:gowdyphenom}.

This conjecture claims that generic singularities of solutions of EFE are
spacelike and locally (i.e.\ pointwise) modeled by the 
family of Mixmaster universes; in particular one believes to find infinite
sequences of Kasner epochs, observable as oscillations, in
the approach to the singularity. Each timelike worldline is supposed
to become
decoupled from all neighboring worldlines and to behave as an individual
``spatially homogeneous'' solution of the type above. This is referred
to as \term{silent singularities}. 
In \Sectionref{sec:gowdyphenom}, we mention only a few more details on
Kasner and Mixmaster 
solutions, so the reader who is not familiar with this should look into
\cite{Wainwright}. Further, the BKL-authors suggest 
that ``matter does not matter'' at the singularity (except
for special cases), i.e.\ the details of the matter model are not
important for the behavior of the solution at the
singularity. However, there
are cases, for instance stiff fluids, which can ``stop'' the BKL-like
oscillations 
such that the solution approaches a, possibly pointwise dependent, Kasner
solution. Such behavior is then called \term{quiescent}
\cite{Andersson00}. However, in many of the special classes of
solutions considered so far,  even in vacuum, the solutions
converge ``only'' to a pointwise dependent Kasner solution
without oscillations
and this behavior is then referred to as \term{asymptotically velocity
  dominated}; a notion introduced first in
\cite{Eardley71} and then extended and applied in
\cite{Isenberg89}.

The
arguments by the BKL-authors are heuristic, formal and essentially
consistency checks. To give an example, the argument for the  ``matter 
does not matter'' conjecture goes roughly as follows. Let us assume
that we have convinced 
ourself that at the singularities, generic solutions converge locally to a
Kasner spacetime, at least for the period of time which can be
considered as one Kasner epoch. Furthermore,
assume for simplicity that the 
matter is a perfect fluid (cf.\ \Eqref{eq:perfectfluid}) with linear 
equation of state 
\begin{equation}
  \label{eq:linEOS}
  p=(\gamma-1)\rho\quad\text{with}\quad\gamma=const.
\end{equation}
For a Kasner
spacetime that is foliated by its symmetry hypersurfaces with a
symmetry adapted frame $\{e_0,e_a\}$ such that $e_0=\partial_t$ and
$t=0$ corresponds to the singularity,
one finds that the frame components of the 
$2$nd fundamental form are $O(t^{-1})$ at the
singularity. Now consider the 
following 
``consistency'' argument. If our solution above
really converges in some sense to a Kasner spacetime along all
timelines at the singularity then it should be allowed to 
substitute some of the
terms in the Hamiltonian constraint\footnote{Substitute $\lambda$
 by the energy density $\rho$ of the perfect fluid in
 \Eqref{eq:Hamconstraint}.} 
\Eqref{eq:Hamconstraint} by 
their Kasner values at the singularity; in 
particular for the Ricci
scalar set $r=0$ and for the $2$nd fundamental form set $O(t^{-1})$. As we
will not discuss here, one knows for
general Bianchi models with a perfect fluid source and the equation of
state above that  
$\rho=O(t^{-\gamma})$ at the singularity. Hence, if
the solution 
converges to Kasner and if $\gamma<2$, the Hamiltonian constraint  
is dominated by the $2$nd fundamental form for $t\rightarrow 0$ and
\textit{not} by the energy density and so ``matter does not
matter''. However, for the stiff fluid $\gamma=2$, this
argument cannot be 
applied and 
indeed, stiff fluids \textit{do} matter, see below. Moreover note, that for all
flat FLRW with the same matter source as above we have
$\rho=O(t^{-2})$ for all $\gamma$ and hence the argument
also fails. But, the cosmological constant given by $\gamma=0$ is
covered by this kind of argument, and the general expectation is that
its influence becomes 
negligible when a singularity is approached. 

It is hard to judge if such kinds of arguments make sense in general
cases, and it is currently hard to say how ``generic''
BKL-singularities really are. Below we will discuss some cases where
the conjecture has been either confirmed rigorously or at least supported
numerically. However, the  
conjecture is still controversial. For example one knows about the
existence of \term{weak null 
  singularities} and it cannot be excluded that those are also
generic in certain classes of solutions. A list of relevant
references together with the problem and its history on this
alternative type of singularities can be found
in \cite{Rendall05,Ori06}.

\subsection{Cosmic no-hair conjecture}
\label{sec:cosmicnohair}
As we described before, according to our current understanding, the
history of our
universe is associated with at least two ``inflationary'' phases of accelerated
expansions; one shortly after the big bang known as \term{inflation}, the
other at the present time. In this thesis we want to restrict to
spacetimes which show such inflationary behavior asymptotically in the future.
Hawking et al.\ in \cite{Hawking82} introduced the
notion of \term{cosmic no-hair}. The underlying natural question is if
in generic inflationary scenarios
cosmological solutions of EFE converge locally to the de-Sitter
spacetime in the future. 

With respect to a given foliation and time coordinate, spacetimes which
obey the cosmic no-hair picture are characterized as follows.
The 
shear quantities become asymptotically small in comparison to
the Hubble scalar, the slices become more
and more homogeneous and isotropic, in particular flat in the
expanding time direction and 
approximate the
exponentially accelerated expanding phase of the de-Sitter spacetime.

There is a large collection of results about this
issue in several distinct 
situations. Indeed, there are cases which do not obey the cosmic no-hair
picture or obey generalized kinds like power-law
inflation.  We do not comment on this further. The class of spacetimes
which we want to restrict to in this thesis is the class of future
asymptotically de-Sitter spacetimes (\Sectionref{sec:fads}). Those
``almost'' obey the cosmic no-hair picture in a nice geometrically motivated
manner, as we will discuss in
\Sectionref{sec:FAdScosmicnohair}. 

\newpage

\subsection{Results in special classes of cosmological solutions}
\label{sec:gowdyphenom}

Here we report briefly on the status of research in important classes of
spacetimes about
the fundamental issues stated in the previous sections. It is
interesting to realize that many of these rigorous results
were actually stimulated by numerical investigations.

\subsubsection{Bianchi solutions}
For Bianchi solutions (\Sectionref{sec:relevantsymmclass}), some
important results are as follows. Regarding the cosmic no-hair
picture, the classic theorem in this setting is due to Wald
\cite{Wald83}. He considers 
Bianchi models with $\lambda>0$ and with matter that satisfies the
strong and dominant energy conditions, and finds that all initially
expanding models, except possibly 
for Bianchi IX, evolve towards the de-Sitter solution in the
future on an 
exponentially rapid time scale. In particular he proves isotropization
and exponential expansion. For Bianchi IX he finds the corresponding
behavior if the spacetime does not recollapse, and a sufficient
condition relating the size 
of the cosmological 
constant with the initial values of other curvature quantities for
this was also found. For 
vanishing $\lambda$, Lin and Wald prove the closed
  universe recollapse theorem in 
\cite{Lin89,Lin90} which states that there are no Bianchi-IX models
that expand forever if the matter fields satisfy the
dominant energy condition and have positive mean principal pressure. 
Results about the future behavior for the other non-tilted Bianchi-A
models with $\lambda=0$ are listed in \cite{Ringstrom00}.
There, the assumption is that the matter is a perfect fluid
with linear equation of state 
\Eqref{eq:linEOS} such that $2/3<\gamma\le 2$. In particular, one
finds that these spacetimes are future
geodesically complete after having chosen the
time orientation accordingly. However, not all of them isotropize
\cite{Wainwright}.  

Let us postpone the discussion of Bianchi VIII and IX solutions for
the moment and let us restrict to the other Bianchi-A models with
perfect fluid source and linear equation of state. Then, one can
prove geodesic incompleteness for generic such spacetimes in the
opposite time direction.
Bianchi I and VII$_0$ with $\tr\chi=0$ on a Cauchy surface
are the only exceptions, namely, these are
geodesically complete in both time directions.  
Even more, Rendall \cite{rendall97d} showed that Bianchi I, VII$_0$
(except for the $\tr\chi=0$ cases), Bianchi II and VI$_0$
vacuum initial data have $C^2$-inextendible maximal Cauchy 
developments with blow up of the Kretschmann scalar.
In the dynamical 
system language with Hubble normalized quantities, the
$\alpha$-limit set is then a
single point of type I (i.e.\ a Kasner point), apart from the
possibility of a flat point of type VII$_0$ for solutions independent
of time. Rendall
was able to confirm that the singularities are asymptotically
velocity dominated (\Sectionref{sec:bklconjecture}) and so showed that
the behavior for these solutions is, in this simple sense, consistent
with the  
BKL-conjecture.  However, he was not able to find similarly strong
results in the Bianchi VIII and IX cases. In fact, at that time,
heuristic and numerical work suggested 
that the past behavior of
these latter solutions is more complicated due to ``chaotic'' oscillations
and that they are not asymptotically velocity dominated. Indeed, this
complicated behavior was the basis for the speculations of the
BKL-authors for generic gravitational singularities. It was Misner 
\cite{Misner69} who as one of the first
analyzed this behavior non-rigorously and he introduced the name
\term{Mixmaster universe}. 
In the vacuum case with vanishing $\lambda$,
Ringström was able to confirm the heuristic picture rigorously in
\cite{Ringstrom99}.
He considered the maximal Cauchy  
development of generic Bianchi VIII and Bianchi IX vacuum initial
data and found that these are 
$C^2$-inextendible so that the Kretschmann scalar is unbounded along all
incomplete causal geodesics in every incomplete direction. The
Taub(-NUT) spacetimes, which are non-generic in the class of all
Bianchi VIII and XI spacetimes due to their additional LRS-symmetry,
are the only exceptions and they have 
smooth extensions larger than the maximal Cauchy developments. 
Ringström \cite{Ringstrom00} was able to extend his results to the
non-vacuum case 
(perfect fluid with linear equation of state and \Eqref{eq:linEOS}
such that $2/3<\gamma\le 2$).
Hence,
strong cosmic censorship holds for the class of Bianchi-A 
models with perfect fluids and linear equation of state.
Ringström
also obtained information about the conjectured 
oscillations at the singularity. Generic Bianchi VIII and IX models
must oscillate indefinitely (in
the variables of Wainwright and Hsu \cite{Wainwright89}) as the
singularity is approached, in particular because the $\alpha$-limit sets
contain more than one point. In the Bianchi IX case
he found a further characterization which
was conjectured before.
On the one hand 
matter becomes unimportant in a precise sense near the
singularity except for the case $\gamma=2$ (stiff fluid); in the
latter case
the matter \textit{is} important and the behavior is quiescent
(\Sectionref{sec:bklconjecture}). For $\gamma<2$ the solutions
generically converge 
to the attractor
\[A:=\left\{\left(\Omega,\Sigma_+,\Sigma_-,N_1,N_2,N_3\right)\in M\,\bigl|\,
  \Omega+|N_1N_2|+|N_2N_3|+|N_3N_1|=0\right\}\]
in 
the variables of Wainwright and Hsu \cite{Wainwright89}
where $M$ is the subset of the state space consistent with the
Hamiltonian constraint. The convergence is such that
\[\lim_{\tau\rightarrow -\infty}(\Omega+N_1N_2+N_2N_3+N_3N_1)=0.\]
Hence Ringström has described rigorously the
character of the mixmaster behavior which can be seen as a first step
in the direction to understand the BKL-conjecture. Note that similar
results about a conjectured attractor of similar form are not
available yet for the Bianchi VIII case.

Further results and a summary of the status of Bianchi-B models can be
found in \cite{andersson04a}.

\subsubsection{Gowdy solutions}
\label{sec:gowdyphenomon_real}
In the following discussion we restrict to Gowdy solutions
(\Sectionref{sec:relevantsymmclass}) in vacuum
with vanishing cosmological constant and 
spatial $\T$-topology if not noted otherwise.

In \Sectionref{sec:comm_field_eqs} we have already given a
representation of $\T$-Gowdy spacetimes in terms of the orthonormal
frame \Eqref{eq:GowdyONF}. However, this is not the standard representation
used in the literature.
From the investigations in \cite{Gowdy73,chrusciel1990}, it follows that
coordinates for generic $\T$-Gowdy spacetimes can be chosen such that
the metric takes the form  
\begin{equation}
  \label{eq:GowdyparMetric}
  g=e^{(\tau-\lambda)/2}(-e^{-2\tau}d\tau^{2}+d\theta^2)
  +e^{-\tau}[e^{P}d\sigma^2+2e^{P}Qd\sigma d\delta+
  (e^{P}Q^2+e^{-P})d\delta^2].
\end{equation}
Here, $\tau\in\mathbb{R}$ and $(\theta,\sigma,\delta)$ are the
standard coordinates
on $\T$, the quantities $P$, $Q$ and $\lambda$ (not to be confused with
the cosmological 
constant) are smooth $2\pi$-periodic functions in $\theta$ and the Gowdy
Killing fields correspond to the coordinate vector fields
$\partial_\sigma$, $\partial_\delta$. The underlying gauge is called areal
gauge or timelike area gauge, cf.\
\Sectionref{sec:comm_field_eqs}. 
The relation of this representation of the Gowdy metric and that in
\Sectionref{sec:comm_field_eqs} is written out explicitly in
\cite{Andersson03} for the choice $\mathcal N_0=-1/2$. 

In this representation above, EFE imply the following
evolution equations 
\nobreak
\begin{subequations}
  \label{eq:Gowdy_equations}
  \begin{align}
    P_{\tau\tau}-e^{-2\tau}P_{\theta\theta}-
    e^{2P}(Q_{\tau}^2-e^{-2\tau}Q_{\theta}^2) & = 0 \\
    Q_{\tau\tau}-e^{-2\tau}Q_{\theta\theta}
    +2(P_{\tau}Q_{\tau}-e^{-2\tau}P_{\theta}Q_{\theta}) & = 0,
  \end{align}
  and the constraint equations are
  \begin{align}
    \lambda_{\tau} & = P_{\tau}^{2}+e^{-2\tau}P_{\theta}^{2}+
    e^{2P}(Q_{\tau}^{2}+e^{-2\tau}Q_{\theta}^{2})\\
    \label{eq:Gowdyconstr2}
    \lambda_{\theta} & = 2(P_{\theta}P_{\tau}+e^{2P}Q_{\theta}Q_{\tau}).
  \end{align}
\end{subequations}
One finds that the integrability condition for $\lambda$ is satisfied
if $P$ and $Q$ fulfill the evolution equation and if the
integral of the right hand side of \Eqref{eq:Gowdyconstr2} vanishes. Then
the constraints decouple from the  evolution equations because
$\lambda$ can be computed from the constraints (up to a constant) as
soon as 
$P$ and $Q$ are determined from the evolution equations.
Hence, for the analysis one can restrict the attention to the
two semi-linear coupled wave equations for $P$ and $Q$ with arbitrary
initial data subject only to that integral condition. Recall that in
\Sectionref{sec:comm_field_eqs} we also found that the main evolution
system is 
unconstrained if the cosmological constant vanishes.

In \cite{Moncrief80}, Moncrief proves global existence of the
solutions of these equations in these areal coordinates and showed
that there is a crushing singularity in the limit $\tau\rightarrow\infty$.
Hence the maximal Cauchy development of generic Gowdy data sets 
can be foliated with areal coordinates. However, this does
not exclude the possibility that there are extensions which are not
covered by these 
coordinates. Geodesic completeness 
in the expanding time direction ($\tau\rightarrow-\infty$) was proven
in \cite{ringstrom04a}. The 
main problem is the shrinking time direction
($\tau\rightarrow+\infty$) and the question of extendibility there.
In \cite{Isenberg89,Chrusciel90}, the polarized case
$Q=0$, where both KVFs can be 
chosen mutually orthogonal everywhere, was studied. Indeed their
techniques applied for
all allowed spatial topologies. Strong cosmic censorship,
asymptotic velocity dominance and hence the BKL-conjecture was
confirmed in this class and a characterisation of those spacetimes was
given which admit $C^2$-extensions in the shrinking time direction.

The non-polarized Gowdy case stayed out of reach. Its phenomenology was
first investigated 
numerically \cite{Berger93,Hern97,Berger97} and the results suggested that the
behavior is non-oscillatory, in fact asymptotically velocity dominated almost
everywhere with ``spiky 
features'' at the exceptional points. It turned
out that the idea of asymptotic expansions,
which define the notion of asymptotically velocity dominated
singularities, is crucial. It was applied to the general Gowdy case in
\cite{Kichenassamy97} where analytic solutions of the Gowdy equations
were constructed 
with prescribed asymptotic
expansions of the form
\nobreak
\begin{subequations}
  \label{eq:asymptexpansionsGowdy}
  \begin{align}
    P(\tau,\theta) & = v(\theta)\tau+\phi(\theta)+u(\theta,\tau)\\
    Q(\tau,\theta) & = q(\theta)+e^{-2v(\theta)\tau}[\psi(\theta)+
    w(\tau,\theta)]
  \end{align}
\end{subequations}
with the ``error'' functions $u$, $w$ satisfying 
$\lim_{\tau\rightarrow\infty}u,w=0$. Such
solutions are indeed asymptotically velocity 
dominated and hence obey the BKL-conjecture. In this ansatz, the functions $v$,
$\phi$, $q$ and $\psi$ are 
considered as freely choosable and the Gowdy equations written for
$u$ and $w$ become a Fuchsian 
system. It turned out that this ansatz yields a well-posed initial
value problem with data  
$v$, $\phi$, $q$ and 
$\psi$
on the
singularity if $0<v(\theta)<1$ for all $\theta$, and such $u$ and $v$
always exist uniquely.
However, if 
in particular
$v(\theta)>1$ somewhere, the
corresponding results could only be proven for $q=const$. All these results
were extended to the smooth case in 
\cite{Rendall00}.
Although
this was a first step in the direction to show that Gowdy
spacetimes are asymptotically velocity dominated, the question remained why
generic Gowdy solutions should  have asymptotic expansions of the form
above. Further, the rigorous description of the spiky features
conjectured to ``happen'' when $v(\theta_0)>1$ at isolated points
remained unclear. 

To clarify these issues, let us introduce the quantity  
\begin{equation}
  \label{eq:hyperbolickinenergy}
  \kappa(P,Q):=P_\tau^2+e^{2P}Q_\tau^2.
\end{equation}
It can be interpreted as the ``geometric
kinetic energy'' of the solution, and its square root is often referred
to as \textbf{hyperbolic velocity}. It
is a geometric quantity because the Gowdy 
evolution equations are ``almost'' wave map equations\footnote{See for
instance the comments in \cite{Rendall01}.} from the
two-dimensional Minkowski spacetime to the $2$-dim.\ hyperbolic space
$\mathbb H^2$ with metric  
$dP\otimes dP+e^{2P} dQ\otimes dQ$, so that the solution $(P,Q)$ can
be considered as a curve in $\mathbb H^2$. In terms of the $\beta$-rescaled
quantities of \Sectionref{sec:comm_field_eqs}, the hyperbolic velocity
takes the following form \cite{Andersson03}
\begin{equation}
  \label{eq:hypvel}
  v=\sqrt 3\sqrt{\Sigma_-^2+\Sigma_\times^2}.
\end{equation}
Indeed, it turns out that isometries of the 
hyperbolic plane map solutions of the Gowdy equations to isometric solutions.
Now, if asymptotic expansions of the form \Eqsref{eq:asymptexpansionsGowdy}
are valid and if naive differentiation is 
allowed then $\lim_{\tau\rightarrow\infty}\kappa=v^2$. Hence $v$ has a geometric
meaning, namely it represents the asymptotic velocity of the solution in the
target space. Ringström \cite{Ringstrom06} was able to prove that
$\lim_{\tau\rightarrow\infty}\kappa(\tau,\theta)$ 
always exists irrespectively of the question if expansions of the form
above are valid. So it makes sense to define 
\[v_\infty(\theta):=\sqrt{\lim_{\tau\rightarrow\infty}\kappa(\tau,\theta)},\]
called \term{asymptotic velocity}. In particular, he showed (Proposition 1.3 in
\cite{Ringstrom06}) that 
\[\lim_{\tau\rightarrow\infty}P_\tau(\tau,\theta)=\pm
  v_\infty(\theta);\] 
the sign depends on the
parametrization of the metric and can be changed by an inversion in the
hyperbolic plane. One finds that if for a given
$\theta_0\in[0,2\pi[$ we have
$v_\infty(\theta_0)\not=1$, then the Kretschmann scalar blows up along
all causal curves ``ending at the singularity at
$\theta_0$''. However, for $v_\infty(\theta_0)=1$ such general
conclusion cannot be drawn;
indeed explicit examples exist with 
bounded curvature in this case. Now, if an asymptotic
expansion of the form \Eqsref{eq:asymptexpansionsGowdy} is valid then
$v_\infty(\theta)=\pm v(\theta)$ with, in general,
pointwise dependent sign. Another main result of
Ringström \cite{Ringstrom06} is that if for a given Gowdy solution we
have $0<v_\infty(\theta_0)<1$ for
one $\theta_0$ and if
$P_\tau(\tau,\theta_0)\rightarrow +v_\infty(\theta_0)$, then the
solution has a smooth 
asymptotic expansion of the form above
in a space 
neighborhood of $\theta_0$. Namely roughly speaking, all
$C^k$-norms of the error function $u$ and $w$ 
converge exponentially fast to zero as $\tau\rightarrow\infty$ in that
neighborhood. This 
means that under this 
condition, knowledge of the behavior of the solution at a given point
on the singularity implies the behavior in a whole
neighborhood, namely yields locally an 
asymptotic expansion of the form above.

However, the numerical studies suggested that such asymptotic
expansions are in general not valid globally on the singularity. We have
to discuss the notion of {spikes}, first 
rigorously defined in \cite{Rendall01} and applied in
\cite{Ringstrom06}, although we cannot go into the 
technical details. The main point is that for making statements about
generic Gowdy spacetimes it is sufficient to consider
two distinct  
classes of spikes, namely {non-degenerate false
  spikes} and {non-degenerate true spikes}. The first ones are
localized ``problems'' in the parametrization of the metric without
geometric meaning while the latter is a localized change in
the geometric behavior of the solution at the singularity. The first
type of spikes can be characterized like this. Let $v_\infty$ be a smooth
function in a neighborhood of a $\theta_0\in[0,2\pi[$ with
$0<v_\infty(\theta_0)<1$.
According to the expansions
\Eqsref{eq:asymptexpansionsGowdy} and the results above, it can happen that
$P$ converges to leading order as $-v_\infty(\theta_0)\tau$ at
$\theta_0$ and as $+v_\infty(\theta)\tau$ in a punctured
neighborhood which leads to a downward pointing ``spiky feature'' in
$P$. It 
is a \term{non-degenerate false spike} if $Q$ has a first 
order pole at $\theta_0$. One can show that an inversion
in the hyperbolic 
plane yields that $P$ and $Q$ are smooth in a neighborhood of
$\theta_0$ (including $\theta_0$) and thus that the solution has
a smooth asymptotic expansion as above and is
asymptotically velocity dominated there. This is the reason for the
name \textit{false} spike.  In contrast to that, \term{non-degenerate
  true spikes} are 
characterized as follows. They occur if at a $\theta_0$, one has
$1<v_\infty(\theta_0)<2$ while $0<v_\infty(\theta)<1$ in a punctured
neighborhood and $P_\tau\rightarrow+v_\infty$ on the whole
neighborhood. This implies that $P$ cannot be ``continuous at  
$\theta_0$'' and so it shows an upward pointing ``spiky feature'' at
$\theta_0$. However, $Q$ can be smooth and if $Q_\theta$ has a
simple zero at $\theta_0$, a true spike occurs.  The discontinuous
behavior in $v_\infty$ is 
geometric and not a consequence of a ``bad'' parametrization. Indeed,
one can show  
that the Kretschmann 
scalar blows up along causal curves ``ending at $\theta_0$'' with a
faster rate dependent on the value of $v_\infty(\theta_0)$ than in a
punctured neighborhood.  

It is a deep result of Ringström
that generic vacuum $\T$-Gowdy spacetimes do not have any further
pathologies. Ringström 
\cite{Ringstrom06} was able to show first that the set of initial data
sets on a Cauchy surface
corresponding to solutions
with $l\in\N$ false and $m\in\N$ true spikes is open in the $C^2\times
C^1$-topology on initial data. Second he showed that the union of 
all these sets is dense in the 
$C^\infty$-topology on initial data \cite{Ringstrom06b}. This is the
precise formulation of ``genericity''; namely generic Gowdy solutions
develop finitely many false and true spikes and are asymptotically
velocity dominated ``in between''. Spikes cannot
accumulate somewhere and so the BKL-conjecture is confirmed. Regarding
the SCC conjecture, note that generic 
Gowdy solutions have 
$v_\infty\not=1$ everywhere. Thus, as was already stated above, the
Kretschmann scalar blows up along all causal 
geodesics in the incomplete direction and the solutions are hence
$C^2$-inextendible. Since the solution is 
geodesically complete in the expanding direction,
strong cosmic censorship is
confirmed within this class of spacetimes.

The evolution of spikes is explained in a non-rigorous manner by the
\term{method of consistent potentials} \cite{Berger97}. Further, the
authors of \cite{Garfinkle03} also discuss the evolution of \term{high
  velocity spikes} 
which can occur when the initial hyperbolic velocity is bigger than
two. Some details on those are 
given in \Sectionref{sec:commNumR}.

After this discussion of the $\T$-Gowdy case, let us make a
few remarks about other $\U\times\U$-symmetric solutions with spatial
$\T$-topology.  
For $\U\times\U$-symmetric spacetimes with non-vanishing twist
constants, there is not such a deep understanding, and we refer the
reader to discussions in
\cite{isenberg99,Berger01,Isenberg03} and references therein. In
particular, 
oscillatory, and not asymptotically velocity dominated singularities
of general BKL-type are believed to be generic. 

Regarding the Gowdy case for the other spatial topologies the only
complete result is due to Isenberg et al.\ in \cite{Isenberg89}
restricted to the polarized case. Asymptotic velocity dominance and
SCC are confirmed. St{\aa}hl \cite{Stahl02} made a
Fuchsian analysis analogous to \cite{Kichenassamy97}
in the $\S$- and $\Stwo\times\mathbb S^1$-cases. 
One
can show that for spatial $\S$- and 
$\Stwo\times\mathbb S^1$-topologies, the function $v$ must be $-1$ or $3$ at
that points where the group orbits become $1$-dimensional. But then,
as in the $\T$-case above, an ansatz for the asymptotic form of the
solutions as in \Eqsref{eq:asymptexpansionsGowdy} does not allow to
control the full set of free 
functions. Hence St{\aa}hl's results have to be considered as incomplete.
Numerical investigations of
the $\Stwo\times\mathbb S^1$-case can be found in \cite{garfinkle1999} and
similar behavior as in the $\T$-case is observed.
The $\S$-case is outstanding even numerically.

The only analytical result for the case of Gowdy spacetimes with
non-vanishing cosmological constant is in \cite{Clausen07}. It is a
result about global foliations with areal coordinates and Fuchsian
analysis. The case of non-vanishing twist constants is
included. However, there are no published numerical investigations of
the outstanding issues yet. See
\Sectionref{sec:ExpectT3GowdyLambda} for further discussions.

\subsubsection{Other cases}
There was also some work on solutions with a spatial
$\U$-symmetry, and we just point here to some important references.
The newest (to my knowledge) analytical results are in
\cite{choquet04,Choquet-Bruhat05} 
where one can also find a summary of the current status and
the relevant references; 
cf.\ also \cite{andersson04a}. Numerical investigations were performed
in \cite{Berger98a,Berger98b}; see also some discussion in
\cite{berger98}.
For spacetimes without any symmetry assumptions there are a number of
interesting results in special settings. One of them is the theorem
about non-linear 
stability of the de-Sitter spacetime that we discuss in
\Sectionref{sec:non-linear stability of de-Sitter}. This restricts to
the case of a positive
cosmological constant. Similar theorems for scalar fields
were recently proved by Ringström \cite{Ringstrom06c}. 
Strong cosmic
censorship and the BKL-conjecture for spacetimes with a scalar field,
which resembles a stiff fluid with its quiescent behavior, have
been studied \cite{Andersson00}. Numerical investigations on those
were carried out in \cite{Curtis05} and in the vacuum case in
\cite{garfinkle04a}.

\section{Future asymptotically de-Sitter spacetimes}
\label{sec:fads}

\subsection{Basic definitions and properties}
\label{sec:fads_def}
We will now define the class of spacetimes which will be considered in
this thesis. For that we have to define the notion of conformal
compactifications and 
conformal boundaries. Our terminology closely follows
\cite{galloway2002}. We shall make the same assumptions on the
considered manifolds as before.

\renewcommand{\labelenumi}{(\roman{enumi})}
\begin{Def}
  \label{def:conform_compact}
  A Lorentz manifold $(\tilde M,\tilde g)$ is said to have a smooth
  \term{conformal compactification} (or smooth \term{conformal
    completion}) if there 
  exists a smooth oriented time oriented causal Lorentz
  manifold-with-boundary $(M,g)$ with 
  boundary $\scri:=\partial M$ and a
  smooth function $\Omega:M\rightarrow\R$ such that
  \begin{enumerate}
  \item there is a diffeomorphism 
    $\Phi:\tilde M\rightarrow M\backslash\scri$ such that
    $\tilde g=\Phi^*\left(\Omega^{-2} g\bigr|_{M\backslash\scri}\right)$,
  \item we have $\Omega>0$ in the interior of $M$, and, $\Omega=0$ and
      $d\Omega\not=0$ on $\scri$. 
  \end{enumerate}
\end{Def}
One should note that the terminology is partly misleading because the
manifold $M$ needs to be neither compact nor complete as we see below.
For brevity, we will identify the manifold $\tilde M$ with the
interior of $M$ by means of $\Phi$ and call $\scri$ 
equivalently its \term{conformal
  boundary} or \term{conformal infinity}.  The metric $g$ on $M$
determines a unique conformal structure $\mathcal C_g$ on $M$ in the
sense of \Sectionref{sec:conf_weyl}. The metric 
$\tilde g$ is in $\mathcal C_g$, however, in contrast to $g$, it does
not extent in a regular manner to $\partial M$. If there is no risk of
confusion we will leave the index of 
$\mathcal C_g$ away and simply write $\mathcal C$. Further, 
when we write $g$ we will always mean an arbitrary
global smooth representative of the conformal structure $\mathcal C$ so that
$\Omega$ always means the 
\term{conformal factor} which relates $g$ to $\tilde g$ on $\tilde M$ by
$g=\Omega^2\tilde g$ according to the previous definition. We will
refer to $(M,g,\Omega)$ or equivalently to $(M,\mathcal C)$ as the
\term{conformal spacetime}. 

In \Sectionref{sec:cfe} we have already mentioned Penrose's original
motivation to study conformal
compactifications of solutions of Einstein's field
equations. Penrose was particularly interested in solutions of EFE
with vanishing cosmological constant to 
describe gravitational radiation.
\Eqref{eq:regular_conformal_field_equations_lambda} shows that
conformal boundaries must be null in this case.
In this thesis, we will assume $\lambda>0$, and then the same equation
implies that 
conformal 
boundaries are spacelike.
If $\scri$ is spacelike, we require additionally that it is the
disjoint union of two sets $\scrip$ 
and $\scrim$
given by
\[\scripm:=I_\pm(\tilde M,M)\cap\scri.\]
We call $\scrip$  the \term{future
  conformal boundary}; analogously the past conformal boundary. Either
of these two components can be empty.
Now, a spacetime $(\tilde M,\tilde g)$, not necessarily a
solution of Einstein's field equations, which has a smooth conformal
completion with spacelike $\scri$ 
with disjoint components $\scrip$ and $\scrim$
is said to be of \term{de-Sitter
  type}. If $\scrip$ is non-empty, then it is called \term{future
  asymptotically de-Sitter} (FAdS); analogously for the past case.
If both components are non-empty the spacetime is
referred to as \term{asymptotically de-Sitter}. Some authors prefer
the term (future, past) asymptotically \textit{locally} de-Sitter; these
names will be motivated below.

\begin{Def}
Under the same conditions as above, let $(\tilde M,\tilde g)$ be
future asymptotically de-Sitter. Then, $(\tilde M,\tilde g)$ is called
\term{future asymptotically simple} if all future
inextendible null curves have a 
future endpoint on $\scrip$. Analogously for the past case.
\end{Def}
To understand how ``global'' the assumption of asymptotic simplicity
is, consider the following fundamental facts proven in \cite{galloway2002}.
\begin{Prop}
\label{prop:structure}
  Let $(\tilde M,\tilde g)$ be future asymptotically de-Sitter with
  smooth future conformal boundary $\scrip$.
  \begin{enumerate}
  \item If $(\tilde M,\tilde g)$ is globally hyperbolic and $\scrip$ is compact
    then $(\tilde M,\tilde g)$ is future asymptotically simple.
  \item If $(\tilde M,\tilde g)$ is future asymptotically simple then
    it is globally hyperbolic.
  \end{enumerate}
  In both cases, the Cauchy surfaces of $(\tilde M,\tilde g)$ are
  homeomorphic to $\scrip$.
\end{Prop}
Examples will be discussed in the following section.
The idea for the proof is to extend the
manifold-with-boundary $M$  to a
slightly larger manifold without boundary which is globally hyperbolic
if and only if $\tilde M$ is globally hyperbolic. In
the first case one can argue that $\scrip$ is a Cauchy surface of this
extension, hence all null curves must hit it and both future asymptotic
simplicity and the existence of the homeomorphism between any Cauchy
surface of $\tilde M$ and $\scrip$ follows. For the second point, one
considers the Cauchy horizon of $\scrip$ in the extended
spacetime. If it were non-empty, its null generators were not allowed
to hit $\scrip$ which is a contradiction to asymptotic simplicity.

In particular, we point out that if we restrict to globally hyperbolic
spacetimes which are 
FAdS with compact $\scrip$, future asymptotic simplicity is automatically
implied and need not to be required additionally. However, situations
where one has to be careful in particular concerning the past
direction will be discussed later. Furthermore note 
that we have not made any further assumptions on the topology of
$\scrip$ so far; for the definitions in this sections there is even no need
to require compactness. More details on our assumptions will given later.

Using a weaker notion of conformal boundary completions, Chru\'{s}ciel
\cite{Chrusciel06} proves the following fundamental result. 
\begin{Theorem}
Every spacetime $(\tilde M, \tilde g)$, not necessarily of solution of
EFE, admits a 
unique, up to equivalence, future conformal boundary completion which
is maximal within the class of all completions with spacelike boundaries.
\end{Theorem}
For the rigorous definitions of the notions used in this theorem, the
reader is referred to Chru\'{s}ciel's article. In his considerations,
he does not 
require that the conformal factor vanishes on the conformal boundary.
However, if we
assume EFE with $\lambda>0$ to hold, then conformal boundaries (in our
stronger sense) must be spacelike and then this theorem implies
uniqueness of the maximal conformal boundary which in particular can be empty.

When we study the conformal properties of spacetimes with symmetries
we need the following trivial further result.
\begin{Lem}
  \label{lem:symmetryconformal}
  Let $(\tilde M,\tilde g)$ be a Lorentz manifold with 
  smooth conformal compactification $(M,\mathcal C)$  with a
  smooth vector field 
  $\xi$ on $M$. Let $g\in\mathcal C$ be a smooth global conformal
  metric and $\Omega$ the corresponding smooth conformal factor such that
  $g=\Omega^2\tilde g$ on $\tilde M$ and $\xi(\Omega)=0$ on $M$. Then,
  $\xi$ is a 
  $\tilde g$-Killing vector field, i.e.\ 
  $\left.\lieder{\xi}{\tilde g}\right|_{\tilde M}=0$, if and only if
  $\xi$ is a $g$-Killing vector field, i.e.\
  $\left.\lieder{\xi}{g}\right|_{M}=0$.
  \begin{Proof}
    We have
    \[\left.\lieder{\xi}{\tilde g}\right|_{\tilde M}=
    \left.\lieder{\xi}{\Omega^{-2} g}\right|_{\tilde M}=
    \left.-2\frac{\xi(\Omega)}{\Omega^3}g+
      \Omega^{-2}\lieder{\xi}{g}\right|_{\tilde M}.\]
    Hence for $\xi(\Omega)=0$, we have $\left.\lieder{\xi}{\tilde
        g}\right|_{\tilde M}=0$ 
    $\Leftrightarrow$ 
    $\left.\lieder{\xi}{g}\right|_{\tilde M}=0$. 
    Additionally, continuity implies that
    $\left.\lieder{\xi}{g}\right|_{\tilde M}=0$ $\Leftrightarrow$
    $\left.\lieder{\xi}{g}\right|_{M}=0$. This proves the claim.
  \end{Proof}
\end{Lem}

\subsection{Important examples}
\subsubsection{De-Sitter spacetime}
\label{sec:deSitter}
\begin{figure}[tb]
  \centering
  \subfloat[Physical representation, \Eqref{eq:dS_standard}]{%
    \label{fig:dS_standard}
    \includegraphics[width=0.49\linewidth]{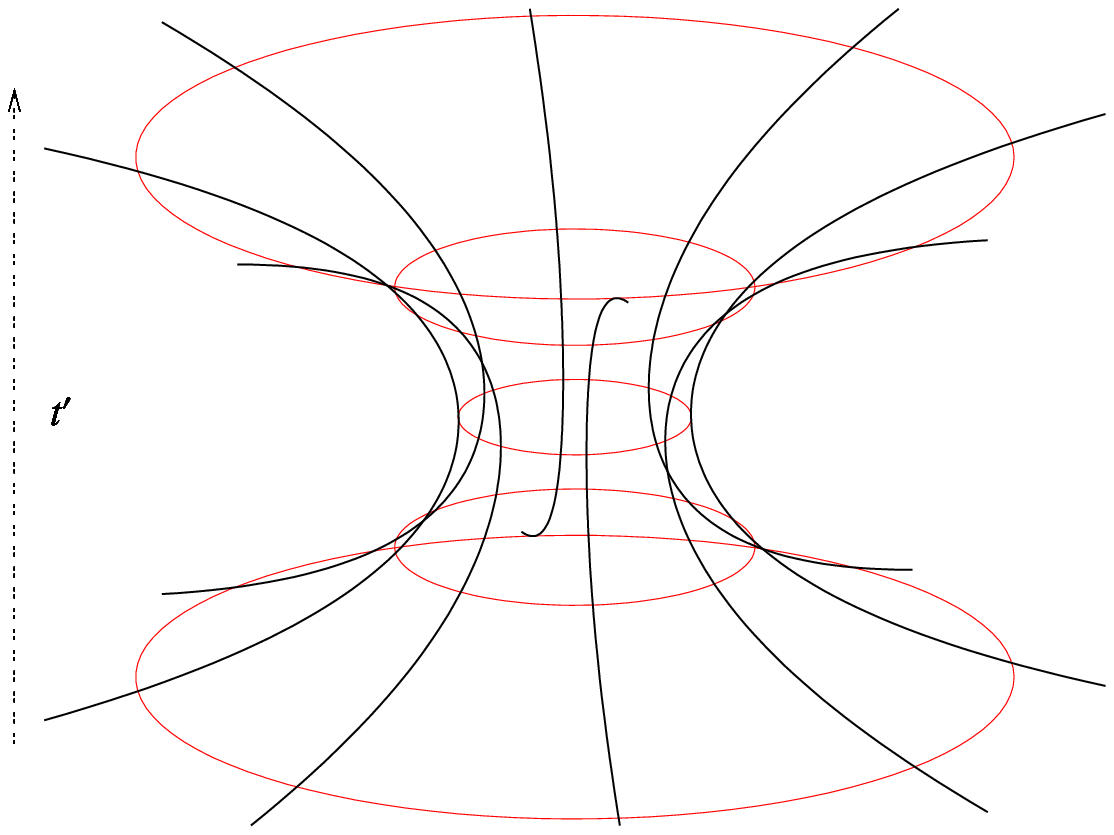}}%
  \subfloat[Conformal representation, \Eqref{eq:dS_conformal}]{%
    \label{fig:dS_conformal}
    \includegraphics[width=0.49\linewidth]{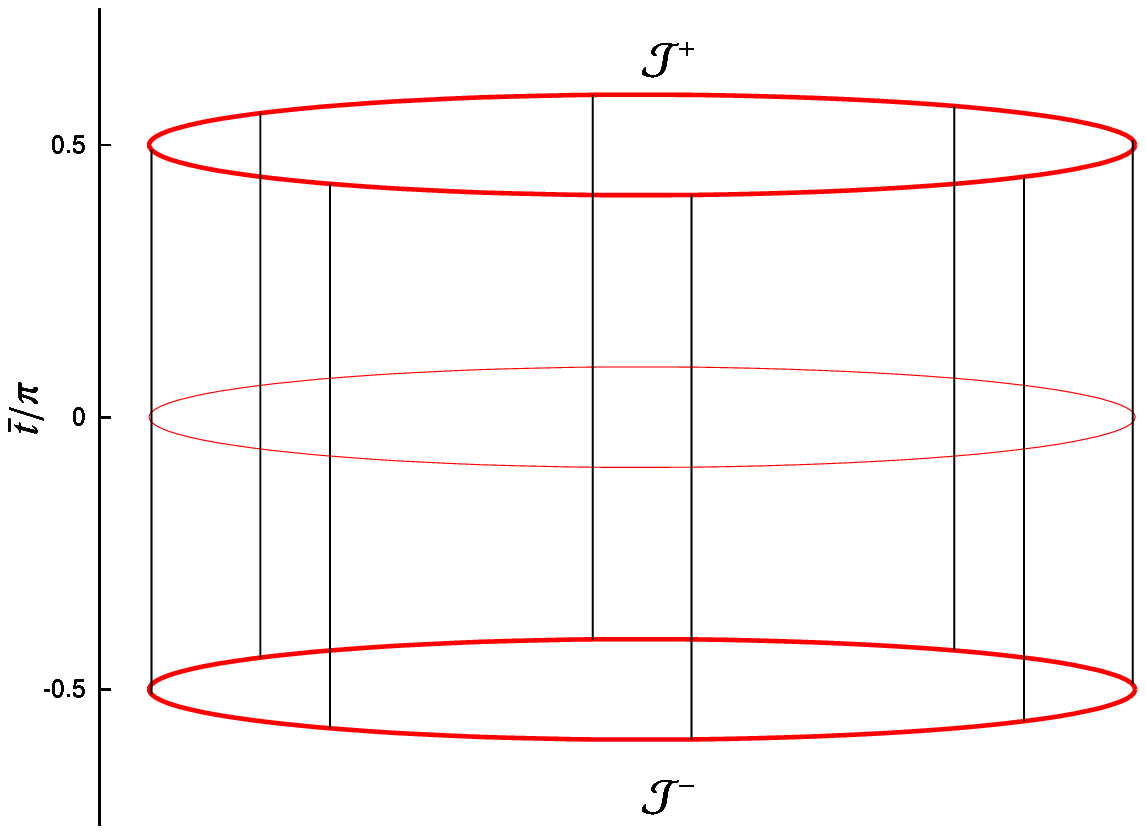}}
  \caption{De-Sitter (dS) spacetime}
\end{figure}
Let the line element on the standard round unit $3$-sphere be
denoted by $d\omega^2$.
After a 
suitable conformal rescaling and choice of coordinates, the line
element of the \term{de-Sitter spacetime} (dS) \cite{hawking} can be
written as  
\begin{equation}
  \label{eq:dS_standard}
  d\tilde s^2=-{dt'}^2+\cosh^2t' d\omega^2
\end{equation}
with $t'\in ]-\infty,\infty[$.
Indeed, this spacetime is a solution of
Einstein's field equations in vacuum with $\lambda=3$.
\Figref{fig:dS_standard} shows this
representation of dS
for a finite time interval; each point represents a
two-sphere in the sense of the Hopf fibration
(\Sectionref{sec:Hopf_fibration}).  

With the coordinate transformation
\[t'=2\text{Arctanh} (t-1)\]
for $t\in]0,2[$, the line element takes the form
\begin{equation}
  \label{eq:dSLCCGG}
  d\tilde s^2=\frac 4{t^2(2-t)^2}
  \left(-dt^2+\frac{\left((t-1)^2+1\right)^2}{4}d\omega^2\right).
\end{equation}
Indeed this is the representation of the de-Sitter spacetime obtained
in the special gauge discussed in
\Sectionref{sec:LCCGG} and will be of relevance for our
investigations. Note that in this form, one can explicitly check that
the conditions for \Defref{def:conform_compact} are satisfied. The
conformal metric is
$g=-dt^2+\frac{\left((t-1)^2+1\right)^2}{4}d\omega^2$, the
corresponding conformal factor is $\Omega=\frac 12{t(2-t)}$ and the
manifold $M$ is $[0,2]\times\S$. So, $t=0$ corresponds to $\scrim$ and
$t=2$ to $\scrip$ (or vice versa in a time-reversed manner). Indeed,
the de-Sitter spacetime is asymptotically de-Sitter both to the future
and to the 
past, and since it is globally 
hyperbolic with compact $\scripm$ it is also asymptotically simple.

The standard ``conformal representation'' of the de-Sitter spacetime
is obtained by the coordinate transformation 
\[\bar t=2\text{Arctan}(t-1)\]
which leads to the form
\begin{equation}
  \label{eq:dS_conformal}
  d\tilde s^2=\frac 1{\cos^2\bar t}
  \left(-{d\bar t}^2+d\omega^2\right)
\end{equation}
with $\bar t\in]-\frac\pi 2,\frac\pi 2[$. This can be interpreted as a
conformal embedding of the de-Sitter spacetime into the Einstein
cylinder,
which is the Lorentz manifold $\R\times\S$ with the standard product
metric; cf.\ \Figref{fig:dS_conformal}.

According to \cite{hawking}, part of the de-Sitter spacetime can also
be foliated by slices of $\R^3$-topology with conformally flat
induced metrics. This foliation covers ``half'' of the finite de-Sitter
cylinder \Figref{fig:dS_conformal} including one conformal
boundary. In cosmology, in particular for questions related to cosmic no-hair,
one usually considers such a foliation to describe a neighborhood of
$\scrip$.  However, one can 
formulate cosmic no-hair equally well with respect to
spherical slices with the difference that the $3$-Ricci tensor
does not vanish asymptotically. In any case, we will show in
\Sectionref{sec:FAdScosmicnohair} that all FAdS ``almost'' obey the cosmic
no-hair picture in a precise sense.

\subsubsection{\texorpdfstring{$\lambda$}{lambda}-Taub-NUT spacetimes}
\label{sec:TaubNUT}
In this section, let us discuss briefly the $\lambda$-Taub-NUT
spacetimes due to Brill et al.\ \cite{Brill1978}, which is a family of
solutions of EFE in vacuum with $\lambda>0$. The corresponding family 
for $\lambda=0$, the Taub-NUT solutions due to Taub and Newman et al.\
\cite{Taub51,NUT63,hawking}, is maybe more familiar to relativists, see
also discussions in \cite{Wainwright,chrusciel93}. However, most of the
well-known phenomena are similar in both families and we
will not say more about the Taub-NUT spacetimes.

Here, we summarize the assumptions to derive the family of
$\lambda$-Taub-NUT spacetimes from
\cite{Brill1978}.  
Consider the standard left invariant coframe
$\{\omega^a\}$ on $\S$ dual to the frame $\{Y_a\}$
(\Sectionref{sec:identification_su2s3}) 
such that $\{Y_a\}$ satisfies \Eqsref{eq:commutator_rel_Y}.
Let us look for solutions of EFE in vacuum with
$\lambda>0$ with Cauchy surfaces of $\S$-topology, and let us make the
following ansatz for an orthonormal coframe of the form
\[\sigma^0=\frac {B_0^2}{A(\tau)}d\tau,\quad
\sigma^{1,2}=B(\tau)\omega^{1,2},\quad
\sigma^{3}=A(\tau)\omega^3,\]
i.e.\ the metric is $g=-\sigma^0\otimes\sigma^0+\sum_{a=1}^3
\sigma^a\otimes\sigma^a$. Here, $\tau\in\R$ is a time coordinate, $A(\tau)$
and $B(\tau)$ are so far undetermined functions and $B_0:=B(0)$.
Such a spacetime is of LRS-Bianchi IX type,
see our related discussion in \Sectionref{sec:bergersphere}. With this
ansatz for the metric, Brill et al.\ solved Einstein's field equations
with $\lambda>0$
and obtained (modulo time translations) 
\nobreak
\begin{subequations}
  \label{eq:TaubNUT}
  \begin{align}
    B(\tau)&=B_0\sqrt{1+\tau^2},\quad
    A(\tau)=B_0^2\frac{\tilde A(\tau)}{B(\tau)},\\
    \tilde A(\tau)&=\sqrt{\frac{\lambda B_0^2}3\tau^4+2(\lambda
      B_0^2-2)\tau^2+C_0\tau+4-B_0^2\lambda},
  \end{align}
\end{subequations}
where $B_0$ and $C_0$ are freely choosable constants.
In
particular, the solution given by $B_0=1$ and $C_0=0$ represents the
de-Sitter spacetime.

The analysis of these solutions can only be sketched here, although the
phenomenology is very interesting for our applications; see also the
discussions in Sections~\ref{sec:SCC}, \ref{sec:dSSR} and
\ref{sec:situation_FADS}. The properties of the solutions are
determined by the root structure
of $\tilde A^2$ which depends on the choice of the parameters $B_0$
and $C_0$. If it has no real roots then the solution is future
and past asymptotically de-Sitter, asymptotically
simple and globally hyperbolic. The limits
$\tau\rightarrow\pm\infty$ correspond to $\scripm$.
One can convince oneself that the geometry of
$\scripm$, in a natural  
conformal gauge, is a Berger sphere
(\Sectionref{sec:bergersphere}); in
\Sectionref{sec:solelectrconstraintscri} we mention how the parameters
$B_0$ and $C_0$ are related to the parameters of the Berger sphere and
the other data.

Now, let $\tilde A^2$ have real zeros. One can interpret such a solution as
consisting of individual pieces separated by the zeros of 
$\tilde A^2$. The geometrical properties of the pieces and their
``boundaries'' depend on the types and positions of the zeros. 
The limit $\tau\rightarrow-\infty$ corresponds, independent on the
root structure of $\tilde A^2$,  to a smooth
$\scrim$ with Berger sphere geometry. The corresponding piece of the
solution in the future is always past asymptotically simple and
globally hyperbolic. The same can be said about the future
limit. If $\tilde A^2$ has a first order zero in the future of
$\scrim$, then the 
``boundary'' surface corresponds to a Cauchy horizon. It is possible
to extend the solution through this surface in non-equivalent ways
such that the extended part has to be identified with the ``next'' piece
of the solution in the future. The extended spacetime is not globally
hyperbolic and possesses closed causal curves. However, if the zero is
of second order\footnote{Comment: it turns out that the
  claims of Anderson \cite{anderson05,Anderson04} about the second
  order case are false. In fact,
  also here, the ``boundary'' is a Cauchy horizon as in the
  previous case, but there are no closed causal curves.}, 
then the piece in the past is
future geodesically complete, i.e.\ the ``boundary'' lies in the
infinite future but does not correspond to a future conformal
boundary. Such a piece is ``infinitely tall'' in the terminology of
\Sectionref{sec:dSSR}. The function $\tilde A^2$ can have at most $4$
real first order zeros; it is straight forward to analyze which
combinations of zeros are possible and hence
characterize all individual pieces and their ``boundaries''. In all
these cases, the curvature is bounded.  

\subsubsection{Schwarzschild de-Sitter spacetime}
\label{sec:SSdS}
Here, we only note that
the family of Schwarzschild de-Sitter spacetimes is a family of
cosmological black and white
hole vacuum solutions with $\lambda>0$, whose Cauchy surfaces have topology
$\Stwo\times\mathbb S^1$, which are both future and past
asymptotically de-Sitter with $\scripm$ of  topology $\R\times\Stwo$
(i.e.\ not compact) and which are not asymptotically simple (neither
to the future nor to the past).
Everything else of interest for this thesis is summarized in
\cite{galloway2004}.

\subsection{Accelerated expansion and cosmic no-hair}
\label{sec:FAdScosmicnohair}
Let
$(\tilde M,\tilde g)$ be a FAdS spacetime and $(M,g,\Omega)$ a
corresponding conformal representation with $\Omega=0$,
$d\Omega\not=0$ on $\scrip$ and $g=\Omega^2\tilde g$. A neighborhood
of $\scrip$ can be foliated by spacelike level sets of $\Omega$. Let $n$ be the
smooth future directed normal to the $\Omega=const$-surfaces with $g(n,n)=-1$
defined at least in a neighborhood of $\scrip$; set 
$\tilde n=\Omega n$ which is a smooth vector field in a
neighborhood of \scrip with $\tilde g(\tilde n,\tilde n)=-1$ in
$\tilde M$.
The $2$nd fundamental form $\tilde\chi_{\mu\nu}$
with respect to $\tilde g$ and $\tilde n$, and the $2$nd fundamental
form $\chi_{\mu\nu}$ 
with respect to $g$ and $n$ are related by the formula
\begin{equation}
  \label{eq:rescal2ndFF}
  \tilde\chi_{\mu\nu}=\Omega^{-1}\chi_{\mu\nu}-\Omega^{-2}n(\Omega)h_{\mu\nu}
\end{equation}
where $h_{\mu\nu}=g_{\mu\nu}+n_\mu n_\nu$ is the induced metric from
$g_{\mu\nu}$ on
the orthogonal complement of $n$, i.e.\ on the
$\Omega=const$-hypersurfaces; similarly define
$\tilde h_{\mu\nu}$ with respect to $\tilde n$ and $\tilde g_{\mu\nu}$. 
Making use of the split \Eqsref{eq:split2ndFF} with respect to the
relevant metrics we obtain that
\[\tilde H=\Omega H-n(\Omega),\quad 
\tilde\sigma\indices{^\mu_\nu}=\Omega\sigma\indices{^\mu_\nu};\]
note that the twist tensor vanishes in both cases because $n$ and
$\tilde n$ are surface orthogonal.
In \Sectionref{sec:initial_data} we will discuss that Einstein's field
equations in vacuum with $\lambda>0$ imply that the
value\footnote{Note that the sign difference to the expression in
  \Sectionref{sec:initial_data} is caused by the assumption
  that $n$ is \textit{future} pointing here.} of
$n(\Omega)$ on $\scrip$ is
$-\sqrt{\lambda/3}$, independent of the conformal gauge. Thus, for
general future  
asymptotically de-Sitter solutions of EFE we find that at $\scrip$
\[\tilde H=\sqrt{\frac\lambda 3}+O(\Omega),\quad 
\tilde\sigma\indices{^\mu_\nu}=O(\Omega).\]
Hence the physical Hubble scalar has the same value as
that of the de-Sitter 
spacetime asymptotically, as one can easily check for $\lambda=3$ with
formula \Eqref{eq:dS_standard}. Further,
the shear tensor, in particular its
eigenvalues, vanish. However, the $\Omega=const$-slices need not to
become homogeneous and isotropic. Since 
$h_{\mu\nu}=\Omega^2\tilde h_{\mu\nu}$ and $\Omega=const$ we find that
the physical Ricci 
tensor of the slices satisfy
\[\tilde r_{\mu\nu}=\text{${}^3$Ricci}[\tilde h]
=\text{${}^3$Ricci}[h]=r_{\mu\nu}\rightarrow
\left.r_{\mu\nu}\right|_{\scrip}.\] 
So the slices need neither become homogeneous, isotropic nor flat,
which is nevertheless
required for the cosmic 
no-hair picture (\Sectionref{sec:cosmicnohair}). Indeed, as we will see in
\Sectionref{sec:initial_data}, the geometry of \scrip, i.e.\ the
$\Omega=0$-surface in our slicing, can be
prescribed almost freely and hence we can produce solutions for which
these 
slices converge to arbitrary geometries; of course in particular also
to the flat geometry (if the topology allows it). 
In any case, the fact that FAdS solutions of EFE obey the cosmic no-hair
picture possibly except for the spatial curvature aspect, gives us the
motivation to state that cosmic no-hair is ``almost'' obeyed. 
Note that although
our slicing based on the level sets of the 
conformal factor does not fulfill all the asymptotic requirements for
cosmic no-hair, there might be
other slices which do. However, the level sets of the conformal factor
form a natural foliation of a neighborhood of \scrip of
FAdS solutions, in particular because all arguments and limits above do not
depend on the particular choice of $\Omega$ and hence the conformal
gauge freedom is respected.

\subsection{Singularity theorems}
\label{sec:singularitytheoremGA}
We have shown in \Sectionref{sec:FAdScosmicnohair} that solutions
of EFE with positive cosmological constant
of de-Sitter type expand exponentially 
when a conformal boundary is approached. The
behavior close to a conformal boundary is well controlled and, additionally, 
such spacetimes are in agreement with the current status of
the observations. However, our knowledge about their behavior away
from the conformal 
boundaries is quite limited. A first step to shed light
on this are singularity theorems worked out in \cite{galloway2002},
some of them which we want to present here. In fact, these theorems
are more general than those versions here since general
matter fields under suitable energy conditions and arbitrary spacetime
dimensions are allowed. We restrict to the $3+1$-case and vacuum (but
$\lambda>0$)  here.

In the notation of \Sectionref{sec:FAdScosmicnohair}, a classical
theorem by Hawking \cite{hawking,oneil,LorentzGeometry} states the
following. 
\begin{Prop}
  Let $(\tilde M,\tilde g)$ be a spacetime satisfying the condition
  $\tilde R_{\mu\nu}V^{\mu}V^{\nu} \ge 0$
  for all unit timelike vector fields $V$.  Suppose that $\tilde M$
  has a smooth 
  compact spacelike Cauchy surface $\Sigma$ with
  Hubble scalar $\tilde H$ satisfying $\tilde H>\beta$ on $N$.  Then, every
  past directed
  timelike curve in $\tilde M$ starting in $\Sigma$ has length $\le 1/\beta$.
\end{Prop}
The proof of this theorem relies in particular on
\Theoremref{th:existence_of_maximal_geodesic}.
Assuming EFE with general matter to hold, the condition on the Ricci
tensor is equivalent to the strong energy condition for the
matter. Thus, a positive cosmological constant considered as part of
the 
matter fields is excluded. To study the case with $\lambda>0$, in particular
spacetimes of 
de-Sitter type, Andersson and Galloway \cite{galloway2002} found the following
modification of Hawking's theorem.
\begin{Prop}
  \label{prop:generalizedHawking}
  Let $(\tilde M,\tilde g)$ be a spacetime satisfying the energy condition
  $\tilde R_{\mu\nu}V^{\mu}V^{\nu} \ge -3$
  for all unit timelike vector fields $V$.  Suppose that $M$ has a smooth
  compact spacelike Cauchy surface $N$ with
  Hubble scalar $\tilde H$ satisfying $\tilde H>1$ on $N$.  Then every
  timelike geodesic in $\tilde M$ 
  is past incomplete.
\end{Prop}
After a conformal
rescaling with a suitable constant conformal factor, we can assume
that $\lambda=3$ if the cosmological constant 
is positive. If the other matter fields fulfill, say,
the strong
energy condition, or as we will assume, simply vanish, then
\Propref{prop:generalizedHawking} has a chance to apply.

For the following discussion we introduce further terminology. From
the investigations of the Yamabe problem (see the review in
\cite{lee87}) one knows that the 
conformal class of 
any smooth Riemannian metric on a compact manifold contains a
metric with constant scalar curvature. Although the value of the scalar
curvature $R$ is not invariant under conformal rescalings, the sign of
the constant $R$ is. Hence, conformal classes of compact
Riemannian manifolds are divided into three disjoint (possibly empty) classes
determined by the existence of a metric in the conformal class with
negative, zero or positive constant Ricci scalar. We
will say that a given conformal class has either negative, zero or positive
scalar curvature. Now, we can formulate the following theorem proved
in \cite{galloway2002}.
\begin{Theorem}
  \label{th:singularityScalarCurvature}
  Let $(\tilde M,\tilde g)$ be a globally hyperbolic solution of EFE in
  vacuum with $\lambda>0$ which is future asymptotically de-Sitter with
  compact $\scrip$.
  Then if $\scrip$ has negative scalar curvature, every timelike geodesic in
  $(\tilde M,\tilde g)$ is past incomplete.
\end{Theorem}
In particular, no past conformal boundary can exist.
To prove this theorem, recall our
result from \Sectionref{sec:FAdScosmicnohair} that $\tilde H=1$ on
$\scrip$ for $\lambda=3$. Further, one can derive easily from the knowledge of
\Sectionref{sec:initial_data} that $n(\tilde H)=0$ on 
$\scrip$, where $n$ is given as in
\Sectionref{sec:FAdScosmicnohair}. Now, the hypothesis in this theorem
is tailored exactly such that 
$n(n(\tilde H))>0$ on $\scrip$, because then the hypothesis of
\Propref{prop:generalizedHawking} holds on any Cauchy surface
arbitrarily close to $\scrip$ and past
geodesic incompleteness is implied. Despite the geometrical elegance
of \Theoremref{th:singularityScalarCurvature}, this suggests that
this result is probably too weak to describe a large class of
interesting past
incomplete FAdS
spacetimes since we cannot expect that general incomplete FAdS
solutions have the 
property that the ``collapse'' hypothesis of
\Propref{prop:generalizedHawking} is satisfied
arbitrarily close to $\scrip$ in the past. It would
be nice to have more general results, but in any case, a few interesting
conclusions can be drawn; see below. The authors of \cite{galloway2002}
discuss a slight 
generalization of \Theoremref{th:singularityScalarCurvature} which
applies
when the scalar curvature of $\scrip$ is zero. Roughly speaking,
except for a special case, all past timelike
geodesics are incomplete also under this hypothesis.

Another result of different nature proved in
\cite{galloway2002} is the following. 
\begin{Theorem}
  \label{th:singularityFundamentalGroup}
  Let $(\tilde M,\tilde g)$ be a globally hyperbolic solution of EFE in
  vacuum with $\lambda>0$ which is future and past asymptotically
  de-Sitter such that $\scrip$ (or $\scrim$) is compact. 
  Then  the Cauchy surfaces
  of $(\tilde M,\tilde g)$ (homeomorphic to
  $\scri^+$ (or $\scri^-$)), have finite fundamental group.
\end{Theorem}

We will discuss a few implications of these singularity theorems with
regard to our applications
in \Sectionref{sec:situation_FADS}. 

\subsection{Friedrich's Cauchy Problem}
\label{sec:initial_data}
In this thesis we want to construct FAdS solutions and analyze their
properties. Friedrich's idea \cite{DeSitter} was to construct FAdS
spacetimes in 
terms of a Cauchy problem. Since, for these spacetimes, $\scrip$ is a
smooth spacelike 
hypersurface in the conformal spacetime it makes sense to try to
formulate the initial value problem with $\scrip$ as the ``initial''
hypersurface using the conformal field equations to integrate the data
backwards in time.  

Let $\Sigma$ be a smooth pseudo-Riemannian $3$-manifold with metric
$h$, an orthonormal frame $\{e_a\}$ and the corresponding Levi-Civita
connection $D$. Then one defines the \term{Cotton tensor} by
\[b_{cab}=D_{[a}r_{b]c}-\frac 14 D_{[a}r\, g_{b]c}\]
where $r_{ab}$ is the Ricci tensor and $r$ the Ricci scalar of
$h$. The Cotton tensor vanishes exactly if $h$ is conformally flat.

Now, we formulate the following theorem about the construction of
initial
data on $\scrip$ for the conformal field equations \Eqsref{eq:gcfe} in
conformal 
Gauß gauge.
Friedrich \cite{DeSitter} proves the general version allowing
general gauges.
\begin{Theorem}
  \label{th:solving_constraints}
  \hypertarget{th:solving_constraints}
  Let $\Sigma$ be a smooth connected orientable $3$-dim.\ manifold (without
  boundary). Let $\{x^\alpha\}$ be local coordinates, $h$ a smooth
  Riemannian metric, $\{c_a\}$ a 
  smooth orthonormal frame and $D$ the covariant derivative of the
  Levi-Civita connection of $h$ with frame connection coefficients
  $\Connectionh abc$, Ricci tensor $r_{ab}$ and Ricci scalar
  $r$. Furthermore, choose a smooth symmetric tracefree
  tensor field $W_{ab}$ on $\Sigma$ that satisfies
  \begin{equation}
    \label{eq:scri_electric_constraint}
    D_a W\indices{^a_b}=0,
  \end{equation}
  a smooth function $k$, a smooth $1$-form $\omega$ on
  $\Sigma$ and the positive value of the constant $\Lambda$. Then the
  following data both constitute an initial data set for 
  the general conformal field equations \Eqsref{eq:gcfe} in vacuum
  with $\lambda=\Lambda$ considering $\Sigma$ as a $\Omega=0$-hypersurface
  in a spacetime with topology $\R\times\Sigma$, and fix a
  conformal Gauß gauge with respect to $\Sigma$: 
  \begin{enumerate}
  \item $e_a=c_a$ and $e_0=\partial_t$ transversal to $\Sigma$ in
    $\R\times\Sigma$,  
  \item $f_a=\omega_a$ and $f_0=0$,
  \item $\ConnectionWeyl abc=\Connectionh abc
    +\delta\indices{^b_a} \omega_c
    +\delta\indices{^b_c}\omega_a-g_{ac}g^{bd}\omega_d$
  \item $\ConnectionWeyl ab0=-k\delta\indices{_a^b}$, 
    $\ConnectionWeyl a0b=-k h\indices{_a_b}$ and $\ConnectionWeyl 0ij=0$
  \item $\hat L_{ab}
    =r_{ab}-\frac 12 h_{ab}\left(\frac r2-k^2\right)
    -D_a \omega_b-\omega_a \omega_b
    -\frac12g_{ab}g^{cd}\omega_c \omega_d$,\\
    $\hat L_{a0}
    =D_a k+k \omega_a$ and $\hat L_{0i}=0$
  \item $E_{ab}=W_{ab}$ and
    $B_{ab}=-\sqrt{\frac3\lambda}\,b_{acd}\epsilon\indices{^c^d_b}$.
  \end{enumerate}
  On $\R\times\Sigma$, we have for the conformal factor, setting $t=0$
  on $\Sigma$, 
  \begin{equation}
    \label{eq:general_omega}
    \Omega(t,x^\alpha)=\frac 12\sqrt{\frac\lambda 3}\, t
    \left(2-k(x^\alpha)\, t\right)
  \end{equation}
  and $d_a$ vanishes identically for all times.
  \begin{Proof}
    Friedrich's version of the theorem involves general
    gauges, but his expressions are given in terms of the Levi-Civita
    connection of $g$. He finds that one must make the following
    identifications. 
    \begin{enumerate}
    \item $\lambda=\Lambda$,
    \item $e_a=c_a$ and $e_0$ transversal to $\Sigma$,
    \item $\Connection abc=\Connectionh abc$
    \item $\Connection ab0=-k\delta\indices{_a^b}$, 
      $\Connection a0b=-k h\indices{_a_b}$
    \item $\Connection 0ba$, $\Connection 00a$ and $\Connection 0a0$
      freely specifiable by the frame transport 
    \item $L_{a0}=D_a k$, 
      $L_{ab}=r_{ab}-\frac 12 h_{ab}\left(\frac r2-k^2\right)$ and
      $L_{00}$ determined by $L\indices{^a_a}$ and an arbitrarily
      chosen gauge source function $R$,
    \item $E_{ab}=W_{ab}$ and
      $B_{ab}=-\sqrt{\frac3\lambda}\,b_{acd}\epsilon\indices{^c^d_b}$
    \item $\Sigma:=\Sigma_0=\sqrt{\frac\lambda 3}$, $\Sigma_a=0$ and
      $s=k\sqrt{\frac\lambda 3}$ (definitions in \Sectionref{sec:cfe})
    \end{enumerate}  
    The main work left for us is to find the
    expressions in a conformal Gauß gauge to be fixed before. 
    Set $f_a \sigma^a=\omega$ and $f_0=0$ on $\Sigma$. This and the frame
    and coordinate choices above fix a conformal Gauß gauge.
    The determination of the Weyl connection quantities corresponding to
    the quantities above
    are
    straight 
    forward using \Eqsref{eq:levi_civita2Weyl1} and
    \eqref{eq:relate_Weyl_L_and_LeviCivita_L}. Further we have to
    determine $\Omega$ according to 
    \Eqref{eq:Omega2ndOrderPolyn} and 
    $d_a$ according to \Eqref{eq:CGG_d}. We will set $t=0$ on
    $\Sigma$ and assume $e_0=\partial_t$. On $\Sigma$ we have
    \[\dot\Omega:=e_0(\Omega)=\Sigma_0=\sqrt{\frac\lambda 3}.\]
    But we also find 
    \[k\sqrt{\frac\lambda 3}=s=\frac 14\nabla_i\Sigma^i
    =\frac 14(e_i(\Sigma^i)-\Connection iij\Sigma_j)
    \quad\Rightarrow\quad
    \ddot\Omega:=e_0(e_0(\Omega))=-k\sqrt{\frac\lambda 3}.\]
    Since $\Sigma$ is a $\Omega=0$-surface,
    this implies the claimed time dependence for the conformal
    factor. On $\scrip$ we have that $d_a=e_a(\Omega)=0$, hence $d_a$
    identically vanishes during evolution.
  \end{Proof}
\end{Theorem}
Note that the conformal gauge is chosen here such that the conformal
geodesics start from $\scrip$ orthogonally. This can be generalized
straight forwardly which has not yet been done for this
thesis. Indeed, according to the 
discussion in \Sectionref{sec:confgeodesicsFAdS} this can be a serious
issue.

The natural next question is if each such initial data set can be
realized as the conformal boundary of a FAdS solution of EFE in
vacuum with $\lambda>0$. 
Consider the 
following theorem also due to Friedrich \cite{DeSitter} which gives a
positive answer to this question. Without loss of generality we can
assume that the conformal boundary corresponding to the initial data
is $\scrip$.
In this case $e_0$ is past
directed and hence the coordinate $t$ increases into the
past. Evidently, we can also 
consider the time dual case with $\scrim$ as initial hypersurface and
$e_0$ future directed which is the convention used in
\cite{DeSitter}.  
\begin{Theorem}
  \label{th:cauchyproblem}
  Assume that we prescribe initial data for the conformal field
  equations as in the general form \cite{DeSitter} of
  \Theoremref{th:solving_constraints} on a compact 
  $3$-surface $\Sigma$ such that the fields are of type $H^k(\Sigma)$
  with $k\ge 4$. Then there is a unique (up to questions of
  extendibility) FAdS solution $(\tilde M,\tilde g)$ of Einstein's
  field equations in vacuum with 
  $\lambda=\Lambda>0$ with a conformal representation $(M,g,\Omega)$ of class
  $H^k(M)$ and with
  a conformal embedding $\Psi:\Sigma\rightarrow M$ with
  $\Psi(\Sigma)=\scrip$
  such that the pull-back of the 
  fields on $\Psi(\Sigma)$ to $\Sigma$ correspond, after a suitable conformal
  transformation, to those on $\Sigma$.
  In  particular, $\Omega_{|\Psi(\Sigma)}=0$,
  $d\Omega_{|\Psi(\Sigma)}\not=0$.
\end{Theorem}
For the proof which is based on Friedrich's conformal field equations,
see again \cite{DeSitter}. Here, $H^k(\Sigma)$ (and in the same way of
$M$) is constructed as follows. Consider the space
$C^\infty(\Sigma,\R^N)$ with the standard Sobolev 
norm
\[\|w\|_m:=\left\{\sum_{k=0}^m\int_{\Sigma}|D^kw|^2d\mu\right\}^{1/2}\]
where we use
standard multiindex notation for the covariant derivatives $D$ and
$d\mu$ is some appropriate measure on $\Sigma$. Then $H^m(\Sigma)$ is
the completion of $C^\infty(\Sigma,\R^N)$ with respect to this
norm. For our considerations it is sufficient 
that the solution is smooth on its existence
interval if smooth initial data are chosen.

This is a ``semi-global'' result since by construction,
solutions exist for all physical times into the future but not necessarily into
the past. Indeed, there are no general global existence results
available for the past except for the special situation discussed in
\Sectionref{sec:non-linear stability of de-Sitter}. 
Analogously to the standard Cauchy problem of EFE discussed in 
\Sectionref{sec:maximaldevelopments}, it makes sense to talk about
maximal Cauchy 
developments of $\scrip$-initial data sets, i.e.\ the maximal
globally hyperbolic solution component connected to $\scrip$. Thus the
standard formulation of the strong cosmic censorship conjecture
(\Sectionref{sec:SCC}) can be
transfered directly to Friedrich's Cauchy problem.

The full conformal gauge
freedom is preserved in the discussion above in the sense that certain
transformations of the initial data on $\scrip$ together with the appropriate
choices of gauge source 
functions lead to a conformally rescaled solution of the conformal
field equations. 
Let us elaborate a bit further on this. In this discussion one should have
the general form of \Theoremref{th:solving_constraints} in
\cite{DeSitter} in mind.
Suppose that the solution $(M,g,\Omega)$ of the conformal field equations
corresponding to a $\scrip$-initial data set
in a conformal gauge
determined by the gauge source function $R$ is given with an
orthonormal frame as in \Theoremref{th:solving_constraints} such that
$e_0$ is orthogonal to $\scrip$ and past directed. Denote the underlying
FAdS spacetime as usual as $(\tilde M,\tilde g)$. Consider
a conformal rescaling of the solution of the
form $\bar g=\Theta^{-2} g$ with $\Theta>0$ smooth. 
The restrictions of
$\Theta$, $e_0(\Theta)$ and $e_0(e_0(\Theta))$ to $\scrip$ 
determine the corresponding transformation of the $\scrip$-initial
data set completely. In particular the function $k$, being the conformal
expansion of $\scrip$ with respect $e_0$, transforms as
\begin{equation}
\label{eq:transfk}
\bar k=\left.(\Theta k-e_0(\Theta))\right|_{\scrip}
\end{equation}
which is analogous to \Eqref{eq:rescal2ndFF}. The value of the gauge
source function $R$ on $\scrip$ is the only initial data component
which is influenced by $e_0(e_0(\Theta))$ on $\scrip$. This
shows that by an appropriate choice of conformal factor $\Theta$ we
can, independently from each other, make an arbitrary conformal
rescaling of the $3$-metric 
of $\scrip$ (and hence change all derived initial data quantities such
that they stay consistent with \Theoremref{th:solving_constraints}),
an arbitrary 
transformation of the function $k$ and an arbitrary transformation of
$R$. This shows that in particular, the conformal class of the initial
$3$-metric and the function $k$ are pure gauge
and have no physical meaning. Indeed, the uniqueness result of
\Theoremref{th:cauchyproblem} implies that solutions corresponding to
initial data 
sets which only differ in this way are isometric up to
questions of extendibility. In any case, note that questions about
extendibility and the gauge can be crucial for practical purposes. In
particular, when we want to stay in the family of conformal Gauß
gauges, a different choice of the conformal class of the initial
$3$-metric and the function $k$ lead
necessarily to different
coordinates which cover different parts of the solution
manifold in general. Thus it is a difficult, but crucial, task to
choose the gauge
such that the solution manifold is covered in an optimal way.

We make a few more comments here. First note that there is no
assumption in \Theoremref{th:solving_constraints} about the topology
of the initial hypersurface. Solutions to the constraints exist in any
case, in contrast to the initial data problem on standard Cauchy
surfaces where the Yamabe type plays a role, see for instance
\cite{isenberg95}. The main reason for this simplification is that
there is no ``\mbox{Hamiltonian}
constraint'' involved here; the only differential equation
\Eqref{eq:scri_electric_constraint} is always solvable because it
admits at least the trivial solution. In
\Theoremref{th:cauchyproblem}, one requires compactness of $\scrip$
but this is rather a technical requirement and could possibly be
generalized. Hence, at least roughly, Friedrich's Cauchy problem is well-posed
independent on the choice of manifold of $\scrip$. However,
compactness of $\scrip$ seems like a natural assumption in the
cosmological setting as discussed before.
Another
comment is that Friedrich's results based on the conformal field
equations are only valid in four spacetime dimensions. Anderson
\cite{Anderson04} succeeded in finding a generalization of these
results here to arbitrary even
spacetime dimensions.

\subsection{Levi-Civita conformal Gauß gauge}  
\label{sec:LCCGG}
In \Sectionref{sec:initial_data} we discussed how to set up the initial
value problem of the conformal field equations with initial data on
$\scrip$ using the conformal Gauß gauge. Here we show that
by a certain choice of the initial gauge functions $k$ and $\omega_a$ we
can arrange that the Weyl-$1$-form $f$ vanishes identically for all
times. 
This simplified subcase of the class of conformal Gauß gauges will be
used in the numerical computations later in this thesis.

\begin{Prop}
  \label{prop:IDLCCGG}
  Let initial data be given for the GCFE as described in
  \Theoremref{th:solving_constraints}, but restrict to the case
  $k=const$ and $\omega_a=0$. Then the corresponding solution of
  the general conformal field equations satisfies $f\equiv 0$, i.e.\
  all quantities reduce to their Levi-Civita versions of the conformal
  metric $g$. With the further conventions $k=1$ and $\lambda=3$, the
  evolution equations simplify to 
  \nobreak
  \begin{subequations}
    \label{eq:gcfe_levi_cevita_evolution}
    \begin{align}
      \partial_t e_a^\alpha&=-\chi\indices{_a^b}e_b^\alpha\\
      \partial_t\chi_{ab}
      &=-\chi\indices{_{a}^c}\chi_{cb}
      -\Omega E_{ab}
      +L\indices{_{{a}}_{b}}\\
      \partial_t\Connection abc
      &=-\chi\indices{_a^d}\Connection dbc
      +\Omega B_{ad}\epsilon\indices{^b_c^d}\\
      \partial_t L_{ab}
      &=-\dot\Omega E_{ab}-\chi\indices{_a^c}L_{cb}\\
      \label{eq:Bianchi1}
      \partial_t E_{fe}&=-2\chi\indices{_c^c}E_{fe}
      +3\chi\indices{_{(e}^c}E_{f)c}
      -\chi\indices{_c^b}E\indices{_b^c}g_{ef}
      +D_{e_c}B_{a(f}\epsilon\indices{^a^c_{e)}}\\
      \label{eq:Bianchi2}
      \partial_t B_{fe}&=-2\chi\indices{_c^c}B_{fe}
      +3\chi\indices{_{(e}^c}B_{f)c}
      -\chi\indices{_c^b}B\indices{_b^c}g_{ef}
      -D_{e_c}E_{a(f}\epsilon\indices{^a^c_{e)}}\\
      \Omega(t)&=\frac 12 t\, (2-t)
    \end{align}
    for the unknowns 
    \begin{equation}
      u=\left(e_a^\alpha, \chi_{ab}, \Connection abc, L_{ab}, E_{fe},
        B_{fe}, \Omega\right).
    \end{equation}
  \end{subequations}
  Moreover, $e_0$ is hypersurface orthogonal, i.e.\ the $2$nd
  fundamental form (cf.\ \Eqsref{eq:2ndFundamentalForm}) is symmetric.
  We refer to this choice of gauge as \term{Levi-Civita conformal Gauß gauge}.
  \begin{Proof}
    Since
    \[d_a=\Omega f_a+e_a(\Omega)\]
    and since $d_a=0$ (\Theoremref{th:solving_constraints}) and
    $e_a(\Omega)=0$ due to  
    \Eqref{eq:general_omega} in a neighborhood of $\scrip$, we have
    $f_a\equiv 0$. Hence $f\equiv 0$.  The evolution equations in
    \Eqsref{eq:gcfe_levi_cevita_evolution} follow from
    \Eqsref{eq:gcfe}. But note that we have written down a reduction
    of the Bianchi system here. It remains to check that all solutions under
    these conditions are really compatible with the Levi-Civita
    connection of $g$ and that $e_0$ is hypersurface orthogonal.
    For instance, to prove that $\chi_{ab}$ is really
    symmetric we can derive an
    evolution equation for its antisymmetric part from the evolution
    equations above. Let $\chi$, $E$,
    $L$ denote the matrices $(\chi_{ab})$, $(E_{ab})$ and $(L_{ab})$. Then
    \[\partial_t\chi=-\chi\cdot\chi-\Omega E+L,\quad 
    \partial_t\chi^t=-\chi^t\cdot\chi^t-\Omega E^t+L^t.\]
    Now let $\chi^S$ and $\chi^A$ be the symmetric and antisymmetric
    parts of $\chi$. Since $E$ and $L$ are symmetric we get
    \[2\partial_t \chi^A=-\chi\cdot\chi-\chi^t\cdot\chi^t
    =-(\chi^S+\chi^A)\cdot(\chi^S+\chi^A)+(\chi^S-\chi^A)\cdot(\chi^S-\chi^A)
    =-2\chi^A\cdot\chi^S-2\chi^S\cdot\chi^A.\]
    Hence, since $\chi^A=0$ initially ($\chi\sim h$), it will vanish
    identically for all times. 
  \end{Proof}
\end{Prop}
Note that \Eqsref{eq:Bianchi1} and \eqref{eq:Bianchi2} constitute
a reduction of the Bianchi system written in terms of $E$ and $B$
(\Eqref{eq:electricmagnetic} for the rescaled Weyl tensor) compatible
with the gauge. Ignoring the tracefreeness 
of $E$ and $B$ in a first step, one can show that this reduction is (apart
from trivial factors) symmetric hyperbolic. Then in a second step, it
is easy to check 
that if the traces of $E$ and $B$ vanish initially, then they vanish for all
times. Hence the complete system
\Eqsref{eq:gcfe_levi_cevita_evolution} is, apart from
these subtleties, symmetric hyperbolic.

In
this gauge, we also write down the constraint equations implied by 
the Bianchi system 
\nobreak
\begin{subequations}
\label{eq:bianchi_constraints}
\begin{align}
  \label{eq:bianchi_electric_constraints}
  D_{e_c} E\indices{^c_e}
  &=\epsilon\indices{^a^b_e}B_{da}\chi\indices{_b^d}\\
  D_{e_c} B\indices{^c_e}
  &=-\epsilon\indices{^a^b_e}E_{da}\chi\indices{_b^d}.
\end{align}
\end{subequations}
The other constraints of the system above are equally important but
are not yet considered in this thesis.

The Levi-Civita conformal Gauß gauge cannot be expected to be a
``good'' gauge choice
in all practical 
situations. However, it simplifies the evolution equations and can
lead to preliminary understanding of situations where no
further a priori insights or expectations exist. Furthermore, note that in this
gauge spatial symmetries are represented in a very simple manner; first,
since the conformal factor 
is constant in space, any Killing vector field of the conformal metric
tangent to the $t=const$ hypersurfaces 
is also a Killing vector field of the physical metric. Second, according to
\Propref{prop:xiconst}, the
coordinate components 
of such KVFs are constant in time. This is not the case
in general 
conformal Gauß gauges because the vector field $e_0$ is not necessarily
hypersurface orthogonal.

\subsection{Conformal geodesics in future asymptotically de-Sitter 
spacetimes}
\label{sec:confgeodesicsFAdS}
Here, we briefly list relevant results from
\cite{ConfGeodesics}. First, for any solution
$(\tilde M,\tilde g)$ of EFE
in vacuum with any cosmological constant, all $\tilde g$-geodesics
are also conformal geodesics of the associated 
conformal structure up to parametrization, in the sense that one can
always construct an associated Weyl connection under these conditions.
The authors of \cite{ConfGeodesics} succeeded in
showing that  the conformal geodesic equations
\Eqsref{eq:conformal_geodesic2} can
be integrated even explicitly for $\tilde f$. In particular, the
resulting Weyl connection is in  
general not the Levi-Civita connection of $\tilde g$.

Another relevant finding is as follows.
Let a future asymptotically de-Sitter solution 
$(\tilde M,\tilde g)$ of EFE in vacuum with $\lambda>0$ be given. Any
timelike conformal geodesic that leaves $\scrip$
orthogonally into the past is, up to parametrization, a 
$\tilde g$-geodesic. Any 
conformal geodesic that leaves $\scrip$ non-orthogonally cannot
represent a physical geodesic. This has indeed important consequences.
The gauge of solutions of GCFE given by initial data as in
\Theoremref{th:solving_constraints} is based on conformal
geodesics which start orthogonally from $\scrip$. Hence it actually
corresponds, up to parametrization, to a ``physical'' Gauß gauge. We
will say more about this later in the discussions of the applications.

\subsection{Non-linear stability of the de-Sitter spacetime}
\label{sec:non-linear stability of de-Sitter}
In \cite{friedrich86}, Friedrich extended his results about FAdS
vacuum spacetimes, cf.\ \Sectionref{sec:initial_data}. He studied
non-linear perturbations of 
the de-Sitter spacetime and concluded that the perturbed spacetimes
behave like the de-Sitter spacetime if they do not deviate too much
at \scrip. He still restricts to vacuum with $\lambda>0$. This is so
far the only non-trivial global existence result for Friedrich's
Cauchy problem. 

We avoid going into too many technical details here, however, some
are necessary for the further understanding. In particular, the Sobolev spaces
$H^m(\S,\R^N)$ with norm $\|\cdot\|_m$ defined as in
\Sectionref{sec:initial_data} are crucial. Note that the following arguments
are actually also used in the proof of
\Theoremref{th:cauchyproblem}. Consider a symmetric hyperbolic
reduction of the conformal field equations.
Friedrich was able to show the following using the theory of symmetric
hyperbolic systems. Suppose that the
solution of the evolution system, symbolically written as $w(t)$ with
$w$ taking values in a 
Sobolev space, corresponding to a choice of data $w_0$ at some
initial time $t_0$ exists in the interval $[t_0,T_0]$ for some
$T_0>t_0$. Then there is a 
neighborhood of initial
data in the $\|\cdot\|_m$-norm at the same initial time such
that all corresponding solutions also exist on $[t_0,T_0]$. On this
interval, these solutions are in the class
$C^{m-2}([t_0,T_0]\times\S)$ in particular. Similar statements can be
made for the 
past time direction. Further, he could show that if a
sequence of initial data (all at the same 
initial time) is given that converges to some initial datum $w_0$ in the
$\|\cdot\|_m$-norm, then the corresponding solutions converge to the
solution corresponding to $w_0$ in the
$\|\cdot\|_m$-norm uniformly in time on the common existence
interval. Friedrich requires for all these statements that
$m\ge 4$; however, he also notes that it is likely to be possible to
get similar results under less regularity assumptions.

Now, consider the de-Sitter spacetime in the conformal representation
given by \Eqref{eq:dS_conformal} depicted in
\Figref{fig:dS_conformal}. The conformal spacetime $(M,g,\Omega)$
corresponding to the de-Sitter solution extends smoothly to all
values of $\bar t$, although the part representing the physical solution
corresponds only to the time interval $]-\pi/2,\pi/2[$. In any case,
with the results quoted above, one can
choose an arbitrary constant $T_0>\pi/2$ and always find a
corresponding neighborhood in the
$\|\cdot\|_m$-norm of the
de-Sitter initial data on $\scrip$ such that all corresponding
solutions exist on $[-T_0,T_0]$. Hence, on the one hand one can choose
this neighborhood so small such 
that the corresponding solutions all exist for an arbitrary long time;
on the other hand, by the continuity property mentioned above
the behavior of the associated conformal factors can be brought arbitrary close
to that of the de-Sitter solution; in particular it can be achieved
that these conformal factors have
non-degenerate zeros within the time interval of existence.  One can conclude
that all solutions in such a sufficiently small
neighborhood are of de-Sitter type, both future and past
asymptotically simple and constitute
$C^{m-2}$-solutions of EFE in vacuum with $\lambda>0$.

In \cite{friedrich91}, Friedrich generalized these results to
spacetimes with Yang-Mills fields. Generalizations to all even
spacetime dimensions can be found in
\cite{Anderson04}.  It is trivial to conclude that cosmic censorship
holds in the class of solutions close to the de-Sitter 
spacetime (or any other reference spacetime) since all
these solutions are geodesically complete and 
hence $C^2$-inextendible. Further note that the arguments above can also be
applied
to prove stability of \textit{other} future and past asymptotically simple
reference solutions than dS.

\subsection{Characterization of the boundary of the de-Sitter
  stability region}
\label{sec:dSSR}
In \cite{anderson05,Anderson04}, Anderson studies the properties of
solutions of Friedrich's Cauchy problem with $\scrip$-initial data sets
on the boundary of the stability neighborhood of
\Sectionref{sec:non-linear stability of de-Sitter}.

Let us assume general even spacetime dimensions and consider FAdS
solutions of EFE in vacuum with $\lambda>0$
which satisfy the hypothesis of Friedrich's stability result with
respect to any future and past asymptotically simple reference
solution; for instance dS. The
union of all these $\scrip$-initial data sets is
open in the topology above, as follows directly from Friedrich's
arguments. Now consider the closure of this set.
Anderson 
found the following 
characterization. The solution corresponding to a $\scrip$-initial data
set on the boundary of that closure has exactly one
of the following properties\footnote{Comment: The following
  characterisation has turned out to be false.  The degenerate
  $\lambda$-Taub-NUT 
  solutions (see \Sectionref{sec:TaubNUT}) are actually
  counterexamples to 1), because they are neither geodesically
  complete, infinitely tall nor always exist of \textit{two} such pieces.}: 
\renewcommand{\labelenumi}{\arabic{enumi})}
\begin{enumerate}
\item It is the union of two spacetimes, one with a regular past
  conformal boundary and empty future conformal boundary, the other
  with a regular future conformal and empty past conformal boundary.
  Each of the two pieces is geodesically complete,
  globally hyperbolic and infinitely tall.
\item It consists of a single complete globally hyperbolic spacetime
  with smooth future conformal boundary and either with a
  partial or empty past conformal boundary.
\end{enumerate}
The physical spacetimes are regular in both cases.
Of course, this result could have been formulated equally well with
respect to $\scrim$.
Here, a spacetime is said to be
\term{partially conformally compact} in the past if the
past conformal boundary is not compact but also
non-empty. Note that in this case the induced metric on the partial
conformal boundary can be complete or incomplete. Let us now discuss
the notion of infinitely tall spacetimes. A spacetime is
called \term{tall} if any observer can ``see'' an entire Cauchy
surface in the past
after sufficiently late times. Hence, in a tall spacetime
all observers far enough in the future have vanishing particle
horizon\footnote{The notion of \term{particle horizons} is visualized
  nicely in Figure~18 in \cite{hawking}.}.
In particular, in a tall cosmological 
spacetime, observers are able to detect the
compactness of the Cauchy surfaces in principle. Gao and Wald
\cite{Gao00} study 
conditions for which a cosmological spacetime is
tall\footnote{However, the authors do not use the word
  ``tall''.}. They find in particular, although the de-Sitter
spacetime just ``barely fails'' to be tall, ``generic'' perturbations
of it have this property. So, all solutions in the de-Sitter stability
neighborhood are tall except for those satisfying the degeneracy
condition by Gao and Wald. Now, a spacetime is \term{infinitely tall},
if \textit{all} 
Cauchy surfaces are visible by observers after late enough times.

According to Anderson, the first possibility is realized for instance
by the degenerate 
$\lambda$-Taub-NUT cases, which has turned out to be false. 
For the second possibility no examples are known. As discussed in
\Sectionref{sec:situation_FADS} some of these spacetimes could maybe
be interpreted as
cosmological black or white hole spacetimes. 

In the following we will call the maximal connected component of
Friedrich's stability neighborhood of dS with respect to dS 
the \term{de-Sitter stability region} (dSSR). 

\subsection{Situation for FAdS solutions}
\label{sec:situation_FADS}
In this section we discuss the situation for FAdS solutions of
Friedrich's Cauchy problem with compact 
$\scrip$. We use the various results
and theorems before to draw a preliminary picture of what we can
expect in this class of spacetimes. For the discussion in the section
here, we  
ignore all the results obtained under special symmetry conditions, in
particular those listed in \Sectionref{sec:gowdyphenom}, despite of
their importance.

We are interested in the
maximal Cauchy developments of $\scrip$-initial data sets
which lead to spacetimes which are globally hyperbolic
FAdS solutions of EFE in vacuum with $\lambda>0$ with
compact Cauchy surfaces homeomorphic to $\scrip$. \Propref{prop:structure}
implies that these solutions are future
asymptotically simple, i.e.\ future causal geodesically complete. As
mentioned before, there is so far only limited a priori information
about the past behavior for given $\scrip$-initial
data sets. \Propref{prop:structure} implies that such a
solution can only be not past 
asymptotically simple, which means that some past directed
inextendible null
geodesics are incomplete due to cosmological singularities, white
holes, Cauchy horizons or whatever, if the solution does not have a
smooth past conformal boundary at all in the maximal Cauchy
development of $\scrip$ or if the past conformal 
boundary is only partial (i.e.\ non-compact).

\begin{floatingfigure}{0.49\linewidth}
  \centering
  \includegraphics[width=0.49\linewidth]{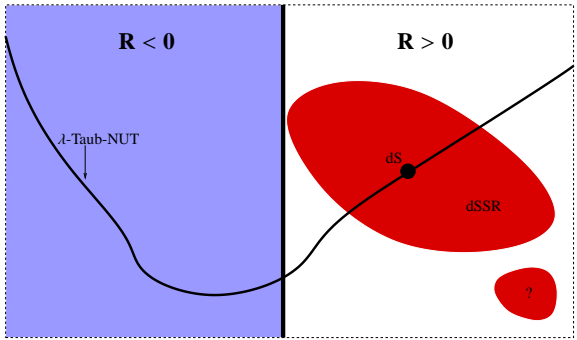}
  \caption{Situation: set of initial data for FAdS solutions with
    $\scrip\cong\S$} 
  \label{fig:situation}
\end{floatingfigure}
After these general remarks, let us restrict now to two cases which
will be studied in this thesis, namely $\scrip$ diffeomorphic to
either $\T$ or $\S$ where
lens spaces are always included implicitly. If under this assumption a
$\scrip$-initial data set has negative scalar curvature
then, according to
\Theoremref{th:singularityScalarCurvature}, all past directed timelike 
geodesics are incomplete and there cannot exist even a piece of
\scrim in the maximal Cauchy development of \scrip. Basically, the
same can be said about initial data of zero scalar curvature with the
exception of a special case \cite{galloway2002}. For $\scrip\cong\T$
being non-simply connected,
this special situation can be excluded from the start because
\Theoremref{th:singularityFundamentalGroup} does not allow a smooth
past conformal boundary (neither compact nor non-compact). But, what
is the physical 
process that prevents such solutions in general from reexpanding into the
past? This is an interesting outstanding problem. 

Let us ignore the special case
addressed in the zero scalar curvature theorem for $\S$-topology for
now, and
let us rather continue with $\scrip$-initial data
of positive scalar curvature. The results related to the Yamabe
problem exclude such data for $\scrip\cong\T$ since $\T$ is of zero
Yamabe type. Hence consider the case $\scrip\cong\S$. Recall that
there is no singularity theorem known for this situation, but
complementary, there is the stability result of the de-Sitter
spacetime. Consider
\Figref{fig:situation} which shows the space of
$\scrip$-initial data sets\footnote{Modulo equivalence transformations, i.e.\ 
those transformations of the initial data which lead to isometric
maximal Cauchy 
developments of $\scrip$.} schematically, divided into several regions
explained in a 
moment. In that figure, the open stability set of dS initial data and
its subset dSSR are marked 
red. Currently, we have no 
understanding if the total set is connected, that is why
\Figref{fig:situation} shows a red region disconnected from dSSR
marked with question marks. Further, we do not know
if it is bounded with respect to the Sobolev norm used in the
stability result and
which parts of the $\{R>0\}$-set are filled up with it. It is
clear from the singularity theorems that $\mathrm{dSSR}\subset \{R\ge 0\}$
but we do not know if the $\{R=0\}$-set is touched somewhere; a bit more
about this in a moment.

The
solutions corresponding to boundary points of dSSR can (maybe) be of the two
types listed in \Sectionref{sec:dSSR}. Particularly interesting are
solutions of type 2) with partial past compactifications, because these
can fail to be past asymptotically simple in a non-trivial manner. If
such a solution is not past 
asymptotically simple but if one finds
at least one causal geodesic connecting \scrip and \scrim,  then this
would maybe give rise 
to the interpretation as a cosmological white hole spacetime since some
regions are collapsing while others are expanding into the
past. Compare this to the 
situation when $\scrip\cong\mathbb S^1\times\Stwo$ motivated by the
Schwarzschild de-Sitter spacetime. Due to the singularity theorem
\Theoremref{th:singularityFundamentalGroup}, the maximal Cauchy development
of $\scrip$ cannot develop a smooth past conformal boundary (neither
compact nor non-compact) at all, hence there cannot be simultaneously
expanding and collapsing regions in the past. Thus such solutions
can never be interpreted as cosmological white hole spacetimes. In
particular, such a solution cannot approach a Schwarzschild de-Sitter
spacetime in the past, as one might have speculated before.

The family of $\lambda$-Taub-NUT spacetimes
interpolates, as indicated in \Figref{fig:situation}, between
some of the relevant regions of the $\scrip$-initial data space
because the corresponding geometries of 
$\scrip$ are 
Berger spheres (\Sectionref{sec:bergersphere}) after suitable
conformal gauge transformations and hence the scalar 
curvature can have any sign (\Eqref{eq:BergerR}). Further, the
de-Sitter spacetime is part of this 
family. Now, one can explicitly check that there are past
\textit{singular}\footnote{Here by singular we mean that the function
  $\tilde A^2(\tau)$ in \Eqsref{eq:TaubNUT} has a zero.}
$\lambda$-Taub-NUT solutions whose initial data have \textit{positive} scalar
curvature, i.e.\ correspond to a point on the right half of
\Figref{fig:situation}. This tells us already that dSSR does not fill
up the $\{R>0\}$-set completely and that indeed, as already suspected,
the singularity theorems described in 
\Sectionref{sec:singularitytheoremGA} are not sufficient to explain the
complete picture.
Note that the past singular $\lambda$-Taub-NUT spacetimes are past
extendible because they form smooth Cauchy horizons except for the
degenerate cases in \Sectionref{sec:TaubNUT}. To understand
more of our picture it will be particularly important to study non-linear
perturbations of these  and to
investigate, 
keeping the strong cosmic censorship conjecture and the results in
\Sectionref{sec:CHcosm} in mind, how the properties of the perturbed
spacetimes change. For this
inhomogeneous singular solutions corresponding to the $\{R>0\}$-set
are constructed numerically later in this thesis.



\chapter{Numerical analysis in general relativity}
\label{ch:numerics}

\section{Introduction}
\label{sec:num_intro}
We stated before that we want to construct and analyze FAdS solutions
numerically. In 
this section we give an overview over the relevant background from
numerical analysis.

Numerical relativity incorporates a wide range
of research problems; particularly important are studies of isolated systems
like stars, black holes and binaries thereof and the calculation of
the gravitational wave signal originating in the dynamical 
processes taking place within 
such systems. Another class of problems is related to cosmology and
this thesis project restricts its attention to those. These and many
research projects in this field require a wide range of approaches on the
one hand for the mathematical formulation of the problem: Cauchy
problems with or without conformal compactifications or initial
boundary value formulations, characteristic formulations, the question of
choosing appropriate reductions of the problem, the choice of gauges,
the way singularities (both in the coordinates and in the curvature) are
treated etc. Further one has to decide if one prefers free evolutions
or if constrained evolutions are more appropriate, and to find a usable
formulation for this.
On the other hand, one has to make a choice for the numerical
techniques including discretization schemes, the optimal use of
computer resources by parallelizations of the code etc.
The reviews \cite{Lehner02,Alcubierre04} try to reflect the status of
numerical relativity and to summarize the most important approaches
and numerical techniques. However, due to rapid progress
these reviews have already become partly obsolete. In particular the
recent 
break throughs in the binary black hole problem 
\cite{Pretorius05,Campanelli05,Baker05,Herrmann06,Scheel06,Bruegmann06,Baker07,Thornburg07}
are not yet covered. Nowadays, one is even
able to perform parameter studies for the binary black hole problem
and has started to collaborate with the experimental gravitational
wave detector community. In this context, numerical studies 
with the conformal field equations play a role which are also
of particular interest for this thesis;
see \cite{Husa02,Husa02b,Frauendiener02} for an overview on the
status. More recent applications using the conformal field equations
are discussed in \cite{Frauendiener02c,Husa05,Zenginoglu07}. 

In this chapter we first introduce some fundamental facts from
numerical analysis, but restrict to spectral discretization schemes;
the mathematical and practical background for finite 
differencing discretizations can be found in \cite{Kreiss}. Among the
spectral methods we restrict further to the collocation method which
will be applied in this thesis and
introduce important associated error quantities.
Then, we discuss the method of
lines, semi- and fully discretized systems and the issue of numerical
stability. Further, we comment on the status of the knowledge of
estimates for the 
error quantities and about convergence. In the development of my
numerical method in this thesis I dealt with two major problems. The
first is the numerical
treatment of 
``non-trivial'' topologies as $\S$ and the second is the approach to
singularities in Gowdy solutions. In
\Sectionref{sec:specialtechniques}, I list relevant techniques in
the literature.

\section{Spectral discretization for time dependent problems}
\subsection{Collocation method -- spatial discretization}
\label{sec:pseudospsectr_background}
A basic
introduction to spectral methods 
can be found in the books 
\cite{orszag,canuto88,Kreiss,boyd} and further references. Here we
restrict to those aspects of the theory which are of importance for
this thesis. 
 
Our interest here lies in initial value problems
and this section is 
devoted to the first step involved, namely spatial discretization. To
discretize the full problem the method of lines is used, see 
\Sectionref{sec:mol}. We consider only the case
with periodic boundary conditions and assume that the functions involved are
$2\pi$-periodic in each spatial direction. A well adapted basis for
such functions is the Fourier basis; for information about other
basis systems, including the very important Chebyshev polynomials, the
reader is referred to the references above.

Consider a system of
PDEs. We want to find solutions taking values in a
Hilbert space $\left(X,\scalarpr{\cdot}{\cdot}\right)$. The idea of spectral 
methods is, on the one hand, to approximate the functions by
orthogonal projections $P_N:X\rightarrow X_N$ onto $N$-dimensional
subspaces $X_N\subset X$ such that for any $f\in X$ we have 
$\lim_{N\rightarrow\infty}\|f-P_N f\|=0$ where $\|\cdot\|$ is the norm
induced by $\scalarpr{\cdot}{\cdot}$. On the other hand, one also projects
the PDEs such that the original PDEs are
recovered for $N\rightarrow\infty$ in some way and such that, for a
given $N$, the 
PDEs are reduced to a finite system of algebraic or ordinary
differential equations for the
finitely many 
parameters describing the projected unknowns. Solving these
equations for each $N$, one hopes that the corresponding sequence of
solutions 
converges to the solution of the original PDEs. The details of this
convergence process have to be studied and depend on the
particular projection method of choice and, in general, on the
considered equations.

Let a smooth function $f\in\mathbb T^1$ be given, i.e.\ $f$ is 
$2\pi$-periodic. Note that many results here can be
formulated under less regularity assumptions but this would be of no direct
relevance for this thesis. According to the
standard theory of Fourier series (which can be seen as an application
of the Peter-Weyl theorem \cite{sugiura}),
$f$ can be written as the  
absolutely and uniformly converging series
$f(x)=\sum_{k=-\infty}^{\infty} f_k e^{ikx}$
with Fourier coefficients $f_k\in\C$ which are rapidly decreasing in
$k$.
The Hilbert space underlying this analysis is $L^2(\mathbb T^1)$ with
its standard scalar product $\scalarpr{\cdot}{\cdot}$ and induced norm
$\|\cdot\|$.
Following the notation of \cite{canuto88}, we set for a given $N\in\N$
\[S_N:=\text{span}\left\{e^{ikx}\,|\,-N\le k< N\right\}\]
and define the projection operator $P_N$  to be the orthogonal
projection into $S_N$, i.e.\
\[\scalarpr{f-P_N f}{\phi}=0\quad\forall\phi\in S_N.\]
Since
$f$ has the Fourier representation above,
\[P_N(f)=\sum_{k=-N}^{N-1} f_k e^{ikx}.\]
We define $\|f-P_N f\|$ as the \term{truncation error}; for smooth
$2\pi$-periodic functions
this error decays faster than any positive power in $1/N$, see
\Sectionref{sec:convergence}.  

Now let us introduce the \term{discrete Fourier transform}. Let
$f_N\in S_N$. Consistently with the conventions used in our
code we write for a real valued function ($N$ odd)
\begin{equation}
  \label{eq:trigpolyn}
  f_N(x)=\frac{a_0}{\sqrt{2\pi}}
  +\frac 1{\sqrt\pi}\sum_{n=1}^{(N-1)/2}(a_n\cos nx+b_n\sin nx)
\end{equation}
with real coefficients $a_k$, $b_k$. Naturally, the expansion
could have been written equally well in terms of the basis
$\{e^{ikx}\}$.
One calls $N$ the \term{degree} of the trigonometric polynomial $f_N$, i.e.\
there are \mbox{$(N-1)/2$} \mbox{$\sin$-terms} and \mbox{$(N-1)/2$}
\mbox{$\cos$-terms} in the 
polynomial. The Fourier coefficients (also
called \term{spectral coefficients}) are given by
\begin{equation}
  \label{eq:FourierCoeff}
    a_0=\int_0^{2\pi}f_N(x) \frac{1}{\sqrt{2\pi}}dx,\,\,\,
    a_k=\int_0^{2\pi}f_N(x) \frac{\cos kx}{\sqrt{\pi}}dx,\,\,\,
    b_k=\int_0^{2\pi}f_N(x) \frac{\sin kx}{\sqrt{\pi}}dx,
\end{equation}
for $1\le k\le(N-1)/2$. These formulas also apply in the limit
$N\rightarrow\infty$ for arbitrary $k\in\N$. 
If $g$ is any continuous function on $\mathbb T^1$, the trapezoidal
rule with $M+1$ points\footnote{This is the trapezoidal
rule with $M+1$ points since it is built with the points $x_k=2\pi k/M$ for
$k=0,\ldots,M$ and the periodicity condition.}
is
\[\int_0^{2\pi} g(x)dx= \frac{2\pi}M
\sum_{k=0}^{M-1}g\left(\frac{2\pi}M k\right)+O(M^{-2}).\]
However, if $g$ is a trigonometric polynomial of degree $N\le 2M-1$
over the domain of $\mathbb T^1$ as in
\Eqref{eq:trigpolyn}, then this formula, without the $O(M^{-2})$-terms,
is exact because it is a Gauß 
quadrature. The points 
\mbox{$x_k={2\pi}k/M$} are called \term{collocation points} (or
\term{quadrature points}). The theory of Gauß quadrature is surveyed in
\cite{boyd}. As one can see, the
collocation points in the Gauß quadrature above 
are equally spaced, however, this is not necessarily the case for
other basis functions. Now we want to compute
the Fourier coefficients $a_0,a_1,b_1,\ldots,a_{(N-1)/2},b_{(N-1)/2}$
of the trigonometric polynomial $f_N$ by evaluating the expressions
\Eqsref{eq:FourierCoeff} with our quadrature formula. If $M$ is chosen
high enough relative to $N$, this evaluation is exact.
How to choose $M$? The
integrands with highest trigonometric polynomial degree occurring  are 
$f_N(x){\cos((N-1)x/2)}/{\sqrt{\pi}}$ and 
$f_N(x){\sin((N-1)x/2)}/{\sqrt{\pi}}$ which are trigonometric
polynomials of degree $2N-1$. Hence, to compute the Fourier
coefficients of a trigonometric polynomial of degree $N$ by means of
the quadrature formula above exactly, we have to choose (at least)
$M=N$. With this choice, we obtain the exact representation
\begin{gather*}
  a_0=\frac{2\pi}N\sum_{l=0}^{N-1}f_N(x_l)
  \frac{1}{\sqrt{2\pi}},\,\,
  a_k=\frac{2\pi}N\sum_{l=0}^{N-1}f_N(x_l)
  \frac{\cos(k x_l)}{\sqrt{\pi}},\,\,
  b_k=\frac{2\pi}N\sum_{l=0}^{N-1}f_N(x_l)
  \frac{\sin(k x_l)}{\sqrt{\pi}},
\end{gather*}
with $x_l:={2\pi}N/l$ for $1\le k\le(N-1)/2$. We can  write
this in the form
\nobreak 
\begin{subequations}
\label{eq:discreteFT}
\begin{equation}
  \label{eq:DFT}
  \vec a=\frac{2\pi}N\Phi\cdot\vec f_N
\end{equation}
with 
\begin{align}
\vec a&:=\left(a_0,a_1,b_1,\ldots,a_{(N-1)/2},b_{(N-1)/2}\right)^T,\\
\vec f_N&:=\left(f_N\left(\frac {2\pi}N\cdot 0\right), 
  f_N\left(\frac {2\pi}N\cdot 1\right),\ldots,
  f_N\left(\frac {2\pi}N\cdot(N-1)\right)\right)^T;
\end{align}
\end{subequations}
the $N\times N$-matrix $\Phi$ can be constructed from above. It turns
out that $\Phi$ is orthogonal up to a factor such that
\[\vec f_N=\Phi^T\cdot\vec a.\]
The map $\Phi$ is called {discrete Fourier transform} (DFT).

Now suppose that for a given $N\in\N$ we knew only the values $\vec f$
of a smooth function $f$ on $\mathbb T^1$ at the $N$ 
collocation points $x_k$. From these values we can compute
corresponding spectral 
coefficients $\vec a$ by the DFT map $\Phi$ which yield a
trigonometric polynomial referred to as $I_N f$. This polynomial
interpolates the 
function $f$ 
such that $(I_N f)(x_j)=f(x_j)$ at each collocation point $x_j$. The
quantity $\|f-I_N f\|$ is called \term{interpolation error}. Note
that in general $I_Nf\not=P_N f$, so the truncation error and the
interpolation error are different. In particular, the
spectral coefficients 
corresponding to $I_N f$ are not the same as those of $P_N f$ which
leads to the notion of \textit{aliasing}, see below. 
Below, we  
list some estimates which show that both errors, truncation and
interpolation, decay in the same way 
for $N\rightarrow\infty$, namely faster than any power in $1/N$. As a
side remark note, that in the same way 
as the operator 
$P_N$ is the orthogonal projection to $S_N$ with respect to the scalar
product $\scalarpr{\cdot}{\cdot}$, $I_N$ is the orthogonal projection
with respect to the discrete scalar product
$\scalarpr{\cdot}{\cdot}_N$, which is defined as
$\scalarpr{\cdot}{\cdot}$ but the integral is substituted by the
quadrature formula above.

Indeed, the fundamental fact that $I_N f$ is a good approximation
for $P_N f$ for large $N$ is the basis for the
\term{collocation method} for solving PDEs. 
Like in finite difference methods at a given time $t$, one
enforces the PDEs only at the collocation (spatial grid) points and computes
multiplications and other non-linear operations from the values of the
functions at these points. The main difference to finite difference
methods is the way spatial derivatives are computed. In the collocation
method, one uses 
the trigonometric interpolants $I_N u$ of the discrete values of the
unknowns $u$
for differentiation. This means that first, one computes the spectral
coefficients 
from the values of $u$ at the collocation points by DFT, next
applies the matrix to the coefficient vector which maps the original
coefficients 
to the coefficients of the differentiated trigonometric polynomial and
finally uses the inverse DFT to transform back to ``collocation
space''. 

The \term{pseudospectral method} is similar to the
collocation method with the only difference that one does not seek
equations in collocation space, i.e.\ the space of collocation (grid)
points, but in spectral space, i.e.\ the space of Fourier
coefficients. Thus the  
collocation and the pseudospectral method are equivalent up to a DFT. In
the remainder of this thesis we will often not distinguish
between these two methods anymore: although we often speak of
pseudospectral methods we will always mean the collocation method. In fact,
it is common in the literature to abuse the terminology in this way.

Consider again the basis $\{e^{ikx}\}$ as at the beginning of this
section. One can show straight forwardly \cite{canuto88} that the
$k$th Fourier 
coefficient $\tilde f_k$ of $I_N f$ is given by
$\tilde f_k=\sum_{l\in\Z} f_{k+lN}$
where $f_n$ is the $n$th Fourier coefficient of $f$.
This is the \term{aliasing effect}. The trigonometric interpolant $I_N f$
cannot represent wavelengths shorter than the grid spacing and higher
frequencies get mapped to lower ones according to this rule. One
defines the aliased part of 
$I_N f$ by the following orthogonal decomposition
\begin{equation}
  \label{eq:aliasing}
  I_N f-f=:(P_N f-f)+R_N f\quad\Rightarrow\quad
  \|f-I_N f\|^2=\|f-P_N f\|^2+\|R_N f\|^2
\end{equation}
and $\|R_N f\|$ as the \term{aliasing error}. More on this can be
found in \Sectionref{sec:convergence}.

A few more comments are in order. DFT is the heart of all
Fourier pseudospectral and collocation methods to solve PDEs. However, in
practice 
it is usually not 
wise to use the matrix representation of the transform above; there
are more efficient 
ways. Of particular importance are \textit{partial summation}
(\Sectionref{sec:myspectralcode}) and the
\term{Fast Fourier Transforms} (FFT), introduced in \cite{FFT}; for 
more modern discussions see
\cite{canuto88,numericalrecipes,boyd}. 

\subsection{Method of lines and time-marching schemes}
\label{sec:mol}
So far, we have only discretized the problem in space. The next step
is to treat the time dependence.
Let $u(t,x)$ represent the vector of unknowns and let the system of
PDEs under consideration be written symbolically as
\begin{equation}
  \label{eq:PDEproto}
  \frac{\partial u}{\partial t}=f(t,x,u)
\end{equation}
where $f$ includes all spatial derivatives, non-linearities etc., and
where we follow the approach in \cite{canuto88}. Let us assume that
the problem, including initial conditions and boundary conditions etc., is
well-posed and all further conditions are satisfied such
that all
following arguments can be justified.
For the \term{method of lines} one assumes that at a time $t$ for
given data $u$ the function $f(t,\cdot,u)$ has been determined
approximately  somehow.  For instance, assume that this has been done by the
collocation method above. In this process, also the unknown $u$ has been
approximated, in the case of the collocation method by projection with
$I_N$. We refer to this approximated unknown by $U$. We yield an
$x$-dependent system of ODEs for $U$ which we symbolically write as
\begin{equation}
  \label{eq:semidiscrete}
  \frac{dU}{dt}=F(t,x,U);
\end{equation}
the spatial coordinate $x$ takes values in the space of grid points.
This system is called the \term{semi-discrete system} of the original
PDE problem because the spatial
derivatives have been approximated completely while the time
derivative is still ``exact''. When we attempt to solve this
{semi-discrete system} numerically, also time has to be discretized
using ODE solution techniques and the resulting system is then referred to
as the \term{fully discrete system}.

As a \term{time-marching scheme} we mean the actual time
discretization scheme.
There are many time-marching schemes discussed in
the references above. Our scheme of choice is the $4$th-order
Runge-Kutta scheme. It is known to be
stable, in the sense below, in a broad class of situations, and in many
fields it is the 
standard scheme of choice. Its representation
formula is as follows. Consider 
general equations 
of the form\footnote{We do not write the $x$-dependence explicitly here.} 
$U'=F(t,U)$ for a given fixed $x$ at time step $n$
($t=t_n)$ with $U=U_n$. Then, at time $t_{n+1}=t_n+h$,
\nobreak
\begin{subequations}
\label{eq:rk4}
\begin{equation}
  U_{n+1}=U_n+\frac 16(k_1+2 k_2+2 k_3+k_4)
\end{equation}
with
\begin{align}
  k_1&=F(t_n,U_n) h\\
  k_2&=F(t_n+h/2,U_n+k_1/2) h\\
  k_3&=F(t_n+h/2,U_n+k_2/2) h\\
  k_4&=F(t_n+h,U_n+k_3) h.
\end{align}
\end{subequations}
This scheme converges in $4$th-order in $h$.
In the references above we find generalizations of this scheme to any
discretization order.

\subsection{Error estimates, stability and convergence}
\label{sec:convergence}
In the previous discussions, we have collected a couple of important
error types for spectral approximations, truncation, interpolation and
aliasing errors. A further 
error is the \term{discretization error} describing the difference between
the solution 
of the (semi-)discretized problem and the solution of the actual
problem. Each of these errors must be controlled; otherwise we cannot
expect that a spectral 
approximation makes sense at all. Truncation, interpolation and
hence aliasing errors are independent of the equations to solve and
are rather determined by 
the choice of basis functions, in our case the Fourier basis, and the
size of the
discretization parameter. There is
quite a number of estimates for these kinds of errors available and
some of them are listed now. 

In Canuto et al.\ \cite{canuto88} it is proven that in $L^2$-Sobolev
spaces of differentiability index $l$
we have the following estimate for the truncation error of a smooth
function on $\mathbb T^1$ with the Fourier basis
\[\|f-P_N f\|_{H^l}\le C N^{l-m}\|f^{(m)}\|_{L^2}\]
for any $m\ge 0$ and $0\le l\le m$ for some $C>0$. From this it
follows that for a 
smooth function the truncation error decays faster than any
positive power in $1/N$.
For the interpolation error, Canuto et al.\ give the same
kind of estimate 
\begin{equation*}
  \|f-I_N f\|_{H^l}\le C N^{l-m}\|f^{(m)}\|_{L^2}
\end{equation*}
for any $m\ge 0$ and $0\le l\le m$.
Note that from this estimate, we
directly get an estimate for the error that is made when spatial
derivatives are computed as in 
the collocation method described in
\Sectionref{sec:pseudospsectr_background}. Hence, we can see that for 
increasing $N$  the
error for
spectral differentiation decays much faster than for finite
differencing differentiation. This is so because the latter
approximation is only accurate to a given polynomial order. These
error estimates are
the rigorous basis of the common parlance that ``spectral methods
converge exponentially'' and that ``they are much more accurate than
finite differencing methods''.
Finally note that both truncation and interpolation errors decay in the same
way for $N\rightarrow\infty$. 
Thus, due to relation \Eqref{eq:aliasing}, also the
aliasing error is controlled.

For practical purposes it is not only important to know how the errors
behave in the limit $N\rightarrow\infty$. Since the constants in these
estimates are not determined, these estimates do not provide 
much information on the actual error for a given finite
approximation. We can estimate the constants roughly by means of
convergence tests. However, in particular the influence of the
aliasing effect on the discrete solutions has been discussed controversially
in the literature \cite{canuto88,boyd}. We can not  elaborate on this
here, but come back to this in \Sectionref{sec:spatial_adaption}.

Now we want to discuss how to control the errors made by discretizing
a system of PDEs.
Assume that we treat time-dependent problems \Eqref{eq:PDEproto} with
energy estimates on the continuum level, 
for instance symmetric hyperbolic problems.
\term{Numerical stability} resembles the notion of continuum energy
estimates.  Roughly speaking one says that a discretization of the
equations  is 
stable if one has a similar estimate for the discretized as
for the non-discretized (i.e.\ continuum) problem with constants
independent (at least 
within certain limits) of the
discretization parameters. In general, the energy estimates available
for non-linear symmetric hyperbolic equations are not sharp enough to
be of practical use,
apart from special situations where sharper estimates can be
obtained. As a strategy in
practice, one considers 
a discretization scheme as a good candidate for a non-linear problem
if it is stable for linear 
problems with constant coefficients.
 
Let us hence restrict to the linear case with constant coefficients
and write our semi-discrete system \Eqref{eq:semidiscrete} as
\[\frac{dU}{dt}=LU, \quad U(0)=U_0\]
where $L$ is a constant matrix and where we assume that $U$ has been
written as a vector. Let $U_n$ be the solution of this system using a choice of
discretization in time for
initial data $U_0$ at time $t_n=n h$ with $h=\Delta t$. Let
the spatial resolution, included in the operator $L$, be fixed for the
moment. The
discretization is called \textbf{stable} if there are positive
constants $\delta$, $C$ and $K$ independent of $h$ such that
$\forall T>0$ we have
\begin{equation}
  \label{eq:discreteenergyestimate}
  \|U_n\|\le C e^{K T}\|U_0\|
\end{equation}
for all $0\le t_n\le T$ and for all $0\le h<\delta$. Here
$\|\cdot\|$ is some ``spatial'' norm. This means that a stable
discretization allows exponential growth, however, this might not be
good enough for practical purposes; consider for example the case when
the original problem has a solution which is bounded for
$t\rightarrow\infty$. Then a stable discretization as above would still
allow errors to grow exponentially. So stability must be considered
as a necessary property of discretization but, dependent on the
problem, it is not a sufficient one. Stronger notions of
stability are introduced in \cite{Kreiss,canuto88} to take this
issue into account. For those, the quantity $\lambda h$ with $\lambda$ an
eigenvalue of $L$ plays the key role. We give no further discussion here.

For the notion of stability just introduced, the spatial
discretization, and hence the operator $L$, was 
fixed. Let us additionally introduce a spatial discretization
parameter $N$, for instance the number of spatial grid points. Then
we require for stability of the fully discrete system in addition to
the above that the constants in \Eqref{eq:discreteenergyestimate} do not 
dependent on $N$, possibly except for $\delta$. The dependence of $\delta$ on
$N$ is called \term{stability limit} of the scheme; if $\delta$ does
not depend on $N$ then the scheme is called \term{unconditionally stable}.

For a Runge-Kutta method one finds that the higher its order the
better its stability 
properties are \cite{canuto88}. In \cite{Kreiss} it is proven 
that the method of lines with Runge Kutta (of any order) is
stable for linear strongly hyperbolic systems with constant
coefficients for sufficiently accurate spatial finite differencing and spectral
discretizations.   

As already said, to consider a numerical
scheme as useful, stability is a necessary property. However, it is not
sufficient because in addition we need to
control that the sequence of solutions 
obtained from the discretized problem convergences to the solution of
the original problem when the discretization parameters go to their
continuum limits. 
In \cite{canuto88}, stability and convergence for the semi-discrete
system of a linear hyperbolic system are proved and estimates for the
discretization error are derived. There does not exist a complete
discussion for a fully discretized 
system. For non-linear equations the situations looks even worse; just
special cases are treated in \cite{canuto88,Kreiss}. However, note that
in analysis, spectral approximations are actually a common technique to
prove the existence of solutions to PDEs. For example, the
proof summarized in \cite{friedrich00} uses a special Garlerkin approximation to
prove convergence of the solutions of the semi-discrete system to the
solution of the continuum problem for
general quasi-linear 
symmetric hyperbolic systems. But note that first, there are, from the
practical point of view, no useful estimates
for the discretization error provided, and
second, there are no
results of this kind for other spectral approximations.

To summarize, in most practical situations, in particular also
regarding the problems discussed in this thesis, one has to live with
the fact that it is not 
rigorously clear if the discretization scheme of choice actually reproduces
the solutions of the original problem in the limit. Boyd \cite{boyd}
formulates the 
empirical ``rule of equal errors'' stating that in most
situations one can expect that truncation, interpolation and
discretization errors 
are of the same 
order of magnitude. Hence, if this were true, letting the
discretization parameters go to 
their continuum limits would lead to a uniform decay of the errors
and the approximate solutions would converge to the actual solution.
Nevertheless, he also discusses situations where this rule does not
hold. The message for the numerical analyst is that each problem
that one likes to study requires careful 
analysis of the properties of the approximated solutions, if possible
either by analytical means but at least by numerical experiments. Such
studies allow one to get ideas for the orders of magnitudes of the
errors. Such 
numerical experiments should involve convergence tests, analyses of
conserved quantities like constraint violations and comparisons of the
results obtained with
other numerical techniques.
It
should be the general attitude to stay
skeptical about numerical results and the
corresponding description of phenomena of
solutions; skeptical in the positive sense that one is always willing
to optimize the numerical technique, to be open for alternative approaches
and, if possible,
to discuss carefully conclusions drawn from the numerical analysis in
the light of known rigorous results. 

In any case,
note, even if we knew rigorously that the rule of equal errors was
valid for a given discretization scheme, we would still be limited by
the actual capacities of
the used computer. In general, even in this optimal case from the
analytical point of view, it is far from being clear if all
important phenomena can be resolved with the maximum resolution that
the machine provides. Hence, from the practical point of view, it is maybe not
always the most crucial point to have a proof that the discretization used is
convergent but rather to find other strong reasons that some of the
important phenomena of the actual solutions are represented correctly in
the approximations.

\section{Further relevant numerical techniques}
\label{sec:specialtechniques}
\subsection{Coordinate pathologies and ``non-trivial''
  topologies in numerical relativity}
\label{sec:nontrivialtopologies}
In this thesis, we want to compute FAdS solutions with
Cauchy surfaces of $\T$- and of $\S$-topology numerically. Further we will
restrict our attention to solutions with Gowdy symmetry, although the
code will be implemented for more general situations. The Euler parametrization
of $\S$ introduced in \Sectionref{sec:geometryS3} is adapted to this
kind of symmetry, however, it constitutes coordinate
singularities. Here now, we summarize the most important numerical techniques
for problems with ``non-trivial'' topologies and singular coordinates.
We do not claim to give the
complete set of references but 
only list a few ones which seem to be particularly relevant. We
restrict to cases where either pseudospectral or finite 
differencing methods are used and ignore, despite of their importance, in
particular finite elements methods. 

There are various approaches to solve Einstein's field equations in
the presence of coordinate singularities. Often the singular coordinate
components of the metric and of its derived quantities are
considered. For example, the finite differencing approaches in
\cite{Garfinkle00,Choptuik03,Rinne05} for axial symmetry in cylindrical
coordinates rescale these coordinate 
components in a clever way to factor out the singular behavior and
derive boundary conditions at the axis which are necessary for
smoothness. Choosing smooth 
initial data, the ``exact'' evolution equations together with
these boundary conditions imply that the solution stays smooth for all later
times. Nevertheless, the discretized versions of the equations do not
respect this property necessarily and this can
lead to problems, in particular instabilities are common.
Choptuik et al.\ in \cite{Choptuik03} report that 
the instabilities in their code
can only be cured by introducing additional dissipative terms
into the equations. On the other hand the authors of
\cite{Garfinkle00} observe a
stable evolution without any additional terms. We compare these
methods a bit further in the light of the results obtained with our method
in 
\Sectionref{sec:comparisonGowdymethods}. 
Another, distinct approach to solve the axis problem in axial symmetry
is the \textit{cartoon method} in
\cite{Alcubierre99}; related problems are addressed in
\cite{Frauendiener02b}.

Another possibility of
dealing with the coordinate problem is to choose spectral
methods; this is indeed very promising since the function space and
its basis can be adapted 
to have the right ``fall-off'' behavior at the coordinate
singularities. For this thesis, such a method 
is worked out and we report on it in the rest of this thesis.

Recently, very advanced multipatch techniques 
were developed. The idea is to cover the computational domain by
more than one \textit{regular}
coordinate patch. The most important two implementations are by
Thornburg \cite{thornburg04} and by Diener et al.\ \cite{tiglio05}. Thornburg's
implementation requires that the patches overlap and information is
shared between the patches by interpolation. His approach has been
used so far in apparent horizon finder codes and in
\cite{Zenginoglu07}. The method discussed in 
\cite{tiglio05} assumes that the patches only touch and information is
shared at the boundaries via characteristics. To distinguish from
Thornburg's approach, this one is called \term{multi-block}
method. 

In numerical relativity one usually deals with spatial ``non-trivial''
topologies. This addresses spatial topologies other than simply
connected subsets 
of $\R^3$ and $\T$. For instance, consider black hole spacetimes
or certain cosmological spacetimes.
In these cases, one cannot avoid 
the problem of dealing either with a single pathological coordinate
patch or with 
multiple regular patches in one or the other way. In the cosmological case,
which will be the case of interest for us, an
important example of a successful numerical implementation is the work in
\cite{garfinkle1999} where Garfinkle uses a single patch method as
above to compute Gowdy solutions with spatial $\mathbb
S^1\times\Stwo$-topology. The coordinate singularity in this case is
similar to that in cylindrical coordinates; more 
details are discussed later. My implementation, which is subject of
this thesis, serves as another example of such an approach. It is
designed in particular for spatial
$\S$-topology and is based on pseudospectral techniques and smooth global
orthonormal frames such that all variables in the equations are
regular everywhere 
except for the coordinate components of the frame. Future
implementations will certainly involve also multipatch techniques.

\subsection{Approach to Gowdy singularities}
\label{sec:techniquesGowdySing}
Over the last 10 years there have been a number of attempts to
simulate Gowdy solutions numerically. Here we list briefly the most
important existing methods in the literature. 
A further list of references on this topic can
be found in the review article by Berger \cite{Berger2002}.
All of the following implementations assume a vanishing cosmological
constant and, with only one exception, restrict to the $\T$-case.

Historically, the first published numerical calculation of Gowdy spacetimes can
be found in
\cite{Berger93}, see also \cite{Berger97}. The method that is used is
based on $4$th-order finite differencing in space and a $4$th-order 
symplectic time integrator; details on this can be found in that
reference. The resolution was fixed in each run, and the authors found
that a given fixed resolution was sufficient to resolve the fine
structure up to a given time. This limit time was reached when the
spatial features became so localized such that the given grid was
not able to represent them anymore.

The first one to apply pseudospectral methods to the Gowdy problem was
van Putten \cite{vanPutten97}.
He used a Yang-Mills type formulation of the field equations,
which is actually very similar to the commutator field equations. His
spatial discretization was done with the collocation method, and in time he
chose a $2$nd-order leapfrog scheme. All his runs were done with fixed
resolution in time and in space.  

In the same year, Hern and Stewart \cite{Hern97}
reported about their application of
their implementation of the Berger-Oliger adaptive mesh refinement (AMR)
algorithm (see references therein) to the Gowdy problem. Adaptive mesh
refinement is indeed 
a very natural approach to study fine scale structure as the Gowdy
spikes. However, as the time-marching scheme the authors used
Lax-Wendroff which is known to involve a quite strong dissipation
component. Although this artificial dissipation vanishes
in a controllable manner by increasing the resolution, it seems to be
responsible for the phenomenon that their well resolved numerical spikes do not
sharpen after some time anymore; 
criticism on their method and also on their interpretation 
of the results can be found in \cite{Berger97b}. It seems that
the implementation of Hern et al.\ cannot give reliable
results for the Gowdy phenomenology. In fact, AMR is not necessary
to study the Gowdy phenomenology because the spikes have a quite
simple behavior. As soon as one knows where they are created,
simple $1$-dim.\ fixed mesh refinement approaches are also
usable \cite{Weaver99}. A clever approach to study
the behavior of a 
single spike with fixed resolution was introduced by Garfinkle and Weaver
\cite{Garfinkle03}. They used
characteristic coordinates that are chosen such that they cover a shrinking
neighborhood of the evolving spike.

The first, and so far apparently only, attempt to study one of the
Gowdy topologies other than $\T$ 
numerically was due to Garfinkle \cite{garfinkle1999}. His observation was that
in the $\Stwo\times\mathbb S^1$-topology case the coordinate
singularity is very similar to the coordinate axis in cylindrical
coordinates and hence a similar
approach as in \cite{Garfinkle00} can be used. For the time evolution,
Garfinkle  used
$2$nd-order finite differencing discretization 
to march forward in time with the iterative Crank-Nicholson
scheme.

\part{Development of my numerical method}
\label{part:treatment}

\chapter{Introduction -- choice of a numerical method}
\label{ch:underlyingquestions}
\markright{Introduction -- choice of a numerical method}

In \Partref{part:preliminaries}, we have given the basic motivations
and collected the necessary
background material. We have selected a class of 
spacetimes, reported on the status of the research and pointed to
some outstanding problems of interest. The main point in this part of this
thesis is to work out our numerical method.

Recall that observations motivated us to restrict our investigations
to the class of 
FAdS spacetimes (\Sectionref{sec:fads}), in particular to
inhomogeneous ones, since these ``almost'' obey the cosmic
no-hair picture in  a very natural way.
As already discussed, we restrict to vacuum because the situation without
matter fields is already complicated enough. 
Although the future behavior of any FAdS solution of EFE is
well understood, it can show complicated and so far not
completely controlled past
behavior; the situation is summarized in
\Sectionref{sec:situation_FADS}. 
In this thesis we want to shed further
light on the possible phenomena of the past behavior 
and on their relation to prescribed properties of $\scrip$ by means of
Friedrich's Cauchy problem.
As we have discussed, we restrict our attention to the cases
when $\scrip$, and hence all Cauchy surfaces, have the topology of
$\S$ or $\T$.

We have decided to do our investigations numerically.
In general,
note the following.
There is a mutual interrelation of the choice of the subject of
investigations and the 
choice of the method which one wants to use; each influences the
other. Because it is sometimes difficult to recognize all
these 
dependencies, in particular when one enters ``new terrain'', 
the process of finding a suitable combination of research
problem and method often has the consequence that, eventually, both 
turn out differently than expected initially. 
This is not problematic in principle, however,
to make this
process as transparent as possible, it is important to collect as much
knowledge as possible before one starts the research. Hence, in this
thesis we try to
discuss our questions of interest always in
correlation with suitable methods, thereby comparing their expected
limitations, advantages and disadvantages.

After this general remark let us turn back to our application
problems; in particular we also want to consider spacetimes with spatial
$\S$-topology. From the 
numerical point of view this topology is
non-trivial and we listed several relevant numerical techniques in
\Sectionref{sec:nontrivialtopologies}. Note that in this thesis, we ignore
completely finite element methods despite their importance for the
solution of problems with complicated topologies.
My decision to write a new code based on pseudospectral methods was
motivated by the following facts. First, spectral methods have
high accuracy compared to 
finite differencing methods as we remarked in
\Sectionref{sec:convergence}. This means that 
it is usually sufficient to use
relatively low resolutions with spectral codes to obtain
the same accuracy as for finite differencing with much higher
resolutions. In many practical situations this means
that spectral codes run faster and do not require so much
computer resources. In 
\cite{boyd} one can find comparisons supporting these statements and a
few practical rules-of-thumb. A
further argument pro spectral methods is that for solutions with
known types of singularities the underlying function space can be
adapted to regularize the problem. The case that we have in
mind are spacetimes with $\S$-topology covered by one dense coordinate
patch, namely the Euler parametrization
(\Sectionref{sec:coordinatesS3}). At the ``boundaries'' of 
this patch one encounters coordinate 
singularities, and since those are of a known type it turns out that
one can 
use a specially adapted basis to make the problem regular. 
In
fact it turns out further that we can consider a map
$\T\rightarrow\S$ such that solutions with $\S$-topologies can be computed
by means of the same underlying spectral infrastructure as in the
$\T$-case. All these issues are discussed in
\Chapterref{ch:treatmentofS3}. 

But, of course, spectral methods also have disadvantages. One of
them is that the
computational cost per grid point is higher 
than for finite differencing. This cost and its scaling with the
discretization parameter $N$
depends strongly on the choice of discrete Fourier transform method. In any
case, this
can mean that when a solution develops spatially localized 
features, it is harder to resolve these features with spectral
methods than with finite differencing. This issue is indeed of
importance in our further investigations.  
Multipatch methods to deal with ``non-trivial''
topologies, in particular those being currently developed
(\Sectionref{sec:nontrivialtopologies}), are certainly
applicable in a quite broad class of such applications;
also for the underlying questions of this thesis.  However,
they are also 
technically involved and so far there is not so much
experience. Furthermore, in spirit of our skeptical philosophy
mentioned at the end of \Sectionref{sec:convergence}, it is desirable
to implement and compare as many very different approaches as
possible to obtain a better feeling for the errors involved.
Spectral methods with their 
geometrical elegance and high accuracy look promising for many of
the applications that we have in mind.
Hence, I
decided to work out a single patch spectral method for the spatial
topologies $\T$ 
and $\S$, and I will discuss it in \Partref{part:treatment} and
\ref{part:analysis} of this thesis.
In any case, our plans for future research certainly involves the
multi-block method \cite{tiglio05}, see further ideas in
\Chapterref{ch:finalchapter}.

In this thesis we restrict the applications to $\lambda$-Gowdy
spacetimes with spatial $\S$- or $\T$-topology.
Gowdy symmetry is also the main motivation to choose
Euler parametrization of $\S$; however, the code is
implemented such that it requires only $\U$-symmetry.
Further the implementation is
done such that generalizations of the code to situations without symmetry
in future work are straight 
forward.
Of particular interest for us in this thesis are numerical approaches to Gowdy
singularities, and it is indeed questionable if spectral methods are suitable
for this situation. The reason is that localized features are common
at singularities
in Gowdy spacetimes. Nevertheless, as stated before, the possibility
of a nice treatment 
of the coordinate 
singularity of $\S$ made me decide to experiment with spectral method in
these situations, and we will discuss whether in our applications the
advantages of 
pseudospectral methods   prevail the disadvantages 
when we analyze our numerical runs in
\Partref{part:analysis} of this thesis. We should point out that such
investigations 
do certainly not allow us to conclude how good or bad our
method performs in other situations. Although the expectation is that
the applicability of spectral methods to approach Gowdy spacetimes is
limited, 
we summarize some ideas for applications, not directly related to singular
Gowdy spacetimes, where we expect that our
method is particularly \textit{well} adapted in
\Sectionref{sec:outluckfuture}. 

To summarize so far, the two main points of attention for this
thesis are first, to find a reliable way to cope with the presence
of the coordinate 
singularities of the Euler parametrization of $\S$ (main focus of
\Partref{part:treatment}), and second to 
optimize the code for approaches to 
Gowdy singularity. Speculations about going ``beyond Gowdy'' and other
interesting applications are discussed at the end of this thesis in
\Sectionref{sec:outluckfuture}. 

After all these thoughts about the numerical method to tackle our
questions of interest we still have to choose appropriate
formulations of EFE (\Sectionref{sec:maximaldevelopments}). Since we
want to apply Friedrich's Cauchy problem 
(\Sectionref{sec:initial_data}) we need 
to deal with Friedrich's conformal field equations
(\Sectionref{sec:conformal_geom}). 
The general formulation of the
conformal field equations as in
\Eqsref{eq:regular_conformal_field_equations} is quite complicated and
although there 
are attempts for their numerical usage (see references in
\Sectionref{sec:num_intro}) there is not so much 
experience. The general conformal field equations
(\Eqsref{eq:gcfe}) represent the conformal field equations
in
conformal Gauß gauge, and the corresponding evolution equations simplify
drastically. For my thesis,
I only considered a further specialization of the conformal
Gauss gauge, namely the Levi-Civita conformal Gauß gauge
(\Sectionref{sec:LCCGG}). Under the conditions of
\Propref{prop:IDLCCGG}, the evolution systems takes the form
\Eqsref{eq:gcfe_levi_cevita_evolution}. 
The main
reason why I did not implement the equations
with their full conformal Gauß gauge freedom is simplicity. On the one
hand there are more unknowns in the equations when the Weyl $1$-form
$f$ does not vanish. On the other hand, for
general Weyl $1$-forms, the spatial frame vector fields $\{e_a\}$
develop a non-vanishing time coordinate component in general, even if it
vanishes initially. Hence, the matrix
in front of the time derivatives in the evolution
equations derived from the Bianchi system does not equal the unit
matrix and has to be 
inverted numerically. Although this is no principal problem, 
I decided to avoid the possible complications related to
this as a first step. Another reason to start with the Levi-Civita
conformal Gauß gauge is that symmetries are 
represented nicely as was remarked in \Sectionref{sec:LCCGG}, but this
is not so in general conformal Gauß gauges.
During my thesis work I obtained
some experience with this system; in particular it turns out that,
although it is suitable to compute \textit{regular} $\lambda$-Gowdy solutions
of $\S$-topology, it
is not so well adapted to approach a Gowdy
singularity (\Sectionref{sec:runsGCFE}) due to problems with the gauge
and constraint growth. 
A ``nicer''
evolution system from this point of view is given by the commutator
field equations 
(\Sectionref{sec:commutatorfieldequations}). Numerical experience by
other people suggests that it is
well-behaved in the situations we have in mind. 
This system relies on the assumption of spatial $\T$-topology and
Gowdy symmetry. It is an outstanding  problem, which has not been solved in
this thesis, if such a system can be
written down also for $\S$-topology.
The $\T$-Gowdy case has been studied often before
numerically (mostly using the system \Eqsref{eq:Gowdy_equations}) with
vanishing cosmological constant, see 
\Sectionref{sec:gowdyphenom}, and we will use the commutator field
equations to obtain 
experience with our spectral infrastructure in particular for
approaching Gowdy singularities. Further, we will study, so far
non-systematically,  the outstanding
case with positive cosmological constant for spatial $\T$-topology by
comparing directly the cases $\lambda=0$ and $\lambda>0$.

\Partref{part:treatment} of this thesis is organized as follows.
We spend most of
the time to explain, discuss and develop our method.
\Chapterref{ch:pseudospectral_impl} is devoted to the description of
our spectral infrastructure including certain adaption techniques.
In \Chapterref{ch:treatmentofS3}, we analyze how $\S$-topology
can be treated within the same spectral infrastructure and how to
deal with the coordinate singularity. Since we have Gowdy symmetry
in mind in this thesis, we further discuss some issues related to
Gowdy symmetry and $\S$-topology.
In \Chapterref{ch:settingupCauchyProblems} we 
construct special classes of initial data which will be used later
in this thesis for numerical computations.

\chapter{Pseudospectral implementation}
\label{ch:pseudospectral_impl}

In \Chapterref{ch:underlyingquestions} we gave the motivation to use
spectral methods to treat the two main problems of interest in
this thesis, namely, the coordinate singularity of
$\S$ and the approach to Gowdy singularities.
This chapter is devoted to the description and discussion of the
spectral infrastructure underlying my
numerical code.
The collocation method with standard Fourier
basis (\Sectionref{sec:pseudospsectr_background}) is particularly
promising and simple to implement for computations of spacetimes
with $\T$-topology. Our way to incorporate  $\S$-topology (and in a
special case also \SoXSt) is described in 
\Chapterref{ch:treatmentofS3}. 
For this thesis work, we decided to use the so called partial
summation method, which we are going to describe in
\Sectionref{sec:myspectralcode}, to compute 
discrete Fourier transformations instead  
of FFT to simplify the implementation. However, nothing prevents us
from switching to FFT for future applications.
In our applications it turns
out quickly that it is problematic to work with fixed resolutions. Hence we
introduce simple adaption techniques, presented in
\Sectionref{sec:spatial_adaption}. 
After that in \Sectionref{sec:implequations}, we make a few
comments about the implementation of the evolution equations. 
The evolution systems that we implement are the 
GCFE in Levi-Civita conformal Gauß gauge (\Sectionref{sec:LCCGG})
and the commutator  
field equations (\Sectionref{sec:comm_field_eqs}).
 
I developed the whole code independently in the programming language
\textit{Fortran 90} \cite{gehrke}. I used the Intel Fortran Compiler
Version 9.0 \cite{Intel} on Intel Pentium 4 processors with compiler
\mbox{options '-O3 -xN'}.

\section{My pseudospectral infrastructure}
\label{sec:myspectralcode}
I implemented the collocation method described in
\Sectionref{sec:pseudospsectr_background} with the same conventions for
the spectral coefficients as given (in one 
dimension) by \Eqref{eq:trigpolyn}. For the method of lines discussed
in \Sectionref{sec:mol} I use the $4$th-order Runge Kutta scheme
\Eqsref{eq:rk4}. As already indicated at the end of
\Sectionref{sec:pseudospsectr_background}, a naive implementation of  
the discrete Fourier transform as given by
\Eqref{eq:DFT} is not practical since, say, in one dimension with $N$
gridpoints, $N$ numerical operations have to be performed for each
spectral coefficient,
giving a total number of $N^2$ computations. In two dimensions this
would already be $N^4$ computations etc. The Fast Fourier
Transform (FFT) algorithm scales as
$N\log_2 N$ in one dimension \cite{numericalrecipes} which is an
improvement in 
particular for high 
$N$. Although the FFT is not so difficult to implement, one could even
use the highly optimized libraries e.g.\ \cite{FFTW}, I decide to
use \term{partial summation} \cite{boyd} which I am going to describe
briefly in a moment because of its simplicity. Substituting this
method by FFT in future work 
promises
speed 
improvements in particular for high spatial resolutions.

The idea of partial summation is the following. Since I will assume
one spatial symmetry and hence my entire code
is so far $2$-dimensional I restrict to the description of the
$2$-dimensional case. To simplify the notation let us further suppose
that we have a grid on $\mathbb T^2$ with equally 
many grid points $N$ in both directions (which I do
not assume in my code)
\[(x_i,y_i)=(i,j)\frac{2\pi} N,\quad i,j\in\{0,\ldots,N-1\}.\]
Let $f:\mathbb T^2\rightarrow\R$ be a function and
$f_{i,j}:=f(x_i,y_i)$. The appropriately normalized Fourier basis functions
are abbreviated as $\{\Phi_i:\mathbb T\rightarrow\R\}$ with
$\Phi_{i,j}=\Phi_i(x_j)$ 
such that, in agreement with 
the discrete Fourier transform formulas \Eqsref{eq:discreteFT}, we have
\[a_{i,j}=\sum_{k,l=0}^Nf_{k,l}\Phi_{i,k}\Phi_{j,l}.\]
Here $a_{i,j}$ is the $2$-dim.\ generalization of the vector $\vec a$
which describes the Fourier coefficients. Computing this sum for each
of the $N^2$ coefficients $a_{i,j}$ involves, as said above, $N^4$
computation steps. The trick is to rewrite this double sum trivially as
\[\sum_{k,l=0}^Nf_{k,l}\Phi_{i,k}\Phi_{j,l}
=\sum_{l=0}^N\left[\sum_{k=0}^Nf_{k,l}\Phi_{i,k}\right]\Phi_{j,l}
=\sum_{l=0}^N\alpha_{i,l}\Phi_{j,l}\]
with $\alpha_{i,l}:=\sum_{k=0}^Nf_{k,l}\Phi_{i,k}$. The computation of
all $\alpha_{i,l}$ coefficients requires $N^3$ computations. Then
computing the coefficients $a_{i,j}$ out of the latter
coefficients takes another $N^3$. So in total we have needed $2N^3$
computations which is advantageous in comparison to $N^4$ in particular
for higher $N$. However, this partial summation method still scales
worse than the FFT 
algorithm.

\section{Adaption methods}
\label{sec:spatial_adaption}
When computing numerical solutions of PDEs one is
always in the following dilemma. On the one hand one should have (more than)
enough spatial (and time) resolution to resolve all features of the
unknowns at a given time. In particular for pseudospectral codes,
aliasing (\Sectionref{sec:pseudospsectr_background}) can play an
important role when the unknowns are
under-resolved. In
\cite{boyd,canuto88}, certain techniques to handle
aliasing are discussed; the most famous one is the so called
\term{2/3-rule}. Our desire is not to run into problems generated by
aliasing but also to avoid the need for such techniques. Then the only
possibility is to make really sure that the spatial 
resolution is always sufficient. However, on the other hand the
resolution should 
also be not too high, otherwise the runs will take too long or one
exceeds the available memory of the machine used. 

Our approach to avoid this dilemma is the following simple global spatial
adaption technique which, in its current form, is designed for Gowdy
symmetry, but which can be generalized easily.
During the run at each time step the program
computes the Fourier transform of one representative unknown; which
one to choose requires some amount of experiments. Then the code determines
how much power this unknown has in the upper third of the
$x_1$-frequency spectrum compared to the total power, where $x_1$ is
that coordinate direction corresponding to the Gowdy inhomogeneity
direction. Usually by 
``power'' 
in accordance with the Parseval equality \cite{sugiura}, one
means the sum 
over the squares of the amplitudes of all frequencies; here, for
various reasons, we sum the absolute values of the frequencies. We
refer to this as the \term{adapt norm} $\normadapt$. A
typical plot of this norm for runs with the general conformal
field equations in Levi-Civita conformal Gauß gauge, which we discuss later,
is shown in \Fignref{fig:typical_adaption_norm} and
\Fignref{fig:S3Gowdyadapt}. Starting from low initial values basically
given by machine precision one typically finds strong exponential
increase of this norm with time. Now, the code is implemented such
that a threshold value can be fixed so that,
as soon as the adapt norm exceeds the threshold value,
the code
stops, interpolates the unknowns with a
higher spatial resolution and continues the run. An
empirically reasonable recipe is to increase the spatial resolution
by $10\%$  when
that happens. In the figures mentioned above, one sees
that then the adaption norm jumps to a lower value. With increasing
time it increases again so that another adaption step will be required
soon. Practically, it turns out to be impossible to keep the
threshold value fixed for the whole run, otherwise the adaption
process becomes instable sooner or later. For instance in
\Fignref{fig:typical_adaption_norm}, the threshold was first at
$10^{-12}$, then at $10^{-11}$ and then $10^{-10}$. It can be seen
from the figure that the time between adaption steps decreases approximately
exponentially  as the singularity at $t\approx 0.9$ is 
approached.

Note that this is a very primitive adaption method since it is
global in space. In particular for spacetimes which develop sharp localized
features, as for instance the spikes in Gowdy spacetimes, a local
adaption method in space would be desirable. Some discussion can be found in
\Sectionref{sec:outluckfuture}.

Another important issue is the right choice of time resolution. One
can expect that typical spacetimes that we want to compute have phases
with only little dynamics on the one hand but which develop strongly
time-dependent features on the other hand when for example a
singularity is approached. It would be nice to have some sort of
automatic adaption 
technique to choose a time step dependent on the current
circumstances of the evolution. Adaptive
Runge-Kutta methods are well known, see for instance
\cite{numericalrecipes}, but these methods also decrease the
integration speed. We can be expect this to be a particular issue when
pseudospectral methods are used since the necessary
trial integration steps are particularly expensive. For the time being the
following primitive time 
adaption methods has turned out to be sufficient; at least for the
conformal field equations (\Sectionref{sec:comm_field_eqs}) which are
conformally invariant. For the 
commutator field equations (\Sectionref{sec:comm_field_eqs}) no time
adaption seems to be necessary because of a ``good'' gauge choice. The right
hand sides of the evolution 
equations \Eqsref{eq:gcfe_levi_cevita_evolution} of the conformal
field equations in Levi-Civita conformal Gauß gauge determine the speed
of the dynamics of the unknowns, i.e.\ their time derivatives. Hence,
say, the $L^1$-norms of all unknowns at a given time step can be
considered as a rough estimate of the speed of the dynamics. The idea
is to keep these norms at or below order unity by conformal rescalings;
this should control the speed of the dynamics to some degree. At
a given time step, the unknowns change under a conformal
rescaling $g\rightarrow \Theta^{-2} g$ with $\Theta$ constant in
space and time in the following way
\begin{equation}
  \label{eq:unknowns_rescale}
  \left(e_a^b, \chi_{ab}, \Connection abc, L_{ab}, E_{fe},
        B_{fe}, \Omega\right)\rightarrow
\left(\Theta e_a^b, \Theta \chi_{ab}, \Theta \Connection abc, 
  \Theta^2 L_{ab}, \Theta^3 E_{fe}, \Theta^3 B_{fe}, \Theta^{-1}\Omega\right),
\end{equation}
and, since we require that $\partial_t$ is orthogonal to the
$t=const$-hypersurfaces and of unit length (with respect to the conformal
metric), we find 
\begin{equation}
  \label{eq:timestep_rescale}
  dt\rightarrow\Theta^{-1} dt.
\end{equation}
Choose $0<\Theta=const<1$,
then the rescaled quantities are decreased according to
\Eqref{eq:unknowns_rescale}. Choosing however a constant time step,
i.e.\ not rescaling the 
time step as suggested by \Eqref{eq:timestep_rescale}, leads to an
effective increase of the time resolution. Hence by such a conformal
transformation with fixed time step one both yields smaller unknowns
and a higher time resolution. Now, the code monitors the orders of
magnitude of the $L^1$-norms of the unknowns and performs a conformal
transformation with $\Theta=1/2$ when some of the unknowns reach the
order of magnitude $10^1$. In fact, so far this is not yet implemented as
an automatic adaption method; those rescalings have to be done rather
manually. 

\section{Implementation of evolution equations and control
  quantities}
\label{sec:implequations}
Suppose we have implemented the underlying infrastructure for discretization on
$\T$ as described in the previous sections.  The next step is to implement
the evolution equations; further geometric and control quantities like
the constraint quantities and the Kretschmann scalar are also required
to analyze and interpret the numerical solutions. I
implemented the general conformal field equations in Levi-Civita
conformal Gauß gauge, see \Sectionref{sec:LCCGG}, and the commutator
field equations discussed in \Sectionref{sec:comm_field_eqs}. 

Let us comment on the frame coefficients $e\indices{_a^\alpha}$ 
given by $e\indices{_a^\alpha}:=\scalarpr{dx^\alpha}{e_a}$
in
\Eqsref{eq:gcfe_levi_cevita_evolution}. If the spatial slices have
$\T$-topology then we can assume that the spatial frame  is globally
smooth and so the functions $e\indices{_a^\alpha}$ with respect
to the standard coordinates on $\T$ are regular everywhere. However,
on $\S$ with the Euler parametrization coordinates
(\Sectionref{sec:coordinatesS3}), the functions $e\indices{_a^\alpha}$
are singular even if the frame is globally smooth. So let us not use
the coordinate components of the orthonormal frame but the components
$e\indices{_a^b}$ 
with respect to a smooth standard frame $\{E_a\}$
given by $e_a=e\indices{_a^b}E_b$
as variables in the
conformal field equations. In the $\T$-case we can set
$E_a=\partial_{x^a}$. In the $\S$-case, however, we will set $E_a=Y_a$
(\Sectionref{sec:identification_su2s3}). 
With this choice, all variables in the equations are
regular functions everywhere. Possible 
coordinate singularities are ``shifted into the frame $\{E_a\}$''
where they have to be controlled, see \Chapterref{ch:treatmentofS3}
for the \S-case.

The implementation of the equations and the other quantities was done
using \Mathematica \cite{Mathematica}. There exist special
packages for \Mathematica which are devoted to make the
representation of geometric quantities in pseudo-Riemannian geometry
as simple as possible; however, I have not used any of them and all
geometric quantities were implemented in \Mathematica manually. Since
this yields more control over the \Mathematica code, this makes it
also straight 
forward to implement the evolution equations and related quantities
and check the expressions carefully. At the end, the
Mathematica package \textit{Format.m} \cite{Format} was used to
generate optimized Fortran 90 code directly from those \Mathematica
expressions.

\chapter{Treatment of \texorpdfstring{$\S$}{S3}-topology} 
\label{ch:treatmentofS3}
\section{Introduction}
As stated before,  our aim is to compute FAdS spacetimes and to focus on
cases with Cauchy surfaces of $\T$- and 
$\S$-topology. In the previous \Chapterref{ch:pseudospectral_impl}, the
discussion was devoted to the description of the underlying spectral
infrastructure which can be used directly to compute spacetimes with spatial
$\T$ topology. Here we show how to incorporate the $\S$-case.
Namely, it turns out that the same infrastructure can be used because $\S$
can be treated as given by a map 
$\T\rightarrow\S$ whose properties we discuss in
\Sectionref{sec:map_Phi}. Because in the applications of this method
in this thesis we will restrict to Gowdy symmetry, we also discuss some
important properties of Gowdy 
symmetric metrics on $\S$ in \Sectionref{sec:GowdyS3}. 

During most of the discussion, we restrict to the case when the fields
on $\S$ are $\U$-symmetric (\Defref{def:U1symm}).  As explained before,
the quotient manifold obtained, when the natural action of this
symmetry group is divided out from $\S$, is $\Stwo$. So  in fact, the
numerical  
computations regarding spatial $\S$-topology with $\U$-symmetry
all have spatial $\Stwo$-topology. Many numerical schemes for 
problems on 
$\Stwo$ are known in the literature; some of
them are presented or at least referenced in \cite{boyd}. However, our
aim is to develop a numerical method such that the $\U$-symmetry
assumption can be dropped at some point. So it would not be wise
to use an algorithm that is restricted to this symmetry. Vice versa, our
``$\S$ point of view to $\Stwo$'' might lead to new approaches for
problems on $\Stwo$.  

For our problems in mind the spatial topology \SoXSt
is as interesting as $\T$ and $\S$. Now, what we said above
means that our method can also be applied to study spacetimes with
$\SoXSt$-topology when there is a symmetry group
acting transitively on the $\mathbb S^1$-part. However, this possibility is
not yet investigated in this thesis.

In \Sectionref{sec:S3GenFourier} we introduced the spin-spherical
harmonics which constitute a basis for the square integrable functions
on $\S$. In principle, this basis can be used to set up a spectral
method for computations involving spatial $\S$-topologies. Some
comments for the $2$-dimensional case, i.e.\ $\Stwo$ with spherical
harmonics, can be found in \cite{boyd}. However, in the $3$-dim.\ case there
seems to be not much experience; in particular no
efficient algorithm for the generalized discrete Fourier transform
is known apparently. Alternatively, one could try to implement a
Garlerkin approach based on spin spherical harmonics, but 
for this one would need to compute generalized convolutions. Such calculations
are expensive numerically because the  
determination of each convolution coefficient involves the sum of the
products of 
all pairs of coefficients. I do not claim that such an approach is not
feasible, however,
these were considerations which motivated me to try the method
which we present in this chapter. 

\section{Numerical treatment of the coordinate singularity on 
\texorpdfstring{$\S$}{S3}}
\subsection{The map \texorpdfstring{$\T\rightarrow\S$}{T3->S3}}
\label{sec:map_Phi}

In this section we define a map $\T\rightarrow\S$ and characterize
functions, vector fields and symmetries compatible with it. By
means of this map, all 
computations on $\S$ involved in the field equations can be performed
in a well-defined sense on $\T$ and hence the numerical pseudospectral
infrastructure for $\T$ can be applied to compute spacetimes with
spatial $\S$-topology.

\subsubsection{Definition and basic properties}
Recall again \Eqref{eq:euler_param} where the Euler
parametrization of $\S$ in terms of the coordinate
$(\chi,\rho_1,\rho_2)$ is defined for
\[(\chi,\rho_1,\rho_2) \in \left]0,\pi/2\right[\times 
\left[0,2\pi\right[\times\left[0,2\pi\right[.\] 
However, it is no problem to extend the domain to $\R^3$. Since the
functions involved are $2\pi$-periodic, this yields a well-defined map
\[\Phi:\T\rightarrow\S\]
with $\T=(\R\text{ mod } 2\pi)^3$. It has the
following properties: 
\renewcommand{\labelenumi}{(\roman{enumi})}
\begin{enumerate}
\item $\Phi$ is surjective and smooth, but not injective.
\item The preimage of $\tildeS$ (defined by \Eqref{eq:definition_tildeS})
  has $4$ connected components\footnote{More details on this are
    given when we discuss invariances of $\Phi$ below.} in $\T$, each of
  which is diffeomorphic to $\tildeS$. For an arbitrary choice of such
  a connected
  component on $\T$, let us fix one of these
  diffeomorphisms
  \[\tilde\Phi:\tildeS\rightarrow\tilde\Phi(\tildeS)\subset\T\]
  by the requirement $\Phi\circ\tilde\Phi=\mathrm{id}_{\tildeS}$.
  This diffeomorphism
  cannot be extended to $\S$ as a continuous map.
\item The function
  $\Phi\circ\tilde\Phi$ can be extended continuously to the 
  identity on
  $\S$. Since $\tildeS$ is a  dense subset of $\S$ this extension is
  the unique continuous extension. 
\end{enumerate}
One can state that $\Phi$ is a covering map in the algebraic, but not
in the topological sense.

Define for each possible $n,i,k$
\[\breve w^n_{ik}:\T\rightarrow\C,\quad \breve w^n_{ik}=w^n_{ik}\circ\Phi\]
which are in $C^\infty(\T)$ with the functions $w^n_{ik}$ defined in
\Sectionref{sec:S3GenFourier}. Their explicit representation is 
given by \Eqref{eq:wnik}. Now, let $f\in C^\infty(\S)$. According to
\Theoremref{th:smooth_fourier} there exist rapidly decreasing coefficients
$a^n_{ik}\in\C$ such that pointwise
\[f=\sum_{n,i,k}a^n_{ik}w^n_{ik}\]
and the convergence is absolute and uniform. Define 
$\breve f:=f\circ\Phi$ which is a smooth function on $\T$ with the
pointwise representation
\[\breve f=\sum_{n,i,k}a^n_{ik}w^n_{ik}\circ\Phi
=\sum_{n,i,k}a^n_{ik}\breve w^n_{ik};\]
the coefficients are rapidly decreasing and the convergence is
absolute and uniform on $\T$. Hence $\Phi$ induces a map from
$C^\infty(\S)$ to the set of functions
\begin{gather}
  \label{eq:definition_breveX}
  \begin{split}
    \breve X&:=\left\{\breve f\in C^\infty(\T),\,\exists\text{ rap.\
        decr.\ coeff. } a^n_{ik}\in\C\text{ such that pointwise }\breve f
      =\sum_{n,i,k}a^n_{ik}\breve w^n_{ik} \right\}\\
    &\subset C^\infty(\T).
  \end{split}
\end{gather}
Note that for any function $\tilde f\in\tilde X$, absolute and uniform
convergence is automatically guaranteed since the coefficients are
rapidly decreasing and the functions $\breve w^n_{ik}$ can be
estimated uniformly according to \Lemref{lem:estimates_basis}.

Now, let vice versa a function $\breve g\in\breve X$ be given by
the pointwise representation 
\[\breve g=\sum_{n,i,k}b^n_{ik}\breve w^n_{ik}\]
with rapidly decreasing coefficients $b^n_{ik}\in\C$. Let $\tilde\Phi$
be the diffeomorphism defined above corresponding to one of the
connected components 
of the preimage of $\tildeS$. We set
\[ g:=\breve g\circ\tilde\Phi\]
which is a smooth function on $\tildeS$ and can also be written as
\[g=\sum_{n,i,k}b^n_{ik} w^n_{ik}\circ(\Phi\circ\tilde\Phi).\]
It is clear that $g$ does not depend on the choice of the connected
component.
The function $\sum b^n_{ik} w^n_{ik}$ is continuous on $\S$ since the
coefficients are rapidly decreasing and due to the uniform estimates
for the basis functions (\Lemref{lem:estimates_basis}). Furthermore,
$\Phi\circ\tilde\Phi$ extends continuously to the identity on
$\S$. Hence, $g$ can be extended in a unique way continuously to the
function $\sum b^n_{ik} w^n_{ik}$ on $\S$ and this extension is also
denoted as $g$. 
Now, according to \Theoremref{th:smooth_fourier}, $g$ is even smooth
on $\S$.  

In summary, the map $\Phi$ induces a bijection between the space
$C^\infty(\S)$ and the space $\breve X\subset C^\infty(\T)$.
It is remarkable that, although $\Phi$ is not diffeomorphism,
we are able transport function from $\T$ to $\S$ (and vice versa) in a
well-defined way, but only if we 
restrict to the function space $\breve X\subset
C^\infty(\T)$. In the following we will use the symbol $\Phi$ for both
the map $\T\rightarrow\S$ that we started with and the just
constructed induced bijection $C^\infty(\S)\rightarrow\breve X$.

Next, we want to check if the bijection $C^\infty(\S)\rightarrow\breve
X$ commutes with the 
evaluation of vector fields in a natural way. This would mean roughly
speaking that the
action of a smooth vector field on $\S$ on a smooth function on $\S$
can be evaluated using the corresponding function in $\breve X$ and a
naturally related (possibly non-smooth) vector field on $\T$. Choose
again one of the 
connected components of the preimage of $\tildeS$ under $\Phi$ with
the corresponding diffeomorphism $\tilde\Phi$ as above and set for a
smooth vector field $V$ on $\S$,
\[\breve V:=\tilde\Phi_* V.\]
This is a smooth vector field on $\mathrm{Im }\tilde\Phi\subset\T$ but
note that it cannot be extended to a smooth vector field on $\T$. Now,
choose $\breve f\in\breve X$ and let $f$ be the corresponding function in
$C^\infty(\S)$. For $x\in\mathrm{Im }\tilde\Phi$ we have
\[\breve V_x(\breve f)
=V_{\Phi(x)}(\breve f\circ\tilde\Phi)=[V(f)]\circ\Phi(x),\]
hence
\[\breve V(\breve f)=[V(f)]\circ\left.\Phi\right|_{\mathrm{Im }\tilde\Phi}.\]
Since $\Phi$ is continuous on $\T$ and $V(f)$ is continuous on $\S$,
$\breve V(\breve f)$ extends to the continuous function $[V(f)]\circ\Phi$, also denoted
by $\breve V(\breve f)$, on $\T$. Because $V(f)$ is even an element of
$C^\infty(\S)$, the function $\breve V(\breve f)$ is the unique
element in $\tilde X$ corresponding to $V(f)$ in the manner above.

To summarize, we have shown the following statement.
\begin{Prop}
  \label{prop:frame_derivatives_on_T3}
  Under the conditions and with definitions above, the following
  diagram is well-defined and commutes:
  \[
  \begin{xy}
    \xymatrix{
      C^\infty(\S) \ar[d]_{\Phi} \ar[r]^{V} & C^\infty(\S) \ar[d]^{\Phi}\\
      \breve X \ar[r]_{\breve V} & \breve X 
    }
  \end{xy}
  \]
  Here, $\Phi$ can be considered as a well-defined bijective map
  $C^\infty(\S)\rightarrow \breve X$. 
\end{Prop}

The map $\Phi$ will be used to construct a pseudospectral method for
spacetimes with $\S$-topological spatial slices. The key point is that we
have shown how to evaluate frame derivatives on $\S$  in a consistent
manner on $\T$ despite of the fact that ``coordinate singularities''
are present, having restricted to the right space of functions 
$\tilde X$. 

The analogous map between $\mathbb T^2$ and $\Stwo$ is discussed in \cite{boyd},
Section 18.8. However, the author does not mention any of these consistency
issues, but rather restricts to the invariance properties which we
discuss now in the $3$-dimensional case.

\subsubsection{Invariances of the map \texorpdfstring{$\Phi$}{Phi}}
\label{sec:invariances_Phi}
The map $\Phi$
is not 
injective. Now we will discuss some of its invariance transformations.

\begin{subequations}
\label{eq:invariances}
The first family of invariance transformations is the standard
translation on $\T$, which in 
principle has already been factored out by the definition of $\T$ but
nevertheless is
listed here for completeness. For all $(k_1,k_2,k_3)\in\Z^3$
the map $\Phi$ is invariant under the transformation
\begin{gather}
\label{eq:invariance1}
  \begin{split}
    \chi&\rightarrow \chi+2\pi k_1\\
    \rho_1&\rightarrow \rho_1+2\pi k_2\\
    \rho_2&\rightarrow \rho_2+2\pi k_3.
  \end{split}
\end{gather}
The orientation of the image of $\Phi$ is naturally preserved by each
of these discrete transformations.
Another invariant orientation preserving transformation is given by
\begin{equation*}
  \begin{split}
    \chi&\rightarrow \chi+\pi\\
    \rho_1+\rho_2&\rightarrow \rho_1+\rho_2+\pi\\
    \rho_1-\rho_2&\rightarrow \rho_1-\rho_2+\pi.
  \end{split}
\end{equation*}
This can be formulated equivalently in terms of the two distinct
possibilities, taking the discrete translations \Eqref{eq:invariance1}
into account, 
\begin{gather}
\label{eq:invariance2}
\begin{split}
  \chi\rightarrow& \chi+\pi\\
  \left\{\begin{aligned}
    \rho_1&\rightarrow \rho_1\\
    \rho_2&\rightarrow \rho_2+\pi
  \end{aligned}\right\}\quad &\text{or}\quad
  \left\{\begin{aligned}
    \rho_1&\rightarrow \rho_1+\pi\\
    \rho_2&\rightarrow \rho_2
  \end{aligned}\right\}.
\end{split}
\end{gather}
The last invariance transformation that we write down is
\begin{equation*}
  \begin{split}
    \chi&\rightarrow \pi-\chi\\
    \rho_1+\rho_2&\rightarrow \rho_1+\rho_2+\pi\\
    \rho_1-\rho_2&\rightarrow \rho_1-\rho_2
  \end{split}
\end{equation*}
and yields the following two distinct possibilities
\begin{gather}
\label{eq:invariance3}
\begin{split}
  \chi\rightarrow& \chi-\pi\\
  \left\{\begin{aligned}
    \rho_1&\rightarrow \rho_1+\frac\pi 2\\
    \rho_2&\rightarrow \rho_2+\frac\pi 2
  \end{aligned}\right\}\quad &\text{or}\quad
  \left\{\begin{aligned}
    \rho_1&\rightarrow \rho_1+\frac{3\pi} 2\\
    \rho_2&\rightarrow \rho_2+\frac{3\pi} 2
  \end{aligned}\right\}.
\end{split}
\end{gather}
These latter transformations are not orientation preserving.
Now, invariances \Eqsref{eq:invariance2} and \eqref{eq:invariance3} imply that
the preimage of $\tildeS$ under $\Phi$ has four connected
components. We will exploit these invariances in
\Sectionref{sec:spectral_analysis_smooth}.  
\end{subequations}

For $\U$-symmetric functions in $\breve X$, i.e.\ functions $\breve f$
with $Z_3(\breve f)=\partial_{\rho_2}\breve f=0$ (\Eqref{eq:Z3Euler}),
we have a continuous 
invariance transformation, namely 
\begin{equation}
  \label{eq:comp_cond}
  \left.\partial_{\rho_1} \breve f\right|_{\chi=k\pi/2}=0,\quad\forall k\in\Z.
\end{equation}
This is so because $Y_3=\partial_{\rho_1}$ and $Z_3=\partial_{\rho_2}$ (as
vector fields on $\S$) are linear dependent at the points 
$\chi=k\pi/2$ ($k\in\Z$) and because $\breve f$ is constant along $Z_3$
everywhere. 

We know that any $f\in C^\infty(\T)$ which is not invariant under any of the
transformations \Eqsref{eq:invariances} cannot be in $\breve X$. However,
the invariance of $f$ 
under these transformations listed there is not sufficient to conclude that
$f\in\breve X$. There are further smoothness requirements. Let us
restrict to the $\U$-symmetric case. One of these requirements
is implied by \Eqsref{eq:wnp_representation_explicit}: any factor
$e^{-2i p \rho_1}$ must be multiplied by a function in $\chi$ which
has zeros of order $p$ at $\chi=k\pi/2$ for all $k\in\Z$. This has
consequences for the Fourier representation of a function in $\breve
X$ but is not yet exploited neither in the
following discussion nor in the code. In the following discussions the
invariance transformations listed above turn out to be sufficient. Possible
further
conditions for $f$ being in $\tilde X$ have not been investigated yet.

From now on we will not distinguish anymore between the function spaces
$C^\infty(\S)$ and $\breve X$, and just write $f\in C^\infty(\S)$
instead of $\breve f\in\breve X$. 

\subsection{Spectral analysis of smooth functions on 
\texorpdfstring{$\S$}{S3} and their frame derivatives}
\label{sec:spectral_analysis_smooth}

\subsubsection{Structure of Fourier series of functions in 
\texorpdfstring{$C^\infty(\S)$}{CInfty(S3)}}
\Eqsref{eq:invariance2} and \eqref{eq:invariance3} give us 
information on the structure of the 
Fourier series of functions  $f\in C^\infty(\S)$. Note that by
``Fourier representation'' we do not mean the series representation
with respect to the basis $w^n_{ik}$ in
\Eqref{eq:standard_repr_smooth_fct_s3}, but rather the expansion
\begin{equation}
  \label{eq:FourierDecInX}
  f(\chi,\rho_1,\rho_2)
  =\sum_{(n,p,q)\in\Z^3}f_{n,p,q}\,e^{i n \chi}e^{i p \rho_1}e^{i q \rho_2}
\end{equation}
with coefficients $f_{n,p,q}\in\C$, after having identified
$C^\infty(\S)$ and $\breve X$. The reality condition is
\begin{equation}
  \label{eq:reality_cond}
  f_{n,p,q}=\overline{f_{-n,-p,-q}}.
\end{equation}
Note furthermore that in this section we do not follow the conventions
for the Fourier transformation given by \Eqref{eq:trigpolyn} to
simplify the notation.

The standard theory for Fourier
series for smooth functions on $\T$ tells us that a series as in
\Eqref{eq:FourierDecInX} converges pointwise 
absolutely and uniformly, and $f_{n,p,q}$ are rapidly decreasing in $n$,
$p$ and $q$, cf.\ \cite{sugiura} for example. 
Smooth functions on $\S$ expressed in Euler coordinates must have
the same invariances as the map $\Phi$. Invariance \Eqref{eq:invariance2}
implies that
\nobreak 
\begin{subequations}
\label{eq:cond_Fourier}
\begin{equation}
  \label{eq:cond_Fourier1}
  f_{n,p,q}=0,\,\text{if $n+p$ odd or $n+q$ odd}.
\end{equation}
From \Eqref{eq:invariance3} it follows that
\begin{equation}
  \label{eq:cond_Fourier2}
  f_{n,p,q}=(-1)^n i^{p+q} f_{-n,p,q}\quad\text{and}\quad 
  f_{n,p,q}=0,\,\text{if $q+p$ odd}.
\end{equation}
\end{subequations}

We will mostly be interested in smooth $\U$-symmetric real functions. For
those we find the following.
\begin{Prop}
  \label{prop:fourier_U1}
  \hypertarget{prop:fourier_U1}
  Let $f$ be a $\U$-symmetric real valued function in 
  $C^\infty(\S)$. Then its Fourier representation reduces to
  \begin{equation}
    \label{eq:writef}
    f(\chi,\rho_1,\rho_2)=F_0(\chi)
    +2\text{Re}\sum_{p=1}^\infty F_p(\chi) e^{2ip\rho_1}
  \end{equation}
  with
  \begin{equation}
    \label{eq:writeF_p}
    F_p(\chi)=
    \begin{cases}
      {\displaystyle 2
        \sum_{n=1}^\infty f_{n,p}\cos 2n\chi+f_{0,p}} &
      \text{for $p\ge0$ even}\\
      {\displaystyle -2 i
        \sum_{n=1}^\infty f_{n,p}\sin 2n\chi} & \text{for $p>0$ odd}
    \end{cases}
  \end{equation}
  where $f_{n,0}\in\R$ for all $n\in\N$ (including zero).
  For even $p>0$, the coefficients satisfy the following compatibility
  conditions
  \begin{equation}
    \label{eq:comp_cond_coeff}
    f_{0,p}+2\sum_{n=1}^\infty f_{2n,p}=0,\quad 
    \sum_{n=1}^\infty f_{2n-1,p}=0,
  \end{equation}
  and hence 
  \begin{equation}
    \label{eq:writeF_p_even}
    F_p(\chi)=2
    \sum_{n=1}^\infty\left\{f_{2n,p}(\cos 4n\chi-1)+
      f_{2n+1,q}\bigl(\cos(4n+2)\chi-\cos 2\chi\bigr)\right\}
  \end{equation}
  for even $p>0$.     
  The coefficients $f_{np}$ are rapidly decreasing in $n$ and in $p$.  
  \begin{Proof}
    We write 
    $f=\sum_{p=-\infty}^\infty F_p(\chi)\exp(2ip\rho_1)$. This
    can be done since for smooth $\U$-symmetric functions there is only the
    $q=0$-mode in \Eqref{eq:FourierDecInX} and hence conditions
    \Eqsref{eq:cond_Fourier1} imply that all modes vanish except for
    $p$ and $n$ even. The functions $F_p$ are given by
    $F_p=\sum_{n=-\infty}^\infty f_{n,p}\exp(2in\chi)$. But note that $n$
    and $p$ here differ by a factor $2$ from those used in
    \Eqref{eq:FourierDecInX} and hence in \Eqsref{eq:cond_Fourier}. Using the
    relation
    $f_{n,p}=(-1)^p f_{-n,p}$ obtained from \Eqref{eq:cond_Fourier2},
    this can be written as
    \[F_p(\chi)
    =\sum_{n=1}^\infty\left(f_{n,p}e^{2in\chi}+f_{-n,p}e^{-2in\chi}\right)
    +f_{0,p}
    =\sum_{n=1}^\infty f_{n,p}\left(e^{2in\chi}+(-1)^pe^{-2in\chi}\right)
    +f_{0,p}.\]
    The same relation implies that $f_{0,p}=0$ if $p$ is odd. The
    reality condition $f_{-n,-p}=\overline{f_{n,p}}$ implies that
    $F_{-p}=\overline{F_p}$, namely
    \[F_{-p}=\sum_{n=1}^\infty f_{n,-p}\left(e^{2in\chi}
      +(-1)^{-p}e^{-2in\chi}\right)+f_{0,-p}
    =\sum_{n=1}^\infty \overline{f_{n,p}}\left(e^{-2in\chi}+(-1)^pe^{2in\chi}\right)
    +\overline{f_{0,p}}.\]
    Thus, $f$ can be written as in \Eqref{eq:writef}. In particular
    $F_0$ is real-valued, hence $f_{n,0}\in\R$. This proves the
    expressions in \Eqref{eq:writeF_p}. The invariance condition
    \Eqref{eq:comp_cond} implies that
    \[F_p(k\pi/2)=0\text{ for all }k\in\Z\text{ and }\forall p\ge 1.\]
    \Eqref{eq:writeF_p} implies that this is automatically
    fulfilled for odd $p>0$. However, for even $p>0$ it leads to the
    conditions \Eqsref{eq:comp_cond_coeff}. Using these compatibility
    condition, $F_p$ can be rewritten as in \Eqref{eq:writeF_p_even}.
  \end{Proof}
\end{Prop}

\subsubsection{\texorpdfstring{$\U$}{U(1)}-symmetric
  functions at the coordinate singularity} 
The aim of this part is to compute finite approximations of the frame
derivatives 
of smooth $\U$-symmetric functions on
$\S$ consistently. Any smooth frame on $\S$ can be considered as a linear
combination of the standard frame $\{Y_a\}$; thus we can restrict to
this special frame here. \Eqsref{eq:coordinate_repr_standard_frame}
tell us what we have 
to do to compute $\{Y_a\}$-derivatives. First, we have to compute the
$\chi$- and 
$\rho_1$-derivatives of the function, and the 
idea of the pseudospectral approach is to do this in spectral
space. Now, in $Y_1$ and $Y_2$ there is the singular factor
$\tan\chi-\cot\chi$. In this section we want to analyze the behavior
of finite approximations of a smooth $\U$-symmetric function on $\S$
being multiplied with this singular factor. First
we need the following simple result. 

\begin{Lem}
  \label{lem:fourier_wnp}
  Let $m\in 2\N$ and $q\in\{-m/2,\ldots,m/2\}$. In the notation of
  \Propref{prop:fourier_U1}, the Fourier coefficients of the
  function $w_{mq}$ vanish for all $n>m$ and for all 
  $p\not=q$. 
  \begin{Proof}
    Consider the explicit
    representation formula given by
    \Eqsref{eq:wnp_representation_explicit}. One sees directly that
    $w_{mq}$ only depends on the $2q$-th $\rho_1$-frequency. To
    determine the highest $\chi$-frequency, write the $\chi$-dependent
    part of its $l$-th summand as
    \begin{equation*}
      \begin{split}
        \cos^{2(l-q)}&\chi\sin^{m-2l}\chi\sin^q\chi\cos^q\chi\\
        &=
        \begin{cases}
          (1-\sin^2\chi)^{l-q}\sin^{m-2l}\chi\sin^q\chi(1-\sin^2\chi)^{q/2}
          &\text{for $q$ even}\\
          (1-\sin^2\chi)^{l-q}\sin^{m-2l}\chi\sin^q\chi(1-\sin^2\chi)^{(q-1)/2}
          \cos\chi&\text{for $q$ odd}.
        \end{cases}
      \end{split}
    \end{equation*}
    In the even case, the highest exponent is $\sin^m\chi$; in the odd
    case $\sin^{m-1}\chi\cos\chi$. Hence, in both cases, the amplitudes of
    $\chi$-frequencies higher than $m$ vanish. This means that in the
    notation of 
    \Propref{prop:fourier_U1}, the series expression for $F_p(\chi)$
    goes only up to $n=m/2$. However, to simplify the following
    discussion, we make nothing wrong when we
    let the series go to $n=m$ by setting all further coefficients to zero.
    In our
    numerical implementation, we make use of the fact that the
    series stops at $n=m/2$.
  \end{Proof}
\end{Lem}

Now, for the analysis of our problem we assume that we
approximate a given smooth $\U$-symmetric function $f$ by smooth
$\U$-symmetric functions\footnote{Be careful to note that in this
  context $f$ with an upstairs 
  $N$  is \textit{not} the $N$th power of $f$,
  but rather a sequence index.} $f^N$ such that 
$f=\lim_{N\rightarrow\infty} f^N$. So each $f^N$ has a representation as in
\Eqref{eq:wnp_representation} with
\begin{equation}
  \label{eq:approxim_assumption}
  f^N:=\sum_{\substack{n=0\\\text{$n$ even}}}^N
  \sum_{p=-n/2}^{n/2} a_{n,p}^Nw_{n,p}
\end{equation}
for all $N\in 2\N$. In particular, $f=\lim_{N\rightarrow\infty} f^N$
absolutely and 
uniformly. 
This convergence assumption is not necessary, however, it turns out to be
helpful in our analysis. In practice this assumption is justified
because we can control the approximation explicitly for the initial
data and then the discretized evolution equations take
care that the 
assumption is fulfilled for all times. However, round-off errors can
spoil this and the code is only stable when we use projections, see
\Sectionref{sec:mymethodS3}. 

Now, we can apply
\Propref{prop:fourier_U1} to $f^N$ together with \Lemref{lem:fourier_wnp}
to find that 
\begin{equation}
  \label{eq:fNFourier}
  f^N(\chi,\rho_1)=F_0^N(\chi)+2\text{Re}\sum_{p=1}^{N/2}F_p^N(\chi)e^{2ip\rho_1}
\end{equation}
with $F_p^N(\chi)$ as in \Eqref{eq:writeF_p} 
\begin{equation}
  \label{eq:FNpFourier0}
  F^N_p(\chi)=
  \begin{cases}
    {\displaystyle 2
      \sum_{n=1}^N f^N_{n,p}\cos 2n\chi+f^N_{0,p}} &
    \text{for $p\ge0$ even}\\
    {\displaystyle -2 i
      \sum_{n=1}^N f^N_{n,p}\sin 2n\chi} & \text{for $p>0$ odd}.
  \end{cases}
\end{equation}
The coefficients
$f^N_{n,p}$ can be computed from the coefficients
$a^N_{n,p}$ by explicitly determining the Fourier coefficients of each
basis function $w_{np}$, but this is not necessary for our
analysis. 
In addition to above, the compatibility conditions hold for even $p>0$
\begin{equation}
  \label{eq:approximation_comp_cond}
  f_{0,p}^N+2\sum_{n=1}^{N/2} f_{2n,p}^N=0,\quad 
  \sum_{n=1}^{N/2} f^N_{2n-1,p}=0,
\end{equation}
and hence
\begin{equation}
  \label{eq:FNpFourier}
  F_p^N(\chi)=2
  \left[\sum_{n=1}^{{N}/2}f^N_{2n,p}(\cos 4n\chi-1)+
    \sum_{n=1}^{{N}/2-1}f^N_{2n+1,q}\bigl(\cos(4n+2)\chi-\cos
    2\chi\bigr)
  \right].
\end{equation}

\begin{Lem}
  \label{lem:singular_trig}
  The following relations are valid for
  $\chi\in\R\backslash(\Z\frac\pi 2)$ and for all $n\in\N$:
  \begin{gather*}
    (\tan\chi-\cot\chi)(\cos 4n\chi-1)
    =2\sum_{k=0}^{n-1}(\sin 4(k+1)\chi+\sin 4k\chi)\\
    (\tan\chi-\cot\chi)(\cos (4n+2)\chi-\cos 2\chi)
    =2\sum_{k=0}^{n-1}(\sin (4k+6)\chi+\sin (4k+2)\chi)\\
    (\tan\chi-\cot\chi)\sin 2n\chi
    =-2\sum_{k=0}^{n-1}\cos2(n-2k)\chi.
  \end{gather*}
  \begin{Proof}
    These identities follow when we write the left hand sides in terms
    of complex exponential functions and manipulate the expressions such that
    the well known formula for geometric sums can be applied.
  \end{Proof}
\end{Lem}

\begin{Cor}
  \label{cor:singtype1}
  Let for a given $N\in 2\N$ the function $F:\R\rightarrow\C$ be given by
  \[F(\chi)=-2i\sum_{n=1}^N c_{n}\sin 2n\chi\]
  where the coefficients $c_n\in\C$ are arbitrary.
  Then, on $\R\backslash(\Z\frac\pi 2)$, we find
  \[F(\chi)(\tan\chi-\cot\chi)=4i\Biggl[
  \mathfrak{b}_{1}+
  \sum_{r=1}^{N/2}\Bigl\{
  (\mathfrak{c}_{r}+\mathfrak{c}_{r+1})\cos(4r-2)\chi
  +(\mathfrak{b}_{r}+\mathfrak{b}_{r+1})\cos 4r\chi
  \Bigr\}
  \Biggr]\]
  with
  \begin{equation}
    \label{eq:def_brcr}
    \mathfrak{b}_{r}:=\sum_{n=r}^{N/2}c_{2n},
    \quad
    \mathfrak{c}_{r}:=\sum_{n=r}^{N/2}c_{2n-1}\quad\text{for }
    r\ge 1.
  \end{equation} 
  Here we understand that $\mathfrak{b}_{r}=\mathfrak{c}_{r}=0$ 
  for $r>N/2$.
  \begin{Proof}
    We have due to \Lemref{lem:singular_trig}
    \[F(\chi)(\tan\chi-\cot\chi)
    =4i\sum_{n=1}^N\sum_{k=0}^{n-1}
    c_{n}\cos2(n-2k)\chi.\]
    Rearranging the terms in this finite sum we find that
    \begin{align*}
      F(\chi)(\tan\chi-\cot\chi)
      =4i&\Biggl[
        \sum_{r=-N/2+1}^{N/2-1}\left\{
          \left(\sum_{n=|r|}^{N/2-1}c_{2n+1}\right)\cos(4r+2)\chi
          +\left(\sum_{n=|r|+1}^{N/2}c_{2n}\right)\cos 4r\chi\right\}\\
        &+\sum_{r=1}^{N/2}c_{2r}\cos 4r\chi
      \Biggr].
    \end{align*}
    Define $\mathfrak{b}_r$ and $\mathfrak{c}_r$ as the sums in the
    brackets. Divide the sums into one for $r<0$, for $r=0$ and for
    $r>0$. The 
    symmetries of the cosine function and the fact that, so far, the
    definitions of $\mathfrak{b}_r$ and $\mathfrak{c}_r$ only involve
    the modulus of $r$ leads to expressions where no negative $r$ is
    present anymore and the modulus can be skipped. Further straight
    forward manipulations involve the shift of indices; in particular
    also $\mathfrak{b}_r$ and $\mathfrak{c}_r$ get redefined
    again appropriately. Eventually the claim is proved.
  \end{Proof}
\end{Cor}

\begin{Cor}
  \label{cor:singtype2}
  Let for a given $N\in 2\N$ the function $F:\R\rightarrow\C$ be given by
  \[F(\chi)=2\sum_{n=1}^N c_{n}\cos 2n\chi+c_{0}\]
  where the coefficients $c_n\in\C$ fulfill
  \[c_{0}+2\sum_{n=1}^{N/2} c_{2n}=0,\quad 
  \sum_{n=1}^{N/2} c_{2n-1}=0.\]
  Then, on $\R\backslash(\Z\frac\pi 2)$, we find
  \[F(\chi)(\tan\chi-\cot\chi)=4\Biggl[
        \mathfrak{c}_{2}\sin 2\chi
        +\sum_{k=1}^{N/2}\Bigl\{
        (\mathfrak{b}_{k}+\mathfrak{b}_{k+1})\sin 4k\chi
        +(\mathfrak{c}_{k+1}+\mathfrak{c}_{k+2})\sin (4k+2)\chi
        \Bigr\}
        \Biggr]\]
   where $\mathfrak{b}_{r}$ and $\mathfrak{c}_{r}$ are defined as in
   \Eqref{eq:def_brcr}.
   \begin{Proof}
     We write the expression for $F$ as in
     \Eqref{eq:FNpFourier}. After having applied 
     \Lemref{lem:singular_trig}, similar manipulations as in
     \Corref{cor:singtype1} lead to the proof.
  \end{Proof}
\end{Cor}

\begin{Prop}
  \label{prop:tancotmultip}
  \hypertarget{sec:tancotmultipl}
  Let $f$ be a smooth $\U$-symmetric function on $\S$ such that
  \[\int_0^{2\pi} f(\chi,\rho_1)d\rho_1=0.\]
  Let $f$ be approximated as in \Eqref{eq:approxim_assumption}, i.e.\
  $f=\lim_{N\rightarrow\infty} f^N$ with $f^N$ smooth $\U$-symmetric functions,
  such that 
  $f^N$ and $F^N_p$ can be expanded as in \Eqref{eq:fNFourier},
  \eqref{eq:FNpFourier0} and  \eqref{eq:FNpFourier} and the
  coefficients fulfill the compatibility condition
  \Eqref{eq:approximation_comp_cond} for even $p>0$.
  Then, on $\tildeS$, we
  can compute the pointwise limit
  \[(\tan\chi-\cot\chi)f(\chi,\rho_1)
  =\lim_{N\rightarrow\infty} (\tan\chi-\cot\chi)f^N(\chi,\rho_1)\]
  by
  \begin{equation*}
    \begin{split}
      &F^N_p(\chi)(\tan\chi-\cot\chi)\\
      &=
      \begin{cases} 
        {\displaystyle
        \begin{split}
           4\Biggl[
            \mathfrak{c}_{2,p}^N\sin 2\chi
            +\sum_{k=1}^{N/2}&\Bigl\{
            (\mathfrak{b}_{k,p}^N+\mathfrak{b}_{k+1,p}^N)\sin 4k\chi\\
            &+(\mathfrak{c}_{k+1,p}^N+\mathfrak{c}_{k+2,p}^N)\sin (4k+2)\chi
            \Bigr\}
            \Biggr]
        \end{split}}
        &\text{for $p>0$ even},\\
      {\displaystyle 4i\Biggl[
        \mathfrak{b}_{1,p}^N+
        \sum_{r=1}^{N/2}\Bigl\{
        (\mathfrak{c}_{r,p}^N+\mathfrak{c}_{r+1,p}^N)\cos(4r-2)\chi
        +(\mathfrak{b}_{r,p}^N+\mathfrak{b}_{r+1,p}^N)\cos 4r\chi
        \Bigr\}
        \Biggr]} &\text{for $p>0$ odd}.
      \end{cases}
    \end{split}
  \end{equation*}
  Here,
  \[\mathfrak{b}_{r,p}^{N}:=\sum_{n=r}^{N/2}f^N_{2n,p},
  \quad
  \mathfrak{c}_{r,p}^N:=\sum_{n=r}^{N/2}f^N_{2n-1,p}\quad\text{for }
  r\ge 1.\]
  \begin{Proof}
    The integral condition implies $F_0^N(\chi)=0$. Now,
    \Corref{cor:singtype1} and \Corref{cor:singtype2} can be applied.
  \end{Proof}
\end{Prop}

It is important to note that we do not claim to have proven uniform
convergence on $\S$ at this point and we do not talk about extending
the analysis to the whole $\S$. In fact, this discussion is not needed
because 
\Propref{prop:tancotmultip} will only be applied to compute the frame
derivatives 
$Y_1(f)$ and $Y_2(f)$ for smooth functions $f$. Since these
derivatives are again smooth functions we know that the
particular combination of formally singular terms involved here has
a well-defined 
smooth extension to $\S$ due to
\Propref{prop:frame_derivatives_on_T3}. With this we also know that the
convergence is uniform on $\S$.

\subsection{Computing \texorpdfstring{$Y_a(f)$}{Ya(f)} and the
  \texorpdfstring{$(\tan\chi-\cot\chi)$}{(tan-cot)}-multiplication} 
\label{sec:discussiontancot}
In \Sectionref{sec:spectral_analysis_smooth} we have performed a
spectral analysis of smooth $\U$-symmetric functions $f$ on $\S$. We used
the properties of the map $\Phi:\T\rightarrow\S$, defined and discussed in
\Sectionref{sec:map_Phi}, to derive the general Fourier representations
of such functions in \Propref{prop:fourier_U1}. With this information we
were able to derive how such functions behave in presence of the
relevant singular factor in \Propref{prop:tancotmultip}. 
Let $f$ be a smooth $\U$-symmetric function on $\S$. To evaluate the
evolution equations using the
variables mentioned in \Sectionref{sec:implequations}, we must calculate
$Y_a(f)$,
cf.~\Eqsref{eq:coordinate_repr_standard_frame}. Having computed the
approximated Fourier representation of $\partial_{\rho_1}f$ pseudospectrally,
\Propref{prop:tancotmultip} enables us to compute 
the Fourier series approximation, i.e.\ the spectral coefficients, of
$(\tan\chi-\cot\chi)\partial_{\rho_1} f$. After having
computed the $\chi$-derivative pseudospectrally and
multiplied with $\cos 2\rho_1$ and $\sin 2\rho_1$ we thus have
determined the pseudospectral approximation of $Y_a(f)$. In this
section we want to note and discuss that this recipe does not yet fix
the procedure completely; moreover we want to introduce a few
alternatives. 

The recipe above requires the use of the results of
\Propref{prop:tancotmultip}; in particular,
the coefficients 
$\mathfrak{b}_{r,p}^{N}$ and $\mathfrak{c}_{r,p}^N$ have to be
computed numerically from the spectral coefficients
of $\partial_{\rho_1}f$. However, for even $p>0$, there are two
ways
of doing this. Although these are equivalent in exact computations
they can be distinct numerically due to round-off
errors. Namely, due to the compatibility conditions 
\Eqsref{eq:approximation_comp_cond} we can write for even $p>0$
\[\mathfrak{b}_{r,p}^{N}:=\sum_{n=r}^{N/2}f^N_{2n,p}
=-\frac 12 f^N_{0,p}-\sum_{n=1}^{r-1}f^N_{2n,p}\quad\text{and}\quad
\mathfrak{c}_{r,p}^N:=\sum_{n=r}^{N/2}f^N_{2n-1,p}
=-\sum_{n=1}^{r-1}f^N_{2n-1,p}.\]
It turns out in practice that there is nearly no
difference between these two ways of determining the coefficients for a
single computation. But having to perform the same 
computation in each time step again and again indeed leads to quite
different behavior as becomes obvious when we discuss numerical
experiments in \Sectionref{sec:comparison}. 

We refer to the first way of computing these coefficients, i.e.\ the
direct use of the definitions, as \term{up-to-down}, since we need
the information of all high frequencies to compute the low
frequency coefficients recursively. In contrast to that the second
variant, i.e.\ taking the 
compatibility conditions into account, is called \term{down-to-up}
since the low frequency coefficients are used to compute the high
frequency coefficients recursively. Recall that 
for odd $p$ there are no compatibility
conditions and hence only the up-to-down method exists.

Both methods, up-to-down and down-to up, are endangered due to the
presence of high-frequency round-off errors. In most practical
situations, the relative round-off errors are larger the higher the
frequencies are because typically the solution has most of its power
in the low frequencies. For the up-to-down method these high-frequency
round-off errors are distributed to the low frequencies and this
might induce instability. For down-to-up these round-off errors stay
at the high 
frequencies but there is the potential risk that they get amplified in
an instable manner
there. However, there seems to be no alternative within this approach
and so one has to 
live with at least one of the two. The influence of round-off errors
in numerical computations is 
difficult to discuss. In finite-differencing approaches
usually the discretization errors are
dominant. This is often not so for pseudospectral
methods and one must take special care of this issue.
We will not make a systematic discussion of this problem 
but instead rely on numerical experiments in
\Sectionref{sec:comparison}. A classic reference for investigations of
this issue based
on statistical analysis is
\cite{Kaneko70}. 

There are further possible variants of our numerical method that could in
principle have enormous impact on the stability and precision properties.
First, we have the possibility of shifting the collocation
points. For instance, the coordinate singularities can be placed on some
collocation points or they can be staggered in between. For our recipe
of computing the frame derivatives above this should not make
a difference, but it is better to be sure and make numerical
experiments. Second, we can try a more naive approach than our recipe
above which I call
\term{direct multiplication}. Assuming that the coordinate
singularities are staggered between the grid points it should be
possible to perform the multiplication of our unknowns with the
singular factor $\tan\chi-\cot\chi$ directly in collocation space
since the unknowns expanded in the basis $\{w_{np}\}$  have the right
fall off behavior there. 
In an exact analytical computation this would be equivalent to our
recipe but it is hard to judge what happens in the discretized
evolution process with round-off errors. 

Numerical
experiments are discussed in
\Sectionref{sec:comparison}. 
I stress that it would be very important to compare the
pseudospectral approaches here to other, particularly finite differencing
ones. However, for this thesis I will only consider pseudospectral
methods and leave these further investigations for future work.

It should be noted that there are also two distinct ways of
doing the multiplications with $\cos 2\rho_1$ and $\sin 2\rho_1$ in
the computation of $Y_1(f)$ and $Y_2(f)$. Namely, one could perform
them  in spectral space or
in collocation space. So far we have only implemented the second more
simple variant; experiments with the first way are under way.

\subsection{Summary of my method for evolution problems with
  \texorpdfstring{$\S$}{S3}-topology}
\label{sec:mymethodS3}

Let us summarize the computational steps involved in our method to
step forward in time in evolution problems with spatial
$\S$-topology.
I restrict here to the evolution system
\Eqsref{eq:gcfe_levi_cevita_evolution} 
with the variables mentioned in
\Sectionref{sec:implequations}
since this is the only system
implemented for spatial $\S$-topology so far. However, I expect that
this method works equally 
well for other evolution systems. Let us assume that the Euler
parametrization coordinates (\Sectionref{sec:coordinatesS3})  have
been chosen on $\S$. 

We give initial data for the variables $u$ at some initial time $t_0$,
mostly $t_0=0$ on $\scrip$,
as expansions in terms of the basis 
$\{w_{np}\}$ using the coordinate expression
\Eqref{eq:wnp_representation_explicit} for these functions. This
ensures that the initial data are smooth functions on $\S$. 
Particular families of such data are constructed in
\Sectionref{sec:initialdataconstruction}. 
Then the code
uses one of the methods in \Sectionref{sec:discussiontancot} to compute
$Y_a(u)$ for all $a=1,2,3$ and all variables $u$. Next, it evaluates
the complete right hand sides of the evolution equations for which
only multiplications and summations done in collocation
space are left. Then it
steps
forward in time by means of Runge-Kutta. Since the spectral
infrastructure was also made for computations with $\T$-slices the
code does not enforce the special properties of the Fourier
series of smooth functions on $\S$ derived previously in this
chapter explicitly. Indeed, it turns out that the code
becomes instable when we just let it run in the way we have just
stated for all of the methods of
\Sectionref{sec:discussiontancot}. However, we found that we can cure this
instability, at least for some of the methods, by projecting those
Fourier coefficients to zero which 
should not be there according to \Eqref{eq:FNpFourier0}. We checked
that then in
particular the compatibility conditions
\Eqref{eq:approximation_comp_cond} behave 
stably. Certainly, one can criticize that the form
\Eqref{eq:FNpFourier0} is only necessary such that the variables can be
considered as smooth functions on $\S$.
But because some of the
methods in \Sectionref{sec:discussiontancot}  have past all tests (see 
\Partref{part:analysis}), in particular, they are stable, convergent and
able to reproduce exact solutions, we believe that these methods
reproduce smooth solutions reliably.
In
future work, I will certainly experiment with further projection
methods.
Of course it would be more
efficient to incorporate the form \Eqref{eq:FNpFourier0} directly into
the DFT method since this would be faster on the one hand and maybe
even avoid
the necessity to project on the other hand; but this has not been done yet. 

Note further that in my implementation of the direct multiplication method of
\Sectionref{sec:discussiontancot}, there is no projection at
all. It is likely that this is 
the reason for the strong instabilities which we observe with this
method in contrast to e.g.\ down-to-up in
\Partref{part:analysis}. However, there might also be a fundamental lack of
understanding and this issue is discussed again in
\Sectionref{sec:comparisonGowdymethods}. 

\section{Gowdy isometries on \texorpdfstring{$\S$}{S3}}
\label{sec:GowdyS3}
\subsection{Gowdy Killing fields}
In this thesis, we will be particularly interested in Gowdy solutions
with spatial \S-topology.
We already mentioned in \Sectionref{sec:relevantsymmclass} what the
topological constraints are when one considers smooth, connected,
compact and orientable $3$-manifolds with a smooth effective isometric
action of
the Gowdy group $\U\times\U$. In particular, such a manifold may be
$\S$ or a lens space or any of the other manifolds listed before; the
lens space case will always be 
included implicitly in our discussion. What is important is that all
such actions 
on one of the admissible manifolds are equivalent and hence we
may choose one representative action. Consider the vector fields
$Y_3$ and $Z_3$; both generate closed curves which correspond to
circles when we consider $\S$ as a subset of $\R^4$. Both vector
fields commute because the first is left and second is right
invariant. Hence, one can convince oneself that they generate an
action of the Gowdy group on $\S$ 
and one can show that this action is effective. Thus, by means of
these two fields we have constructed our representative Gowdy group
action on $\S$. So, the requirement that a Riemannian smooth metric on
$\S$ is Gowdy symmetric is equivalent to the statement that, up to a
diffeomorphism of $\S$ to itself, $Y_3$ and
$Z_3$ are Killing vector fields. 

The basis $(Y_3,Z_3)$ of the Killing
algebra is not yet the canonical one introduced in
\cite{chrusciel1990}. Chru\'{s}ciel requires that on the degenerate orbits,
i.e.\ those two exceptional orbits which are $1$-dimensional, one of
the basis Killing vector fields vanishes and that everywhere the basis Killing
fields are normalized so that the affine length of their closed integral
curves is $2\pi$. This canonical basis is given by 
$(Y_3+Z_3,Y_3-Z_3)/2$. In fact, using the definitions for the
coordinates $\lambda_1$, $\lambda_2$ from \Eqref{eq:lambda12} we find
that
\[Y_3=\partial_{\rho_1}=\partial_{\lambda_1}+\partial_{\lambda_2},\quad
  Z_3=\partial_{\rho_2}=\partial_{\lambda_1}-\partial_{\lambda_2}\]
and hence $(\partial_{\lambda_1},\partial_{\lambda_2})$ is the
canonical basis mentioned above.

Under time evolution, we will always assume that the gauge is chosen
such that the coordinate components of the Killing vector fields are
constant in time; cf.\ \Sectionref{sec:isometry_transport}. Then the
action of the Gowdy group on a $t=const$-hypersurface induces in the
canonical way the action of the Gowdy group on the spacetime. 

\subsection{Orthogonality of the Killing vector fields and
  orbit volume density}
\label{sec:S3orbitvoletc}
To characterize a Gowdy invariant metric on $\S$, the scalar
product $g(\partial_{\lambda_1},\partial_{\lambda_2})$ of the
canonical Killing basis, 
related in particular to the Gowdy
quantity $Q$, see the Gowdy line element \Eqref{eq:GowdyparMetric},
and the area form of 
the orbits $\sqrt{\det g}\, d\lambda_1\wedge d\lambda_2$ play an important
role. Here $\det g$ is the determinant of the matrix
$(g(\partial_{\lambda_A},\partial_{\lambda_B}))$ with $A,B=1,2$. 

In the following we will use the following matrix notation for tensors
with two indices. For all
tensor objects with first index down and second index up, e.g.\
$e\indices{_a^b}$, the matrix $(e\indices{_a^b})$ is given by the
convention 
that $a$ is the row index and $b$ is the column index. For tensor
objects with both indices down, e.g.\
$g(Y_a,Y_b)$, the matrix
$(g(Y_a,Y_b))$ is given by the
convention 
that $a$ is the row index and $b$ is the column index.

In our time evolution formulation of the GCFE on $\S$, part of the set of
the unknowns are the components $e\indices{_a^b}$ of an orthonormal
frame $\{e_a\}$ with respect to the standard frame 
$\{Y_a\}$. Writing $e$ for the matrix
$(e\indices{_a^b})$ and $G$
for the matrix $(g(Y_a,Y_b))$, the orthogonality condition becomes
\[e\cdot G\cdot e^T=\idmatrix \quad\Leftrightarrow\quad
G=(e^T\cdot e)^{-1}.\]
This formula can be used to compute the matrix $G$ from the matrix
$e$.
By means of \Eqsref{eq:expressZbyY} we have
\[Z_3=-\cos(\lambda_1+\lambda_2)\sin 2\chi Y_1
+\sin(\lambda_1+\lambda_2)\sin 2\chi Y_2+\cos2\chi Y_3\]
such that the matrix $(g(V_A,V_B))$ with $V_2=Y_3$ and $V_3=Z_3$ can
be determined from $G$. Now, from this matrix it is straight forward to compute
the required matrix $(g(\partial_{\lambda_A},\partial_{\lambda_B}))$.

In \cite{chrusciel1990} it is discussed that the area density of the
orbits $\sqrt{\det g}$ has a $\sin^2 2\chi$ dependence on each spatial
slice and hence this factor can be divided out. Indeed, this rescaled
quantity is what we will monitor in our numerical runs in
\Sectionref{sec:S3singularGowdy}.

\subsection{Group invariant frames on \texorpdfstring{$\S$}{S3}?}
\label{sec:nogroupinvariantframeS3}
In the torus case, smooth global Gowdy group invariant
frames exist. The curvature quantities
of an invariant metric expressed with respect to such a frame are
constants along the Killing orbits. This
simplifies the analysis drastically since, when in an evolution problem
of EFE
we introduce
coordinates adapted to the Killing fields, the problem is reduced from
$3+1$ to
$1+1$. For instance,
the
choice of a group invariant frame is a key ingredient for the
derivation of the commutator field equations in
\Sectionref{sec:commutatorfieldequations}. The natural question is now
if a global smooth Gowdy invariant frame also exists on $\S$. The
answer is no and the simple reason is that the Gowdy
group has a non-vanishing isotropy subgroup at the ``axes'' where
$Y_3$ and $Z_3$ are linearly dependent. Namely, a frame cannot be
invariant under a non-trivial isotropy subgroup.

Although this simple argument is sufficient, we give another
proof of this statement based on the basis 
functions introduced in \Sectionref{sec:S3GenFourier} in
this section. We do this because the proof gives detailed information
how the frame degenerates at the axes. This information has not been
used yet in this thesis but might become important in studies
involving singular frames on $\S$.

\begin{Theorem}
  There are no global smooth Gowdy invariant frames on $\S$.
  \begin{Proof}
    Assume there was such a frame $\{e_a\}$ which we can decompose as
    $e_a=a\indices{_a^b}Y_b$
    with $(a\indices{_a^b}): \S\rightarrow\mathrm{GL}(3,\R)$
    smooth. The group invariance means
    \[[Z_3,e_a]=0,\quad [Y_3,e_a]=0\quad\forall a=1,2,3.\]
    The commutator relations \Eqsref{eq:commutator_rel_Y} imply the
    two equations
    \[Z_3(a\indices{_a^c})=0,\quad 
    2a\indices{_a^b}\epsilon\indices{_b_3^c}-Y_3(a\indices{_a^c})=0.\]
    The first equation implies the existence of rapidly
    decreasing coefficients (in $n$) $A\indices{_a^c_{np}}\in\C$ such that
    \[a\indices{_a^c}
    =\sum_{n\in 2\N}\sum_{p=-n/2}^{p=n/2}A\indices{_a^c_{np}}
    w_{np},\]
    cf.\ \Corref{cor:expansion_U1symmetric}. Since the frame is smooth
    we have
    \[Y_3(a\indices{_a^c})
    =\sum_{n\in 2\N}\sum_{p=-n/2}^{p=n/2}A\indices{_a^c_{np}} Y_3(w_{np})
    =-2i\sum_{n\in 2\N}\sum_{p=-n/2}^{p=n/2}p A\indices{_a^c_{np}} w_{np},\]
    due to
    \Propref{prop:swap_sum_derivative} and
    \Eqsref{eq:frame_on_basis}. This together with the Parseval
    equality (i.e.\ a smooth function vanishes
    identically if and only if all its coefficients vanish), the
    second identity above leads 
    to the following 
    series of algebraic equations (for all $n\in 2\N$,
    $p\in\{-n/2,\ldots,n/2\}$)
    \[2A\indices{_a^b_{np}}\epsilon\indices{_b_3^c}
    +2ip A\indices{_a^c_{np}}=0.\]
    This can be written as
    \begin{align*}
      A\indices{_a^2_{np}}\epsilon\indices{_2_3^1}
      +ip A\indices{_a^1_{np}}=0\\
      A\indices{_a^1_{np}}\epsilon\indices{_1_3^2}
      +ip A\indices{_a^2_{np}}=0\\
      ip A\indices{_a^3_{np}}=0,
    \end{align*}
    hence
    $A\indices{_a^3_{np}}=0$, $\forall p\not=0$, and
    $A\indices{_a^2_{np}}+ip A\indices{_a^1_{np}}=0$,
    $-A\indices{_a^1_{np}}+ip A\indices{_a^2_{np}}=0$
    such that $A\indices{_a^1_{np}}=0=A\indices{_a^2_{np}}$ 
    $\forall |p|\not=1$. Thus such a frame can be written as
    \begin{align*}
      e_a=&\sum_{\substack{n=2\\n\text{ even}}}^\infty\left[
        \left(A\indices{_a^1_{n,-1}}w_{n,-1}+A\indices{_a^1_{n,1}}w_{n,1}\right)Y_1
        +\left(A\indices{_a^2_{n,-1}}w_{n,-1}+A\indices{_a^2_{n,1}}w_{n,1}\right)Y_2
      \right]\\
      +&\sum_{n\in 2\N}A\indices{_a^3_{n,0}}w_{n,0}Y_3.
    \end{align*}
    Now, the functions $w_{n,-1}$ and $w_{n,1}$ become 
    zero simultaneously at all $\chi=k\pi/2$ ($k\in\Z$) so that the frame
    degenerates there. This is a contradiction to the assumption that
    $\{e_a\}$ is a smooth global frame.
  \end{Proof}
\end{Theorem}

This fact means that a formulation of the field
equations for $\S$-Gowdy spacetimes built on smooth global
frames cannot be reduced to a $1+1$-formulation directly. So far
we can only do a $2+1$-reduction. Some discussions and further ideas
on this issue are 
listed 
in \Sectionref{sec:implementationOfS3Gowdy}.

\subsection{Numerical implementation of the evolution problem in the
  \texorpdfstring{$\S$}{S3}-case with Gowdy symmetry}
\label{sec:implementationOfS3Gowdy}
In the last section we showed that there are no smooth global Gowdy group
invariant frames on $\S$; hence there is no direct reduction of the evolution
problem to $1+1$ as we mentioned there. 

Let us discuss this more carefully. In the evolution problem it
depends very much on the gauge choice how the Killing vector fields
``behave'', as was discussed in
\Sectionref{sec:isometry_transport}. Suppose in the following that the
gauge is chosen such that the coordinate components of the spacelike KVFs
are constant. This is true for the commutator field equations
in \Sectionref{sec:commutatorfieldequations} and the general conformal
field equations in Levi-Civita conformal Gauß gauge in
\Sectionref{sec:LCCGG} according to \Propref{prop:xiconst} and
\Lemref{lem:symmetryconformal}. 

Under these conditions, it is true that the evolution problem cannot
be reduced to $1+1$ because there is no smooth global Gowdy
group invariant frame on $\S$. However, we can reduce it at least to
$2+1$. For this, let us only
consider one
factor of the Gowdy group, say the one generated by $Z_3$. There are
smooth global frames that are invariant with respect to the associated
group $\U$;
for instance all left invariant frames or, more generally, all frames
whose frame coefficients with respect to a left invariant reference
frame are constant along $Z_3$. For such an orthonormal frame all curvature
components are constant
along $Z_3$. So the problem is reduced to $2+1$, since all 
unknowns under these conditions are constant along $\rho_2$ and
effectively only the coordinates $(\chi,\rho_1)$ have to be
considered. 

In this manner, the current implementation of the code for
the conformal field equations on $\S$ only employs this
$2+1$-reduction, and all simulations of Gowdy spacetimes in
\Sectionref{sec:S3singularGowdy}  are
done this way. Of course with such an implementation we cannot expect
to reach high spatial resolutions. Moreover, it could turn out
that the symmetry along the vector field $Y_3$ in a Gowdy simulation,
that in this 
method is ``freely propagated'', drifts due to numerical
errors. In particular, if the equations
are strongly instable with respect to Gowdy symmetry  already on the
continuum level, such 
numerical solutions would certainly be useless to make
statements about the 
class of Gowdy spacetimes itself. We come back to this in
\Sectionref{sec:S3singularGowdy}. 

However, despite this fundamental issue that there are no smooth
global Gowdy group invariant frames on $\S$, here is an idea to get
around the necessity of doing $2+1$-evolutions in the Gowdy case. The
point is that the 
dependence of the unknowns on $\rho_1$ in an orthonormal frame
formulation of the 
field equations under the assumption of $\S$-Gowdy symmetry
is caused by the fact the frame was chosen to be
not group invariant exclusively. All ``dynamics'' in
$\rho_1$-direction stem from  
the bad choice of the frame. Let $T$ be a vector field which is $Y_3$-invariant
and choose an orthonormal frame which is $Z_3$-invariant as above. One
can easily write down the explicit 
expressions that relates the $Y_3$-derivative of any tensor component
$T_a$ and the $Y_3$-derivatives
of the frame components $e\indices{_a^b}$ with respect to the
standard frame 
$\{Y_a\}$. This means that one only needs to know the $\rho_1$-derivative of the
frame components to compute the $\rho_1$-derivatives of the components
of any tensor field
which is $Y_3$-invariant. Now, it appears possible to add the functions
$Y_3(e\indices{_a^b})$ as new variables to the evolution system such
that this extended evolution system is still symmetric hyperbolic. By
this, no $Y_3$-derivatives have to be evaluated anymore during the
evolution and the computations only have to be done for one
$\rho_1$-value. Thus, this can be considered as an effective reduction
to $1+1$ in the $\S$-Gowdy case.
I will
mention no further details here since this idea has not been worked out
completely yet.

\chapter{Construction of initial data}
\label{ch:settingupCauchyProblems}

\section{Introduction}
In this section we want to construct families of simple explicit
$\scrip$-initial data sets for the general conformal field equations,
cf.~\Theoremref{th:solving_constraints}, as well as initial data sets
for the commutator 
field equations (\Sectionref{sec:comm_field_eqs}) on standard Cauchy
surfaces. These data will be used in  
\Chapterref{part:analysis} to compute numerical solutions. I point out
that these class of data 
here are neither 
claimed to be generic nor physically motivated. The
key arguments for their choice are practicability and feasibility, while
keeping contact 
to our main underlying motivational questions. The construction and
investigation of more general classes of initial data are under way.

Note that in the case of the conformal field equations, we always
construct initial data for the
Levi-Civita conformal Gauß gauge (\Sectionref{sec:LCCGG}) for the time
being. In particular this means that we choose the conformal expansion of
$\scrip$ to be $k=1$, the initial value of the Weyl $1$-form
$\omega=0$ and we normalize $\lambda=3$
according to \Propref{prop:IDLCCGG}.

All classes of initial data that we derive here for the conformal
field equations in the case of spatial $\S$-topology are based on the
\term{Berger sphere} geometry 
(\Sectionref{sec:bergersphere}). With zero (or constant) data
for the components of the electric part of the rescaled Weyl tensor,
the corresponding 
solutions are the $\lambda$-Taub-NUT spacetimes
(\Sectionref{sec:TaubNUT}). 
We will write down the relation explicitly.
However, when the initial data of the
electric part, subject to the constraint
\Eqref{eq:scri_electric_constraint},
are not constant, a quite general class of 
inhomogeneous FAdS solutions 
can be constructed (\Sectionref{sec:solelectrconstraintscri}) which,
for sufficiently small data for $W_{ab}$, can be considered as
non-linear perturbations of the $\lambda$-Taub-NUT family. Since the
$\lambda$-Taub-NUT family has very peculiar properties like existence of Cauchy
horizons and causality violations, it will be particularly interesting
to study the behavior of those pathologies under such
perturbations. Moreover, there is an interesting connection to the 
singularity theorems by Andersson and Galloway
(\Sectionref{sec:singularitytheoremGA}) because the Ricci
scalar of the induced
geometry on $\scrip$ for this initial data class can have any sign. We
concentrate here on the case of data which are invariant under the 
group $\U$ or even under the Gowdy group $\U\times\U$. In the
following \Sectionref{sec:otherID} we briefly comment on other
reasonable classes of initial data for the $\S$-case. 

Since we also want to use the general conformal field equations for
spacetimes with spatial $\T$-topology we give a simple class of
initial data for that case in \Sectionref{sec:T3initialdata}.

Finally we also comment on the construction of initial data for the
commutator field equations. 

\newpage

\section{Initial data construction on \texorpdfstring{$\scrip$}{J+} 
  for GCFE}
\label{sec:initialdataconstruction}
\subsection{Berger sphere}
\label{sec:bergersphere}
In this section we construct a family of $3$-metrics on $\S$ that is
known in the mathematical literature
as the \term{Berger sphere} family.  
The underlying requirement is that these metrics are of type
LRS-Bianchi-IX (\Sectionref{sec:relevantsymmclass}), i.e.\ the Killing
algebra is 
$4$-dim.\ with the Lie algebra of $\SU$ as a subalgebra.
The Lie algebra of $\SU$ can be identified with the span of the frame
$\{Z_a\}$ defined in \Eqsref{def:YZ}. With the ansatz 
$e_a=e\indices{_a^b}Y_b$
for an orthonormal frame with a smooth $\text{GL}(3,\R)$-valued
function 
$(e\indices{_a^b})$ on $\S$, this
requirement 
is equivalent to $Z_c(e\indices{_a^b})=0$
and hence $e\indices{_a^b}=const$. A fourth basis element of the
LRS-Bianchi IX
Killing algebra is $Y_3$. Hence, in some sense,
LRS-Bianchi-IX-invariant metrics
are simultaneously $\SU$ and Gowdy group invariant.

Let $(f\indices{_a^b})$ be the inverse
matrix of $(e\indices{_a^b})$, i.e.\ 
$e\indices{_a^b}f\indices{_b^c}=\delta\indices{_a^c}$. Together with the
commutator relations \Eqsref{eq:commutator_rel_Y}, the Killing
equation \Eqref{eq:killing_eq_onf} for $Y_3$  leads to
\begin{align*}
  0=K\indices{_a_b}&:=
  e\indices{_a^c}\epsilon\indices{_3_c^d}f\indices{_d^e}g_{eb}
  +e\indices{_b^c}\epsilon\indices{_3_c^d}f\indices{_d^e}g_{ea}\\
  &=-e\indices{_a^2}f\indices{_1^e}g_{eb}
  +e\indices{_a^1}f\indices{_2^e}g_{eb}
  +e\indices{_b^1}f\indices{_2^e}g_{ea}
  -e\indices{_b^2}f\indices{_1^e}g_{ea}
\end{align*}
with $K_{ab}$ a symmetric matrix.
Making use of the freedom to perform a $\text{O}(3)$-transformation of the
frame (or equivalently by a Gram-Schmidt Orthonormalization of the
frame $\{Y_a\}$) it can be
arranged that $(e\indices{_a^b})$ is an upper triangular
matrix\footnote{We use the same matrix conventions as in 
\Sectionref{sec:S3orbitvoletc}.}, i.e.\
$e\indices{_a^b}=0$ for $b<a$,  with
non-vanishing positive diagonal elements. The inverse matrix
$(f\indices{_a^b})$ is hence also an upper triangular matrix with
positive diagonal elements. Then we find in a first step that
\begin{alignat*}{3}
  K_{33}&=0,& &\quad& K_{32}&=-e\indices{_2^2}f\indices{_1^3}\\
  K_{31}&=e\indices{_1^1}f\indices{_2^3}-e\indices{_1^2}f\indices{_1^3},& & &
  K_{11}&=2(-e\indices{_1^2}f\indices{_1^1}).
\end{alignat*}
The condition $K_{ab}=0$ implies that $(e\indices{_a^b})$ and
$(f\indices{_a^b})$ are diagonal. With this the residual components
are
\[K_{12}=e\indices{_1^1}f\indices{_2^2}-e\indices{_2^2}f\indices{_1^1},
\quad
K_{22}=0.
\]
This tells us that
$e\indices{_2^2}=e\indices{_1^1}$. Hence, under
the symmetry assumption above, we can find a
$\mathrm{O}(3)$-transformation such that the orthonormal frame takes
the form  
\begin{equation}
  \label{eq:BergerFrame}
  (e\indices{_a^b})=\text{diag}(a_1,a_1,a_3)
\end{equation}
with $a_1,a_3>0$.

With this it is straight forward to compute that the Ricci tensor is
\[(R_{ab})=\text{diag}\left(2a_1^2\left(2-\frac{a_1^2}{a_3^2}\right),
2a_1^2\left(2-\frac{a_1^2}{a_3^2}\right),2\frac{a_1^2}{a_3^2}\right)\]
and for the the Ricci scalar we get
\begin{equation}
  \label{eq:BergerR}
  R=2a_1^2\left(4-\frac{a_1^2}{a_3^2}\right).
\end{equation}
It is of interest for us to have found a family of smooth metrics on
$\S$ such that all signs of the Ricci scalar can be obtained. Compare
this to the discussion in \Sectionref{sec:situation_FADS}.
Berger's original motivation to investigate these metric was that these metrics
have finite curvature even if the direction corresponding to $Y_3$ (i.e.\
the fiber in the Hopf fibration, cf.\ \Sectionref{sec:Hopf_fibration})
collapses so that, in some sense, $\S$ becomes $\Stwo$. Namely, in this
case, $a_3\rightarrow\infty$ while $a_1$ and hence the Ricci tensor
stay bounded. This is the process which happens also at the Cauchy
horizons of $\lambda$-Taub-NUT (in the same way for $\lambda=0$) solutions.

\subsection{Solutions of the electric constraint on the Berger sphere}
\label{sec:solelectrconstraintscri}

The family of Berger spheres constructed in the previous section with
vanishing (or constant) components of the electric part of the rescaled Weyl 
tensor can be used as initial data for the general conformal field
equations and leads to the family of the $\lambda$-Taub-NUT solutions.
Since
these spacetimes have a 
quite large symmetry group the aim of this section is to built
initial data with less symmetry. Here, we construct an explicit family
of solutions of the 
electric constraint on $\scrip$ \Eqref{eq:scri_electric_constraint} in
\Theoremref{th:solving_constraints} assuming $\scrip$ to be a Berger
sphere. 

We want to find solutions of 
\[g^{ab}D_a W_{bc}=0\]
on $\S$ with a metric given by the Berger family of the previous
section determined by the orthonormal frame \Eqref{eq:BergerFrame}
such that $W_{ab}$ is a smooth symmetric tracefree tensor field
corresponding to the initial data for the electric part of the
rescaled Weyl tensor. Then, the
components $W_{ab}$ are smooth functions on $\S$. The
tracefree condition will always be realized by requiring that
$W_{33}=-W_{11}-W_{22}$. Without loss of
generality we can assume that $a_1=1$ after a suitable conformal
transformation of the initial data, cf.\ the discussion at the end of
\Sectionref{sec:initial_data}.
 
Let $\Connectionh abc$ be the
connection coefficients of a Berger metric $h$ with respect to the associated
orthonormal frame \Eqref{eq:BergerFrame}. Then it is easy to
check that on the one hand $g^{ac}\Connectionh abc=0$
and on the other hand one finds for the following one-indexed quantity
\[\Bigl(g^{ab}\Connectionh adc W_{bd}\Bigr)
=\frac{2(a_3^2-1)}{a_3}\,\Bigl(W_{23},-W_{13},0\Bigr).\]
So the constraint becomes
\nobreak
\begin{subequations}
\begin{align}
  \label{eq:1}
  e_1(W_{11})+e_2(W_{12})+e_3(W_{13})-\frac{2(a_3^2-1)}{a_3}\,W_{23}&=0,\\
  \label{eq:2}
  e_1(W_{12})+e_2(W_{22})+e_3(W_{23})+\frac{2(a_3^2-1)}{a_3}\,W_{13}&=0,\\
  \label{eq:3}
  e_1(W_{13})+e_2(W_{23})+e_3(-W_{11}-W_{22})&=0.
\end{align}
\end{subequations}
We make the simplifying assumption that
$\lieder{Z_3}{W}=0$, meaning that $W$ is $\U$-invariant which implies
that its components $W_{ab}$ are $\U$-symmetric functions. Hence they have the
decomposition 
\[W_{ab}=\sum_{n\in 2\N}\sum_{p=-n/2}^{n/2}(W_{ab})_{n,p} w_{np}\]
with rapidly decreasing coefficients. Since the equations above are
linear with constant coefficients, \Eqsref{eq:frame_on_basis} hold and
\Propref{prop:swap_sum_derivative} can be applied,
each $n$-mode decouples from the others. Look first at
\Eqref{eq:1}. For each $n$ we find due to 
\Propref{prop:swap_sum_derivative} 
\begin{align*}
  0=\sum_{p=-n/2}^{n/2}&\Biggl\{(W_{11})_{np}Y_1(w_{np})
    +(W_{12})_{np}Y_2(w_{np})+a_3(W_{13})_{np}Y_3(w_{np})\\
    &-\frac{2(a_3^2-1)}{a_3}\,(W_{23})_{np}w_{np}\Biggr\}\\
  =\sum_{p=-n/2}^{n/2}&\Biggl\{
  (W_{11})_{np}(-i)\left(C_{n,p}w_{n,p-1}+C_{n,-p}w_{n,p+1}\right)\\
  &+(W_{12})_{np}\left(C_{n,p}w_{n,p-1}-C_{n,-p}w_{n,p+1}\right)\\
  &+a_3(W_{13})_{np}(-2ip)w_{np}
  -\frac{2(a_3^2-1)}{a_3}\,(W_{23})_{np}w_{np}\Biggr\}
\end{align*}
with
\[C_{n,p}:=\sqrt{\left(\frac n2+p\right)\left(\frac n2-p+1\right)}.\]
Similar equations follow from \Eqsref{eq:2} and \eqref{eq:3}.

Let us, for simplicity, only analyze the equation for the cases $n=0$
and $n=2$.  
For $n=0$ we find 
\[0=(W_{23})_{0,0}=(W_{13})_{0,0} \text{ (otherwise
arbitrary) if } a_3\not=1;\]
if $a_3=1$ there is no condition on $(W_{ab})_{0,0}$ at all. 
For $n=2$ we get
\begin{align*}
  0&=w_{2,1}\left[(-2i)a_3(W_{13})_{2,1}+(-i)\sqrt 2(W_{11})_{2,0}
    -\sqrt 2(W_{12})_{2,0}-\frac{2(a_3^2-1)}{a_3}(W_{23})_{2,1}\right]\\
  &+w_{2,0}\Bigl[(-i)\sqrt{2}(W_{11})_{2,1}+\sqrt 2(W_{12})_{2,1}
    +(-i)\sqrt{2}(W_{11})_{2,-1}\\
    &\quad\quad\quad-\sqrt 2(W_{12})_{2,-1}
    -\frac{2(a_3^2-1)}{a_3}(W_{23})_{2,0}\Bigr]\\
  &+w_{2,-1}\left[-(-2i)a_3(W_{13})_{2,-1}+(-i)\sqrt 2(W_{11})_{2,0}
    +\sqrt 2(W_{12})_{2,0}-\frac{2(a_3^2-1)}{a_3}(W_{23})_{2,-1}\right];
\end{align*}
thus each of the algebraic brackets has to vanish identically. Similar
expressions are obtained from \Eqsref{eq:2} and \eqref{eq:3}. The
reality condition \Eqref{eq:reality_wnp} are consistent with the fact
that the complex conjugate of the third bracket corresponds to the
negative of the first bracket. Writing
\[(W_{ab})_{2,1}=u_{ab}+i v_{ab},\quad (W_{ab})_{2,0}=U_{ab}, \quad
(W_{ab})_{2,-1}=-u_{ab}+i v_{ab}\]
with real valued symmetric tracefree tensorial functions $u_{ab}$, $v_{ab}$ and
$U_{ab}$, we 
obtain, having split into real and imaginary parts,
\begin{align*}
  2a_3v_{13}-\sqrt{2}U_{12}-\frac{2(a_3^2-1)}{a_3}u_{23}&=0,\\
  2a_3u_{13}+\sqrt{2}U_{11}+\frac{2(a_3^2-1)}{a_3}v_{23}&=0\\
  \sqrt{2}v_{11}+\sqrt{2}u_{12}-\frac{a_3^2-1}{a_3}U_{23}&=0
\end{align*}
for \Eqref{eq:1}.With the same calculations we get from \Eqref{eq:2}
\begin{align*}
  2a_3v_{23}-\sqrt{2}U_{22}+\frac{2(a_3^2-1)}{a_3}u_{13}&=0,\\
  2a_3u_{23}+\sqrt{2}U_{12}-\frac{2(a_3^2-1)}{a_3}v_{13}&=0\\
  \sqrt{2}v_{12}+\sqrt{2}u_{22}+\frac{a_3^2-1}{a_3}U_{13}&=0
\end{align*}
and from \Eqref{eq:3}
\begin{align*}
  2a_3(-v_{11}-v_{22})-\sqrt{2}U_{23}&=0,\\
  2a_3(-u_{11}-u_{22})+\sqrt{2}U_{13}&=0\\
  \sqrt{2}v_{13}+\sqrt{2}u_{23}&=0.
\end{align*}
Hence we have $9$ linear equations for the $15$ unknowns $u_{ab}$,
$v_{ab}$ and $U_{ab}$. It turns out that only $8$ of them are linear
independent. Thus we get a $7$-parameter
family of solutions which can be represented as follows ($a_3>0$)
\nobreak
\begin{subequations}
\label{eq:n2solelectrconstraint}
\begin{align}
  & u_{11}=\frac{a_3}{\sqrt 2}C_3+C_7, &
  & u_{12}=\frac{a_3}{\sqrt 2}C_1+C_6, &
  & u_{13}=-\frac{a_3}{\sqrt 2}C_2-a_3C_5, & \\
  & u_{22}=\frac{a_3^2-1}{\sqrt 2\,a_3}C_3-C_7, &
  & u_{23}=a_3C_4, & & &\\
  & v_{11}=-\frac{1}{\sqrt 2\,a_3}C_1-C_6, &
  & v_{12}=C_7, &
  & v_{13}=-a_3 C_4, & \\
  & v_{22}=C_6, &
  & v_{23}=\frac{a_3}{\sqrt 2}C_2-a_3C_5, & & &\\
  & w_{11}=C_2+\sqrt{2}(2a_3^2-1)C_5, &
  & w_{12}=\sqrt{2}(1-2a_3^2)C_4, &
  & w_{13}=C_3, & \\
  & w_{22}=C_2+\sqrt{2}(1-2a_3^2)C_5, &
  & w_{23}=C_1. & & &
\end{align}
\end{subequations}
with seven real parameters $C_1,\ldots,C_7$.

Now let us construct that subspace of solutions which is Gowdy
invariant. This means that additionally to the requirement
$\lieder{Z_3}W=0$ above we demand $\lieder{Y_3}W=0$. The analysis of
this latter equations is similar but more lengthy than the first
one. Again, the tensor field $W$ is expanded in terms of the
orthonormal frame $\{e_a\}$ which in turn is expanded in terms of the
basis $\{Y_a\}$. Using the commutator relations
\Eqsref{eq:commutator_rel_Y} and expanding the component function in
terms of the basis $\{w_{np}\}$ as before, we eventually obtain the
following results. For
$n=0$ we must satisfy
\[(W_{12})_{0,0}=(W_{13})_{0,0}=(W_{23})_{0,0}=0,\quad
(W_{11})_{0,0}=(W_{22})_{0,0}.\]
For $n=2$ we find the condition that all $C_i=0$ except for $C_2$
which can be arbitrary. Hence, the $7$-dim.\ space of solutions of the
electric constraint in the $n=2$-case has a $1$-dim.\ subspace of
Gowdy invariant solutions.

Does this Gowdy subspace lead to polarized Gowdy solutions, i.e.\ are
the Killing vector fields orthogonal everywhere? In Levi-Civita
conformal Gauß gauge (\Sectionref{sec:LCCGG}), the leading order
dynamics close to the initial hypersurface $\scrip$ corresponding to
$t=0$ is in general
\begin{equation}
\label{eq:leadingorderdynamics}
e(t)=\left[\idmatrix+\idmatrix t
+\frac 12(2\idmatrix+L^*)t^2
+\frac 16(6\idmatrix-6L^*+2W)t^3\right]\cdot e^*+O(t^4).
\end{equation}
By $e^*$ we mean the initial value of the matrix\footnote{We use the
  same matrix conventions as in  
\Sectionref{sec:S3orbitvoletc}.}
 $e=(e\indices{_a^b})$
and by $L^*$ the initial value of the matrix $(L\indices{_a^b})$. For these
data here, we have 
\[e^*=\mathrm{diag}(1,1,a_3)\quad\text{and}\quad
L^*=\mathrm{diag}(5a_3^2-3,5a_3^2-3,5-3a_3^2)/(2a_3^2).\]
As mentioned in \Sectionref{sec:S3orbitvoletc}, the matrix
$G=(g(Y_a,Y_b))$ satisfied $G=(e(t)^T\cdot e(t))^{-1}$. By means of
\Eqref{eq:expressZ3byY} we can now get the leading order expression
for the scalar product of the Killing fields $g(Y_3,Z_3)$ and find
\begin{equation*}
  g(Y_3,Z_3)(t,\chi)=\frac{\cos 2\chi}{{{a_3}}^2} 
  -\frac{2\,\cos 2\chi}
  {{{a_3}}^2}t 
  -\frac{\left({{a_3}}^2-5\right) \,
    \cos 2\chi}{2\,
    {{a_3}}^4}{{t}}^2+O(t^3).
\end{equation*}
The third order term is also explicitly known and involves the matrix
$W$, but is too lengthy to write
down here. In any case, there are Gowdy initial data in this family
which correspond to polarized solutions.

As mentioned before, one can show that one can find a conformal
representation of the $\lambda$-Taub-NUT spacetimes with the
parameters $B_0$, $C_0$ (\Sectionref{sec:TaubNUT}) such that $\scrip$
is a Berger sphere. The 
corresponding $\scrip$-initial data set is given by
$a_1=a_2=1$, $a_3=B_0^{-1}$, $W_{11}=W_{22}=-B_0C_0/2, W_{33}=B_0C_0$ and all other
components of $W$ vanish. Here we assume that $\lambda=3$. Hence, by
means of these data and Friedrich's
Cauchy problem, one can compute the maximal Cauchy development of
$\scrip$ of the $\lambda$-Taub-NUT spacetimes.

Let us interpret the solutions corresponding to the data we have just
constructed briefly. If we choose $a_3=1$, i.e.\ the standard sphere, the
corresponding solutions obey the cosmic no-hair picture because in the
slicing given in \Sectionref{sec:FAdScosmicnohair} the spatial
curvature becomes that of the
de-Sitter spacetime in spherical slicing. This is so even
globally. Now, when we choose the
parameter $a_3$ a little smaller than $1$, the corresponding
solutions will have a little anisotropy even asymptotically but are at
least homogeneous in the limit. This
shows that it is not difficult to produce solutions which are
anisotropic even though there is inflation. Hence little anisotropies
of our universe should not be excluded from the start and one should
think about further ways of measuring them observationally.

\subsection{Other families of data on \texorpdfstring{$\S$}{S3} on 
  \texorpdfstring{$\scrip$}{J+}}
\label{sec:otherID}

In the previous sections, special families of initial data for the general
conformal field equations in the case of spatial $\S$-topology were
constructed. These data can be considered as close to corresponding
$\lambda$-Taub-NUT data if the relevant initial data parameters are
chosen small enough. The particular motivation for the construction of
these data was that it is possible to find
explicit representations. One can expect
that initial data corresponding to explicit
solutions of the constraints induce less errors for the time evolution
than in particular numerically obtained data. 
An alternative approach in this direction is the 
following. We are allowed to pick any smooth perturbation of a Berger
$3$-metric as the $3$-metric on $\scrip$ since there is no constraint
that restricts the choice. Then, when
we prescribe the matrix $W$ to vanish, the constraint
\Eqref{eq:scri_electric_constraint} is solved trivially, and this
yields non-trivial $\scrip$-initial data sets. However, to
construct the complete initial data, i.e.\ the curvature
quantities of the 
$3$-metric, for instance the Cotton tensor
$B_{abc}$, up to $2$nd derivatives of the $3$-metric have
to be computed. A strategy to avoid the necessity to compute these
numerically is to fix a parametrized family of such perturbations
and to
obtain the parametrized explicit expressions for the initial data quantities
beforehand, say with \Mathematica.
So far, this possibility has not been tried
systematically. 

In any case, it would be nice to construct a family of initial data
such that the polarized Gowdy case is included. For those, a lot more
analytical results are known (at least for vanishing cosmological
constant), and their study would be a good further test for the code.

Eventually, we want to construct families of initial data that can, in
some sense, be 
considered as ``generic''. A prototype argument for the genericity of
a given class of initial data is in \cite{Berger97}.

\subsection{Class of initial data on 
  \texorpdfstring{$\scrip$}{J+} for the \texorpdfstring{$\T$}{T3}-case }
\label{sec:T3initialdata}
Similar to the initial data construction in the $\S$-case in the
previous section, 
we derive a simple family of data for the $\T$-case in this section.
The basic simplifying assumption is that
the $3$-metric on $\scrip$ is flat. As usual, we suppose that the
data choice 
corresponds to the Levi-Civita conformal Gauß gauge as before.
Let some orthonormal 
frame and coordinates $(x^1,x^2,x^3)$ such that $e_a=\partial_{x^a}$
be given; then
the only relevant equation to 
solve, namely \Eqref{eq:scri_electric_constraint}, reduces to 
\[\partial_{1}W\indices{_1_b}+\partial_{2}W\indices{_2_b}
+\partial_{3}W\indices{_3_b}=0.\]
Let us further restrict to the case of Gowdy symmetry with
$\partial_2$ and $\partial_3$ the 
associated Killing vector fields. Then the previous equation yields
\[\partial_{1}W\indices{_1_1}=0,\,\partial_{1}W\indices{_1_2}=0,
\,\partial_{1}W\indices{_1_3}=0,\]
in other words $W\indices{_1_a}=const$.
The other components of this symmetric tracefree tensor are not
constrained at all.

As in \Sectionref{sec:solelectrconstraintscri} we ask the question
under which conditions these data correspond to polarized Gowdy
solutions. \Eqref{eq:leadingorderdynamics} also holds here, but this
time we have
\[e^*=\idmatrix\quad\text{and}\quad
L^*=\frac 12\idmatrix\]
so that\footnote{We use the
  same matrix conventions as in  
\Sectionref{sec:S3orbitvoletc} and in
\Sectionref{sec:solelectrconstraintscri}.} 
\[e(t)=\idmatrix+t\idmatrix
+\frac 34t^2\idmatrix+\frac 16t^3(3\idmatrix+2W)+O(t^4)\]
where $W$ is the matrix $(W_{ab})$. From this we find that the scalar
product of the two Killing vector fields $\partial_2$ and $\partial_3$ is
\[g(\partial_2,\partial_3)=-\frac 23 W_{23}t^3+O(t^4).\]
Hence a necessary condition for these data to yield polarized Gowdy
solutions is $W_{23}=0$. It turns out that a sufficient condition for
polarization is to choose the matrix $W$ diagonal. Further
criteria concerning polarization can be derived but are not discussed
here. 

\section{Initial data for the commutator field equations}
\label{sec:IDcommutFE}
In the previous section we have constructed some families of initial
data for the general conformal field equations in Levi-Civita
conformal Gauß gauge. In this section, we list some issues for the initial
data construction for the commutator field equations, cf.\
\Sectionref{sec:comm_field_eqs}. 

It is not very difficult to construct explicit initial data for the
commutator field equations on a standard Cauchy
surface since, in particular, for the main system there is only one constraint
\Eqref{eq:constrLambda} with $r$ given by \Eqref{eq:ralb} and with
$A=0$. For example, all initial data that we will use in
\Sectionref{sec:RunsCosmFE} have the 
property that the initial value of $r$ is zero so that the initial value
for $\Omega_{\Lambda}$ must be constant.

However, our aim is to construct FAdS solutions. Prescribing data on a
standard Cauchy surface as just mentioned does not enable us to
control the time asymptotics a priori; 
for instance the corresponding solution might collapse in both time
directions. Attempts to 
regularize the commutator field equation system on $\scri$ and hence to
formulate a Cauchy problem with respect to $\scri$ as for the
conformal field equations have failed so 
far and are maybe not possible at all. However, in principle, such a
regularization is not needed since 
we can use the conformal 
field equations with initial data on $\scrip$ to integrate a bit into
the past and then start a standard Cauchy problem there for the
commutator field equations. Either one can try to 
find a gauge for the conformal field equations such that the final
Cauchy surface computed with the conformal field equations can be used
directly as the initial surface 
for the commutator field
equations, i.e.\ these gauge conditions are satisfied: $A=0$ and the
area density 
of the orbits is constant. However, the Levi-Civita conformal Gauß gauge does
not fulfill this requirement and it is currently not clear how to do this with
other gauges.  Another approach to solve this
``transfer gauge problem'' is to construct a Cauchy surface
in the solution obtained with
the conformal field equations which satisfies these gauge
requirements and use it as
the initial surface for the commutator field equations. However, this
has not been 
investigated so far.

\part{Applications and their analysis}
\label{part:analysis}
\chapter{Introduction}
In the previous \Partref{part:treatment} we have introduced our
underlying questions and problems and developed our method. 
In this following part of the thesis we start by testing our method
in \Chapterref{ch:numexperiments}; this involves checks for formal
errors in the implementation but also experiments to see how well our
approach for the coordinate singularity on $\S$ behaves. For this we
compare the various  possibilities to do this pseudospectrally
mentioned in \Sectionref{sec:discussiontancot}. In particular, these
tests involve computations with explicitly known solutions of the
linearized system but also with fully non-linear
regular $\lambda$-Gowdy spacetimes. In \Chapterref{ch:singulargowdy}
we compute singular $\lambda$-Gowdy spacetimes. Here, the emphasis
lies again on the study of the behavior of the code; but these
considerations can already be seen as first preliminary, though non-systematic,
investigations of some of our underlying questions. In
\Sectionref{sec:comparisonGowdymethods} we discuss the properties and
expectations for our method in the light of the
other numerical methods for singular Gowdy spacetimes that exist in the
literature. Then in
\Sectionref{sec:outluckfuture} and \ref{sec:summary}, we summarize,
name the open problems 
and collect a 
list of projects for future research. 

In fact, one should note that almost all applications which we present
here 
belong to the Gowdy class. On the one hand the problem often
simplifies technically under this symmetry assumption; in particular
the symmetry reduces the problem to a lower number of spatial
dimensions. Additionally 
in the Gowdy class, singularities can be expected to be
non-oscillatory. On the other hand, 
this class of spacetimes has been studied most extensively and one has
quite a good overview where the open issues lie. 

Another remark is the following. During our analysis we use the norms
$L^p(\S)$. Usually these are defined with respect to the standard
measure on $\S$. However, we define the $L^p$-norm
of a function $f\in C^\infty(\S)$ to be the $L^p$-norm of the
corresponding function $f\in\breve X\subset C^\infty(\T)$
(\Sectionref{sec:map_Phi}) with respect to the
standard measure on $\T$.

\chapter{Numerical experiments}
\label{ch:numexperiments}

\section{Tests with explicit solutions}
\label{sec:experimentslinearized}

\setcounter{totalnumber}{1}

\subsection{Explicit solutions of the
  spin-2-system on the de-Sitter background}
\label{sec:linearized_solutions}
In this section we want to show results from experiments with
solutions with 
$\S$-topology of the linearized general conformal field equations on the
de-Sitter background. The purpose is
to check whether our numerical algorithm to compute frame derivatives
on $\S$ that we summarized and deepened in
\Sectionref{sec:discussiontancot}, works and how well it performs. 

Consider again the general conformal field equations in Levi-Civita
conformal Gauß gauge \Eqsref{eq:gcfe_levi_cevita_evolution}. The
de-Sitter solution in this gauge takes the form \Eqref{eq:dSLCCGG}
with conformal factor $\Omega=\frac 12t(2-t)$ from which the unknowns
of this system can be derived. In particular, this solution is
conformally flat, i.e.\ all components of the rescaled conformal Weyl
tensor vanish identically. 
A solution of the
Bianchi system on this given background solution can be considered as the
solution of the
linearization of the conformal field equations with respect to that
background solution.
In \cite{penroserindler}, where also the
original references are listed, a consistency conditions is derived
which is necessary such that Bianchi system on a given background $M$
has a solution at all. This consistency condition is satisfied on any
conformally flat background. 
On a given background, the Bianchi system is
often called \term{spin-2-system} and the rescaled Weyl tensor
$W\indices{^i_j_k_l}$ \term{spin-2-field}.  Related discussions and
further terminology can 
be found in \cite{penroserindler}. Formally one
gets the linearized evolution equations by setting $\Omega=0$ and
$\dot\Omega=0$ 
in \Eqsref{eq:gcfe_levi_cevita_evolution} which means that the
``background part'' of the equations get decoupled from the
spin-$2$-part and we solve for the conformally flat background and the
spin-$2$-field on this background simultaneously. The ``background
part'' consists only of ODEs and the spin-$2$-part is a linear
symmetric hyperbolic system (apart from the small subtleties mentioned
in \Sectionref{sec:LCCGG}). 

Now we briefly summarize the steps to derive explicit solutions of the
linearized equations 
on the de-Sitter background. The main idea is that the spin-$2$-system
is conformally invariant. By a suitable conformal transformation and a
corresponding change in the time coordinate (cf.\ the derivation of
\Eqref{eq:dS_conformal}) we can bring the de-Sitter
solution to the static Einstein cylinder. In this gauge, the evolution
equations derived from the 
spin-$2$-system have constant
coefficients. Expanding the unknowns in terms of the basis functions
$\{w^n_{ik}\}$ we obtain a linear system of ODEs which is decoupled
for each $n$-mode but coupled among the $i$- and $k$-modes. At least
for $n=0$ and $n=2$, this
system can be solved explicitly by diagonalizing the evolution
matrix; the case $n>2$ has not been considered yet. 
We do not write down the lengthy general expressions for the solutions
here since they are not of further interest for us. 

In any case, it is
not sufficient to just solve the evolution equations, also the
constraints \Eqsref{eq:bianchi_constraints} of the spin-$2$-system have
to be satisfied. This means we 
have to choose initial data consistent with the constraints which are
then satisfied for all later times because on
the conformally flat background the constraints propagate. Choosing
the initial data for the magnetic part of the spin-$2$-field to vanish, 
the only constraint left is
the electric constraint
\Eqref{eq:bianchi_electric_constraints}. How to solve this equation
on a general Berger sphere, in particular on $\scrip$ of
de-Sitter, has been discussed in 
\Sectionref{sec:solelectrconstraintscri}. 

\subsection{Numerical solutions of the linearized equations}
For our numerical experiments we choose the data of
\Sectionref{sec:solelectrconstraintscri} with  $a_3=1$,
$C_3=\sqrt{2}$ which means that initially
\[E_{11}=2\, \text{Re}(w_{21})=-E_{33},\quad E_{13}=\sqrt{2}\, w_{20}\]
and all other components of the spin-$2$-field vanish. As a side
remark, note that these initial data are the only data in this whole thesis
which are not Gowdy symmetric.
The numerical
solutions that we compute are compared 
to the corresponding exact solution of
the linearized equations constructed above whose $E_{11}$-component reads
\[E_{11}=-2\,\frac{\sin (10\arctan(t-1))}{(1-t+\frac 12 t^2)^3}\,
\text{Re}(w_{21}).\]

Let us define two error norms. The 
deviation of the numerical solution from the exact solution
is
\[\normdiffexact:=
\left\|E_{11}^{(\mathit{num})}-E_{11}^{(\mathit{exact})}\right\|_{L^1(\S)},\]
and the violation of the electric constraint
\Eqref{eq:bianchi_electric_constraints} is
\begin{equation}
  \label{eq:electric_error_norm}
  \normelec:=\left\|D_{e_c} E\indices{^c_e}
    -\epsilon\indices{^a^b_e}B_{da}\chi\indices{_b^d}\right\|_{L^1(\S)}.
\end{equation}
For tensorial quantities, as in the second definition, we always
assume summation over all components. Note however, that
$\normdiffexact$ takes into
account only one component of the solution. In general, this should
not be done because one could think 
about frames which leave special components roughly regular 
while the other components are problematic. For the analysis of this
simple test case in this section, we are 
convinced that this is sufficient, but
later on, we will try to avoid
such error norms.

We use the methods down-to-up (referred to as ``D2U''
in the plots) with staggered coordinate singularity
and the direct multiplication method (referred to as
``DirMul.'') with staggered coordinate singularity; both are
described in \Sectionref{sec:discussiontancot}. More
thorough comparisons between various pseudospectral methods are done
in \Sectionref{sec:comparison} in the non-linear case.
The runs here were done with various
resolutions referred to as ``lSlT'' (\textit{low space low time}),
``lSmT'' (\textit{low space medium time}) etc. The specific
resolutions are given below.

\begin{figure}[tb]
  \centering
  \subfloat[$\normdiffexact$ absolute (mT)]{
    \includegraphics[width=0.49\linewidth]{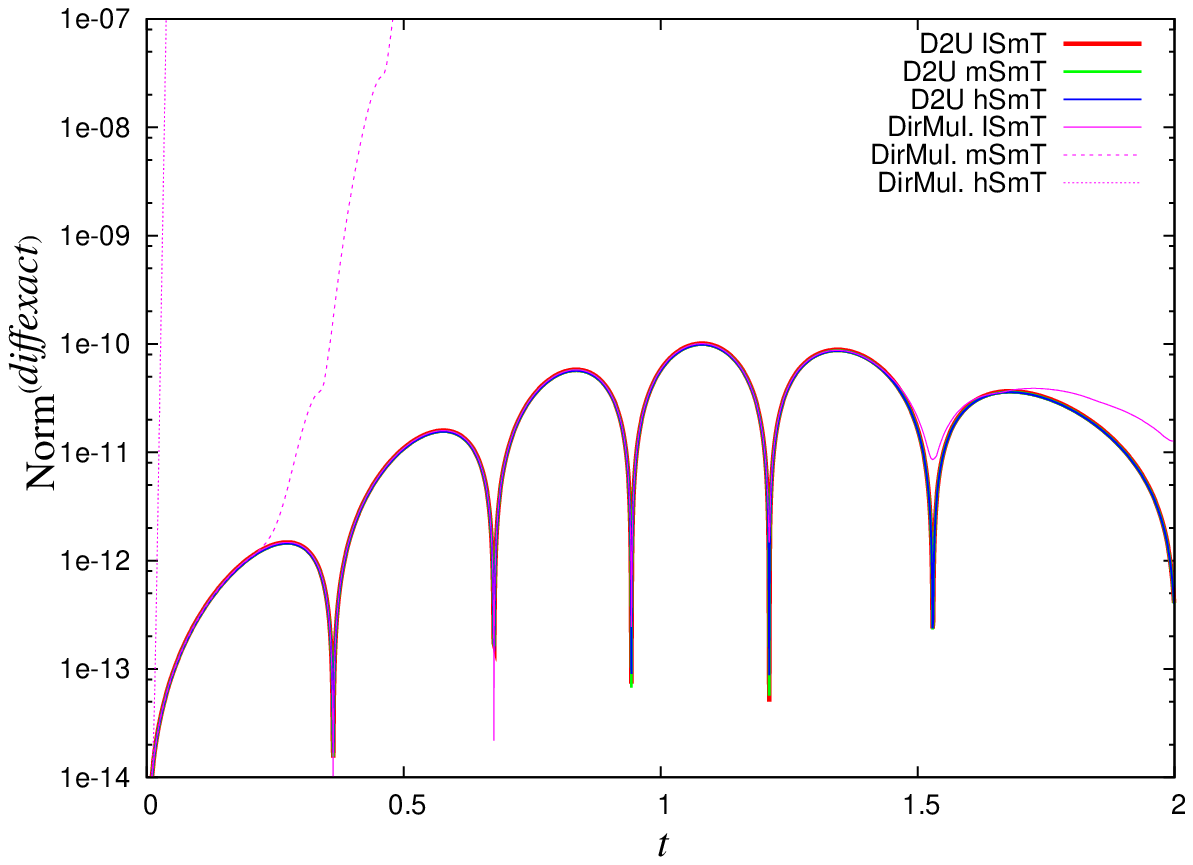}}%
  \subfloat[$\normdiffexact$ convergence]{
    \includegraphics[width=0.49\linewidth]
    {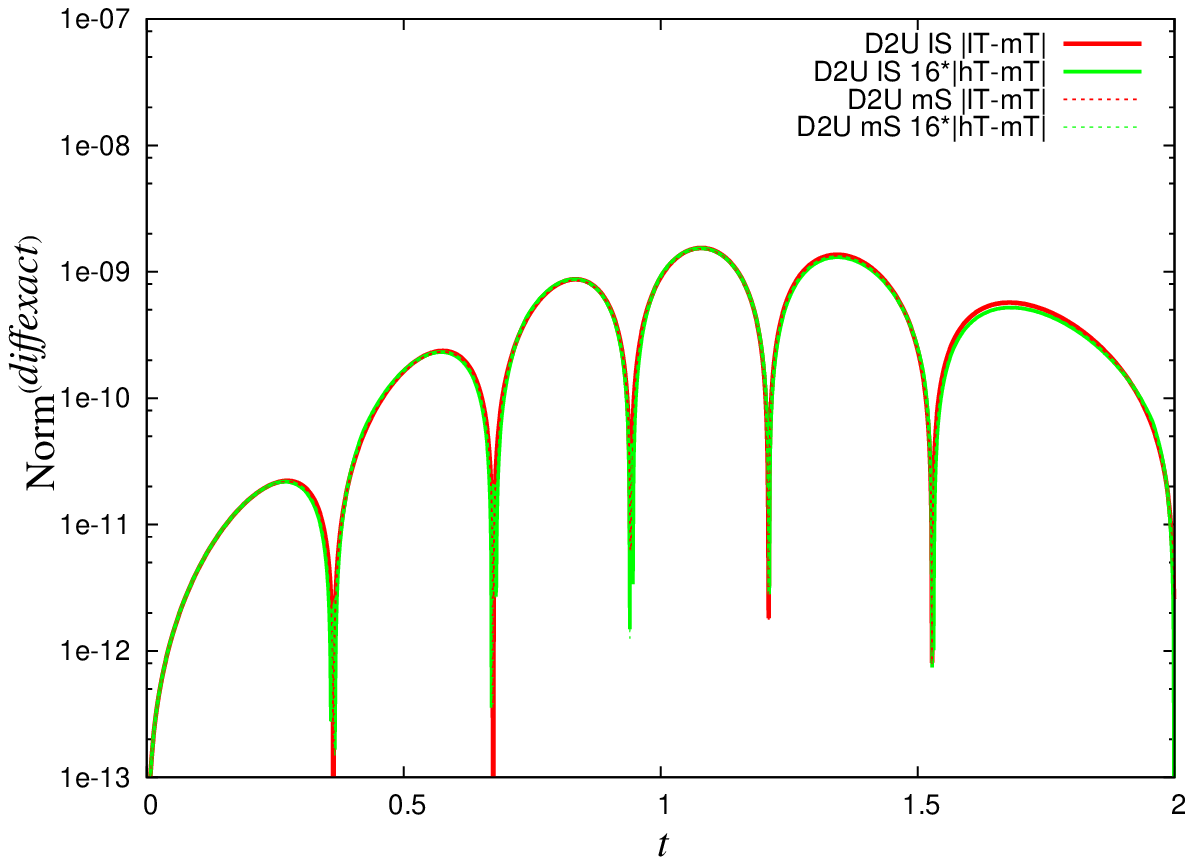}}
  \caption{Deviation from the exact solution}
  \label{fig:deviation_linearized}
\end{figure}
\begin{figure}[tb]
  \begin{minipage}[t]{0.49\linewidth}
    \centering
    \includegraphics[width=\linewidth]{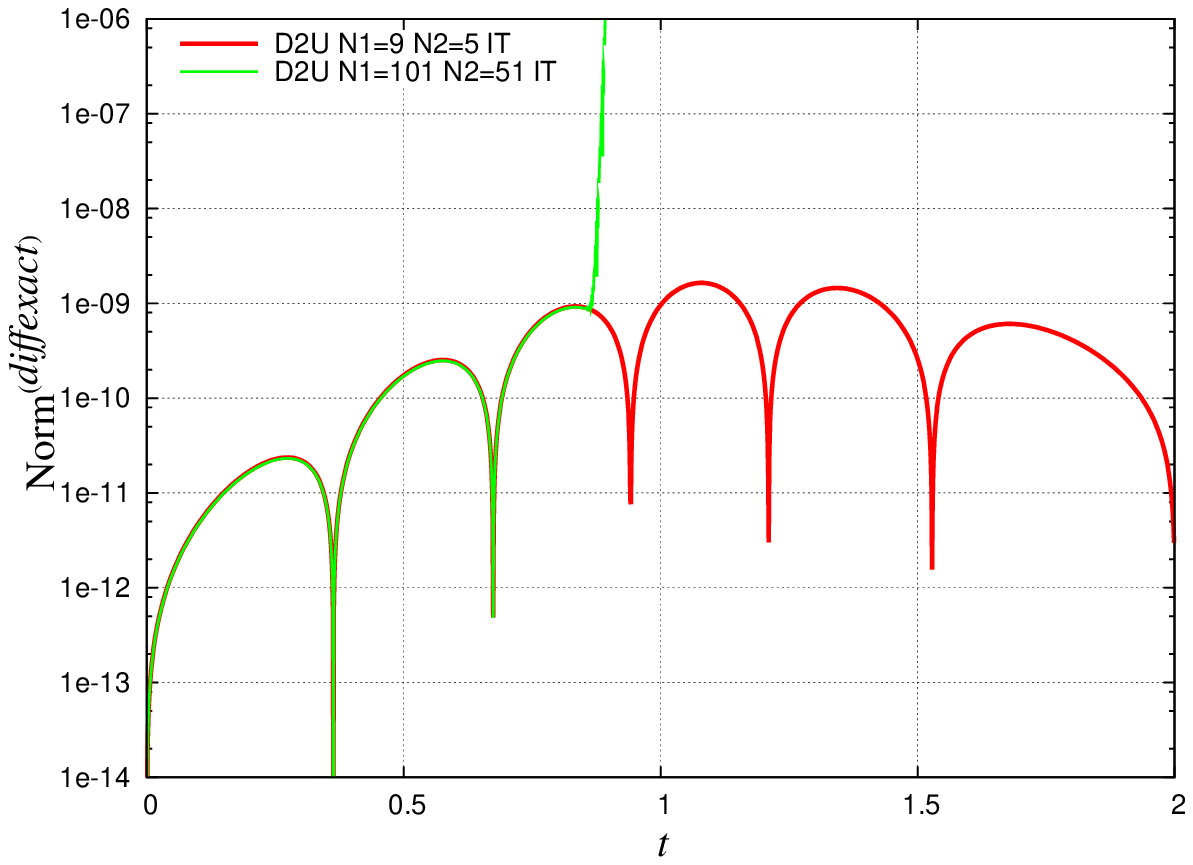}
    \caption{Instability for too high spatial resolutions}
    \label{fig:linearized_instability}
  \end{minipage}
  \begin{minipage}[t]{0.49\linewidth}
    \centering
    \includegraphics[width=\linewidth]{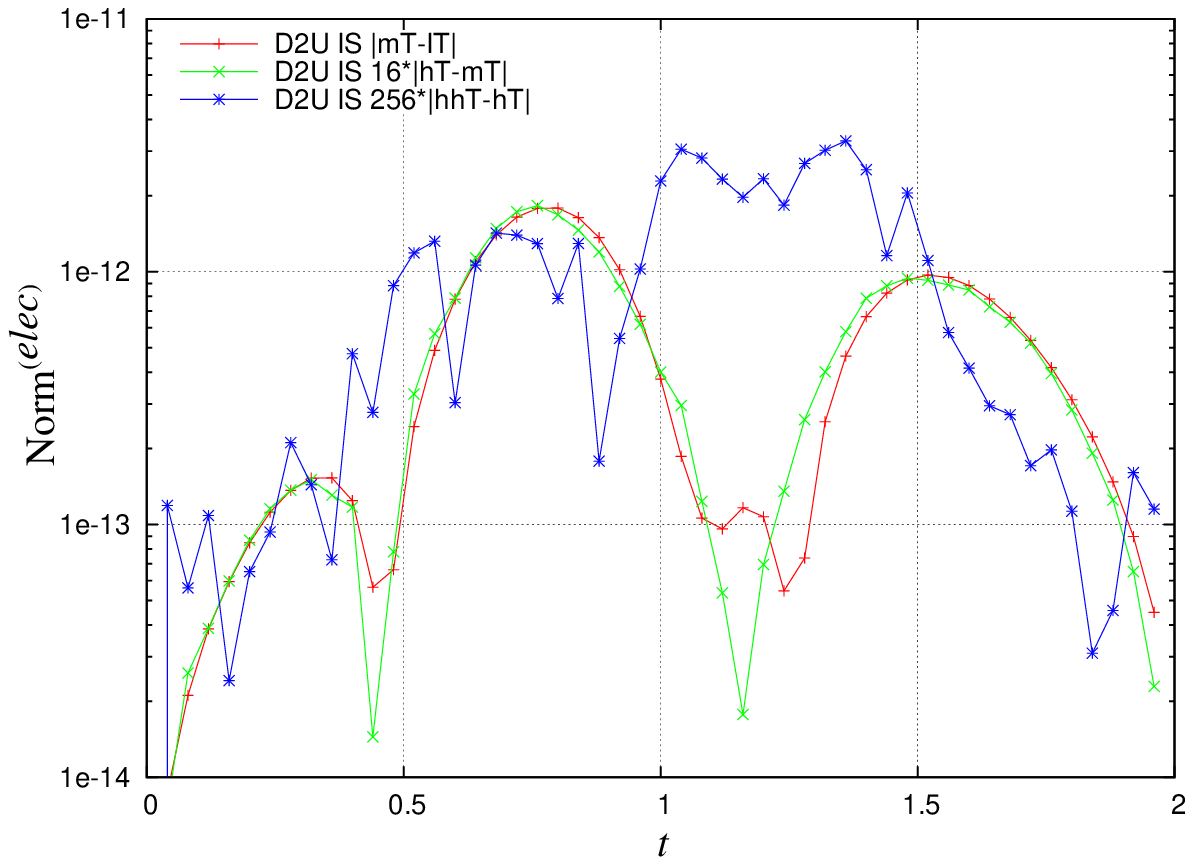} 
    \caption{Convergence of the constraint violations (lS)}
    \label{fig:linearized_convergence_constraint}
  \end{minipage}
\end{figure}

Consider \Figref{fig:deviation_linearized} where we plot the deviation
of the numerical from the exact solution for various
resolutions and for the two different methods mentioned above. The
abscissa represents time $t$ with $t=0$ corresponding to $\scrip$ and
$t=2$ to $\scrim$
while the ordinate shows $\normdiffexact$. The left plot is devoted to
show these errors for a given time resolution mT, namely with time step
$h=5.0\cdot 10^{-4}$, but for varying spatial resolutions lS
$N_1=9$, $N_2=5$, mS $N_1=25$, $N_2=13$, hS $N_1=77$, $N_2=39$. Here
$N_1$ is the number of collocation points in the $\chi$-direction and
$N_2$ the number in the $\rho_1$-direction. 
Note that the ``spiky features'' in the plot just represent the
oscillatority behavior of the exact solution.
In the right plot we show
convergence for varying time resolutions; here in addition to mT above
we have lT $h=1.0\cdot 10^{-3}$ and hT $h=2.5\cdot 10^{-4}$. In both
plots it is obvious that the down-to-up method works very well. The
absolute agreement with the exact solution is of the
order $10^{-10}$. The left plot implies that there is basically no
difference when the spatial resolution is changed which is clear since the 
solution consists only of basis functions $\{w_{np}\}$ with $n=2$ and
is hence, up to round-off errors,  
represented exactly by all three spatial resolutions, cf.\
\Lemref{lem:fourier_wnp}. Indeed, higher
spatial resolutions here can only make the numerical solution worse
because higher round-off errors are introduced. However, the influence
of round-off errors is not notable yet and everything is very stable
including our treatment of the coordinate singularity. To drive the
method instable, much higher spatial resolutions have to be used, see
\Figref{fig:linearized_instability}. The right plot demonstrates nice
$4$th-order convergence  
of the code with down-to-up in time. This means that the errors are
dominated by the 
time discretization for both the low and medium (and in fact also for
the high) spatial resolutions and then the $4$th-order Runge-Kutta method
enforces $4$th-order convergence. What is
interesting about the left picture of
\Fignref{fig:deviation_linearized} is that the direct
multiplication method, i.e.\ the naive way of treating the coordinate
singularity, is strongly instable\footnote{Here, by ``instable'' we do not
  mean the rigorous notion of \Sectionref{sec:convergence} but a
  numerical solutions which deviates from its expected behavior and
  blows up strongly.}. Higher resolution strengthen the 
instability which can be explained because the spatial points get
closer to the coordinate 
singularity.

In \Figref{fig:linearized_convergence_constraint} we also demonstrate
convergence of the constraint violations of the linearized
solution. This is actually not so easy because the initial
constraint violations are of the order of the machine round-off errors
and it stays like that during the evolution; hence we cannot
expect to find a converging constraint violation since round-off
errors spoil convergence. To see convergence of the constraints at all
we introduced  an artificial
constraint violation of the order $10^{-8}$ for the runs underlying
\Fignref{fig:linearized_convergence_constraint}. The constraint
propagation system implies that the constraint violations are
oscillatory in this simple case. However, even for that we find that
the difference in the constraint violation for the various resolutions
above is also of the order of the round-off error and still no
convergence can be observed. Hence, these runs (and only the runs for
this plot) are done at \textit{very} low time resolutions; here lT
$h=4.0\cdot 10^{-2}$, mT $h=2.0\cdot 10^{-2}$, hT $h=1.0\cdot
10^{-2}$, hhT $h=0.5\cdot 10^{-2}$ while spatial resolutions are as
above. Then, we are able to observe as shown in the plot, that there
is $4$th-order convergence but only for the three lower time
resolutions. For the highest time resolution hhT at least the orders of
magnitude are still correct. For even higher resolution no convergence
at all would be visible.

This already shows a problem for the interpretation of our numerical
results. Since in some applications pseudospectral methods are so
accurate that round-off errors dominate, the standard
convergence tests, that work very nicely for finite differencing
methods in particular, have to be handled with care.

\begin{figure}[tb]
  \centering
  \includegraphics[width=0.49\linewidth]{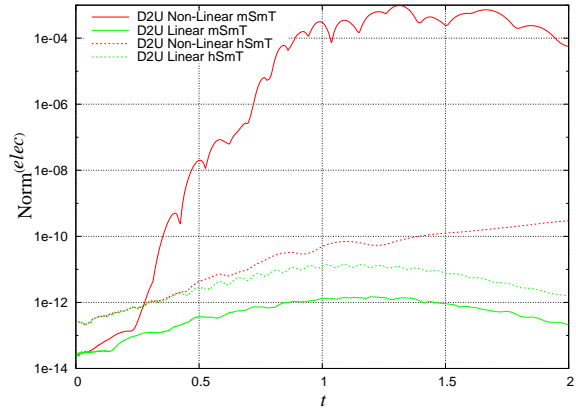}
  \caption{Comparison of constraint violations in the linearized and in the
    non-linear case}
  \label{fig:linearized_nonlinear}
\end{figure}
Finally consider \Figref{fig:linearized_nonlinear} where we compare
the propagation behaviors of the constraint in the linear and
in the non-linear case, i.e.\ when the coupling between the
``background part'' and the spin-$2$-system is switched on again. We
do this for two different spatial resolutions but with one time
resolution. One can expect that the non-linear case is much more
severe since the non-linear coupling induces fine structure
formation. In particular, the medium spatial resolution is only
sufficient at early times in the non-linear case while for the high
resolution there is not very much difference between the violations in
the linear and the non-linear case which means that hS is sufficient
to resolve all fine structure formed. Hence, in generic non-linear
runs one can expect that the demand for spatial resolution increases
with time and hence it will be crucial to have some method which keeps
track of this demand. A possible, and so far working approach to this
is the spatial adaption method introduced in
\Sectionref{sec:spatial_adaption}. This will be used intensively in
the computation of singular spacetimes in
\Chapterref{ch:singulargowdy}. Note another typical feature from
\Fignref{fig:linearized_nonlinear}: the initial constraint violation
is higher the higher the spatial resolution is chosen initially due to
the influence of round-off errors. This
has a some impact for later discussions.

\subsection{Other explicit solutions}
To further check the implementations of the equations I have also
reproduced successfully the explicitly known $\lambda$-Taub-NUT family
\Sectionref{sec:TaubNUT} in the way mentioned at the end of
\Sectionref{sec:solelectrconstraintscri}. We will not discuss
this further here.

\section{Regular \texorpdfstring{$\lambda$}{lambda}-Gowdy 
spacetimes on  \texorpdfstring{$\S$}{S3}}
\label{sec:comparison}

\setcounter{totalnumber}{2}

In this section we make numerical experiments with the methods
introduced and discussed in \Sectionref{sec:discussiontancot} with the
full non-linear evolution equations, namely 
the general conformal field equations in Levi-Civita conformal Gauß
gauge as described in \Sectionref{sec:LCCGG}. As initial data on
$\scrip$ we
choose a Gowdy
initial data set  on a Berger sphere as derived  
in \Sectionref{sec:solelectrconstraintscri} that turns out to be in
the de-Sitter 
stability region dSSR (\Sectionref{sec:non-linear stability of de-Sitter}).
The initial data parameters are
\[a_3=1,\,a_3=0.92,\,(E_{11})_{0,0}=(E_{22})_{0,0}=0,\,C_2=0.5,\]
cf.\ \Eqsref{eq:n2solelectrconstraint}; all other $C_i$ vanish
because we demand Gowdy symmetry. The corresponding solution turns out
to be
both future and past asymptotically de-Sitter, but which nevertheless has 
some non-trivial amount of inhomogeneity. 

In this section, results from five different pseudospectral methods to
compute frame derivatives on $\S$ (\Sectionref{sec:discussiontancot})
are presented and compared.
The first method, referred to as 
``D2U Stag.'' in the following plots, is the down-to-up method
introduced in \Sectionref{sec:discussiontancot} with staggered coordinate
singularities, i.e.\ the collocation points are shifted such that all
coordinate 
singularities are exactly in the middle between two grid
points. Correspondingly, we make the same computation with  the method 
``D2U Non-Stag.'' where the coordinate singularities coincide
with the relevant grid points. We also show results of the method
``D2UMod Stag.'' which is just a reimplementation of the
down-to-up method above but which is arranged slightly
differently such that the distribution of round-off errors is not
the same. We expect that the results of this method are the same as for
``D2U Stag.'' but the influence of round-off errors is hard to
predict; eventually it turns out that there are indeed differences,
see below.
Furthermore, we did runs with the method ``U2D Stag.'' and ``DirMul.'', cf.\
\Sectionref{sec:discussiontancot}. For the direct multiplication
method ``DirMul.'' the 
coordinate singularities are staggered between the grid points.

The runs were performed with various
resolutions referred to as ``lSlT'' (\textit{low space low time}),
``lSmT'' (\textit{low space medium time}), ``lShT'' 
(\textit{low space high time}), ``mShT'' 
(\textit{medium space high time}) etc. The low resolution in time
corresponds to a time step $h=2\cdot 10^{-3}$, the medium one to
$h=1\cdot 10^{-3}$ and the high one to $h=5\cdot 10^{-4}$. Note that
$t=0$ corresponds to $\scrip$ and $t=2$ to $\scrim$ thus the complete
spacetime is computed. The
low resolution in space is $N_1=25$, $N_2=13$ where $N_1$ is the
number of collocation points in $\chi$-direction and $N_2$ the number
in $\rho_1$-direction. The medium resolution in space is given by
$N_1=41$, $N_2=32$.

It is important to note
that here we do not have explicit solutions to compare with;
other measures of the numerical errors have to be found. We use
the violation norm of the electric constraint 
\Eqref{eq:electric_error_norm} and 
the following measure of the violation of Einstein's vacuum field
equations \Eqref{eq:EFE} 
\[\normeinstein
:=\left\|(\tilde R_{ij}-\lambda \tilde g_{ij})/
\Omega(t)\right\|_{W^{1,1}(\S)}.\]
By the $W^{1,1}$-norm of a quantity $u$ we mean the sum of the
$L^1$-norm of $u$ and the $L^1$-norms of all $\{Y_a\}$-derivatives of
$u$. For the tensorial quantities we additionally sum over all
components as we will do in all similar expressions in the following. 
The physical Ricci tensor and the physical metric are 
projected onto the physical orthonormal frame. Since this expression
is $O(\Omega)$ close to $\scri$ with $\Omega$ the physical conformal
factor, this factor is divided out. Another norm that monitors the
behavior of the numerical code is  
$\normadapt$ introduced in \Sectionref{sec:spatial_adaption}; however,
in this section this norm is not yet used for automatic spatial adaption.
A further norm is
\[\normweyl
:=\left\|E_{ab}\right\|_{L^1(\S)}+\left\|B_{ab}\right\|_{L^1(\S)}\]
which gives us the order of magnitude of certain relevant components
of the solution at a given
time. Another error quantity measures the quality of Gowdy symmetry. As
explained in \Sectionref{sec:implementationOfS3Gowdy}, the symmetry
along $Y_3$ is not enforced during time evolution and
there is the possibility that the solution, although initially fully
Gowdy-symmetric, strongly deviates.  To check this we thus define the
norm
\[\normkilling
:=\left\|K_{ab}\right\|_{L^1(\S)}\]
where the Killing operator $K_{ab}$ is defined by
\[K_{ab}:=g([Y_3,e_{(a}],e_{b)}),\]
cf.\ \Eqref{eq:killing_eq_onf}; note that
$[Y_3,e_0]=0=g([Y_3,e_a],e_0)$ by the choice of gauge.

\begin{figure}[tb]
  \centering
  \subfloat[lSmT $N_1=25$, $N_2=13$, $h=1.0\cdot 10^{-3}$]{
    \includegraphics[width=0.49\linewidth]{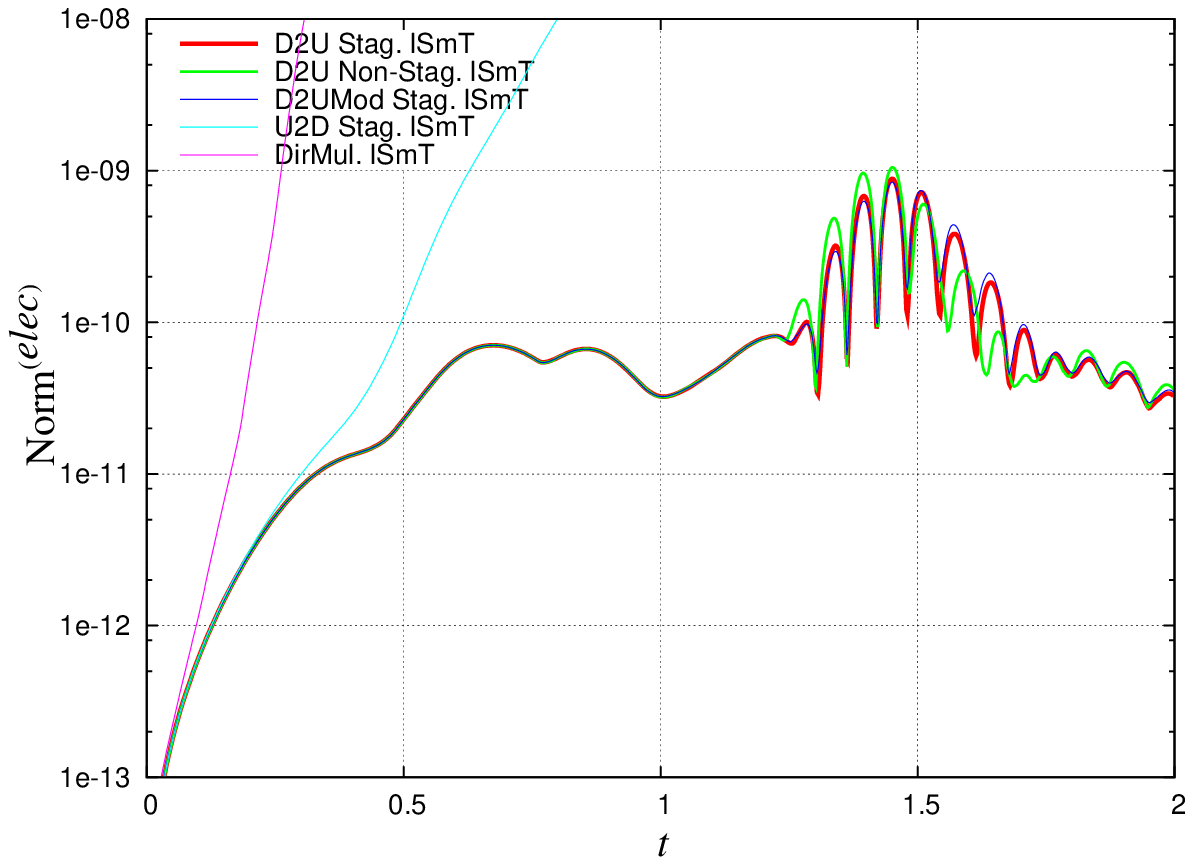}}%
  \subfloat[lShT $N_1=25$, $N_2=13$, $h=0.5\cdot 10^{-3}$]{
    \includegraphics[width=0.49\linewidth]{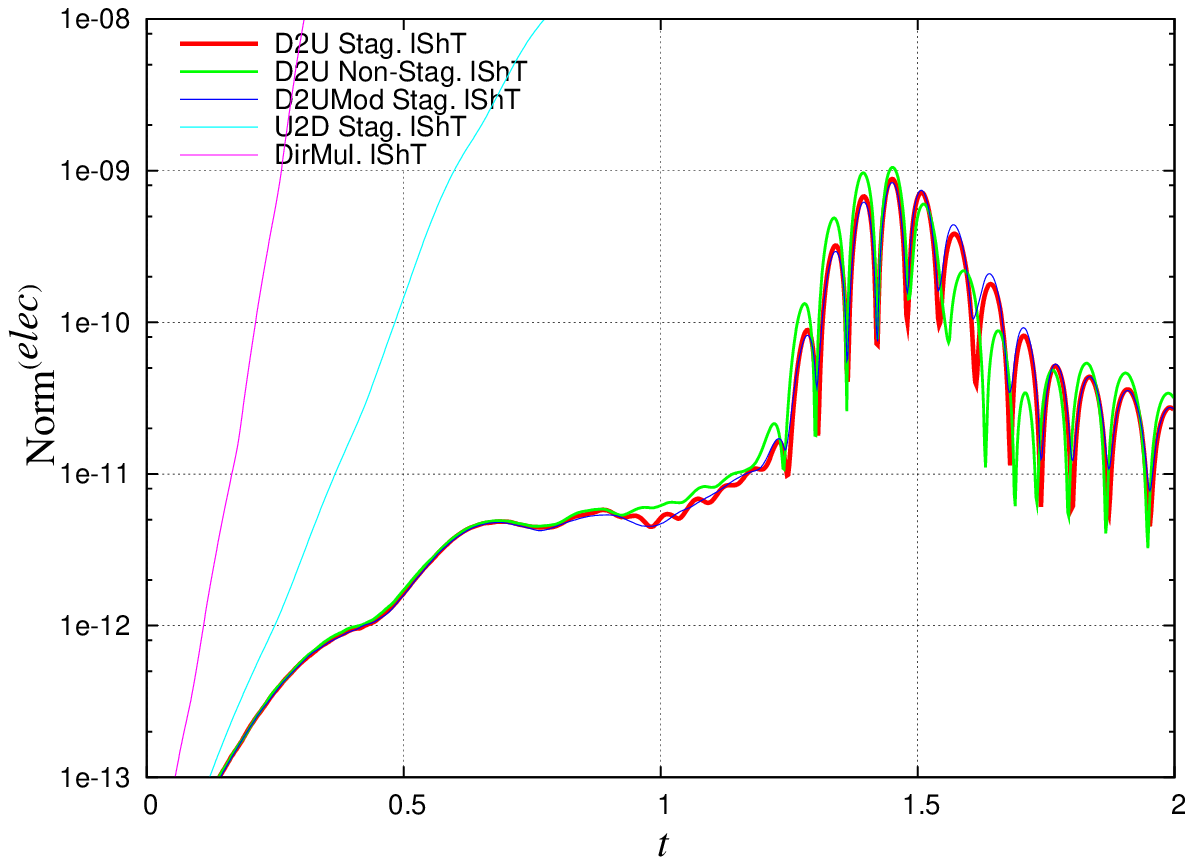}}\par
  \subfloat[mSmT $N_1=41$, $N_2=32$, $h=1.0\cdot 10^{-3}$]{
    \includegraphics[width=0.49\linewidth]{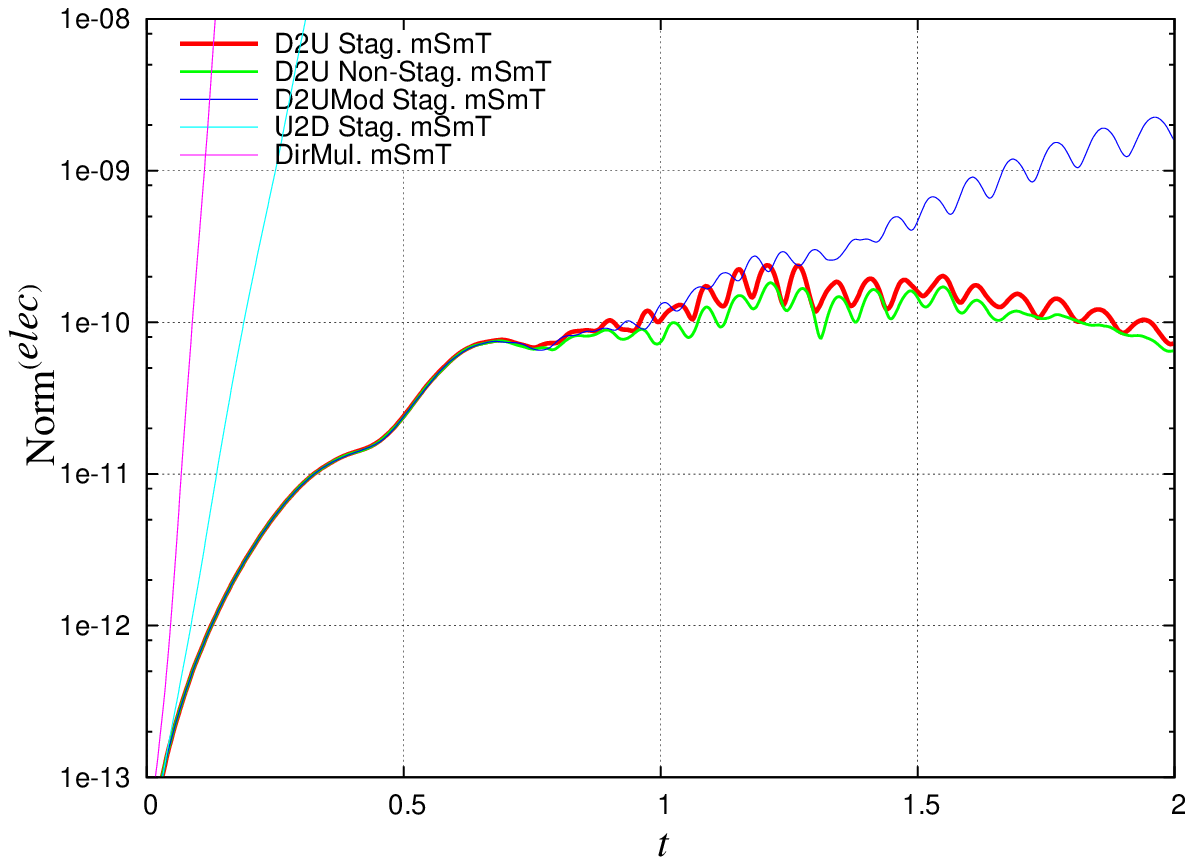}}%
  \subfloat[mShT $N_1=41$, $N_2=32$, $h=0.5\cdot 10^{-3}$]{
    \includegraphics[width=0.49\linewidth]{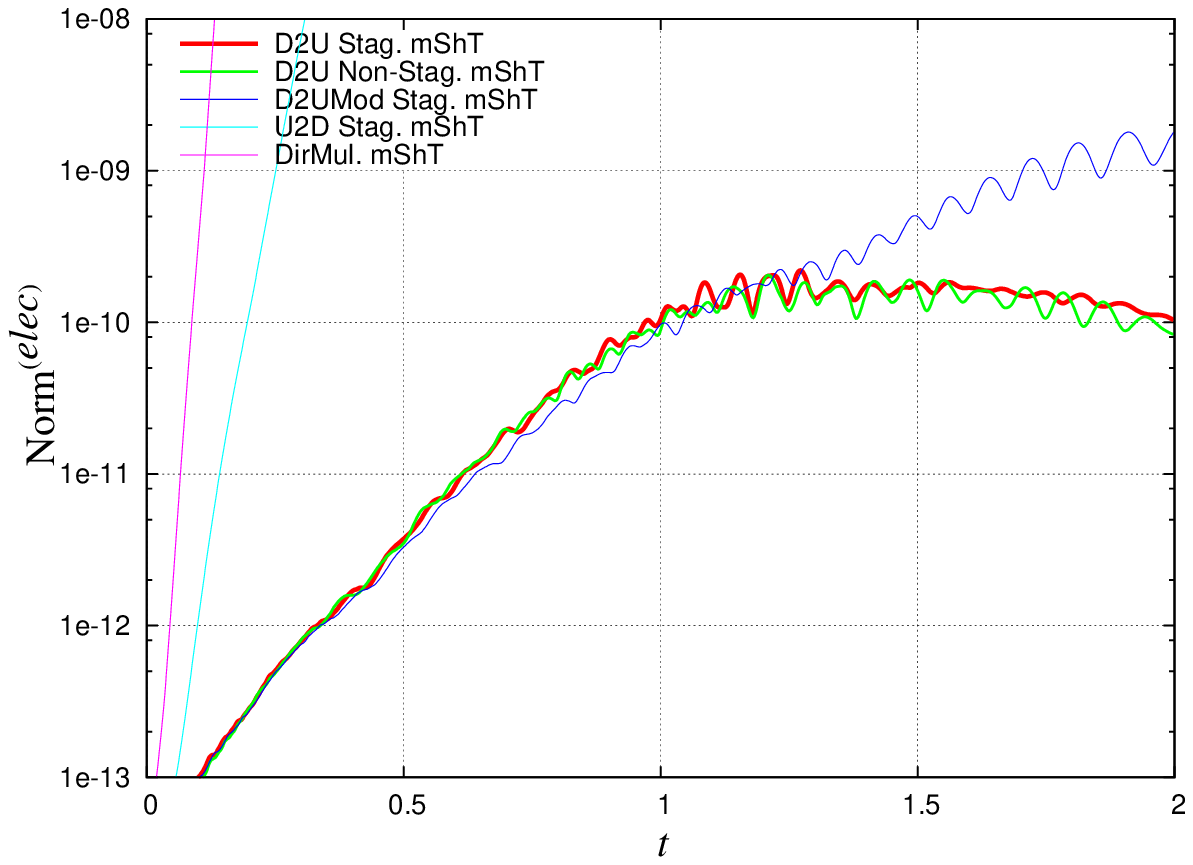}}
  \caption{Violation of electric constraint for the methods discussed
    in the text} 
  \label{fig:electric_viol_all_methods}
\end{figure}

\begin{figure}[tb]
  \centering
  \subfloat[lS $N_1=25$, $N_2=13$]{
    \includegraphics[width=0.49\linewidth]{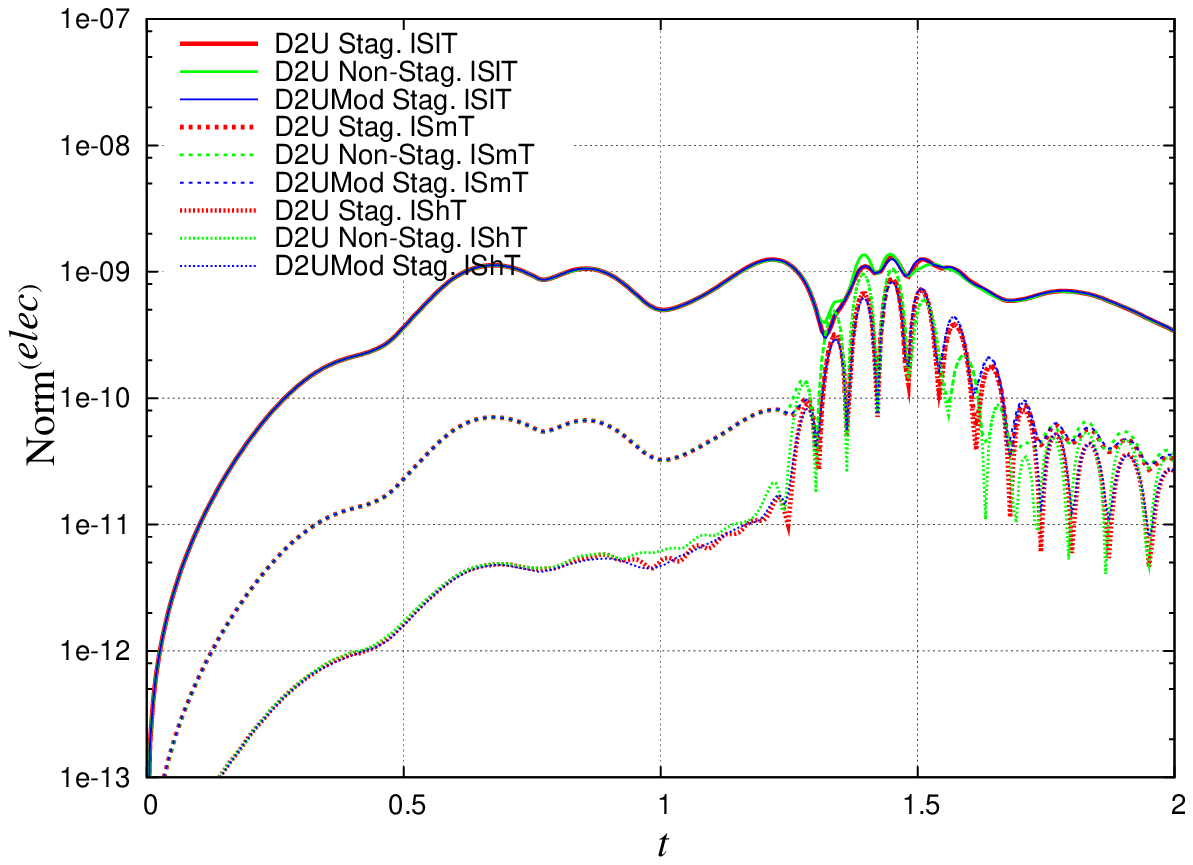}}%
  \subfloat[mS $N_1=41$, $N_2=32$]{
    \includegraphics[width=0.49\linewidth]{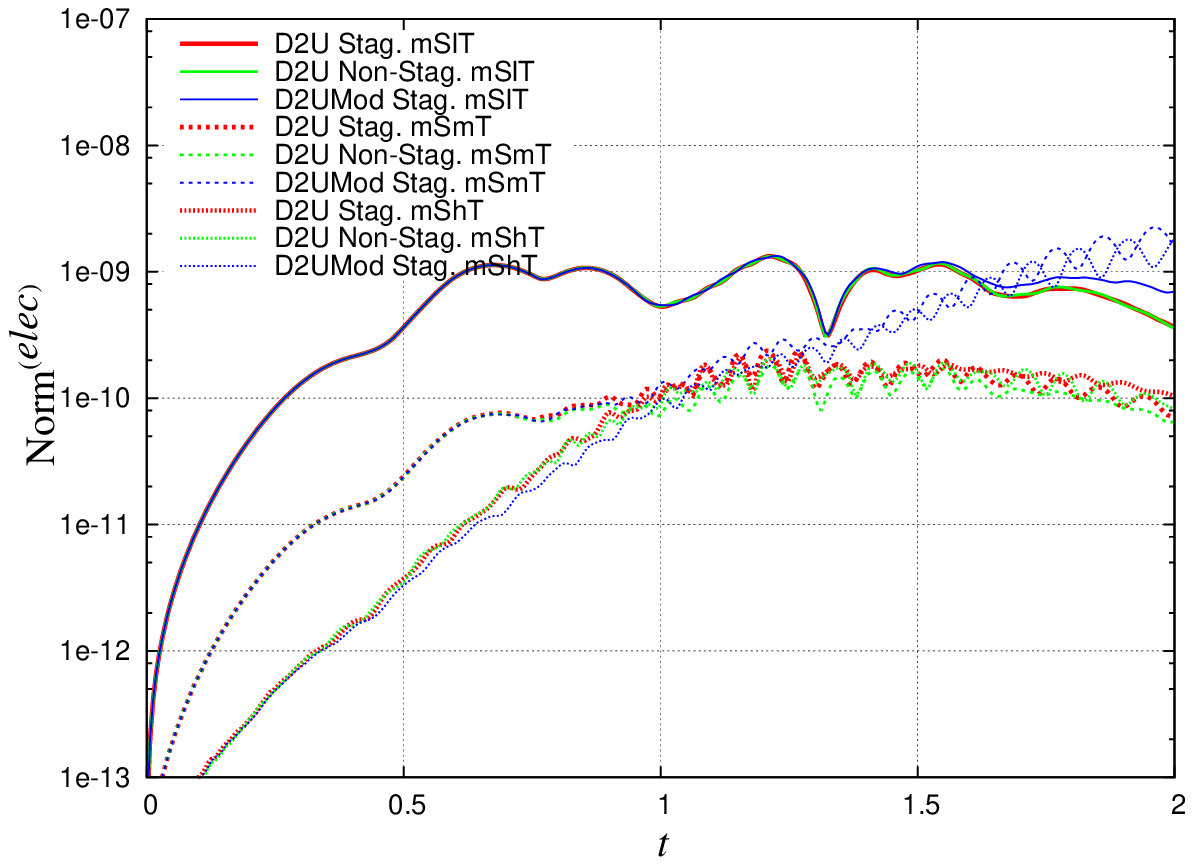}}
  \caption{Violation of electric constraint for the three stable methods}
  \label{fig:electric_viol_3_methods}
\end{figure}

Consider \Figref{fig:electric_viol_all_methods}. Here we plot the
violation of the electric constraint vs.\ time for the methods above
where each 
picture shows a fixed resolution. Exponential constraint growth, as we
observe here, can be
expected in most free evolutions so  the questions
are rather, first, how strong this growth is, second, if the
violations stays small 
enough for the relevant time interval such that the
solution can be trusted, and third, what the main reason for the growth
is. The down-to-up variants all work 
very well and are stable. There is no difference if the coordinate
singularity is staggered or not as expected. For the highest
resolution the modified down-to-up method ``D2U Mod'' is a bit worse
which shows that the influence of round-off errors for different
implementations of the same thing can behave quite
differently. However, this method can also be considered as
stable. Definitely instable are the direct multiplication and the
up-to-down methods. In particular, the instability is stronger the
higher the spatial resolution is. For the direct
multiplication method this can be explained because the grid points
get closer to the coordinate singularity for higher resolutions. For
the up-to-down method, it seems to be true what we suspected already
in \Sectionref{sec:discussiontancot}: the round-off errors which are
relatively strong in high frequencies get distributed to the low
frequencies and this drives instability. Hence, these two methods can
be considered as 
unusable. Increasing the time resolution yields a smaller constraint
violation at least for low times; we check for convergence
in \Figref{fig:electric_viol_3_methods}. As we have already seen in
\Figref{fig:linearized_nonlinear}  
the errors for later times can be dominated by a lack in spatial
resolution, in particular for such fixed resolution runs. This results in
oscillations and we can observe in
\Fignref{fig:electric_viol_all_methods} that their amplitudes are
smaller the higher the spatial resolution is. Comparing the late time
behavior of the third and the fourth plot it can be seen that the
mean magnitude of the constraint violations  does not 
change although the time resolution increases. This gives us
another hint that error quantities in pseudospectral
codes must sometimes be interpreted differently than in
finite-differencing approaches. The problem is the following. Constraint
violations of the Bianchi system are propagated by means of an
homogeneous 
symmetric hyperbolic system of equations, the subsidiary system
(\Sectionref{sec:maximaldevelopments}). In particular, if the 
constraint violations vanish initially, they will vanish for all
times. But in a pseudospectral code, the initial constraint violation
is of the order of the machine precision, which is not zero, and hence
the corresponding
solution of the subsidiary system does not vanish. In fact,
depending on the properties of the evolution system, the constraints
do typically grow exponentially in time. Hence, from some resolution on, the
constraint violations cannot be decreased anymore by increasing the
resolution since the ``true'' solution of the subsidiary system with
initial data of the order of machine 
precision gets resolved. In fact then, increasing the resolution, increases
the initial 
data for the constraint quantities which yields an even higher
constraint
violations as the corresponding solution of the subsidiary system.  In
any case, since practically we cannot increase the machine 
precision\footnote{The machine precision can indeed be increased but
  only with a significant loss of performance.}, the only possibility
that is left is to decrease the actual 
solution of the constraint propagation system by changing the
properties of the evolution system. In the most 
prominent formulations of Einstein's field equations, there are attempts
to introduce so called \term{constraint damping terms} into the
evolution system such that the corresponding constraint propagation
system drives the constraint violations to zero; for instance
\cite{Brodbeck98,Gundlach05}. However, similar techniques have not been applied
to the Bianchi system yet although there are analyses of the
constraint propagation in \cite{Frauendiener04}.
In any case, what we can see here is that
given high enough resolution, our method is able to approximate the actual
solution of the subsidiary system quite accurately. With
such resolutions the constraint 
errors and the round-off errors, but not the discretization error,
dominate the errors of the solution. 

The discussion of \Figref{fig:electric_viol_3_methods} is
related. Here we plot again the behavior of constraint violations
versus time, but we restrict to the three stable methods. In each of
the two plots the spatial resolution is fixed while we vary the time
resolution. Although it is not explicitly checked in this plots, we
find $4$th-order convergence for low times. The same phenomena
observed as before, that
for later times the spatial discretization contributes to the errors,
can be observed and this results in oscillations.
\begin{figure}[tb]
  \centering
  \includegraphics[width=0.49\linewidth]{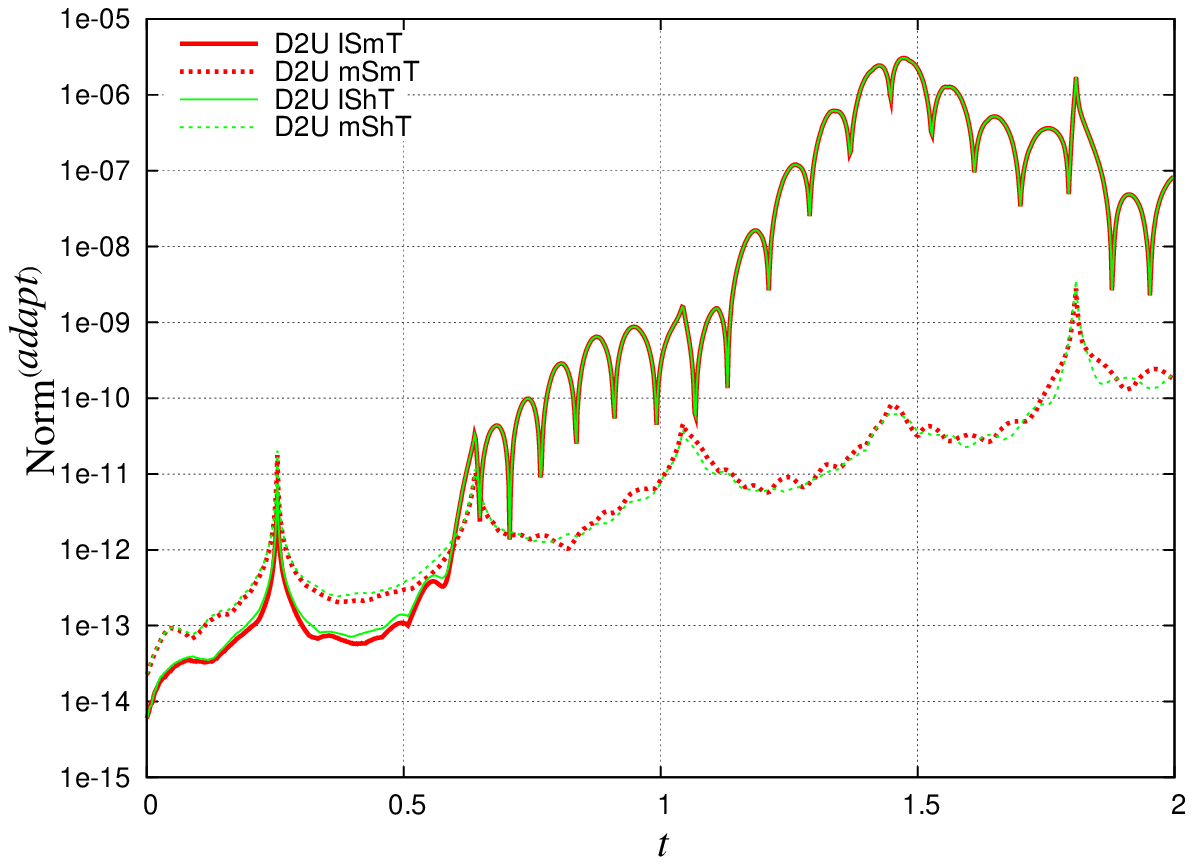}
  \caption{$\normadapt$ (\Sectionref{sec:spatial_adaption}) for some
    resolutions} 
  \label{fig:adapt_norm}
\end{figure}
Related to this is \Figref{fig:adapt_norm} where we show the behavior
of the adaption norm. Recall again that adaption is not used in these
runs; hence the apparent dynamics in this norm is caused exclusively
by the same 
oscillations which we have also seen in the other norms. We see, as
expected, that for low spatial 
resolutions, approximately independent of the time resolution, the
amount of power in the high frequencies is quite high at later
times and the 
aliasing effect (\Sectionref{sec:pseudospsectr_background}) is then
responsible for these oscillations. For higher spatial
resolutions, the power in the high frequencies is correspondingly
smaller and hence the aliasing effect is not dominant.

\begin{figure}[tb]
  \centering
  \subfloat[lS $N_1=25$, $N_2=13$]{%
    \includegraphics[width=0.49\linewidth]{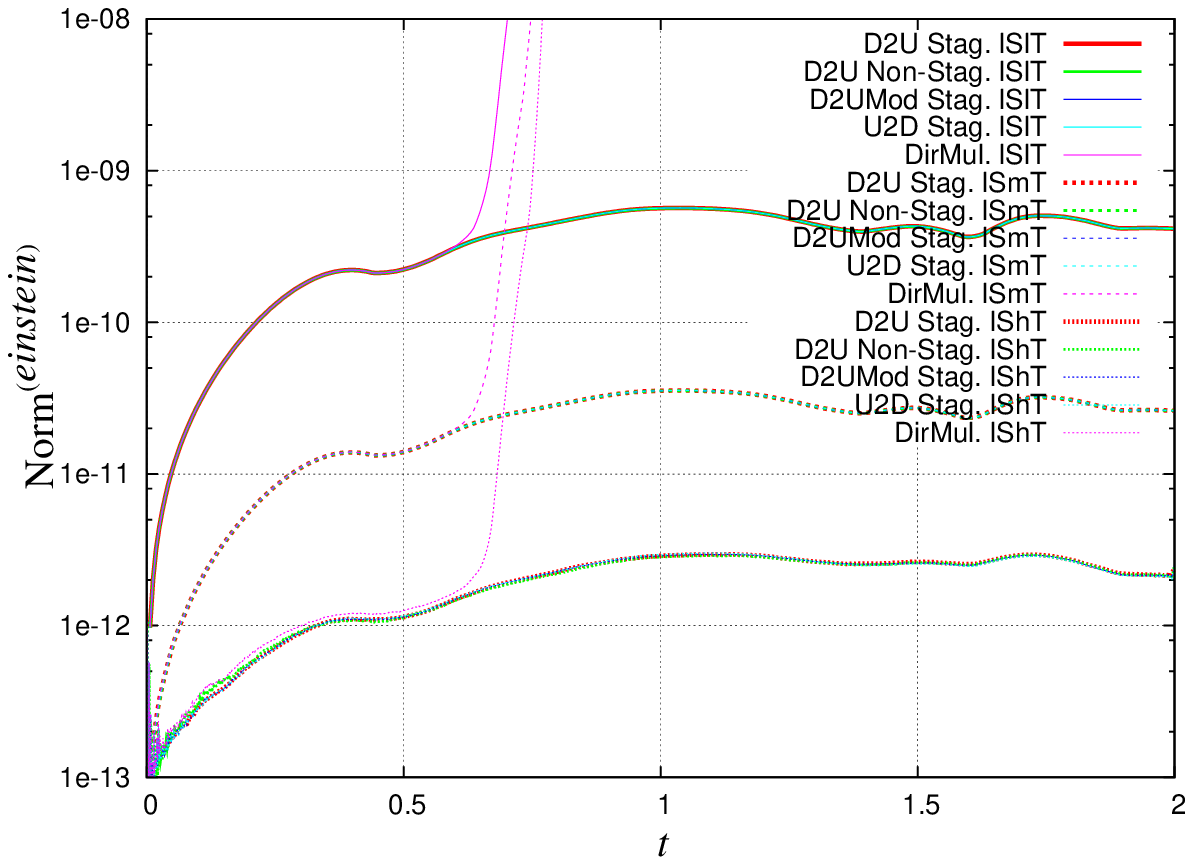}}
  \subfloat[mS $N_1=41$, $N_2=32$]{%
    \includegraphics[width=0.49\linewidth]{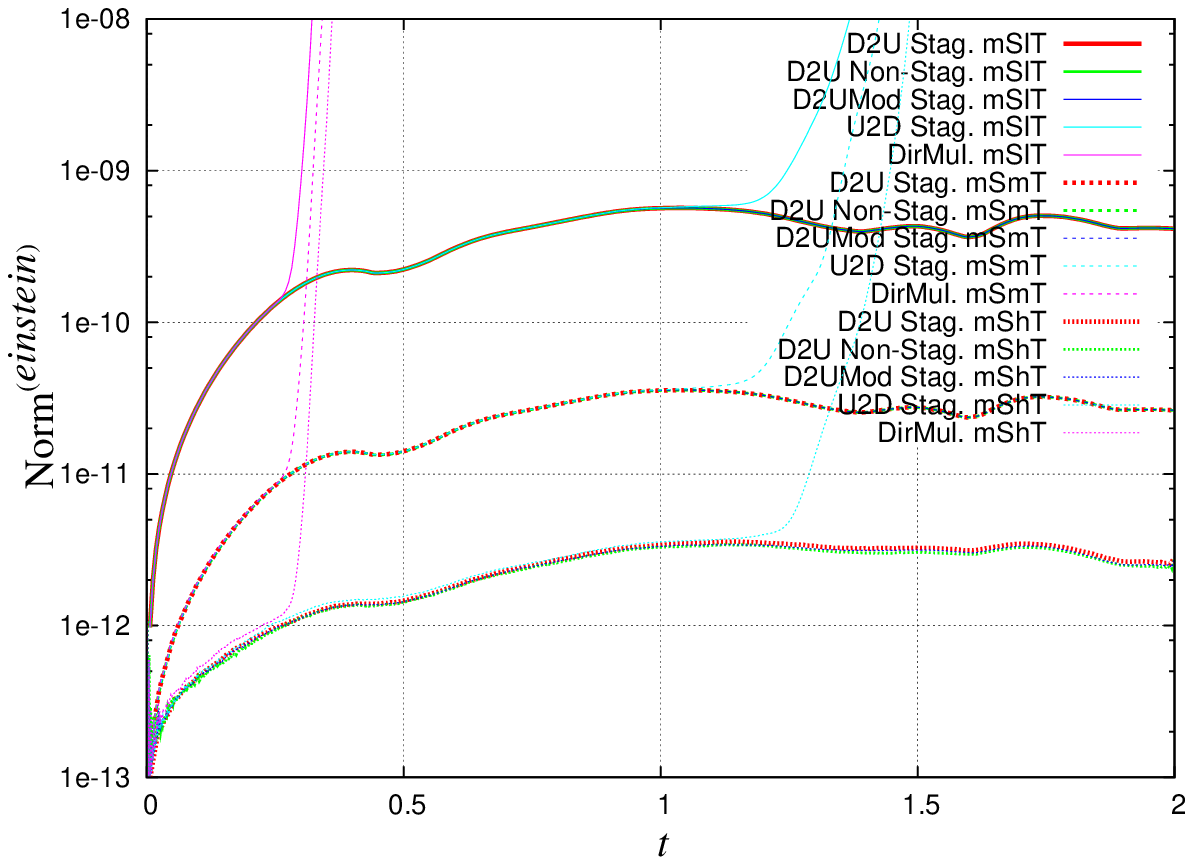}}
  \caption{Violation of Einstein's vacuum equations for various
    resolutions and methods} 
  \label{fig:einstein_all_methods}
\end{figure}
As we discussed before, the ability to use $\normelec$ as an error monitor
quantity has limitations. That is why we also discuss the behavior of
$\normeinstein$ for all the methods above in
\Fignref{fig:einstein_all_methods}. This error 
quantity measures the error in the standard ADM evolution and constraint
equations (\Sectionref{sec:maximaldevelopments}). In fact, the
electric constraint corresponding to the 
quantity $\normelec$ is only one of a large 
number of constraints in our formulation of the field equations, and
it is
one differential order higher than the standard ones included in
$\normeinstein$. 
It is difficult to understand how all these constraints somehow sum up
to the standard constraints. To measure these higher order errors, the
first spatial derivatives of the deviations from Einstein's field
equations are included by the $W^{1,1}$-norm. Now, have a look at 
\Fignref{fig:einstein_all_methods}. One sees that all methods agree
for some time then suddenly, dependent on the spatial resolution, the
direct multiplication method and later also the up-to-down method
strongly deviate. What is surprising is that in the plots before we saw
the two instable methods to deviate much earlier and much stronger.
\begin{figure}[tb]
  \centering
  \includegraphics[width=0.49\linewidth]{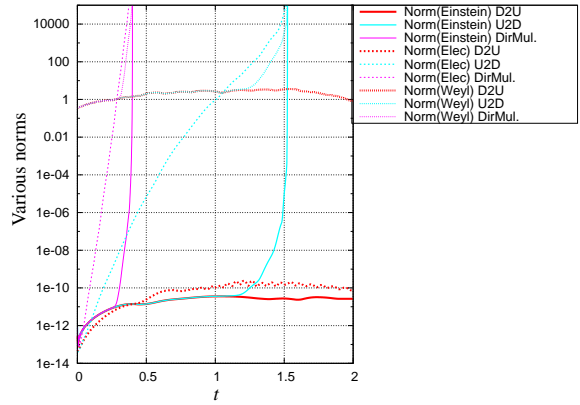}
  \caption{Comparison of the orders of magnitude for the different
    norms (mSmT)}
  \label{fig:orders_magnitude}
\end{figure}
To obtain some deeper understanding look at
\Figref{fig:orders_magnitude}. There we see that the deviation of
$\normeinstein$ happens at that time when $\normelec$ gets higher than the
order of magnitude of the solution itself, represented by
$\normweyl$. Hence, in this sense, $\normeinstein$ is on the one hand
much less sensitive to errors than $\normelec$ but on the other hand
indicates when the errors really begin to dominate the
solution. This should be of general interest since the standard constraints
included in $\normeinstein$ are the typical error monitors in
numerical relativity.

\begin{figure}[tb]
  \begin{minipage}[t]{0.49\linewidth}
    \centering
    \includegraphics[width=\linewidth]
    {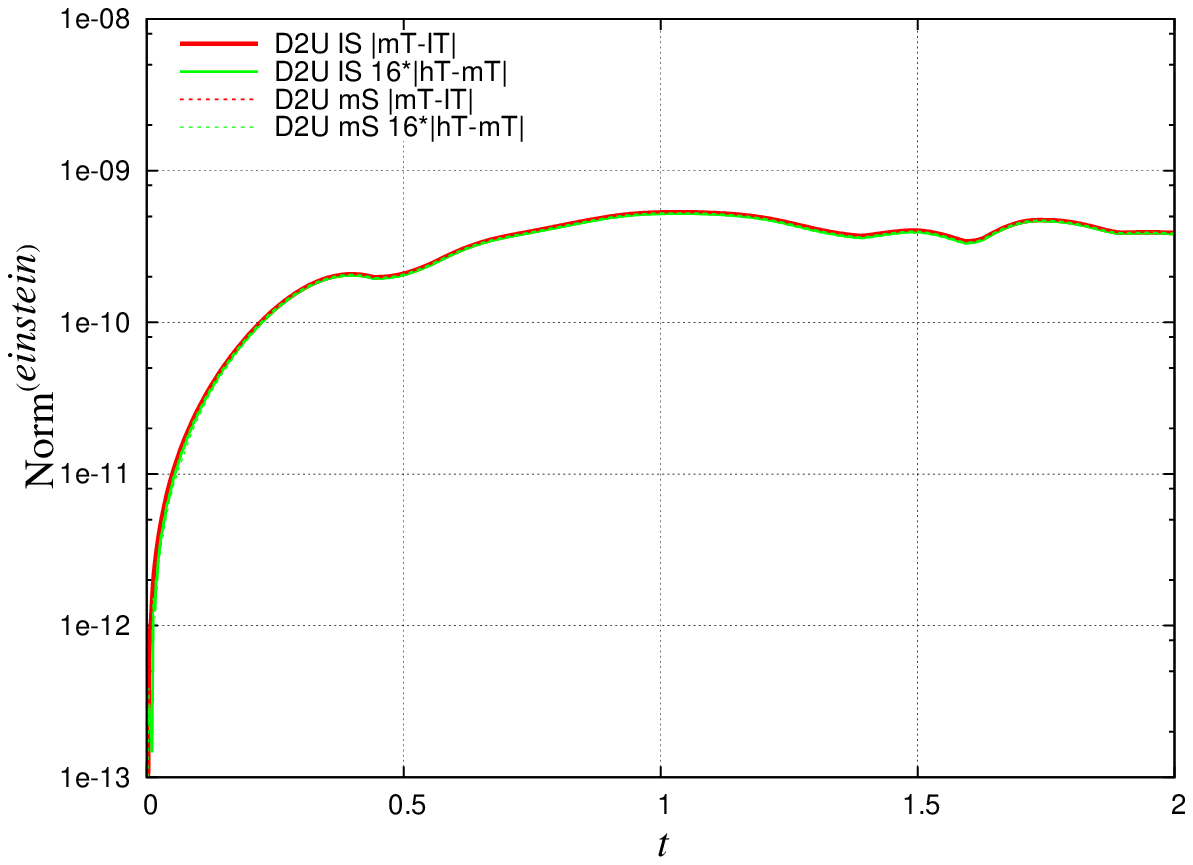}
    \caption{Convergence of $\normeinstein$}
    \label{fig:einstein_convergence}
  \end{minipage}
  \begin{minipage}[t]{0.49\linewidth}
    \centering
    \includegraphics[width=\linewidth]{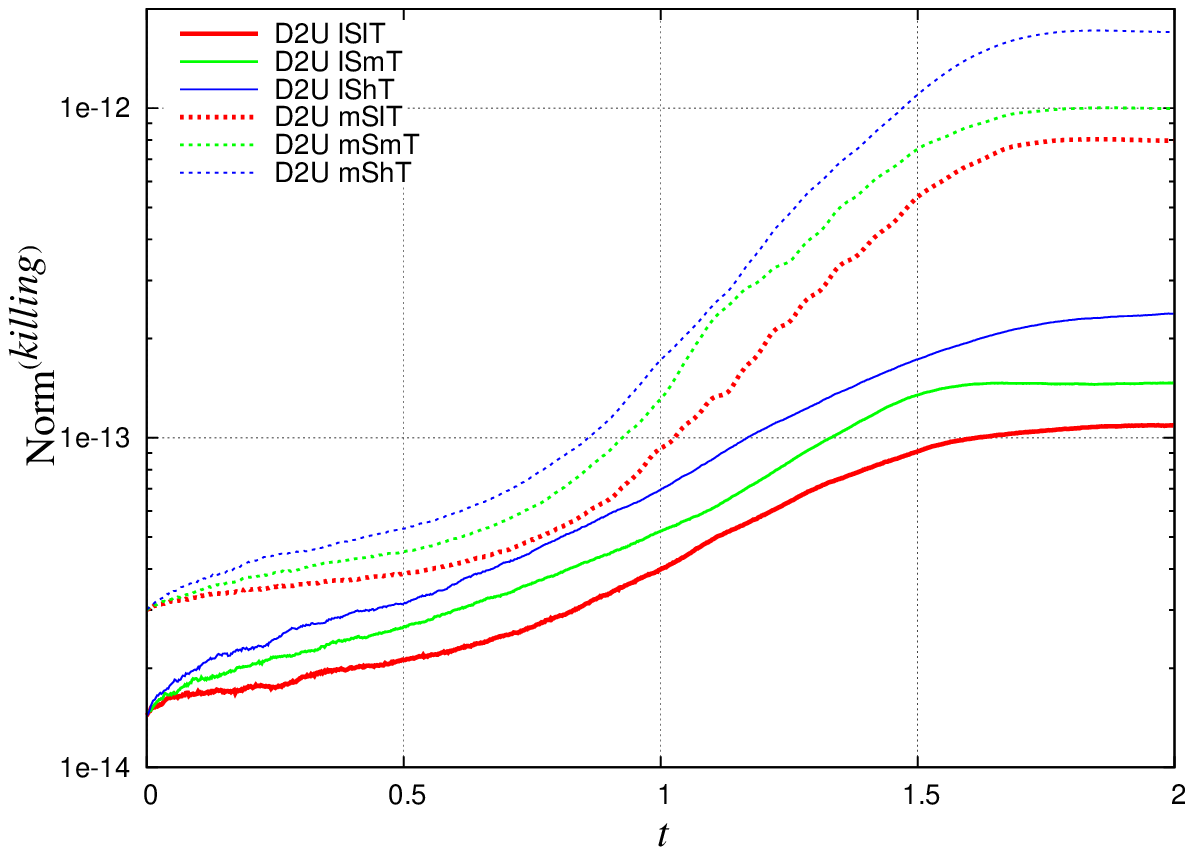}
    \caption{Violation of Gowdy symmetry}
    \label{fig:killing_norm}
  \end{minipage}
\end{figure}
In \Figref{fig:einstein_convergence} we show convergence of
$\normeinstein$. The errors induced by time discretization seem to
play the dominant role for $\normeinstein$ and we find $4$th-order
convergence for all times. 

In \Figref{fig:killing_norm} we show the behavior of
$\normkilling$. We have already explained above that the code does not
explicitly enforce the invariance along $Y_3$, and one cannot exclude that
numerical errors 
together with possible non-linear instabilities of the continuum evolution
equations drive the solution away from this symmetry. This plot shows
that this does not seem to be the case. The
violation of the symmetry is a bit stronger when the resolution is
higher but this is to some degree also caused by the higher round-off
errors in the spatial derivative that one has to take to compute
$\normkilling$. Although these investigations here cannot be
considered as systematic this result gives us indeed a hint that there
is not a strong non-linear instability of the class of Gowdy solutions
within the class of 
$\U$-symmetric solutions. We come back to this later.

Another interesting experiment, which could have been done, is to
give data on $\scrip$, compute, as we did here,  for the
corresponding solution the
data on $\scrim$ backward in time, next use that
data as new initial data on $\scrim$ and finally compute the
corresponding data on $\scrip$ forward in time. The difference of the
original data 
and the resulting data on $\scrip$, which would be zero in an exact
solution, can be interpreted as another 
error measure. Further it would be interesting if those
oscillations, visible for instance in $\normelec$ in
\Figref{fig:electric_viol_3_methods} caused by a lack of spatial
resolution, would start close to $\scrim$ and would stop when $\scrip$ is
approached in the same way as we have observed here. 

In \Sectionref{sec:comparisonGowdymethods} we discuss some implications
of our results here
and of the following chapter in comparison with the existing
numerical methods 
for the Gowdy class of spacetimes.

\chapter{Singular \texorpdfstring{$\lambda$}{lambda}-Gowdy 
  spacetimes}
\label{ch:singulargowdy}

In \Sectionref{sec:comparison} we computed regular $\lambda$-Gowdy
spacetimes with spatial $\S$-topology. In particular, these spacetimes
are geodesically complete and there is no curvature singularity.
Here we want to compute
and analyze FAdS $\lambda$-Gowdy spacetimes that are past
singular by modifying the initial data parameters such that the
de-Sitter stability region (\Sectionref{sec:dSSR}) is left. While the
regular class 
excludes the $\T$-topology, see the 
singularity theorems in \Sectionref{sec:singularitytheoremGA}, in the
singular class both $\T$- and $\S$-topologies are allowed and will be
studied here.

As
the reader can 
imagine, the singular class is much more difficult to study
technically than the regular class. Our results should be considered
to a large degree as tests of our method, both the numerical approach and the
formulations of the field equations, to find out the strengths and
limitations. However, we also present some preliminary 
results of fundamental interest together with their investigations and
discussions.   
In the first section we show numerical results for the conformal
field equations in the Levi-Civita 
conformal Gauß gauge (\Sectionref{sec:LCCGG}) with $\T$-topology. Here
the Gowdy 
symmetry is used to reduce the 
evolution equations to a $1+1$-form explicitly. In the second section, similar
investigations are presented for the case 
of $\S$-topology; note again that up to now the code is reduced only
to $2+1$ as explained before.  
$2+1$-simulations in
the $\T$-case have 
not been tried yet but are expected to behave similarly as in the
$\S$-case.

In the \S-case the
solutions can be considered as non-linear perturbations of
$\lambda$-Taub-NUT spacetimes (\Sectionref{sec:TaubNUT})
and there are, in the special case considered, indications for an
interesting stability of the Taub-NUT Cauchy horizon; however, the
investigations have not been thorough enough yet. Further we reconsider the
non-linear stability issue of the class of
Gowdy spacetimes within the 
class of $\U$-symmetric spacetimes, which already came up in
\Sectionref{sec:comparison}. 
 Afterwards,
we do numerical investigations with the 
commutator field equations introduced in
\Sectionref{sec:commutatorfieldequations}.
Note again, that this
system is currently restricted to $\T$-topology. 

\section{Runs with the GCFE in Levi-Civita conformal Gauß gauge}
\label{sec:runsGCFE}
\subsection{Runs with \texorpdfstring{$\T$}{T3}-topology}
\label{sec:singularT3}
This section is devoted to the study of numerically generated
singular $\lambda$-Gowdy spacetimes with spatial
$\T$-topology. We make use of the initial data constructed in
\Sectionref{sec:T3initialdata} with the following two choices:
\begin{enumerate}
\item non-polarized data 
  \[(W_{ab})=
  \begin{pmatrix}
    10^{-4} & 0 & 0\\
    0 & 0 & \sin x_1\\
    \sin x_1 & 0 & -10^{-4}
  \end{pmatrix},\]
\item polarized data
  \[(W_{ab})=
  \begin{pmatrix}
    10^{-4} & 0 & 0\\
    0 & \sin x_1 & 0\\
    0 & 0 & -10^{-4}-\sin x_1
  \end{pmatrix}.\]
\end{enumerate}
As before, we assume that the other initial data quantities are given to induce
the Levi-Civita conformal Gauß gauge, with
$t=0$ the initial hypersurface $\scrip$. The time $t=2$ would correspond to
$\scrim$ but due to the singularity theorems in
\Sectionref{sec:singularitytheoremGA}, a smooth $\scrim$ cannot exist.

\begin{figure}[tb]
  \begin{minipage}[t]{0.49\linewidth}
    \centering
    \includegraphics[width=\linewidth]
    {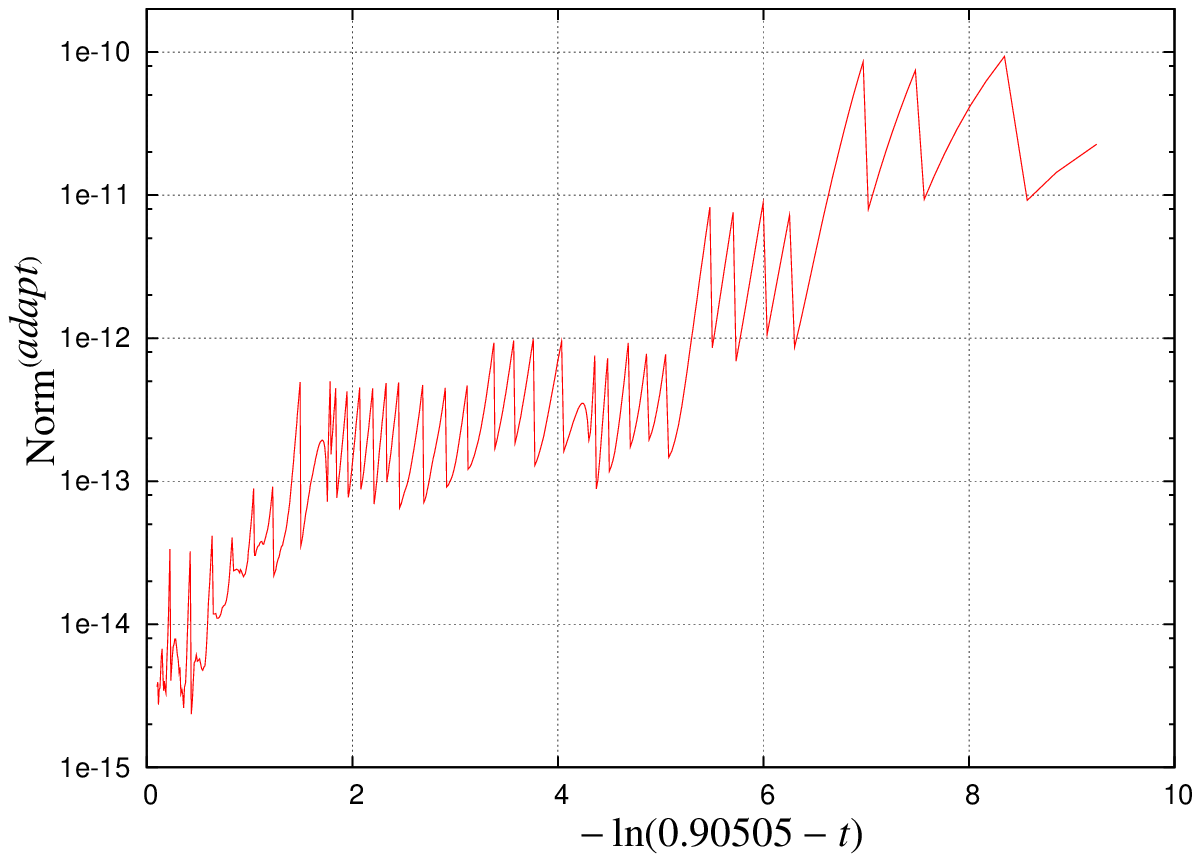}
    \caption{$\normadapt$ for the non-polarized case}
    \label{fig:typical_adaption_norm}
  \end{minipage}
  \begin{minipage}[t]{0.49\linewidth}
    \centering
    \includegraphics[width=\linewidth]
    {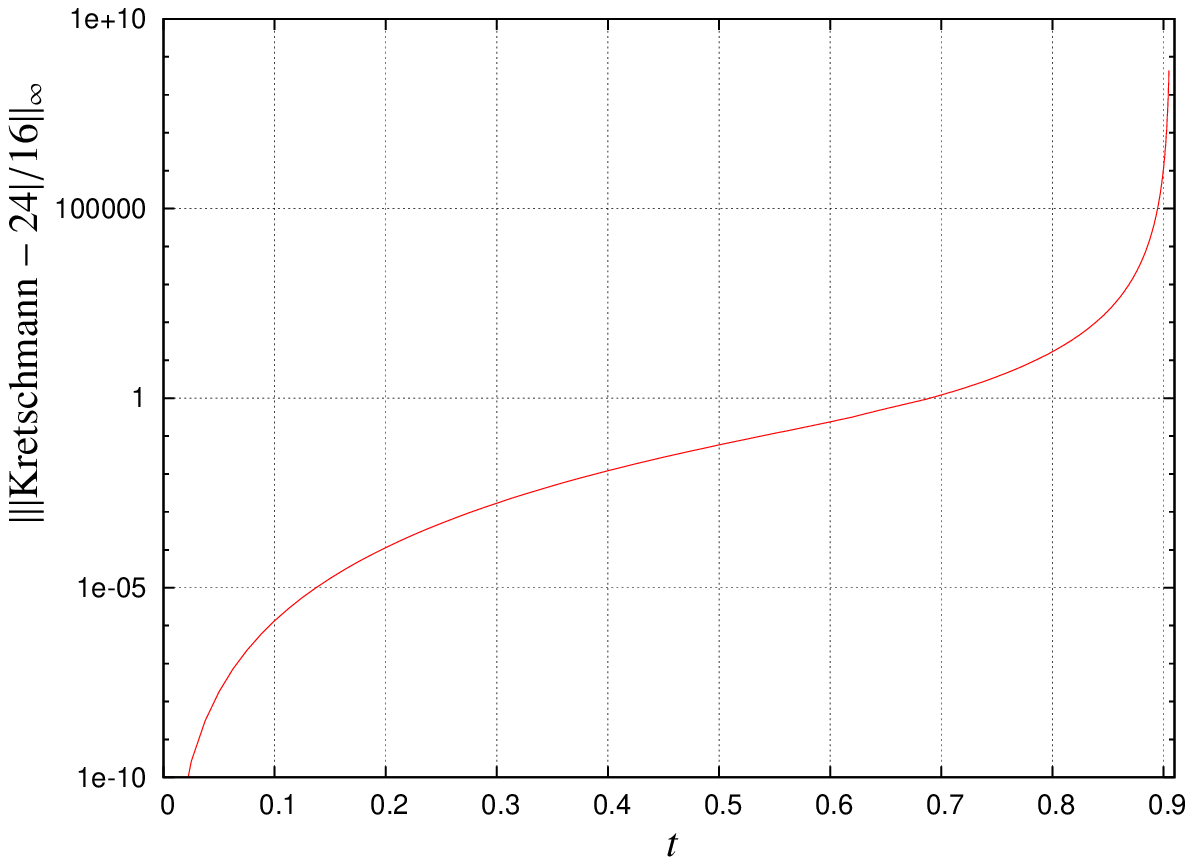}
    \caption{Kretschmann scalar for the non-po\-lar\-ized case}  
    \label{fig:KretschDiv}
  \end{minipage}
\end{figure}
Let us start with the spacetime corresponding to the non-polarized
initial data. It turns out that at
$t\approx 0.9$ the variables in the equations blow up. It is a
curvature singularity because the Kretschmann scalar is divergent, see
\Figref{fig:KretschDiv}. In these runs, we employ 
our  
adaption techniques explained in \Sectionref{sec:spatial_adaption} as
one can see in \Figref{fig:typical_adaption_norm}. There, the time axis is
exponentially stretched and one sees that the demand for spatial resolution
increases almost exponentially in time. The runs were done with
several resolutions. The resolution, which we
refer to as hShT, starts from $N=11$ and 
$h=6.25\cdot 10^{-5}$ and stops at the final time with $N=615$ and 
$h=4.6875 \cdot 10^{-7}$ making use of the adaption methods described
before. For the other resolutions, the automatic 
adaption mechanism is switched off and the adaption history of the
hShT run is copied, on the one hand with half of the
spatial resolution for the lShT run, and other other hand with half
the time resolution for the hSmT run etc. Here
$N$ is the number of collocation 
points in $x_1$ direction. Recall that these runs are $1+1$.

\begin{figure}[tb]
  \centering
  \subfloat[Behavior of $\normelec$]{
    \includegraphics[width=0.49\linewidth]{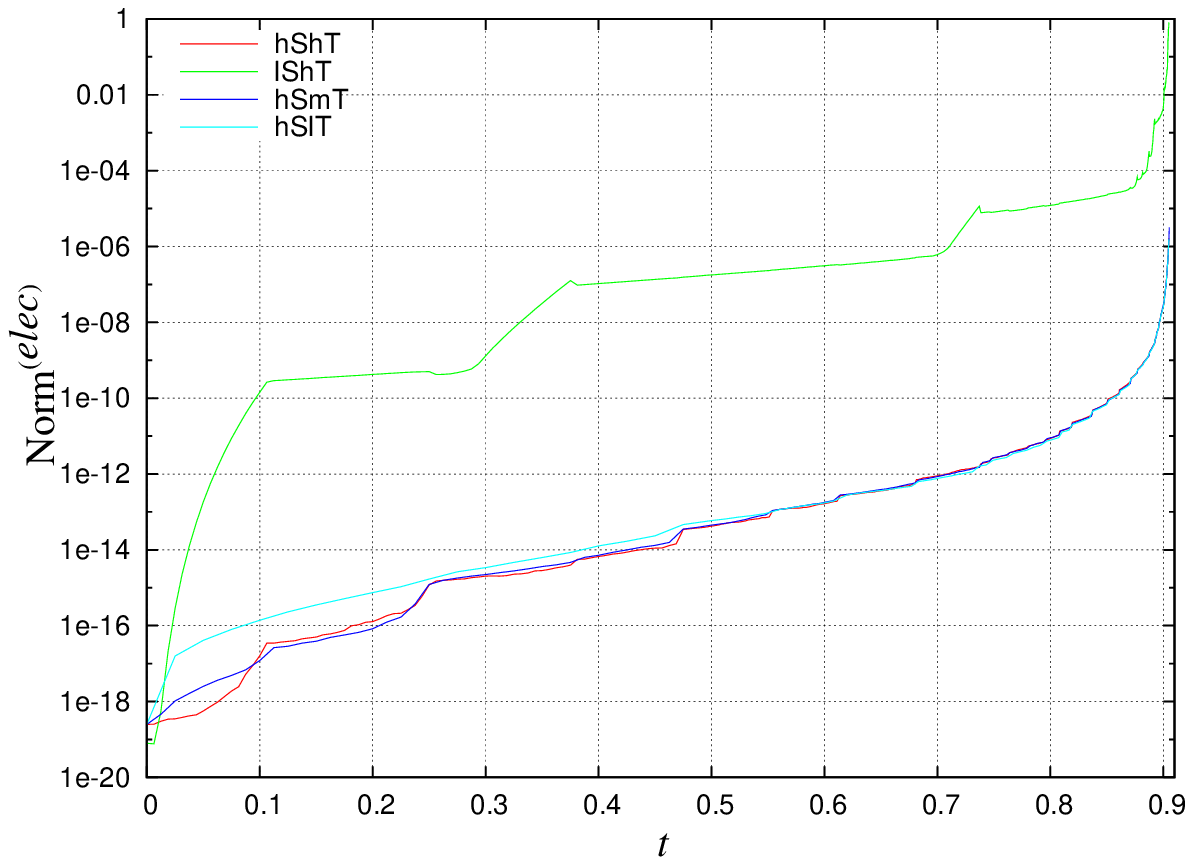}}%
  \subfloat[Behavior of $\normeinstein$]{
    \includegraphics[width=0.49\linewidth]
    {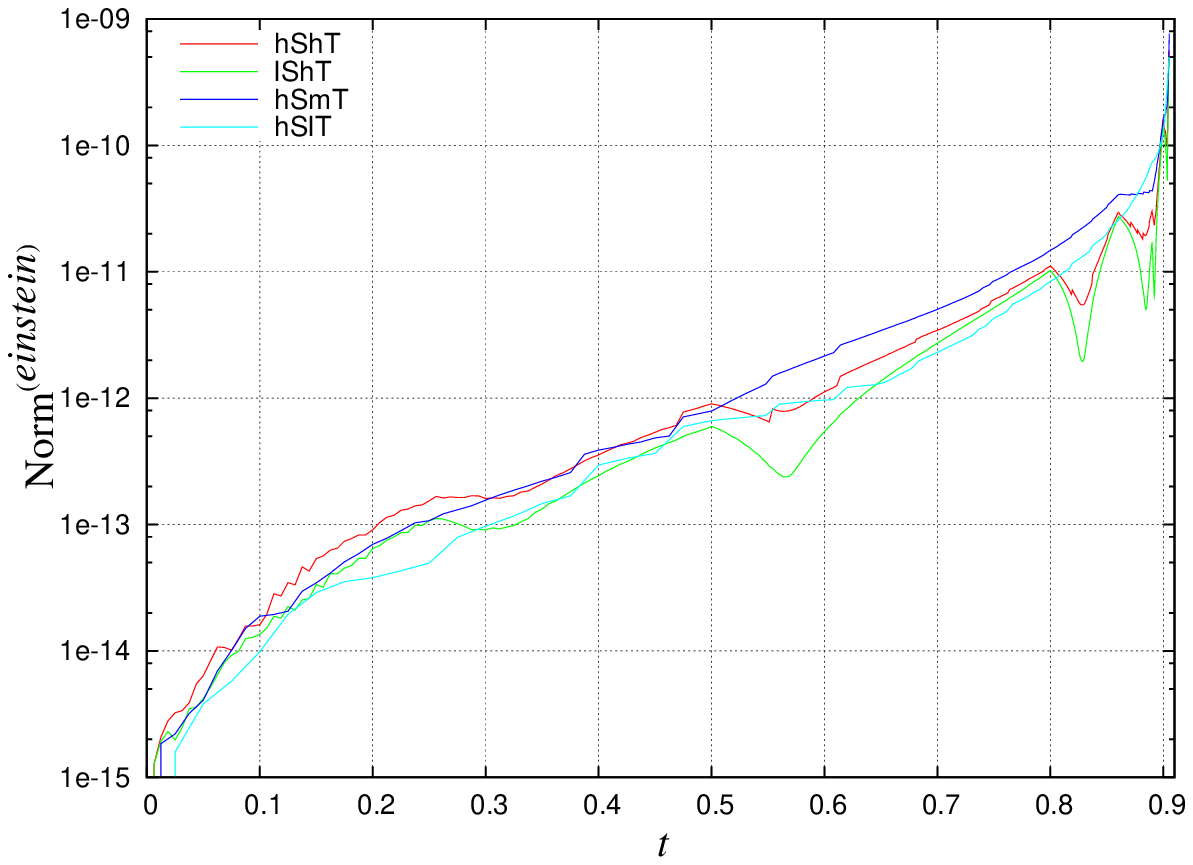}}
  \caption{Some error norms for the non-polarized case}
  \label{fig:T3Gowdyerror}
\end{figure}
\Figref{fig:T3Gowdyerror} shows error norms for the runs for the
resolutions above. On the left $\normelec$ and on the
right $\normeinstein$ are plotted vs.\ time. From the left plot we can
deduce that for the high spatial
resolution  the error is dominated by the time
discretization for low times as before. For larger times, increasing
the time 
resolution stops to make a difference. This is so because
on the one hand spatial discretization plays a bigger role for later
times despite of the adaption; however, apparently the aliasing effect
is not dominant since no oscillations in the error norms can
be observed. On the other hand, the constraint error cannot be
made smaller than the actual solution of the subsidiary
system as we have discussed before.
Hence, so far, everything is in agreement with the results we found
before. The low spatial resolution is surely not enough and we see
large errors in the constraint although the code is still stable. 
For the other runs, one
should note that although the constraint violation gets to the
order $10^{-6}$ at the final time, the relevant
quantities of the unknowns are of the order $10^4$ (as we do not show
here), and this 
means that our accuracy is of the order $10^{-10}$. This is in
agreement with $\normeinstein$ at the final time. As before,
$\normeinstein$ is much less sensible for errors caused by spatial
discretization. 

\begin{figure}[tb]
  \centering
  \subfloat[Orbit volume density]{%
    \label{fig:T3GowdygeometricquantitiesOVol}
    \includegraphics[width=0.49\linewidth]{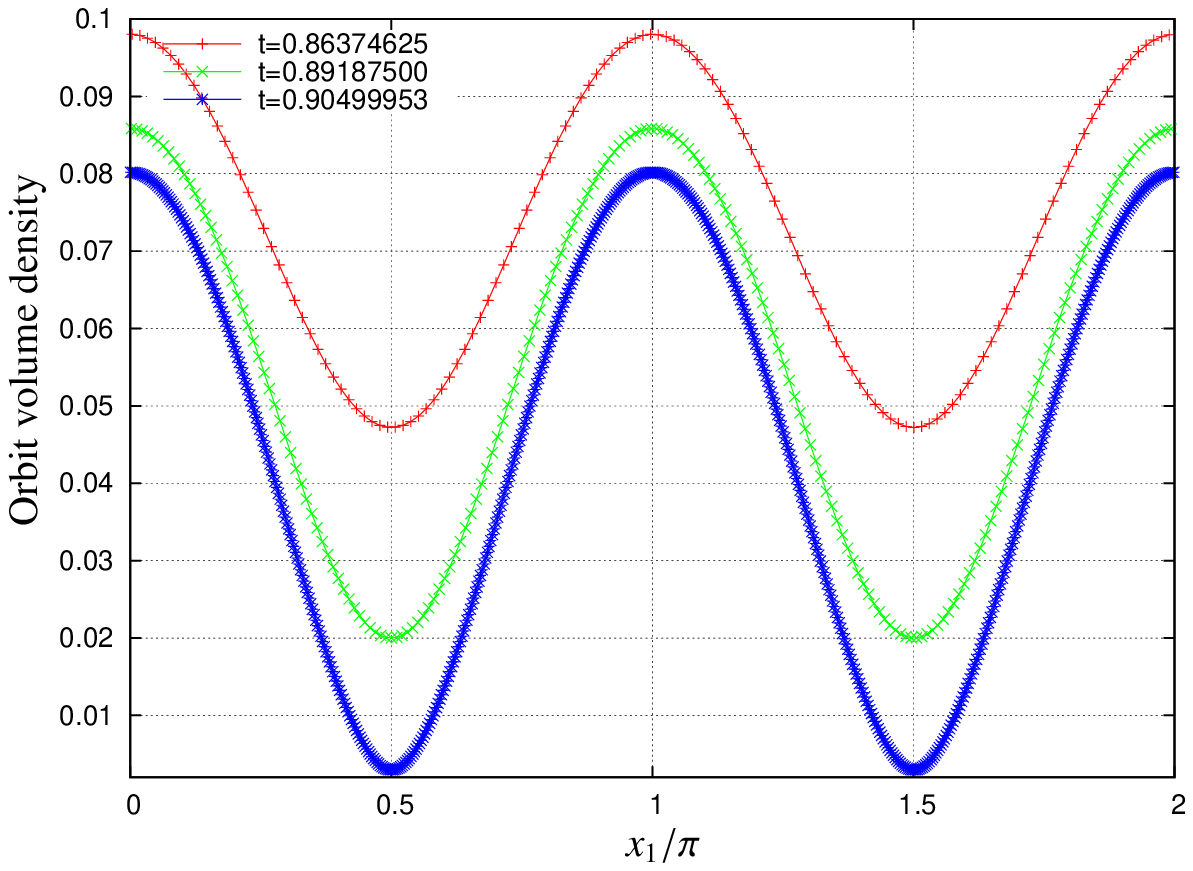}}%
  \subfloat[Kretschmann scalar]{%
    \label{fig:T3GowdygeometricquantitiesKretsch}
    \includegraphics[width=0.49\linewidth]
    {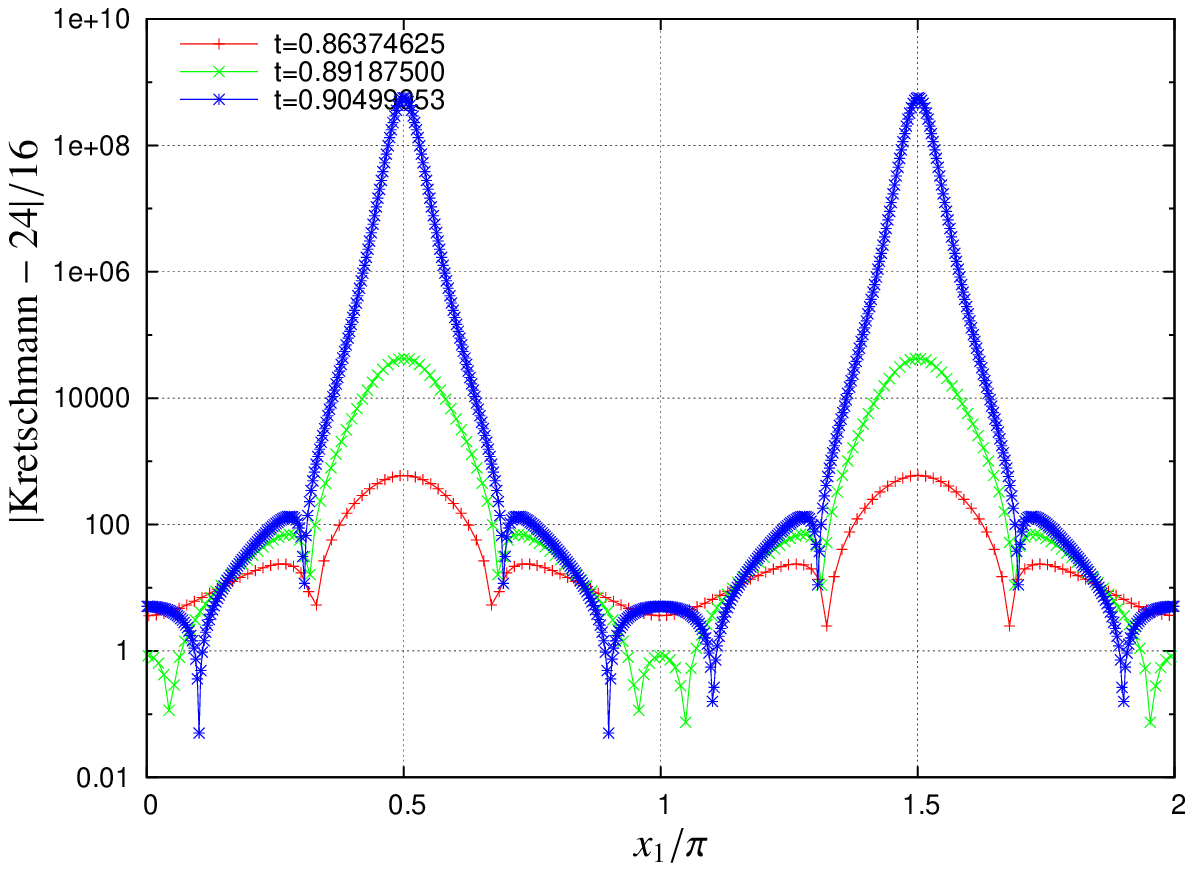}}\par
  \subfloat[Scalar product of the Killing vector fields]{%
    \label{fig:T3GowdygeometricquantitiesScalarpr}
    \includegraphics[width=0.49\linewidth]
    {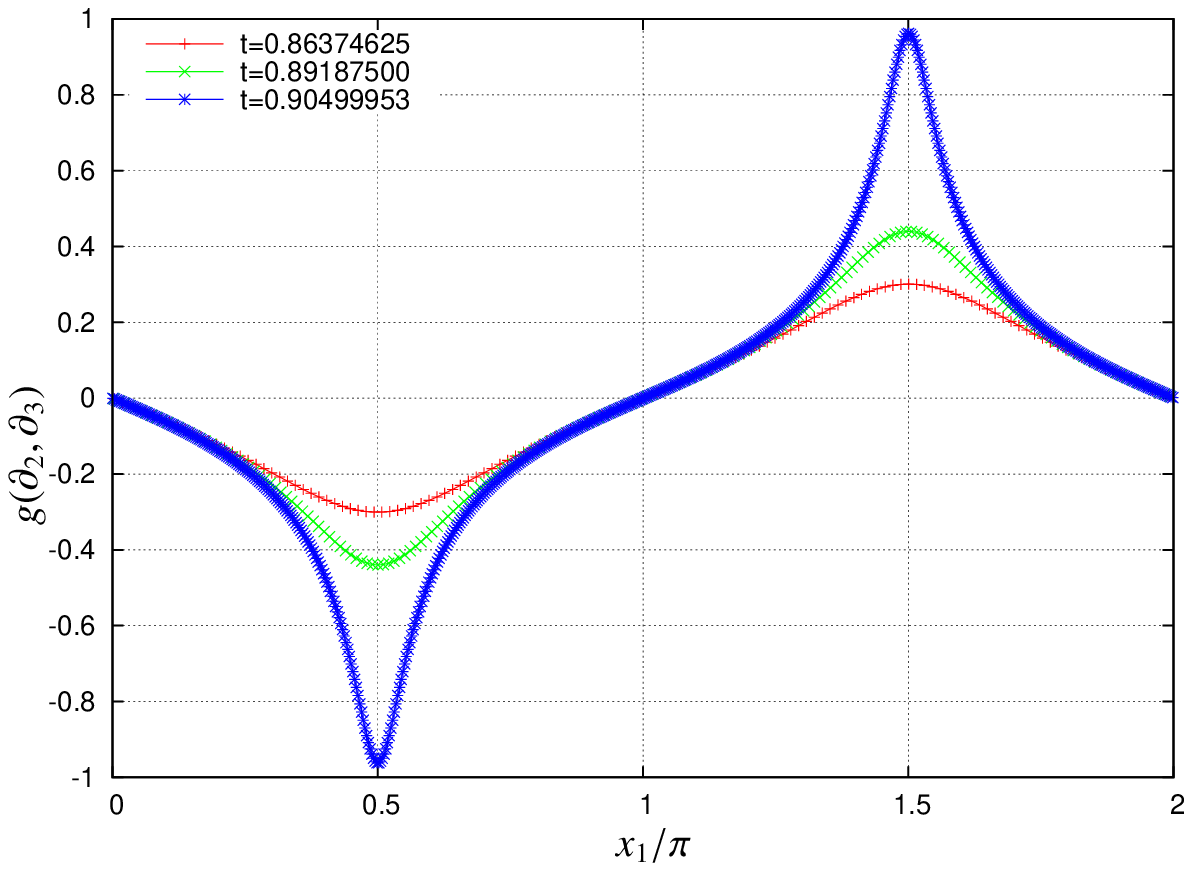}}%
  \subfloat[Hyperbolic velocity]{%
    \label{fig:T3GowdygeometricquantitiesHypVel}
    \includegraphics[width=0.49\linewidth]{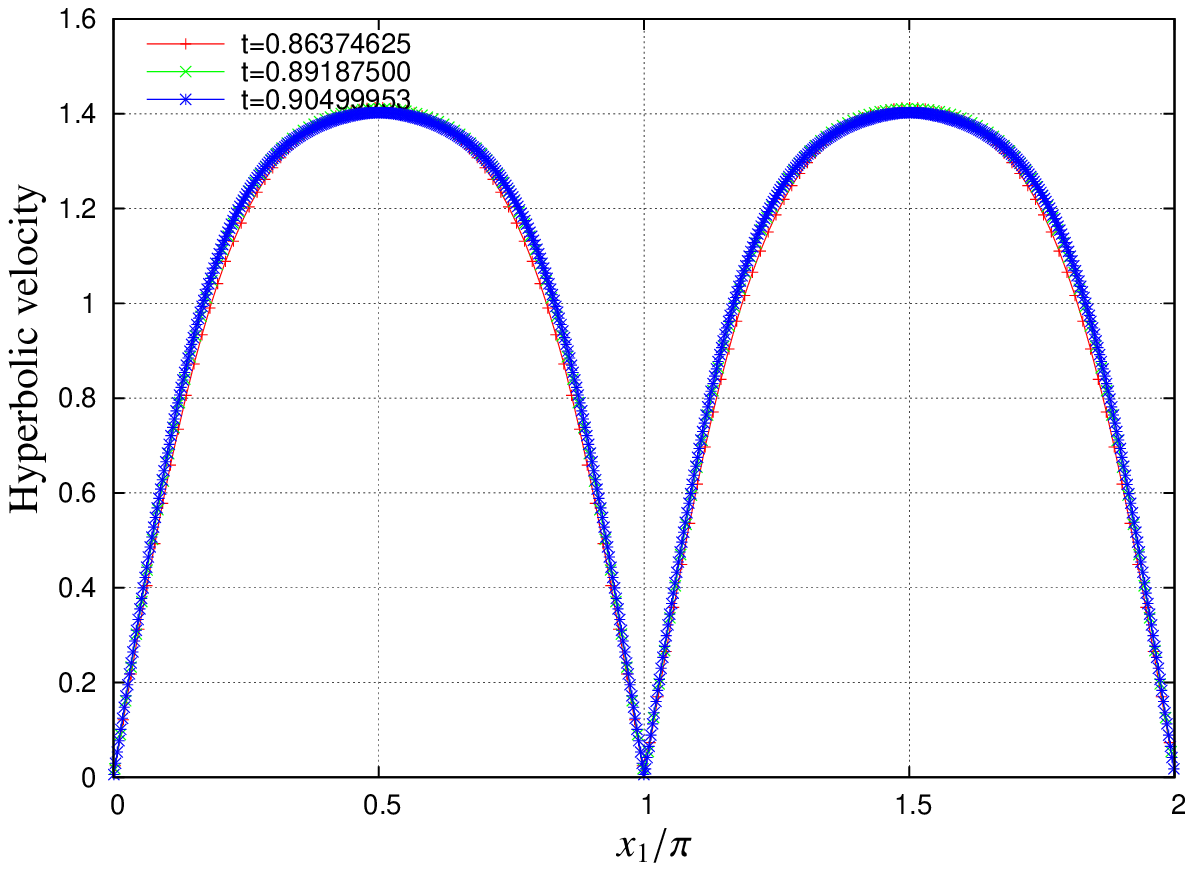}}
  \caption{Spatial behavior of geometric quantities for the
    non-polarized case}
  \label{fig:T3Gowdygeometricquantities}
\end{figure}
In \Figref{fig:T3Gowdygeometricquantities} we plot the spatial
distribution of some geometric
quantities for the singular $\T$-Gowdy solutions for three
different times close to the singularity. In
\Fignref{fig:T3GowdygeometricquantitiesOVol} we see how the 
orbit volume density behaves. The orbit volume density is defined as the
square root of the determinant of the matrix
$(g(\partial_A,\partial_B))$ with $A,B=2,3$, i.e.\ it is, to be
precise, the 
\textit{conformal} orbit volume density; but note that close to 
$t\approx 0.9$ there is not much difference between conformal and physical
quantities. Here one sees one particular drawback of the Levi-Civita
conformal Gauß gauge compared to areal gauge. In the latter gauge, the
orbit volume density is a constant on each $t=const$ slice and
hence the singularity is approached in a ``homogeneous'' way. This is
not the case for our gauge here and this ``inhomogeneity'' is even increased
the further the singularity is approached. Some points, namely those
where intuitively gravity is stronger, are pulled
faster to the singularity than other points. In any case, such a
behavior can be 
expected from a Gauß like gauge as ours. In
\Fignref{fig:T3GowdygeometricquantitiesKretsch}, where we plot the
physical Kretschmann scalar according to
\Eqref{eq:physKretschConfQuant}, we see that exactly at those points,
which approach the singularity most quickly, the Kretschmann scalar
blows up fastest. To avoid confusion note that the downward
pointing ``spiky features'' in 
this plot are caused by the fact the I plot the absolute value of
the Kretschmann scalar and the ordinate is logarithmic. The plot
\Fignref{fig:T3GowdygeometricquantitiesScalarpr} shows 
the scalar product of the killing vector fields, i.e.\
$g(\partial_2,\partial_3)$ which are related to the quantity $Q$ in
the standard parametrization of the Gowdy metric. Hence one sees
that this solution is definitely not polarized. Finally, in
\Fignref{fig:T3GowdygeometricquantitiesHypVel} we show the 
hyperbolic velocity, cf.\ \Sectionref{sec:gowdyphenom}. This is
computed from \Eqref{eq:hypvel} by first computing from our unknowns
an orthonormal frame as in \Sectionref{sec:comm_field_eqs} and then by
computing the relevant rescaled quantities. One sees that the velocity
nearly becomes constant in time which is a hint that we are not
approaching spiky features in the sense of
\Sectionref{sec:gowdyphenom}.

\begin{figure}[tb]
  \centering
  \includegraphics[width=0.49\linewidth]
  {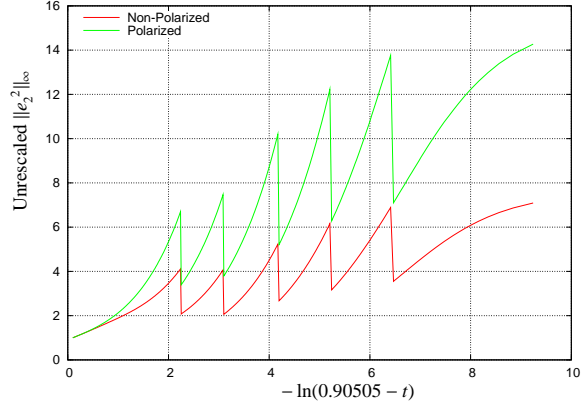}
  \caption{Evolution of non-rescaled $e\indices{_2^2}$ for a polarized and
    non-polarized solution}
  \label{fig:T3Gowdye22}
\end{figure}

There are further reasons to believe that the upward pointing ``spiky''
features in 
\Fignref{fig:T3GowdygeometricquantitiesKretsch} are just artifacts of
the gauge.
A simple argument is the
following. Polarized Gowdy spacetimes in areal gauge cannot develop
spikes which would be visible in the Kretschmann scalar due to the
results in \cite{Isenberg89}, meaning that the Kretschmann scalar
blows up uniformly when the singularity is approached. 
However, the Kretschmann scalar of the polarized
solution corresponding to the polarized initial 
data given at the beginning of this section is nearly
indistinguishable from
\Fignref{fig:T3GowdygeometricquantitiesKretsch}, although individual
variables are very different, see for instance\footnote{Note that the
  jumps in this plot are 
produced by our time adaption method  described in
\Sectionref{sec:spatial_adaption}, since we plot the unrescaled
quantities here, and hence are not geometric.} \Figref{fig:T3Gowdye22}.
The relative deviation is of the order
$10^{-2}$ at the final time. In particular, the same upward pointing
features in the 
Kretschmann scalar can be observed. This is a hint that the cause of
these features is the ``inhomogeneous'' approach to the
singularity in our gauge. However, we also cannot exclude the possibility
that at later times, even in this gauge, ``real'' spikes will be
visible. 
In any case, it would be hard to
distinguish those ``real'' spikes from the effects that are caused by
the gauge.  

I should note that when I stopped
the runs there were no principal numerical problems with the
solutions. Indeed, the evolutions could have continued closer to
the singularity.
The only limiting factor so far is the constraint growth. As
discussed before, I suspect that this is not
caused primarily by my discretization scheme but is rather a problem
either of the evolution equations at the continuum level or of the
gauge. We will discuss this problem again later.

\subsection{Runs with \texorpdfstring{$\S$}{S3}-topology}
\label{sec:S3singularGowdy}

In this section we report on similar investigations in the case of
$\S$-topology as before. Maybe one should note that these are the first
published attempts to study $\S$-Gowdy singularities numerically.
Recall from \Sectionref{sec:implementationOfS3Gowdy} that these are $2+1$ runs 
in contrast to the $\T$-case and that the ideas to reduce to $1+1$ 
have not
been implemented yet. This means that we have higher practical
constraints on the spatial resolution now than in the
previous section. The
solutions constructed here also have to be seen in relation to those
in \Sectionref{sec:comparison} where we chose initial in the de-Sitter
stability region and hence obtained solutions which are both future
and past asymptotically de-Sitter. Here now, we want to leave the
stability region such 
that the solutions become singular in the past. All runs in this
section have been done with the ``D2U Stag.'' method
(\Sectionref{sec:comparison}).  

Two sets of initial data as constructed in
\Sectionref{sec:solelectrconstraintscri} are considered: Gowdy data
with 
\begin{enumerate}
\item ``small inhomogeneity''
  \[a_3=1,\,a_3=0.7,\,(E_{11})_{0,0}=(E_{22})_{0,0}=0,\,C_2=10^{-4},\]
\item ``large inhomogeneity''
  \[a_3=1,\,a_3=0.7,\,(E_{11})_{0,0}=(E_{22})_{0,0}=0,\,C_2=10^{-1}.\]
\end{enumerate}
As always so far, we assume that we are in 
Levi-Civita conformal Gauß gauge.

\begin{figure}[tb]
  \centering
  \includegraphics[width=0.49\linewidth]
  {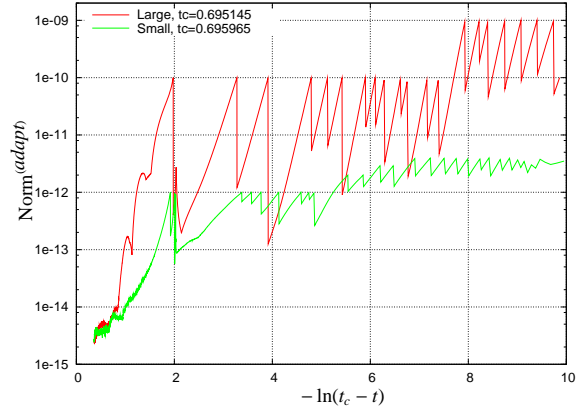}
  \caption{Behavior of $\normadapt$ for large and small 
    inhomogeneity}
  \label{fig:S3Gowdyadapt}
\end{figure}
It turns out that both cases appear to become singular at 
$t\approx 0.7$ because some variables blow up. The behavior of the
Kretschmann scalar and other quantities is discussed in a moment. As
expected, the 
``singular'' times are a bit different in the two
simulations. The runs
where done with the adaption mechanisms described in 
\Sectionref{sec:spatial_adaption};
\Figref{fig:S3Gowdyadapt} shows the behaviors of
$\normadapt$ in both cases. The
$t$-axis has been stretched exponentially such that one can see the
exponentially increasing dynamics close to the singularity in both
cases. For the large inhomogeneity run, the
adapted resolution, referred to as hShT, starts with $N_1=13$, $N_2=7$ and 
$h=2.5\cdot 10^{-4}$ and ends with $N_1=157$, $N_2=79$ and 
$h=3.90625\cdot 10^{-7}$. The resolutions lShT etc.\ are derived from
that as
above. Here $N_1$ is the number of collocation points in
$\chi$-direction and $N_2$ in $\rho_1$-direction.
For the small inhomogeneity case, the resolutions hShT starts with $N_1=13$,
$N_2=7$ and $h=2.5\cdot 10^{-4}$ and ends with $N_1=469$, $N_2=235$ and 
$h=3.75\cdot 10^{-6}$. The reason that the resolution for the small
inhomogeneity case ended up higher than for the large inhomogeneity
case -- the other way around would have certainly made more sense -- was my
unskilful choice of representative variable to compute
$\normadapt$, namely, $E_{11}$ in both runs. From 
\Eqsref{eq:n2solelectrconstraint} we see that the initial data
parameter $C_2$ controls the magnitude of the initial values of
$E_{11}$. Although the initial value of $\normadapt$ was almost
identical in both cases because of the definition of the norm, the
consequence of this was that the time behavior of $\normadapt$ 
was very different in both cases. I tried to compensate this by giving
different threshold values for the adaption, see again
\Fignref{fig:S3Gowdyadapt}, however, the undesired result was that
the low inhomogeneity run was done with higher resolution than the
high inhomogeneity run. The fact, as discussed below, that the error
quantities in the two runs
are almost of identical size suggests on the one hand, that the low
inhomogeneity run did not really require so much resolution, but on the
other hand, that the code is also not instable when the resolution is
too high (at least so far). One could have worried about this issue, cf.\ the
investigations related to
\Figref{fig:linearized_instability}.

\begin{figure}[tb]
  \centering
  \subfloat[Behavior of $\normelec$]{
    \includegraphics[width=0.49\linewidth]{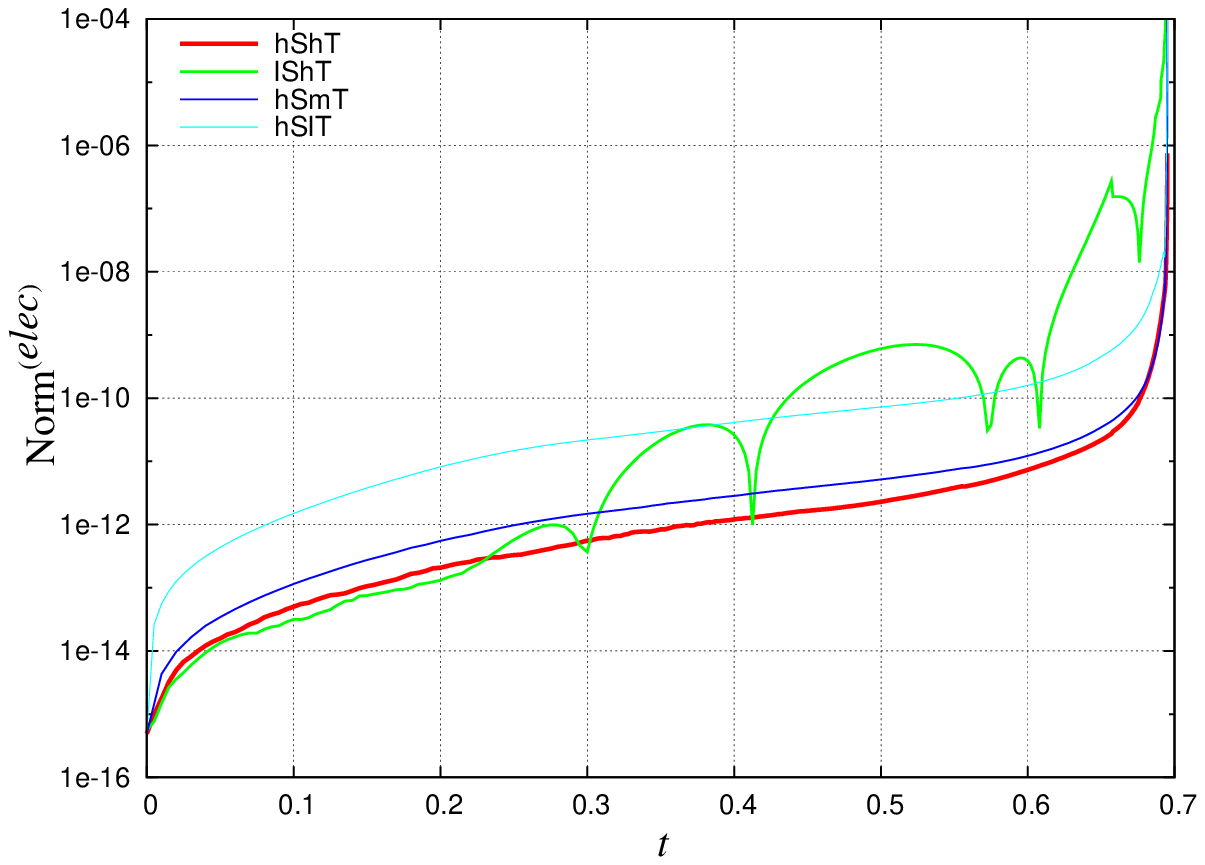}}%
  \subfloat[Behavior of $\normeinstein$]{
    \includegraphics[width=0.49\linewidth]
    {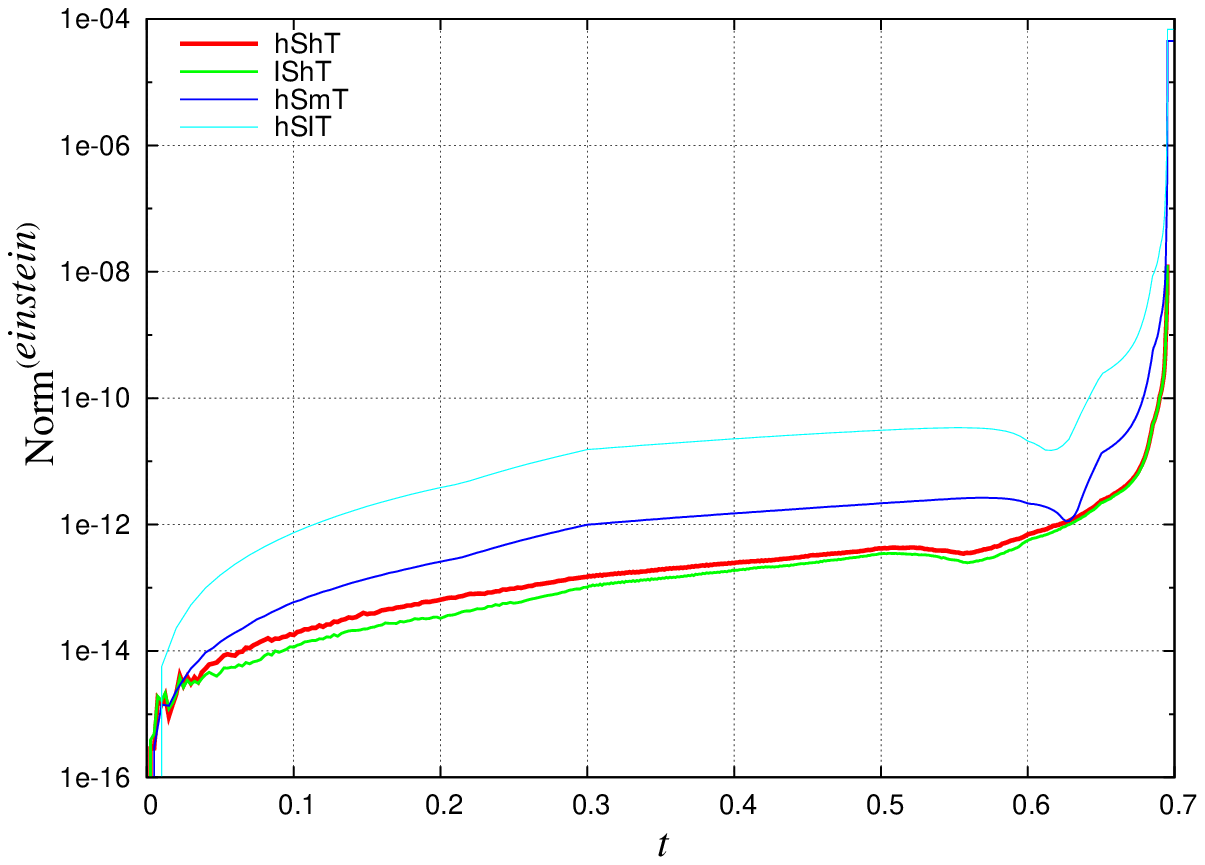}}\par
  \subfloat[Behavior of $\normkilling$]{%
    \label{fig:S3GowdyerrorKilling}
    \includegraphics[width=0.49\linewidth]{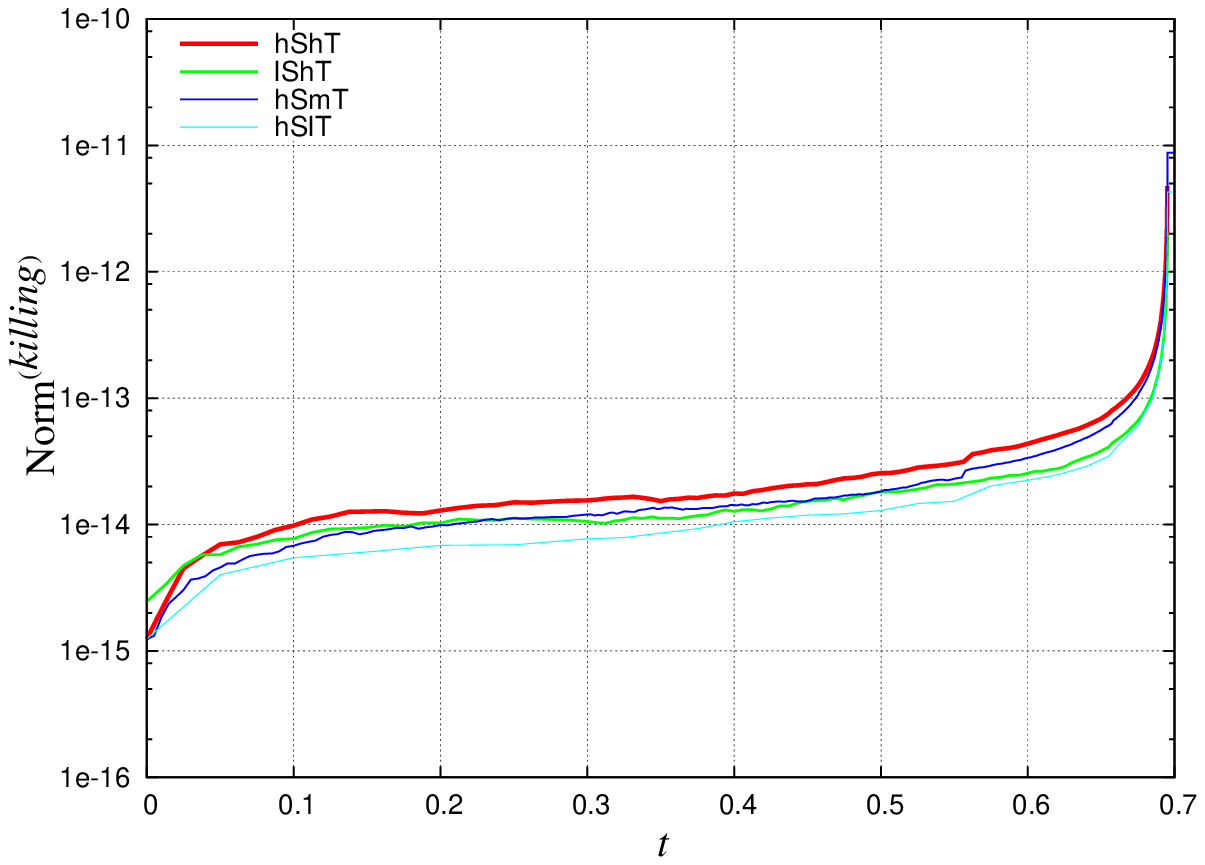}}
  \caption{Some error norms for the large inhomogeneity case}
  \label{fig:S3Gowdyerror}
\end{figure}
In \Figref{fig:S3Gowdyerror} we see the behavior of some error norms
vs.\ time in the large inhomogeneity case. The behavior is
analogous to the $\T$-case before. The errors are
moderate until very close to the singularity; at the end of the run
$\normelec$ is of the order $\sim 10^{-6}$ and $\sim 10^{-8}$ for
$\normeinstein$ for the hShT run. Note however that
here, in contrast to the $\T$-case before, the relevant components of
the solutions are of the order $10^1$ at the final time (as we do not
show here) and hence,
the relative errors indeed grow close the singularity, but are still
acceptable. The suspected 
reason for this is the limited spatial resolution since these are
$2+1$-runs in contrast to $1+1$-runs before. Further, it is
interesting that $\normkilling$ in \Fignref{fig:S3GowdyerrorKilling}
is very stable until the errors in the solution start to grow
more rapidly close to the singularity. We come back to this at the end
of this section.
As mentioned already
above, the corresponding error
quantities for the small inhomogeneity case behave very similar and thus
are not plotted here.

\begin{figure}[tb]
  \centering
  \subfloat[Orbit volume density]{%
    \label{fig:S3GowdygeometricquantitiesLargeOVol}
    \includegraphics[width=0.49\linewidth]{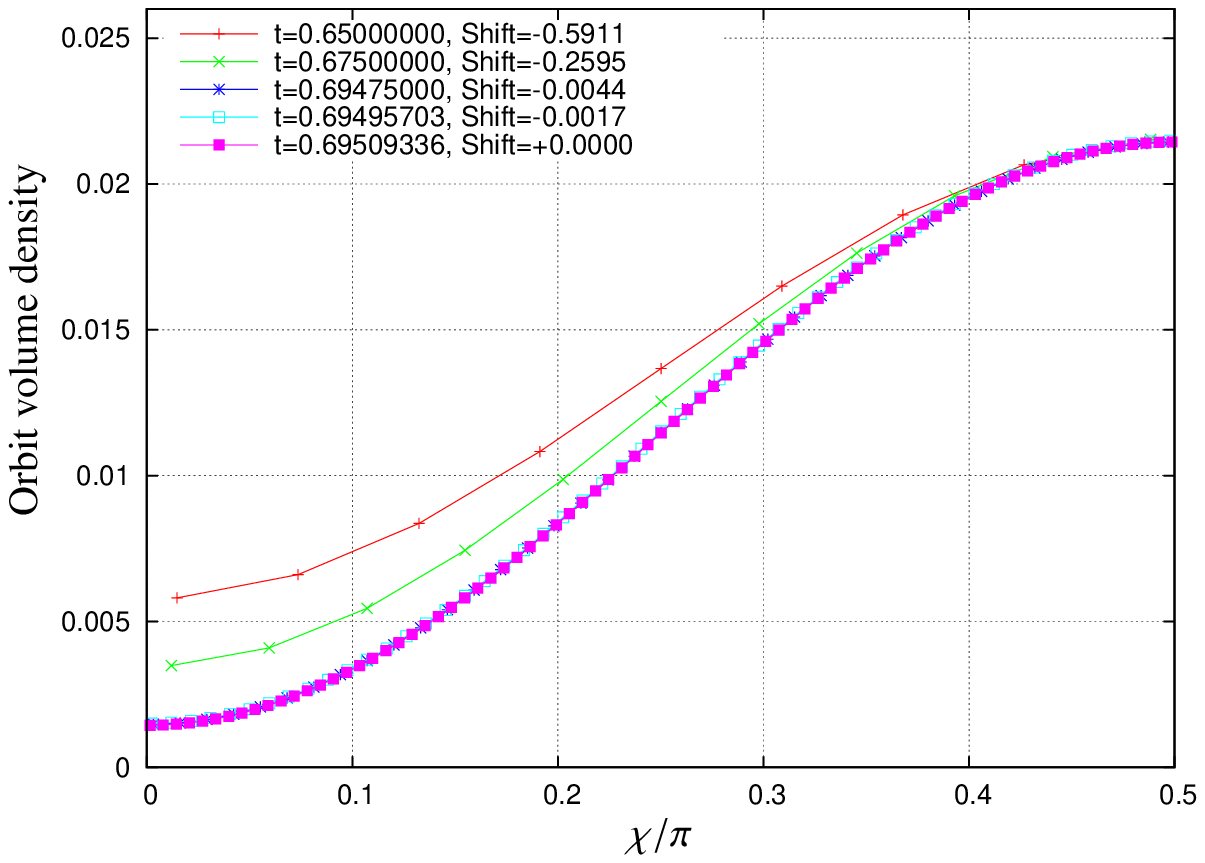}}%
  \subfloat[Kretschmann scalar]{%
    \label{fig:S3GowdygeometricquantitiesLargeKretsch}
    \includegraphics[width=0.49\linewidth]
    {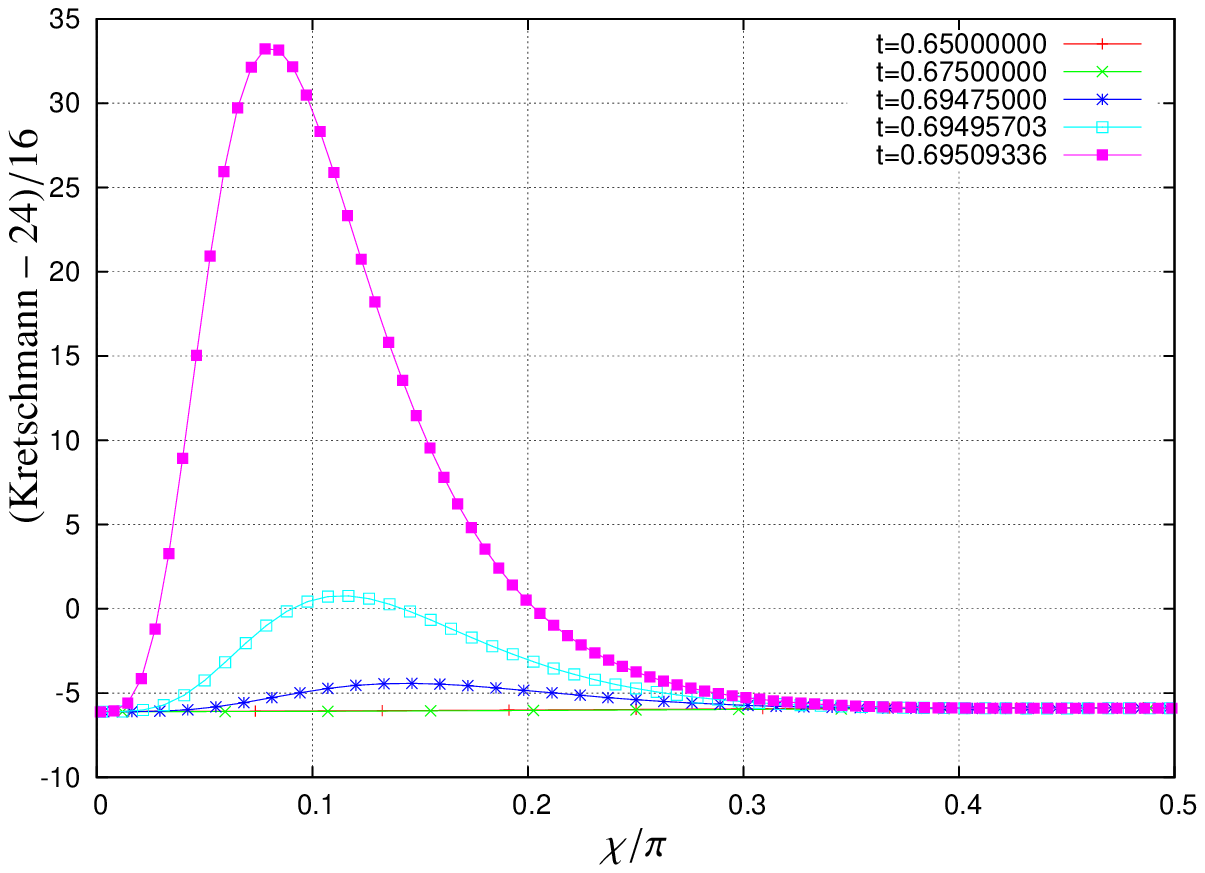}}\par
  \subfloat[Scalar product of Killing vector fields]{%
    \label{fig:S3GowdygeometricquantitiesLargeScalarprKV}
    \includegraphics[width=0.49\linewidth]
    {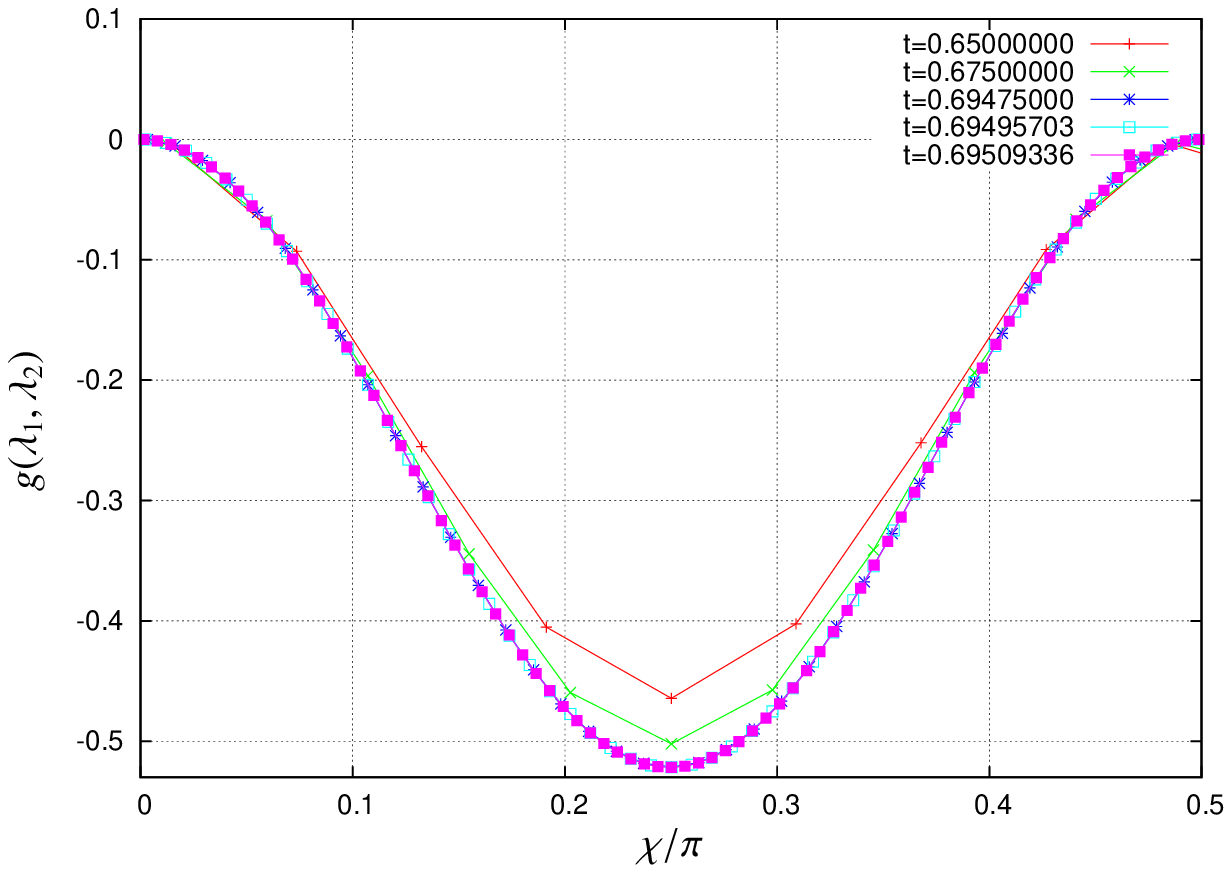}}
  \caption{Spatial behavior of certain geometric quantities for large
    inhomogeneity}
  \label{fig:S3GowdygeometricquantitiesLarge}
\end{figure}
\begin{figure}[tb]
  \centering
  \subfloat[Orbit volume density]{%
    \label{fig:S3GowdygeometricquantitiesSmallOVol}
    \includegraphics[width=0.49\linewidth]{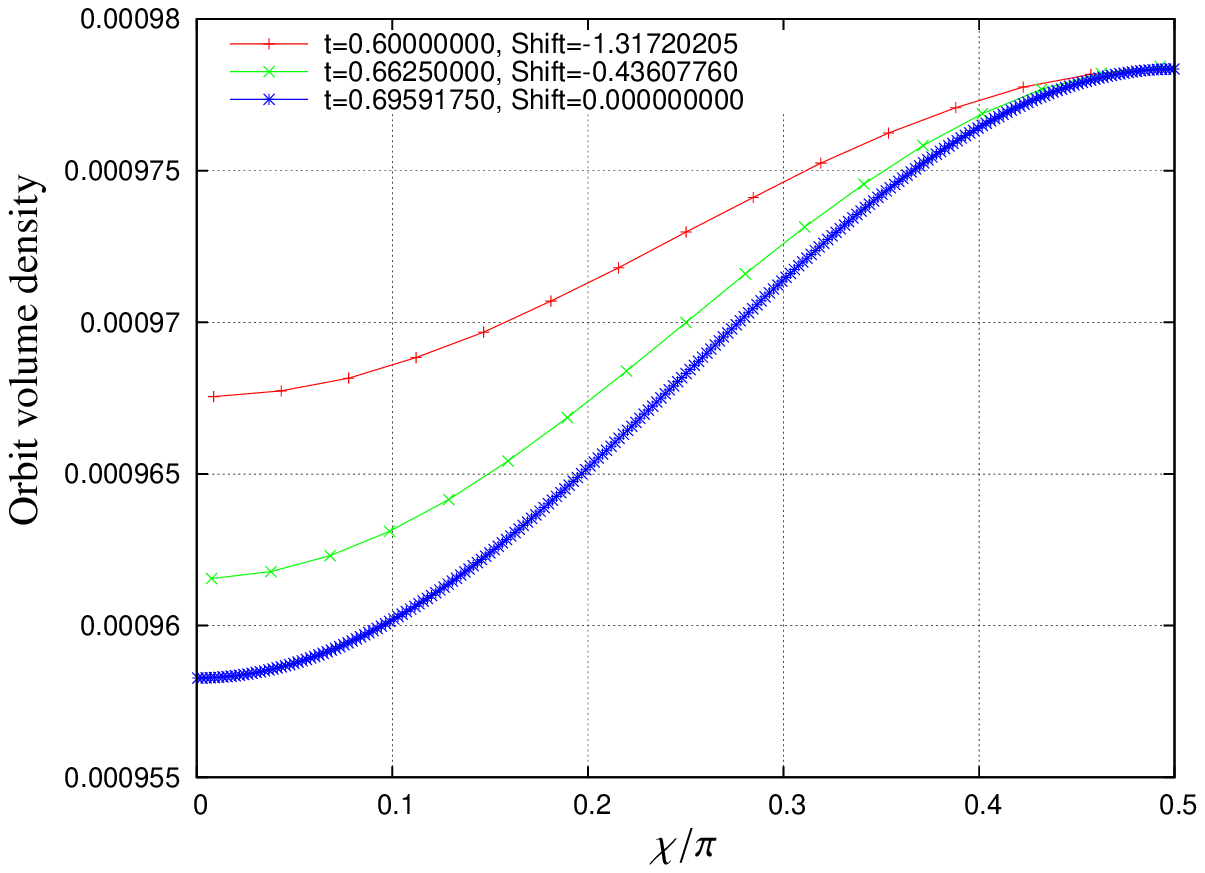}}%
  \subfloat[Kretschmann scalar]{%
    \label{fig:S3GowdygeometricquantitiesSmallKretsch}
    \includegraphics[width=0.49\linewidth]
    {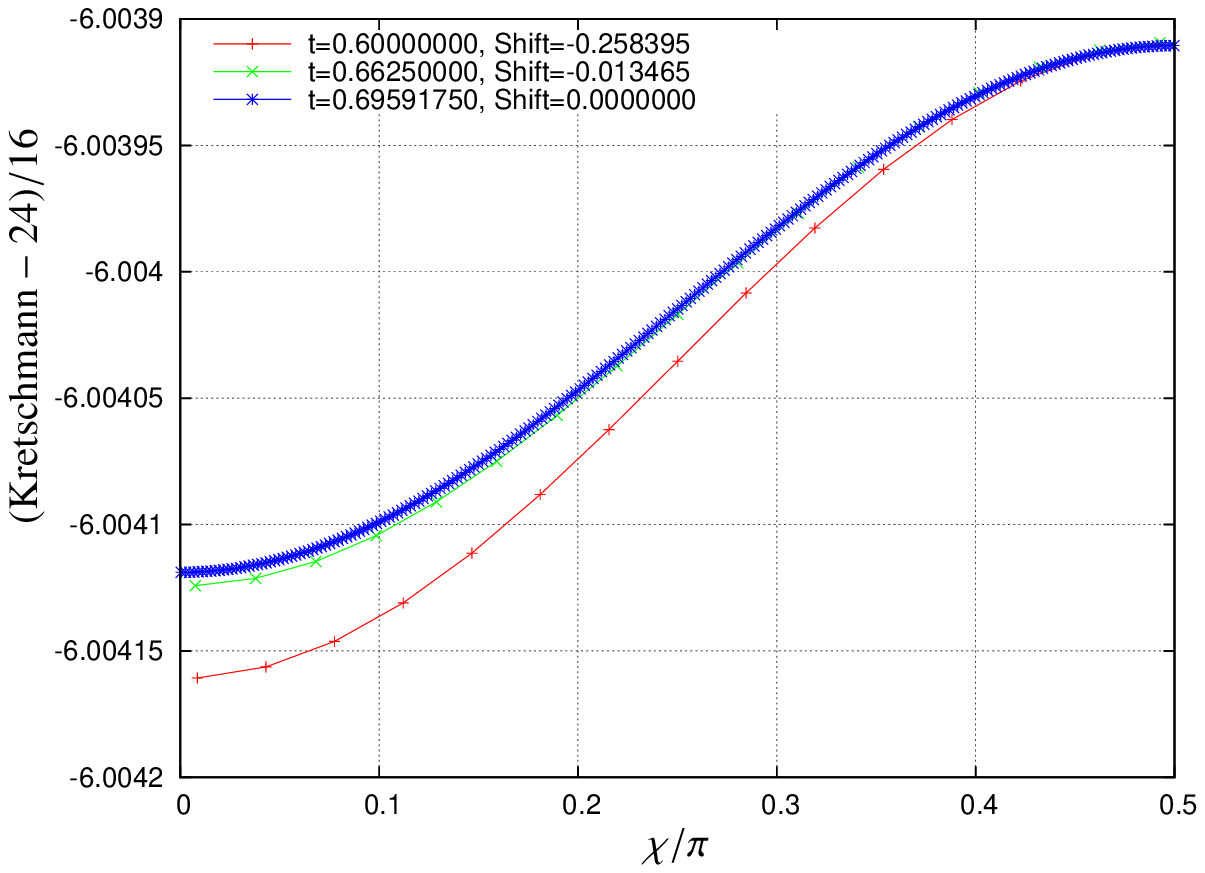}}
  \caption{Spatial behavior of certain geometric quantities for small
    inhomogeneity}
  \label{fig:S3GowdygeometricquantitiesSmall}
\end{figure}
Next consider \Figref{fig:S3GowdygeometricquantitiesLarge} where we
plot the spatial dependency of certain geometric
quantities for five different times close to the singularity for
the large inhomogeneity case. Corresponding plots for the small
inhomogeneity case can be found in
\Figref{fig:S3GowdygeometricquantitiesSmall}. Let us start with the
rescaled orbit volume density\footnote{By ``rescaled'' we mean
that the quantity is divided by $\sin^22\chi$ as explained in
\Sectionref{sec:S3orbitvoletc}.} 
in
\Fignref{fig:S3GowdygeometricquantitiesLargeOVol}. Note, as in the
$\T$-case, that this 
quantity would be constant on each $t=const$ slice in areal gauge,
however in our gauge it is not. In
the plot we shift the values of the functions such that
the most right point agrees for all curves
so that we can study the deformations of the curves
on the way to the ``singularity''. Similar to the $\T$-case, these curves
deform such that points which are closer to the singularity
move faster towards it which is caused by the Gauß like gauge. A
similar behavior can be recognized for the small 
inhomogeneity case
\Figref{fig:S3GowdygeometricquantitiesSmallOVol}. Now consider 
\Figref{fig:S3GowdygeometricquantitiesLargeKretsch} which shows the
spatial dependence of the physical Kretschmann scalar. We see that
with increasing time a localized feature develops as we observed also
in the $\T$-case in
\Fignref{fig:T3GowdygeometricquantitiesKretsch}. However, nothing
like this is visible in the small inhomogeneity case
\Fignref{fig:S3GowdygeometricquantitiesSmallKretsch}. In fact, the contrary
seems to be the case there, namely the curves seem to become flatter
with increasing time. 

What is happening here? When the
inhomogeneity parameter $C_2$ of the initial data, which has the value
$10^{-4}$ in the small inhomogeneity case and $10^{-1}$ in the large
inhomogeneity case, is turned to zero, the corresponding solution is a
$\lambda$-Taub-NUT spacetime with a Cauchy horizon in the past, and
expressed in our variables and gauge,
some quantities, in particular the trace of the $2$nd fundamental form,
blow up there. This is so since the leaves approach a null
surface. However, the Kretschmann scalar stays bounded.
The behavior of the small inhomogeneity solution here shows very 
similar behavior indeed; for instance, the trace of the $2$nd fundamental
form blows up (which we do not show here), but the Kretschmann scalar seems
to stay bounded as indicated by
\Figref{fig:S3GowdygeometricquantitiesSmallKretsch}. Now, we could  
speculate that the small inhomogeneity case also develops a Cauchy
horizon in the past. In contrast, the large
inhomogeneity case shows first signs that the Kretschmann
scalar blows up, see
\Figref{fig:S3GowdygeometricquantitiesLargeKretsch}. If this
speculation was true then the Cauchy horizon of the 
$\lambda$-Taub-NUT spacetime given by zero inhomogeneity parameter
$C_2$ would be stable under small inhomogeneous Gowdy perturbations of our
type. Then, the small inhomogeneity case, provided the generators of
the null-hypersurface are closed, should fit into a
generalization of the family of solutions in \cite{moncrief84}, and it
would be interesting to find 
out how the relation is. There is no simple a priori way of bringing
the two pictures 
together, since Moncrief studies the problem as a singular initial value
problem with the Cauchy horizon as initial hypersurface while we start
from $\scrip$. 
In any case, all this is just speculation since so far our results are
not conclusive. In fact it might
turn out that other curvature quantities than the Kretschmann scalar,
which we do not monitor currently, blow up in the small inhomogeneity
run. On the other 
hand, because the inhomogeneity is so small it might also be the case that
the Kretschmann scalar blows up at just a little later time. In any case, a
systematic study 
of these issues is in order. Even if it turns out that the small
inhomogeneity 
case does not 
correspond to a spacetime with Cauchy horizon or that, even more, the
Cauchy horizon of the corresponding $\lambda$-Taub-NUT spacetime is
not stable under these kind of perturbations at all, it is still interesting
to study the transition from our inhomogeneous spacetime family with
curvature singularities to a
$\lambda$-Taub-NUT spacetime with Cauchy horizon. Further one should
investigate other classes of perturbations. The perturbations
considered so far are in the Gowdy class. According to the results
listed in \Sectionref{sec:CHcosm} there must not be smooth Cauchy
horizons in classes without symmetry. It would be interesting to study
what happens when we systematically reduce the symmetry assumptions
for our perturbations. 

There is another remarkable aspect of
\Fignref{fig:S3GowdygeometricquantitiesLargeKretsch}. The maximum of
the Kretschmann scalar at the latest considered times does not
correspond to the place which is closest to the singularity in
contrast to the $\T$-case. We could speculate if this is a spike
in the sense of \Sectionref{sec:gowdyphenom} but further
investigations of this are clearly necessary. Finally,
\Fignref{fig:S3GowdygeometricquantitiesLargeScalarprKV} 
shows the scalar product of the Killing fields (similar for the
small inhomogeneity) proving that the spacetime is not
polarized. 

Unfortunately, considerations about the interesting non-linear stability
issue of the class of $\S$-Gowdy spacetimes within the class of
$\U$-symmetric spacetimes had a slightly lower priority in the
presentation here. What we find numerically in the
singular class studied in this section, as
indicated in \Fignref{fig:S3GowdyerrorKilling} for the large
inhomogeneity case, is consistent with what we found in the
regular class in \Sectionref{sec:comparison}. Namely, the Gowdy symmetry of
the numerical solutions is quite stable until the errors close to the
singularity become dominant. A
strong non-linear 
instability of the continuum equations would most likely have been
visible in the 
numerical results. Hence, our findings indicate that such an
instability is not present. Certainly, for reliable conclusions,
further investigations are necessary.

\section{Runs with the commutator field equations}
\label{sec:RunsCosmFE}

\subsection{Introduction}
In the previous sections we presented some numerical calculations of
singular $\lambda$-Gowdy spacetimes in the two cases of $\T$- and
$\S$-topology using the conformal field equations in Levi-Civita
conformal Gauß gauge. It turned out that this gauge is not
well adapted to studies of the corresponding singularities because
on the one hand the singularities are approached in a non-homogeneous way
which makes interpretation of the results difficult; on the other hand
the solution demands more and more resolution on time scales which
become exponentially shorter when approaching the singularity. 
Since from previous experience we know that such
problems do not 
occur so strongly
with the commutator field
equations in timelike area gauge (\Sectionref{sec:comm_field_eqs}),
this
motivates us to start numerical experiments. 
Numerical results with similar equations based on finite differencing
methods can be 
found in \cite{Andersson03} in the 
case of a non-vanishing cosmological constant. Our first aim in this section
is to get a feeling how well our pseudospectral approach can cope
with the demands of this class of spacetimes under these ``better''
gauge conditions. Another aim  
is to deepen the fundamental understanding of the Gowdy case with non-vanishing
$\lambda$ by direct comparisons with the $\lambda=0$-case. Indeed, the
investigations in the following are, to my 
knowledge, the only published numerical attempts to treat the Gowdy
case with $\lambda\not=0$. However, due to time 
constraints in this thesis work, I only computed one numerical result
regarding this issue which I present here,
and otherwise just elaborate on my expectations.  We start off by
discussing suitable error analysis and error monitoring for this
system. 

Unfortunately the
formulation of the commutator field equations is restricted to the
case of $\T$-topology and 
the modification to $\S$-topology is outstanding.

Note that the solutions considered in this section are the only ones
in this thesis which are not necessarily FAdS. This is so because we
give data on standard Cauchy surfaces, and this yields no direct control about
the evolution behavior. For $\lambda=0$, corresponding solutions cannot
be FAdS anyway. In the following let us choose the time orientation
such that the solutions collapse into the future, as determined by the
timelike area gauge with positive $\mathcal N_0$ and increasing time $t$.

\subsection{Error analysis and monitoring}
\label{sec:erroranalcommruns}
In \Chapterref{ch:numexperiments} we obtained some preliminary
experience on the numerical behavior of the conformal field equations and fixed
a few error quantities that we monitored in the
discussion which followed. Since the situation is a bit different for
the commutator 
field equations, this section is devoted to the discussion of some error
quantities for these equations.

Let us comment on how to judge the size of the errors involved in our
numerical runs. First, this can be done as usual by convergence
tests but, as explained before, their interpretation can be quite
different for pseudospectral than for finite differencing codes due to
the potentially strong influence of round-off errors in the first case.
Second, one can monitor error quantities, in
particular constraint violations.
We have from \Eqref{eq:constrLambda}
\[{\mathcal C}_{\Lambda}:=(E_1-2r)(\Omega_\Lambda)\]
with $r$ obtained from \Eqref{eq:ralb} with $A\equiv 0$
\[r = - 3\,(N_\times\,\Sigma_--N_-\,\Sigma_\times).\]
Furthermore, we have from \Eqsref{eq:constrEresidual}
\begin{align*}
  {\mathcal C}\indices{_2^2}&:=(E_1-\sqrt{3} N_\times-r)(E\indices{_2^2})\\
  {\mathcal C}\indices{_3^2}&:=(E_1+\sqrt{3} N_\times-r)(E\indices{_3^2})
  +2\sqrt{3} N_-E\indices{_2^2}\\
  {\mathcal C}\indices{_3^3}&:=(E_1+\sqrt{3} N_\times-r)(E\indices{_3^3}).
\end{align*}
The Gauß constraint \Eqref{eq:constrGauss} is solved identically since
$\Sigma_+$ is obtained from it. However, the evolution equation for
$\Sigma_+$ \Eqref{eq:evolSigmaPlus} 
\[3\,E_0(\Sigma_+)
=  -3\,(q+3\Sigma_+)\,(1-\Sigma_+)
+ 6\,(\Sigma_++\Sigma_-^{2}+\Sigma_\times^{2})- 3\,\Omega_\Lambda 
-E_1(r)\]
might be violated numerically. 
How do we measure the violation of this equation at a given time step?
There is no unique 
way to evaluate the partial derivatives in this equation
numerically. The cleanest way, i.e.\ avoiding further errors which are just
caused by the numerical evaluation of the partial derivatives, is to
substitute the relation for $\Sigma_+$ from the Gauß constraint and
the relation for $r$ above into this equation. Then, the equation
involves time derivatives of other unknowns. With the same
argument as above, it is natural to determine
these values from the corresponding evolution equations whose
violations at a given time step cannot be measured without introducing
further distinct numerical methods. It turns out that all terms cancel and
from this point of view the evolution equation for $\Sigma_+$ above is
satisfied identically. This is no surprise since it is just a
reformulation of the statement that the Gauß constraint propagates. In
summary, we cannot measure the violation of the evolution equation of
$\Sigma_+$ at a given time step without introducing further numerical
methods to estimate the partial derivatives involved. In general, it is a
principle that ``we cannot measure errors with methods that rely on
the same errors''. The usage of other numerical methods to evaluate
partial derivatives introduces further errors and it is hard to
distinguish if the corresponding results really represent the true
error quantity or rather the errors of these further numerical
methods. For our purposes, we thus ignore the evolution equation of
$\Sigma_+$.

Now we comment on violations of the integrability condition 
\Eqref{eq:integrabbeta} for $\beta$ and define its violation as
$\Phi_{int}$. After 
the same kind of manipulations as before we find 
\[\Phi_{int}=\frac{3}{2}{\mathcal C}_{\Lambda}.\]
Hence, this error quantity is non-trivial, but it
is explicitly determined by ${\mathcal C}_{\Lambda}$ and so it is 
sufficient to monitor the quantity ${\mathcal C}_{\Lambda}$.

Further, we would like to check if the
orbit area density
$\mathcal A$ given by \Eqref{eq:orbitareadensity} is really constant
on each time slice as required by the underlying gauge conditions. We
define the following error quantity 
\[\mathcal C_A:=E_1(\mathcal A)
=E_1(\beta^2 E\indices{_2^2}E\indices{_3^3}),\]
and want to check if it is zero.
The same kind of manipulations as above lead to
\[\mathcal C_A=\mathcal A^{-1}(
{\mathcal C}\indices{_2^2}+{\mathcal C}\indices{_3^3}).\]
Hence, monitoring the quantities ${\mathcal C}\indices{_2^2}$ and
${\mathcal C}\indices{_3^3}$ is sufficient to estimate the error
quantity $\mathcal C_A$.

In summary, it is sufficient to monitor the following ``constraint
violation'' quantity in our computations which is
the sum of the $L^1$-norms of the quantities
${\mathcal C}_{\Lambda}$, ${\mathcal C}\indices{_2^2}$,
${\mathcal C}\indices{_3^2}$ and ${\mathcal C}\indices{_3^3}$.

For the following runs, it turns out that no time adaption is
needed. This can be expected 
since the gauge is chosen such that the singularity lies at
$t\rightarrow\infty$ in an ``exponential manner''; see the discussion
associated with the choice of lapse in \Eqref{eq:choicelapse}. A
typical plot of the adaption norm is \Figref{fig:CosmBothadapt}, where
one can see  
that the need for spatial resolutions indeed increases in time, but
not on shorter and shorter time scales as in the runs before.

\subsection{Numerical Results}
\label{sec:commNumR}
As initial data for the $\beta$-normalized quantities
(\Sectionref{sec:comm_field_eqs}) we pick
\[E\indices{_1^1}=-2, \quad\Sigma_-=-\frac 5{\sqrt 3}\cos x,
\quad \Sigma_\times=0,\quad N_-=\frac 1{\sqrt 3}\sin x,\quad N_\times=0,\]
which so far agrees with those in \cite{Andersson03} and which is in the
same family of data that was already investigated in \cite{Berger93}. However,
the residual choices are $\Omega_\Lambda=1$ (refer to as
``$\lambda>0$'') and 
$\Omega_{\Lambda}=0$ (refer to as ``$\lambda=0$'') which both satisfy
the constraint \Eqref{eq:constrLambda}. Furthermore, the initial data
for the residual decoupled part is
\[E\indices{_2^2}=-2,\quad E\indices{_3^2}=2 \cos x,\quad
E\indices{_3^3}=-2\]
which is in agreement with the constraints
\Eqsref{eq:constrEresidual}. Note that these data imply the initial
hyperbolic velocity $v=5|\cos x|$ according to
\Eqref{eq:hypvel}. Further recall that
due to our specific choice of $\mathcal N_0=-1$, our
time coordinate $t$ is related to the time coordinate $\tilde t$ in
\cite{Andersson03} by $t=\tilde t/2$.

\begin{figure}[tb]
  \begin{minipage}[t]{0.49\linewidth}
    \centering
    \includegraphics[width=\linewidth]
    {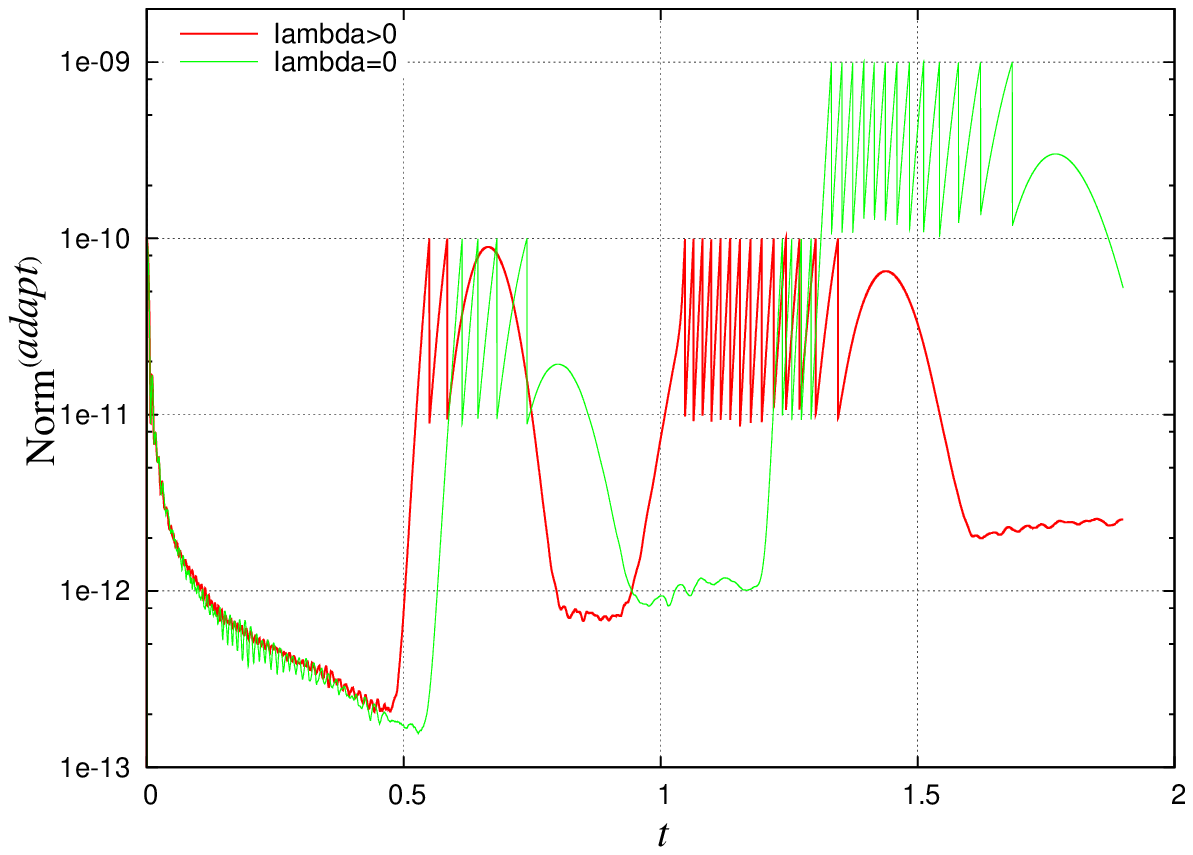}
    \caption{Adapt norm for the $\lambda>0$ and $\lambda=0$ cases}
    \label{fig:CosmBothadapt}
  \end{minipage}
  \begin{minipage}[t]{0.49\linewidth}
    \centering
    \includegraphics[width=\linewidth]
    {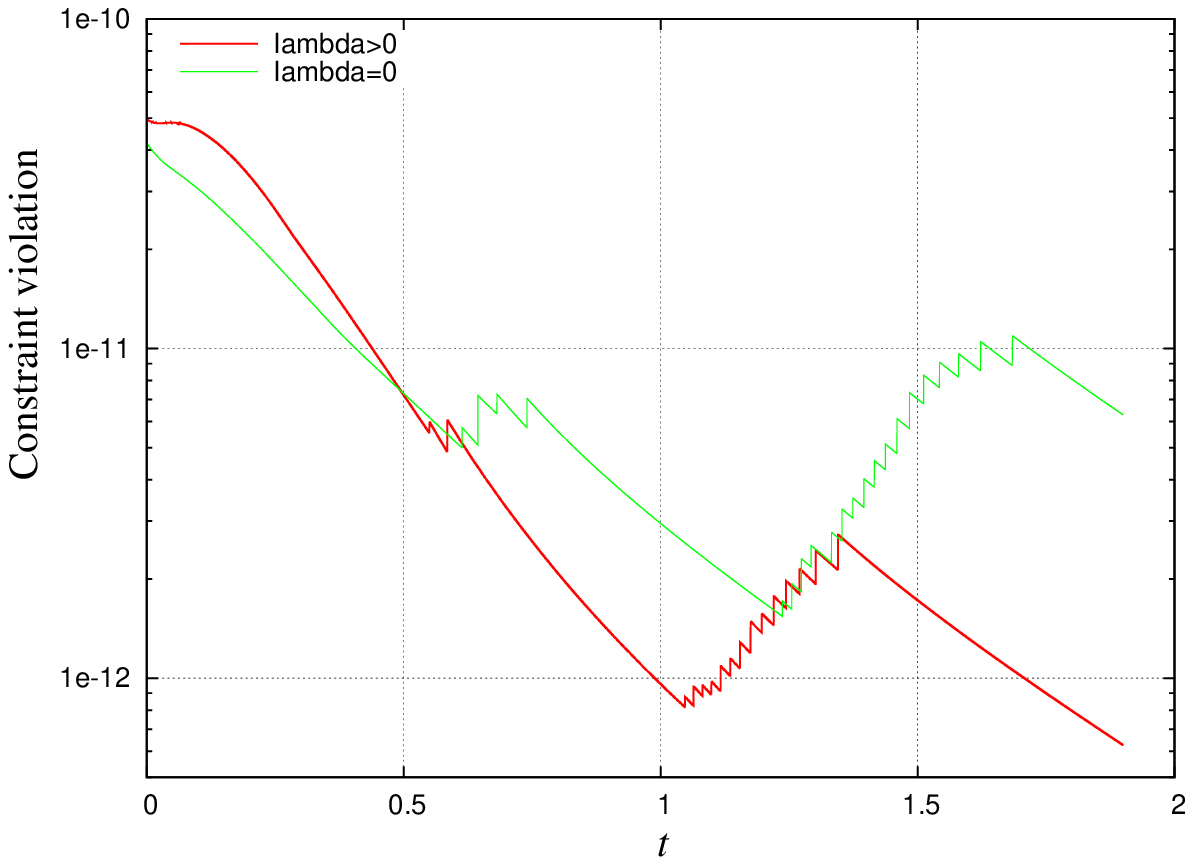}
    \caption{Behavior of the constraint violation for the $\lambda>0$
      and $\lambda=0$ cases}  
    \label{fig:Cosmconstraint}
  \end{minipage}
\end{figure}
In \Figref{fig:CosmBothadapt} we see the behavior of $\normadapt$. As
indicated before due to the choice of gauge, there is no need to
stretch the time axis. Certainly
this subserves the quality of the numerical calculations since no
``artificial'' time adaption is necessary. 
In the following we show only results obtained with one resolution and
in particular no convergence plots; the code is convergent similarly as
we observed before.
The time resolution is
$h=10^{-4}$ in all runs shown here and fixed; the spatial resolution starts with
$N=511$ (number of collocation points in $x$-direction) in both runs
and ends with $N=2133$ in the $\lambda>0$ case and with $3779$ in the
$\lambda=0$-case.
 In both cases,
the final time $t=1.8999$ corresponds 
to $\tilde t=3.7998$ (time coordinate in \cite{Andersson03}). This is
not very close to the singularity; note, however that there was no
principal obstacle to let the runs continue.

In \Figref{fig:Cosmconstraint} one can see the behavior of the
constraint violation as introduced in
\Sectionref{sec:erroranalcommruns}. We see that the constraints are
surprisingly well behaved, 
namely they decay approximately exponentially in time. Note that these jumps in
\Fignref{fig:Cosmconstraint} are produced by the spatial
adaption. Namely, the
individual $L^1$-norms of the violations
are not exactly equal before and after an interpolation step. In any
case, this is a big
difference to the results which we obtained with the conformal field
equations. 

\begin{figure}[tb]
  \centering
  \subfloat[Behavior of $\Sigma_\times$]{
    \includegraphics[width=0.49\linewidth]{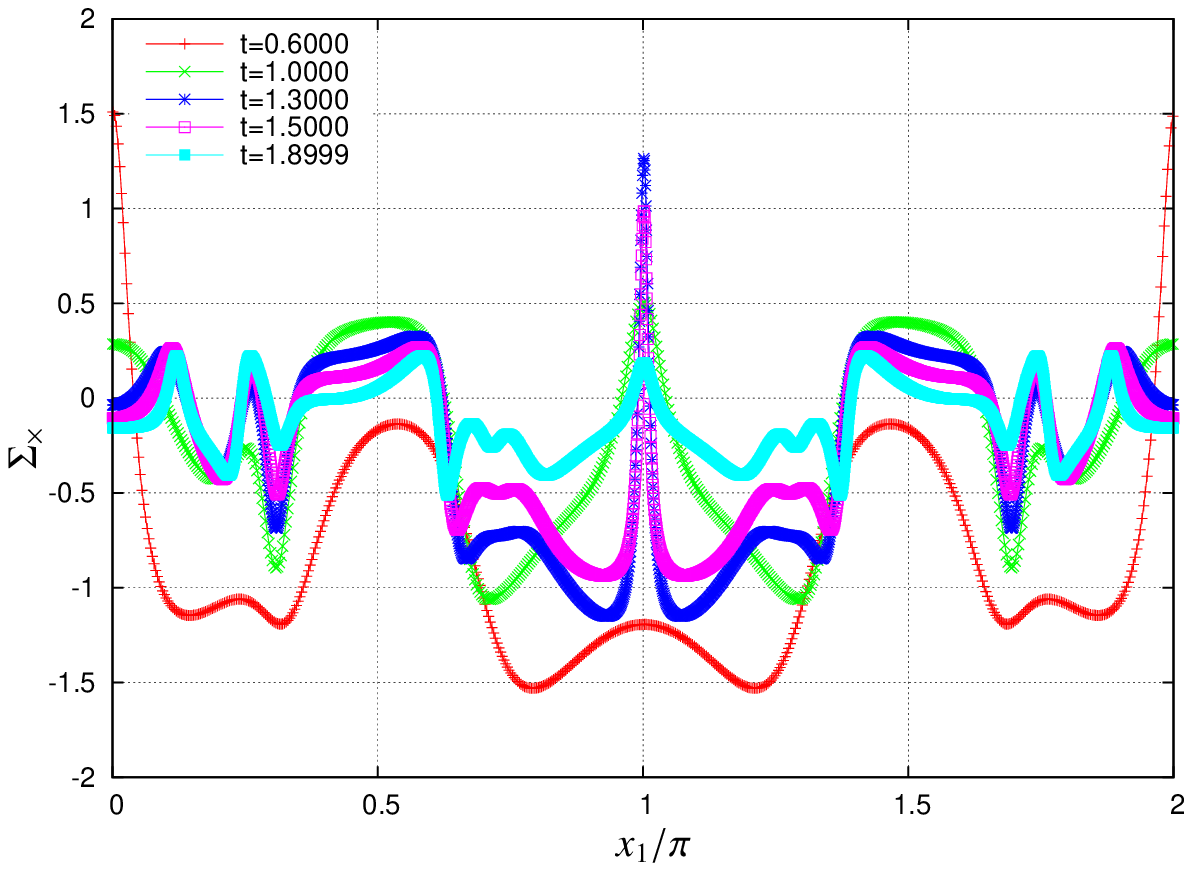}}%
  \subfloat[Behavior of $N_-$]{
    \includegraphics[width=0.49\linewidth]
    {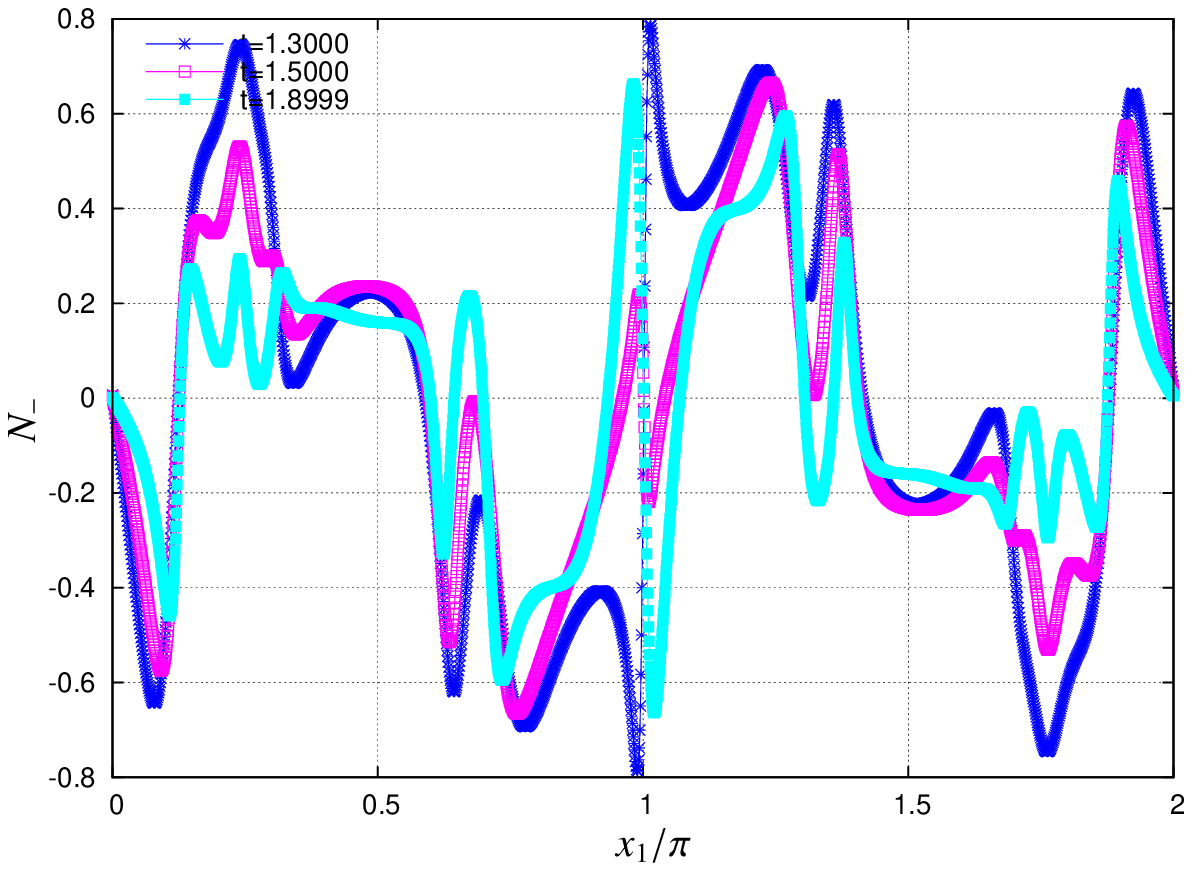}}\par
  \subfloat[Behavior of $\beta$]{%
    \label{fig:CosmNonvanishGeomBeta}
    \includegraphics[width=0.49\linewidth]{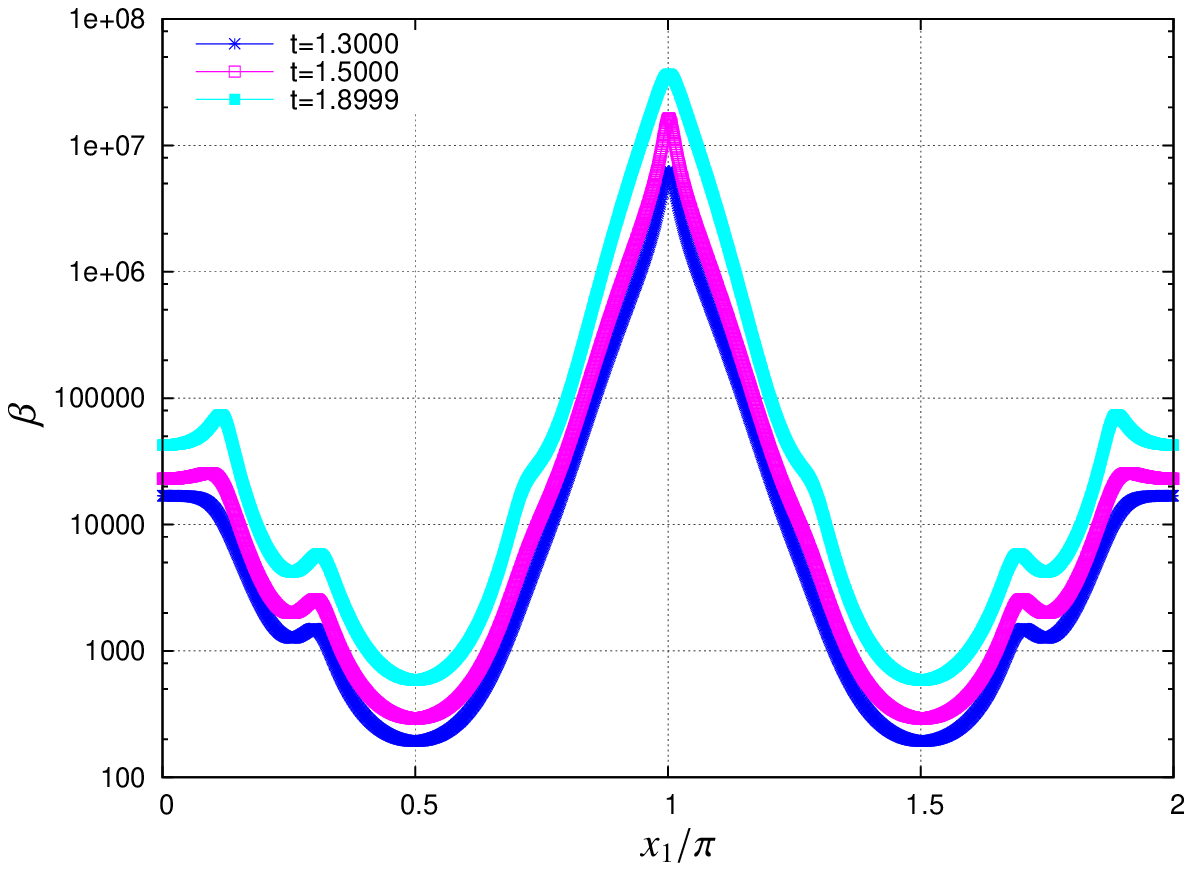}}%
  \subfloat[Behavior of hyperbolic velocity]{
   \includegraphics[width=0.49\linewidth]{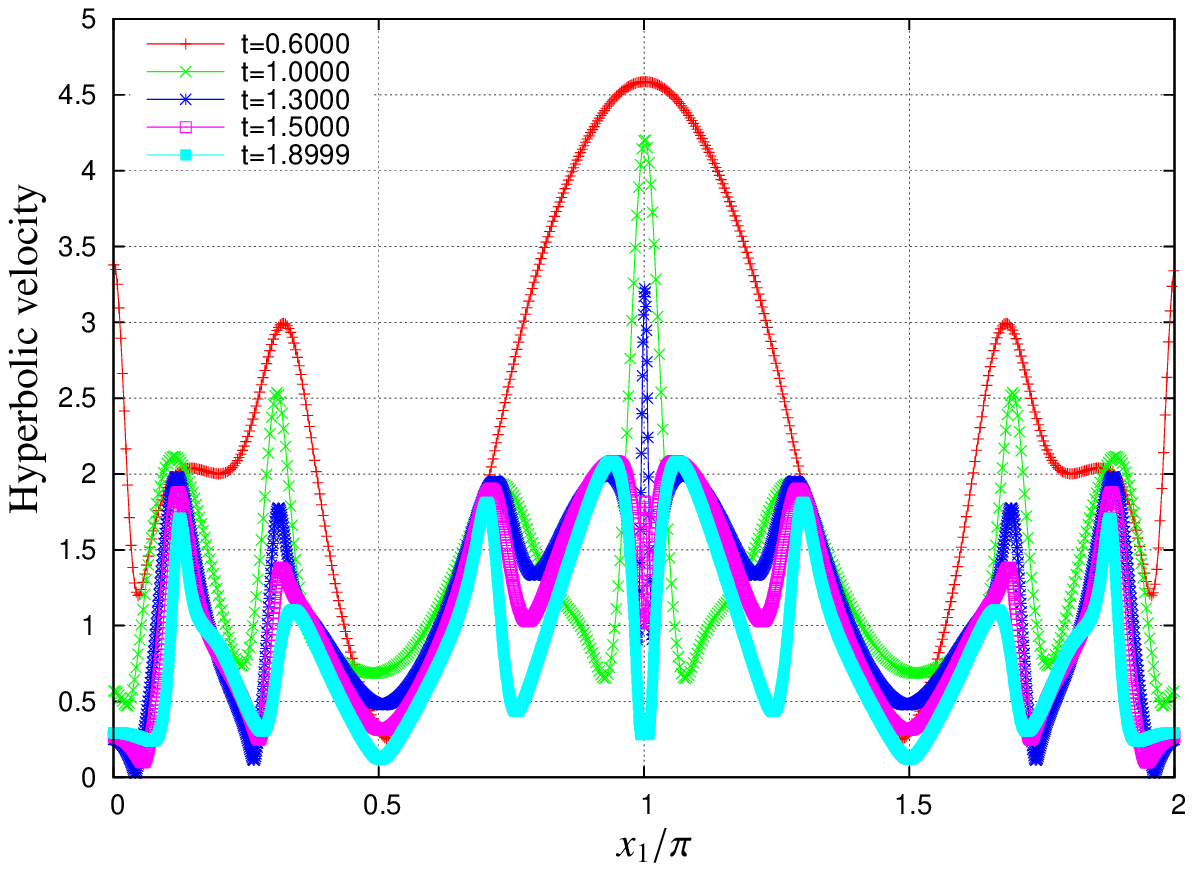}}
  \caption{$\lambda>0$ case: geometric quantities for $5$
    different times}
  \label{fig:CosmNonvanishGeom}
\end{figure}
\begin{figure}[tb]
  \centering
  \subfloat[Behavior of $\Sigma_\times$]{
    \includegraphics[width=0.49\linewidth]{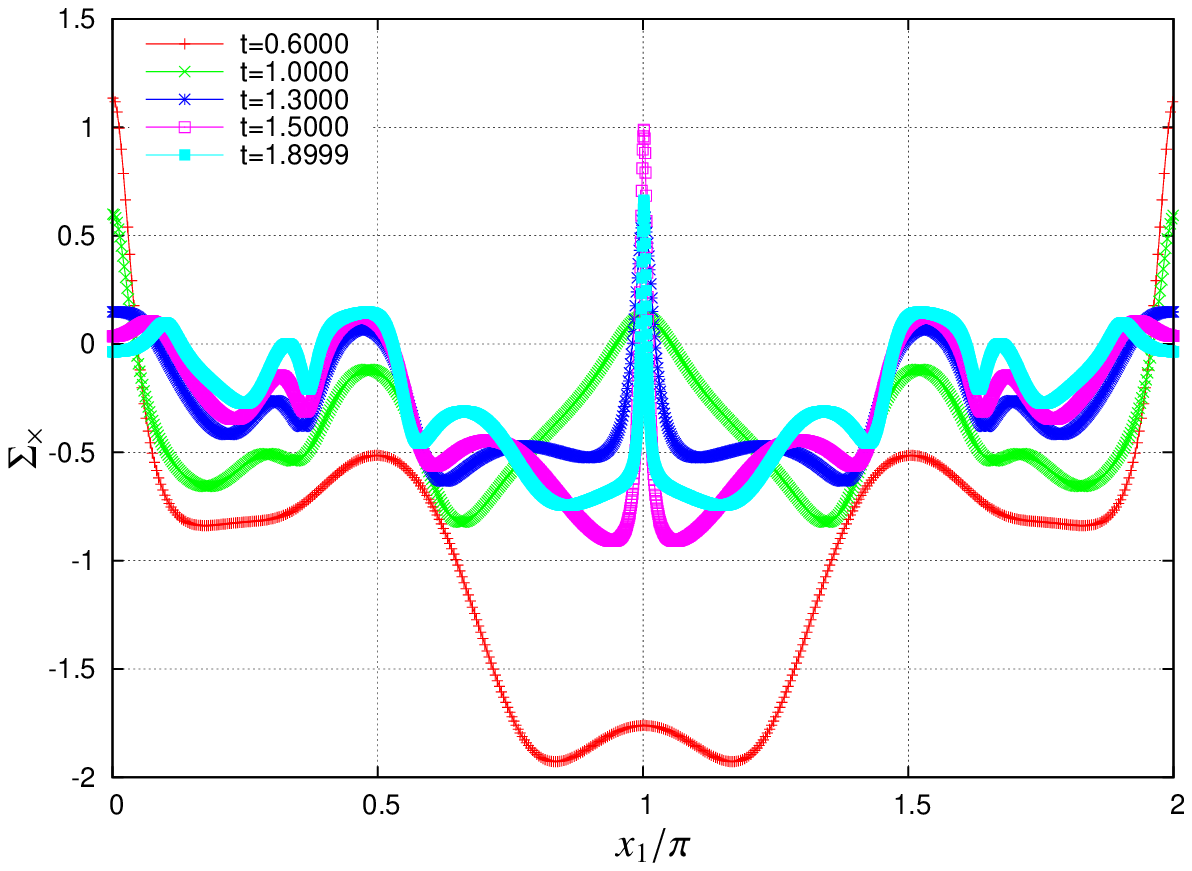}}%
  \subfloat[Behavior of $N_-$]{
    \includegraphics[width=0.49\linewidth]
    {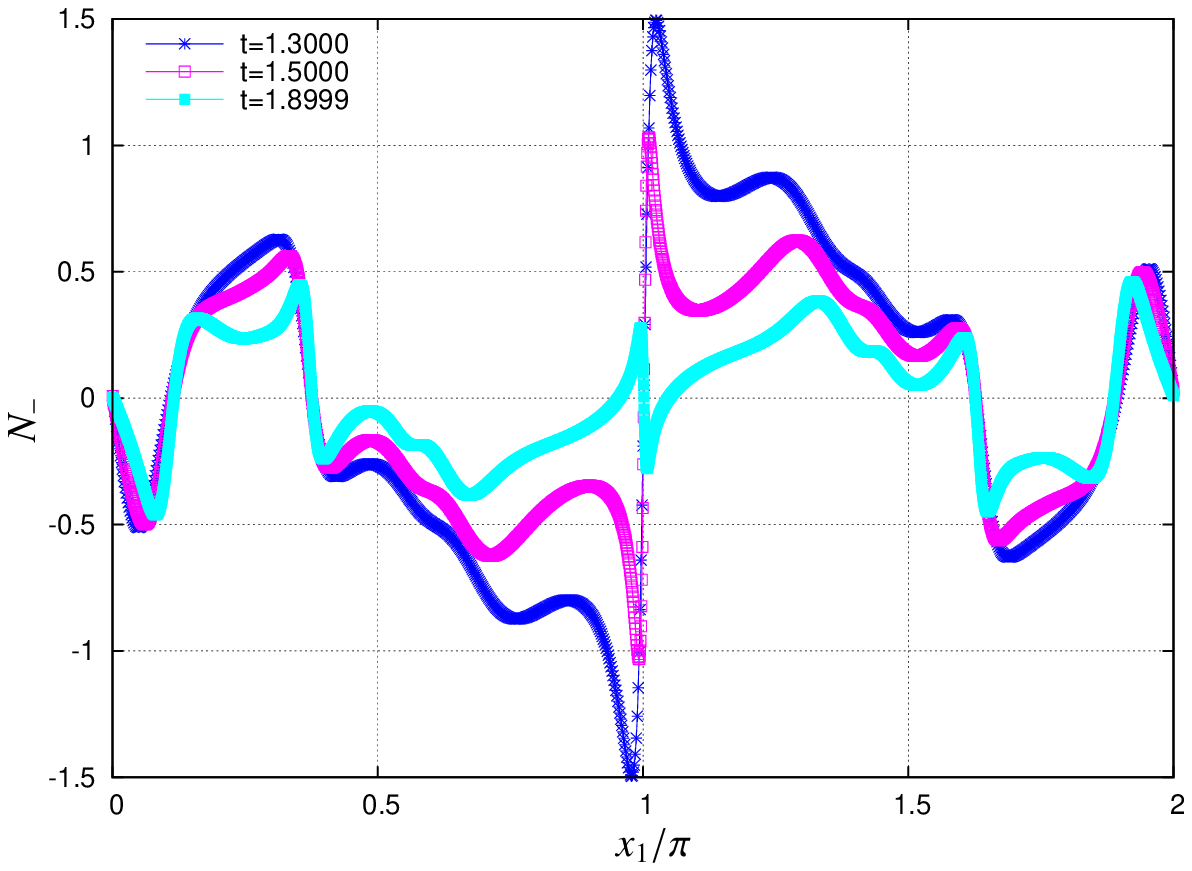}}\par
  \subfloat[Behavior of hyperbolic velocity]{
   \includegraphics[width=0.49\linewidth]{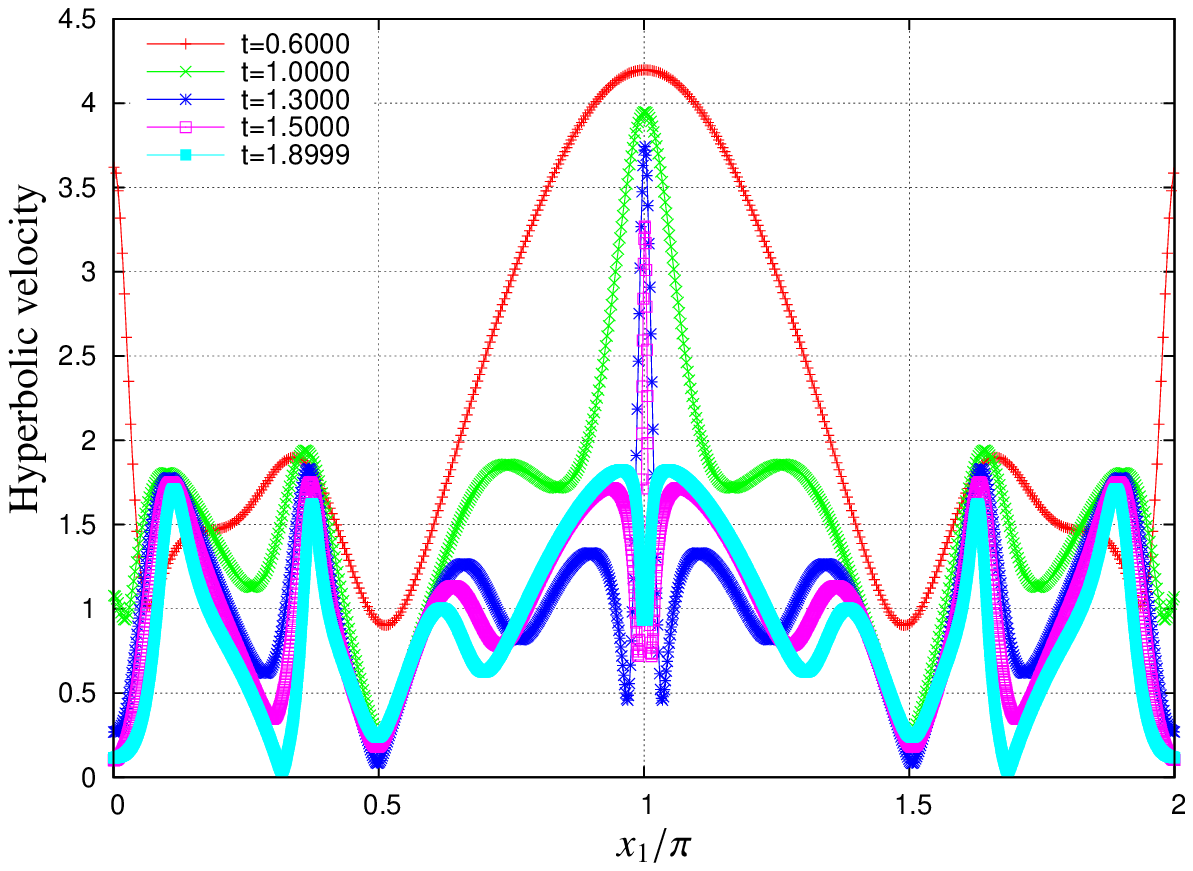}}
  \caption{$\lambda=0$ case: geometric quantities for $5$
    different times}
  \label{fig:CosmvanishGeom}
\end{figure}
Now consider \Figref{fig:CosmNonvanishGeom} and
\Figref{fig:CosmvanishGeom} where we plot the spatial dependence of
some geometric quantities for a few time steps for the $\lambda>0$-
and the $\lambda=0$-case respectively. From these plots we can follow
how the solution develops localized features from smooth data. In both
cases a particularly sharp feature develops at $x=1$, particularly
visible in the $\Sigma_\times$ plots, but which then decays
again. Comparing the $\lambda=0$ plots to Fig.~10 of
\cite{Andersson03}, 
which shows\footnote{But note that the
authors of \cite{Andersson03} use Hubble-normalized quantities so that
the plots cannot be compared directly.} 
$\Sigma_\times$ at $t=5$ ($\tilde t=10$), we see
that there is indeed no ``final'' spike at $x=1$. Those can be
observed more on the left and 
symmetrically on the right of $x=1$ and we can speculate that our
plots show the early stages of such in both cases
$\lambda>0$ and $\lambda=0$. Consider also
\Fignref{fig:CosmNonvanishGeomBeta} 
which has little peaks at those expected positions. But why does the
feature at $x=1$ decay in both cases?  It is a ``high velocity spike''
\cite{Garfinkle03} since its initial hyperbolic velocity is bigger
than $2$. The phenomenology of these features can be described roughly
as follows. The evolution 
equations drive them to lower velocity while some of them (as in our case)
decay completely. In \cite{Garfinkle03}, $\lambda=0$ is assumed but in
this special case here we see that the presence 
of $\lambda$ does not change this behavior very much. However, the
cosmological constant seems to lead to faster decay. An
intuitive argument for this is that the repulsive forces of the
cosmological constant blow up the localized features. Consider
\Fignref{fig:CosmBothadapt} again. Here we can follow the formation and
decay of this intermediate spiky feature in frequency space. We see
that the fine structure is 
built in short phases, shorter for $\lambda>0$, with relaxation phases
in between. This is consistent with the investigations of
\cite{Garfinkle03} where the conclusion was drawn that the high
velocity spikes decay by bouncing from one velocity regime to the next
lower, while in between there are phases of relatively weak
dynamics. In any case, we can see that plotting $\normadapt$ is not only
interesting to study, how the code itself behaves, but also how the
solution develops. This is so since building fine structure requires higher
resolutions and this is represented well in Fourier space. Indeed, one
can view this as an advantage 
of pseudospectral methods compared to other methods because the role
of frequency space, in which part of the relevant information about fine
structure can be read off directly, plays a fundamental role for the
method itself.

These intermediate
``spikes'' demand quite high spatial resolutions already after a short time.
But after that, when these spikes have decayed again, the resolution
is not needed anymore at least for some 
time. The current implementation of my spatial
adaption method cannot cope with 
this situation very well; in particular it is not able to reduce the
resolution when it is not needed 
anymore. However, it should be straight forward to modify the method
to make this possible. In any case, these intermediate spikes make my
runs quite slow already after short time. Although the runs are not
yet impractically slow I stopped them so that I was not able to
study the final distribution of the spikes. Another reason for
stopping the 
runs is that the current implementation of the
code produces much output data, of the order of $10$ GByte, and
the hard disk that I was using ran out of free space.

\begin{figure}[tb]
  \centering
  \subfloat[Behavior of $N_-$]{
    \includegraphics[width=0.49\linewidth]{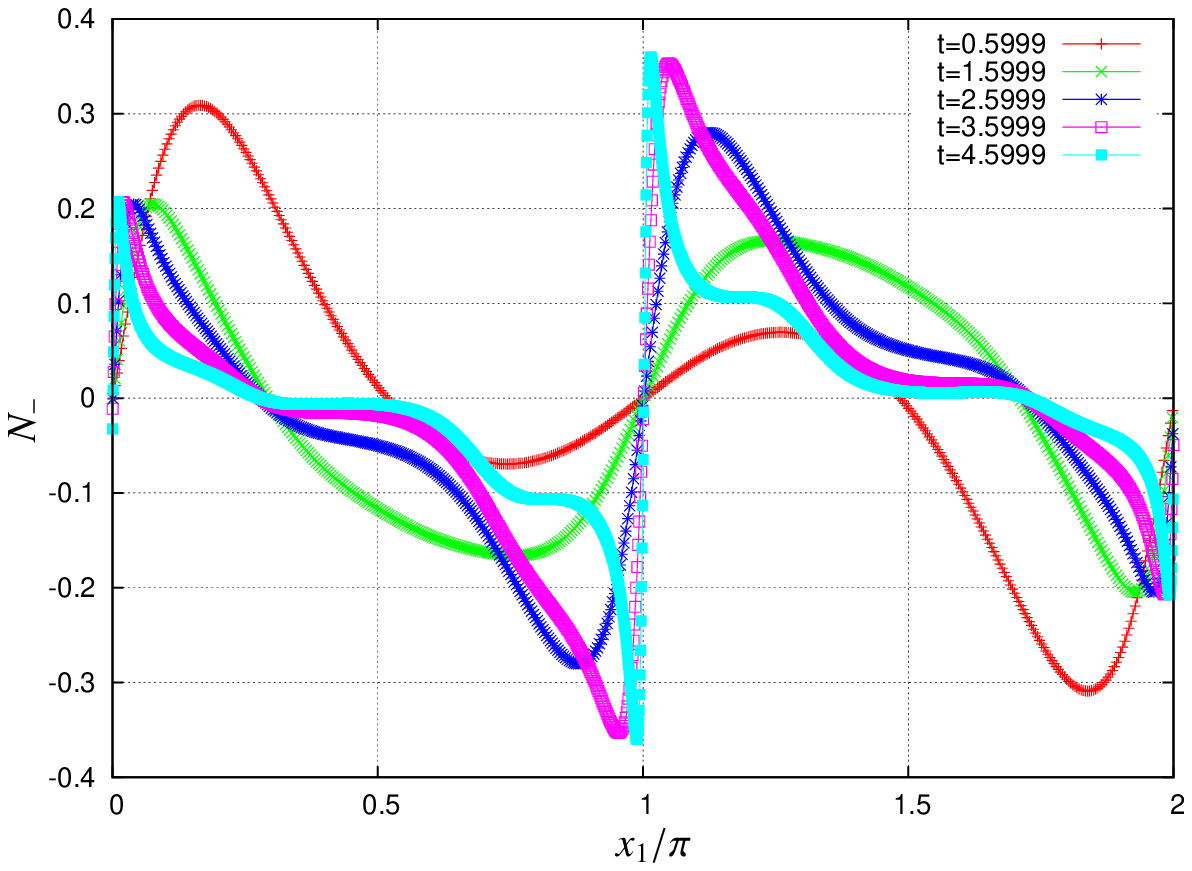}}%
  \subfloat[Behavior of Hyperbolic velocity]{
    \includegraphics[width=0.49\linewidth]
    {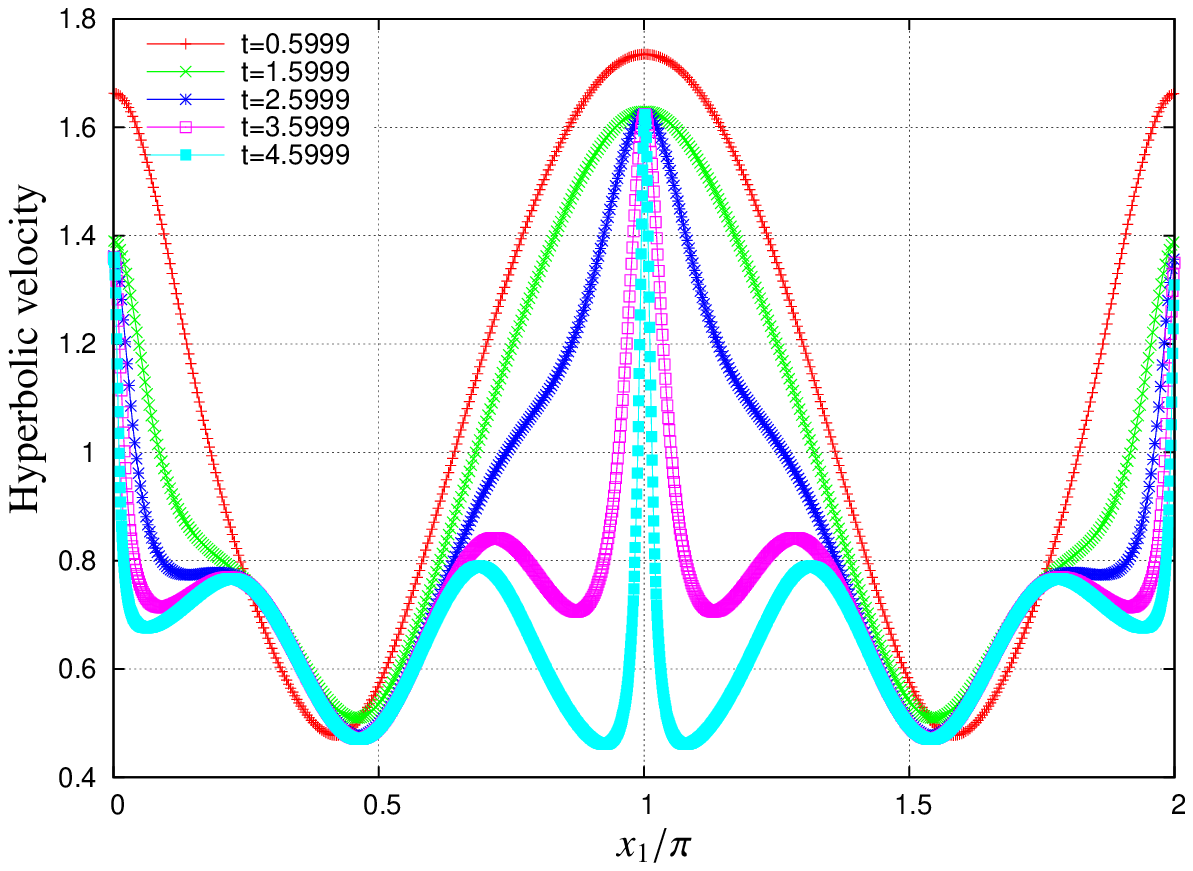}}
  \caption{Low velocity run}
  \label{fig:CosmvanishGeomLowVel}
\end{figure}
However,
since the intermediate spikes are caused by the high
velocity of my choice of initial data sets, I also
computed the solution of another choice of 
initial data with lower initial velocity.
This choice is
the same as above but $\Omega_{\Lambda}=0$ and 
$\Sigma_-=-\frac 2{\sqrt 3}\cos x$, so that the initial hyperbolic velocity is 
$v=2|\cos x|$. In the corresponding \Figref{fig:CosmvanishGeomLowVel}
we see that there is actually a spike at $x=1$ (and also at $x=0$)
which does not decay and we can see its development exemplary for
$N_-$. Note, that the time scales are different for these data than
before and we stopped the code at $t=4.5999$. The
final spatial resolution was $N=2133$.

\subsection{Expectations regarding \texorpdfstring{$\Omega_{\Lambda}$}
{Omega(Lambda)}}
\label{sec:ExpectT3GowdyLambda}
The numerical studies before cannot be considered as systematic
investigations of the influence of the cosmological
constant in solutions with Gowdy symmetry. The only aspect we could
see is that maybe $\lambda>0$ supports the decay of high velocity
spikes. But what about low velocity spikes? Could there be an extreme
value of $\Omega_\Lambda$ such that all spikes decay before the singularity? 

The only analytical results known so far for $\T$-Gowdy with
$\lambda>0$  are due to
Clausen and Isenberg \cite{Clausen07} who prove that the maximal
Cauchy development of any smooth Gowdy initial data on a
standard Cauchy surface is globally
covered by areal coordinates where the orbit area lies in the interval
$]c,\infty[$ for some undetermined constant $c\ge 0$. Furthermore,
by means of Fuchsian methods as in
\cite{Kichenassamy97} they obtain  that one can construct solutions which are
asymptotically velocity dominated in the analytic case. However, one
does not obtain 
control over the full set of free functions.

We can say a little more than this, although many of the following
arguments are heuristic. \Eqref{eq:qgauge} for $q$  tells
us that  
\[q>\frac 12-\frac 32\Omega_\Lambda,\] 
when we exclude the fixpoint solution
$\Sigma_-=N_\times=\Sigma_\times=N_-=0$ (de-Sitter spacetime).
Then, with $\mathcal N_0=-1$, \Eqref{eq:OmegaLambdaDot} implies that
\[\dot\Omega_\Lambda<3(\Omega_\Lambda-1)\,\Omega_\Lambda.\]
Consider the following initial value problem
\[\dot y(t)=3(y(t)-1)y(t),\quad y(0)=\eta.\]
A unique solution exists which has the explicit form
\[y(t)=\frac 1{1-C e^{3t}}\]
with $C=1-1/\eta$. The solution exists for all $t\ge 0$ if and only if
$\eta\le 1$. If $\eta<1$, $y(t)=O(e^{-3t})$ for
$t\rightarrow\infty$. Now since, $\Omega_{\Lambda}$ is a subsolution
of this problem it follows from the standard theory of ODEs that for
$\Omega_{\Lambda}(0)<1$ we have $\Omega_{\Lambda}(t)=O(e^{-3t})$ for
$t\rightarrow\infty$. If $\Omega_{\Lambda}(0)=1$ we can only deduce
that $\Omega_{\Lambda}(t)<1$ for all $t$. Now for 
$\Omega_{\Lambda}\ll 1$ for late $t$ we can expect from \Eqref{eq:E11dot} that
$E\indices{_1^1}=O(e^{-2t})$ for 
$t\rightarrow\infty$ and hence $\Omega_{\Lambda}$ decays exponentially
faster than $E\indices{_1^1}$. Since roughly speaking, the decay of
$E\indices{_1^1}$ is responsible for bringing the solution into the
asymptotically velocity dominated regime and also for spiky features
we can expect the same phenomenology as in the $\lambda=0$-case, at
least if $\Omega_{\Lambda}(0)<1$. 

What about very large $\Omega_{\Lambda}$? If it is initially large
compared to the other unknowns then 
$q\approx -\frac 32\Omega_{\Lambda}$. 
Using this in the evolution equation \Eqref{eq:OmegaLambdaDot} 
implies 
\[\dot\Omega_\Lambda\approx 3 \Omega_\Lambda\Omega_\Lambda,\]
which has a solution that is unbounded after finite time. Let us assume
that the other unknowns are so small initially such that in particular
the quadratic terms can be neglected. Then all of the evolution
equations are of the form  
$\dot u/u\approx 3\Omega_\Lambda$ for a generic unknown $u$. Comparing
this with the equation 
for $\Omega_\Lambda$ this could mean that all these
quantities increase with the same strength such that the approximation
$q\approx -\frac 32\Omega_{\Lambda}$ is still valid for later
times. Then the whole 
solution could blow up after finite time. However, it is not clear in
particular how the non-linear terms in the evolution equations behave
in this situation. If
it turns out that the other unknowns grow faster initially than
$\Omega_{\Lambda}$, then $q$ would become more and more positive
and the grow of $\Omega_{\Lambda}$ would be damped such that possibly the
solution stays finite for all times. In any case, what I want to say
by means of 
this heuristic discussion is that is not easy, without usage of more difficult
arguments, to exclude a blow up after finite time if the
$\Omega_{\Lambda}$ is 
high enough initially. Such a blow up would imply a drastic change in the
dynamics for $t\rightarrow\infty$ compared to the well known
$\lambda=0$ case. This possibility is maybe related to the outstanding
issues in the 
theorems by Clausen et al.\ explained above. Can the constant $c$
always be chosen to be 
zero? As we said, for $\lambda$ ``small enough'', this can be
done. But if not, what kind of singular behavior is there? 
Note, that if the solution is, say, past
asymptotically de-Sitter (past $\leftrightarrow$ decreasing the
$t$-coordinate), then
all timelike
geodesics are future incomplete in this case, at least when the scalar
curvature of $\scrip$ is negative, due to the singularity theorems of
\Sectionref{sec:singularitytheoremGA}. 
But do they run into curvature
singularities or Cauchy horizons? In the first case: are the solution
in some sense asymptotically velocity dominated? Recall, that Clausen
et al.\ were not able to prove that \textit{all} solutions  are
asymptotically velocity dominated, not even
in the analytic 
case.

\chapter{Conclusions, projects for future research and summary}
\label{ch:finalchapter}

Having described the ideas underlying my method and its implementation
  in \Partref{part:treatment} and discussed the analysis of several application
  problems in \Partref{part:analysis}, we now want to conclude and summarize.
We start this chapter with
  \Sectionref{sec:comparisonGowdymethods}, where we
  conclude about the technical   properties of my 
  method with emphasis on the
  restricted set of applications in this thesis. For these discussions
  the technical results presented in the previous chapters are taken
  into account. 
Then, in \Sectionref{sec:outluckfuture}, we describe open
  problems of both technical and physical-mathematical nature. On the
  one hand, we reconsider the  
  preliminary results about properties of the solution space of EFE obtained in
  this thesis, and, on the other hand, we discuss expectations
  about other 
  applications of interest which particularly go beyond the limited set of
  applications considered in this thesis.
At the end of this chapter, in \Sectionref{sec:summary}, we summarize.

\section{Conclusions about the numerical method}
\label{sec:comparisonGowdymethods}
The purpose of this section is to compare my
method to those listed in
\Sectionref{sec:specialtechniques}  
and to draw
some conclusions. 
In agreement with our previous analysis, we focus on two aspects
here, namely  first, the
way my code is able to cope with the presence of the coordinate
singularities of the Euler parametrization of $\S$, and second, how
well it allows to approach
Gowdy singularities. It should be clear that for $\S$-Gowdy spacetimes
these aspects cannot be 
discussed independently. Recall the definitions of the
various pseudospectral techniques to deal with the coordinate singularity in
\Sectionref{sec:discussiontancot}.

In \Sectionref{sec:linearized_solutions} we analyzed the linearized
GCFEs on the de-Sitter background and found that the down-to-up
treatment of the coordinate singularity is stable and
reliable. Similar results were obtained in \Sectionref{sec:comparison}
where we considered non-linear regular $\lambda$-Gowdy solutions. In
particular, there
was no notable difference between staggered and non-staggered
coordinate singularities; both performed equally well. 
Further investigations of singular $\lambda$-Gowdy
spacetimes with $\S$-topology in \Sectionref{sec:S3singularGowdy}
supported the conclusion that the coordinate singularity treatment
introduced in
this thesis works quite well.

However, in these investigations we also
found that both our direct multiplication and up-to-down
methods are 
instable. Although it is in principle clear that round-off
errors are responsible for that, in particular for the direct
multiplication method which is implemented without any projection, we
have no real understanding of the 
mechanisms. Conversely, the same can be said about the
\textit{stability} of the down-to-up method. This issue is not restricted
to our situation here. Indeed, similar observations are made for the
axis singularities of cylindrical coordinates in $\R^3$. Note that the
coordinate singularity on $\S$ resembles these singularities.
In the case of axis singularities,
important contributions are the
implementations (\Sectionref{sec:nontrivialtopologies}) by Garfinkle
et al.\ \cite{garfinkle1999,Garfinkle00}, 
the one by Choptuik et al.\ \cite{Choptuik03} and that by Rinne and
Stewart \cite{Rinne05,Rinne05b}. With the exception of 
\cite{garfinkle1999}, all those assume axisymmetry. The
implementations by Garfinkle et al.\ and Choptuik et al.\ are similar
to our direct multiplication. Namely,  
at the staggered coordinate singularity, one multiplies ``directly''
functions with a zero, in 
particular some of the variables possibly after clever rescalings, 
with functions having a pole, in particular singular terms
in the equations. 
If the initial data are chosen appropriately, the (exact) evolution
equations imply that the variables decay with the 
right order at the singular points at all times such that the limit of such
products is finite.
However, the discretized equations can be instable. 
This is the reason why in the methods above the problem is formulated as an
initial boundary value problem, and boundary
conditions necessary for smoothness are
prescribed to yield at least some sort of control there. 
However, there is still the possibility 
that the solution of the discretized equations behaves in an undesired
manner at the
boundary. Indeed, Choptuik et al.\ find 
that even with such boundary conditions, their code is instable at the 
axis, and they have to use Kreiss-Oliger dissipation to cure this. In
contrast to that the approaches by Garfinkle et al.\ are reported to
be stable. Since the reason for this different behavior is not known,
we can state that the issue of axis instabilities 
is not understood in general.  The implementation by Rinne and Stewart
is a bit 
different because they were able to find a formulation of the equations 
which is completely free of singular terms in the axisymmetric
case. In particular their
evolutions are claimed to be stable at the axis. However, it is
unclear if their treatment can be generalized to cases without axisymmetry.
In general, one can 
expect that, at least to some degree, the choice of the
time-marching scheme has an impact on this stability issue because
certain schemes 
have higher implicit dissipation than others, but to my knowledge this has not
been investigated systematically yet.
In particular it would be interesting to experiment
with other time integrators than Runge-Kutta in my code.

Indeed, there is a further motivation for experiments with other
time-marching schemes which is particularly relevant for spectral  
methods. Since the 
evaluation of the spatial derivatives of all variables on a given time
slice is 
relatively expensive for spectral methods, one should look for 
time-marching schemes which take as few trial time steps as
possible for each time step. In particular, it may be possible to find a better
compromise between speed and stability than the $4$th-order Runge
Kutta scheme used in this thesis.

To conclude, if we want to treat a situation with a coordinate axis
singularity numerically, 
not necessarily with axisymmetry, in which
spectral methods are appropriate then 
the experience in this
thesis suggest
that an approach based on
our down-to-up 
method is highly accurate, stable and reliable. However, when
applications are studied for which spectral methods 
cannot be expected to be optimal then one might have to consider
something else; for instance one of the other techniques mentioned here.
In any case, a description of the mechanisms driving instabilities derived
from first principles is outstanding.

This suggests that for the discussion of our second focus, namely the
approach to the Gowdy singularity, we must clarify to which extend 
spectral methods are appropriate,
or if finite differencing are favorable. For a
fair comparison with other treatments of Gowdy spacetimes, we should
point out that the problems which we observed in
\Sectionref{sec:S3singularGowdy}
to approach the 
singularity in $\S$-Gowdy spacetimes are believed
to be caused to a large degree by 
the unsuitable choice of gauge. Further, the constraint growth in our
reduction of the
conformal field equations was very strong. However, we are optimistic
that one can find a 
modification of the commutator field equations
for the $\S$-case and a $1+1$-reduction according to the ideas in
\Sectionref{sec:implementationOfS3Gowdy}, such that we can use our 
techniques for the $\S$-Gowdy problem in timelike area gauge.
Our experience with the spectral
code to treat the $\T$-Gowdy case with the commutator field equations in
\Sectionref{sec:RunsCosmFE} 
shows that a few thousand grid points are in reach of the method. This
is the order of magnitude of grid points that was also used by
Garfinkle in \cite{garfinkle1999} to treat the $\SoXSt$-case. He
was able to achieve  
accuracies of the order $10^{-3}$ and 
shows results up to time $\tau=10$ where in his convention the orbit
area density is proportional to $\sin t$ and $\tau=-\ln\tan t/2$. With
our pseudospectral approach and with the expected formulation of the
problem as above, one could hence reach at least as much accuracy in
the $\S$-case; it is likely that the accuracy is even higher due to
spectral discretization. However,
we can also expect that resolutions of this order of magnitude would
constitute a limit for 
our method, at 
least without further technical devices like local adaption or
parallelization. 


We have seen that for our method, adaption is crucial to approach the
Gowdy singularity. In the timelike 
area gauge, as we experienced in \Sectionref{sec:RunsCosmFE}, adaption
is not needed in 
time but certainly in space. Fixed resolution codes, as for example
the one by
Berger and Moncrief \cite{Berger93}, have to use very high resolution
right from the 
beginning; for example they made runs with $20000$ grid points in
space and were 
able to keep the errors below the order $10^{-4}$. However, first, such high
number of grid points can probably be never reached with spectral
methods, and second, even if one could reach it, one would eventually
start to lose spikes when they
become smaller than the grid spacing. Hence,
adaption is particularly important for spectral methods, however, note
that the second aspect is surely an 
issue for finite differencing as well. Our adaption
method described in \Sectionref{sec:spatial_adaption}, as primitive as
it may seem on the first sight, is 
particularly nice to track the smallest scales relevant for the solution
due to our analysis in frequency space.

A further argument pro spectral methods for the approach to Gowdy
spacetimes is the following. As argued in \cite{boyd}, it is a
rule-of-thumb that in general spectral methods have lower
implicit dissipation than most finite differencing
discretizations. Too high dissipation can
influence the evolution of localized features drastically as been
experienced by 
Hern et al.\ \cite{Hern97} with their Lax-Wendroff scheme. Since this
feature of 
pseudospectral methods is dependent on the choice of the time
integrator, we have thus found another motivation
to experiment with other time-marching schemes for optimizing the
dissipative character of the method.

For further conclusions one should start direct comparisons
of all relevant techniques systematically. In the single patch case,
changing from pseudospectral discretization to finite differencing 
is in fact straight forward
the way my code is implemented. For multi-block techniques, it is
likely that a new code 
has to be
created based on the already existing infrastructure of
\cite{tiglio05}. However, these 
methods are particularly promising not only due to their 
generality, but in particular also because the
existing infrastructure offers efficient parallelization on
super computers and fixed mesh refinement.

The hope that motivated me to use pseudospectral methods
is so far well supported by the numerical experiments done in this thesis.
Coordinate singularities can be treated in a clean way without having
to rely on complicated multipatch methods. Indeed, it is realistic to
believe that 
sufficient resolutions to treat the $\S$-Gowdy case are in reach of
the method which could give us all
advantages which a spectral method has to offer. For this, it is  
crucial to find a reliable way of formulating the timelike area gauge
in a $1+1$-fashion. As soon as this is possible, the method presented
in this thesis should be able to treat the so far outstanding case of
Gowdy solutions with $\S$-topology with resolutions as in Garfinkle's
work about the $\SoXSt$-case. I should point out that the current
implementation of my code is not yet 
optimized for efficiency, on the one hand because I am not yet using the FFT
algorithm, and on the other hand because I am not yet taking into account
the special structure of the Fourier series in the $\S$-case. These
possibilities make it even more likely that sufficient resolutions  
can be reached.  However, we can also expect that my method, in its
current form, is not able to reach more resolution than this; further
discussions on this aspect can be found in the next section.

It should be clear that
these results of this thesis, which focused on two main aspects, do
not allow to draw real
conclusions about the applicability  of our method, neither in the
positive nor in 
the negative sense, when one wants to go
``beyond'' Gowdy or treat other problems of interest. However,
certainly one can formulate some expectations. A discussion on
modifications which are expected to become necessary can be found in
\Sectionref{sec:outluckfuture}. In particular, for Gowdy solutions the
problem is simplified by the fact that 
a 
nicely adapted gauge is known, namely the timelike area gauge. But which
gauge to choose in more general 
situations? See \cite{garfinkle04a} and references therein for some ideas.
Further, in  Gowdy solutions one observes no oscillations close to the
singularity. In general, 
according to the BKL-conjecture, such difficult phenomena can be
expected and the numerical method must be adjusted to be able to cope
with these. Some ideas on this can be found in \cite{Berger96,Berger2002}.
Further it might turn out that in general we find pathologies even
more severe than spikes. Those would certainly require even more
sophisticated local adaption techniques. For other applications than
the cosmological singularities we list further ideas in the next
section.

\section{Outstanding issues and future research projects}
\label{sec:outluckfuture}
I will now summarize problems that were left open in this thesis,
further interesting application projects and possibly necessary modifications
of my method. All these aspects 
could be among my future research topics. 
In accordance to what we have stated before, we discuss the
questions of interest in 
connection with our expectations about the required techniques in the
following. 

\paragraph{Gravitational singularities}
In \Sectionref{sec:comparisonGowdymethods} we have already drawn
some conclusions on the properties of the current implementation of
my code regarding the two main issues of this thesis, namely the
coordinate singularity of $\S$ and the approach to
Gowdy singularities, both in comparison with other methods on the
market. Here we continue this discussion with the particular emphasis
on future research possibilities.

Certainly, an important future research project will be to continue the
investigations of past singular $\lambda$-Gowdy solutions with
$\S$-topology started in this thesis. In a first step one can ignore
the issue of constructing FAdS solutions. 
Recall that it is conjectured that the singularity is asymptotically
velocity dominated with similar types of spikes as
in the $\T$-case. For the $\SoXSt$-case, this conjecture is supported by the
numerical results in \cite{garfinkle1999}. The only
relevant analytical results of \cite{Isenberg89,Stahl02} cannot be considered
as complete. A bit farther in the future, we also want to study issues like
cosmic censorship and the BKL-conjecture for more general cases,
for example in $\U$-symmetry. In principle my code in its current form
could be applied in this situation. In an even farther future, we
would like to 
give up all symmetry assumptions to
study ``generic singularities'' and continue the research program
started in \cite{garfinkle04a}. In that work, the topology is
restricted to $\T$ and it would be interesting if similar statements
can be obtained in the $\S$-case in particular.
Up to now in the
$\S$-case, my code was implemented assuming $\U$-symmetry. In order to
use it for  
studies of spacetimes without any 
symmetries, our analysis of the 
coordinate singularity  has to be
generalized. I expect no principal problems with this, only that all
expressions in \Chapterref{ch:treatmentofS3} become more complicated.

I have already stated that I
am optimistic to reach sufficient 
resolutions with my code to study the $\S$-Gowdy case, if it turns out
to be possible to
formulate the equations in a 
$1+1$-fashion with timelike area gauge.
For this problem it seems worthwhile to try to adapt
the commutator 
field equations to the $\S$-case
since this system seems to be well-behaved in our experiments in
\Sectionref{sec:RunsCosmFE}. We are also optimistic that the problem
can be reduced to $1+1$ even in the $\S$-case due to the ideas in
\Sectionref{sec:implementationOfS3Gowdy}. 
That general optimism is particularly based on the expectation that my global
spatial adaption 
technique (\Sectionref{sec:spatial_adaption}) is
sufficient to study Gowdy singularities. However, for future studies,
either when  
the localized features become more complicated than Gowdy spikes in
more general situations, or when computer resources have to be
handled more efficiently, one must start to think about local adaption
techniques. For instance, I expect that without such techniques my
current implementation is limited practically to 
a few thousand spatial grid points in $1+1$. With 
finite differencing,
implementing such local adaption
techniques is not as difficult as in spectral methods. Nevertheless,
pseudospectral multidomain approaches, which 
can be used for fixed mesh refinement, have been
implemented successfully in numerical relativity; 
for elliptic equations see
\cite{Pfeiffer02,Ansorg06}, for the
hyperbolic ``generalized harmonic'' formulation of EFE applied to the
binary black hole problem see \cite{Scheel06} and for a special mixed
elliptic-hyperbolic problem see \cite{Lau07}.

In any case, there are several motivations to compare the method
worked out in this thesis to other numerical methods directly, both in
the $\T$-case 
without coordinate singularities and in the $\S$-case. 
First, it is
always important to compare the differences of the results obtained from
distinct numerical 
techniques to get further insights into the errors
involved. Second, despite the better accuracy property, high
resolution spectral approaches are usually more expensive than
finite differencing methods with the same number of grid points. This
is the basis for the expectation that finite 
differencing discretizations may cope better with small scale features,
as those close
to gravitational singularities.
To switch our code to finite 
differencing is, at least on $\T$, straight forward. The issue how finite
differencing is able to deal with the
coordinate singularity on $\S$ is outstanding. 
In systematic comparisons of different methods the multi-block
approach \cite{tiglio05}, see 
\Sectionref{sec:nontrivialtopologies}, should play a major role.
This should be so not only because of its generality but also because
the existing numerical infrastructure of these codes, involving efficient
parallelization routines to put runs onto supercomputer and fixed mesh
refinement, may become necessary as has already been argued
above. Codes based on pseudospectral methods and FFT can also be 
parallelized in principle, for instance the implementation \cite{FFTW} of FFT
supports this, but it cannot be expected that this is as efficient on
computer clusters of standard type as in the finite differencing case.

In any case, our current implementation is not optimized for
efficiency. On the one hand
the  
partial summation method (\Sectionref{sec:myspectralcode}) should be
substituted by FFT,
either by using one of the freely available highly optimized libraries
(e.g.\ \cite{FFTW}) or by implementing it individually. The code
will benefit from this in particular for the high spatial
resolutions needed at gravitational singularities.  Further, in the
$\S$-case we should make use of the special properties of the Fourier
series involved to reduce the amount of computations in each time
step. Another important aspect is that my code
does not yet handle output data in an efficient way.
The runs which I presented before in this thesis produced a
couple of
$10$ GBytes of data in total which caused
hard disk space problems.

If the BKL-conjecture is true, the Gowdy singularity is exceptional
since it is 
asymptotically velocity dominated.
For other more general applications involving gravitational
singularities, one can expect that the solutions behave oscillatory. 
This means that we must pay special attention to the implementation of
the time integrators so that they can cope with this difficult
problem. Ideas for this can be found in 
\cite{Berger96,Berger2002}. Some results on this issue are available  in
the class of 
$\U\times\U$-symmetric spacetimes with non-vanishing twist constants
(which is only possible on $\T$). Also, for $\U$-symmetry, velocity
dominated singularities are expected in the ``polarized'' case
\cite{Berger98a}, while oscillatory ones show up in the
``non-polarized'' case \cite{Berger98b}. 

With evolution systems similar to the commutator field equations to
study the $\S$-Gowdy singularities, we are forced to set up the Cauchy
problem with respect to a standard Cauchy surface because, as noted
before, there seems to be no way of regularizing the equations on
$\scri$. Hence we have the problem that we cannot decide a priori from
the data if the corresponding solution is FAdS.
Hence, to study the
singularities in FAdS $\S$-Gowdy spacetimes, one should explore the
ideas to transfer data from the conformal field equations to the commutator
field equations  in 
\Sectionref{sec:IDcommutFE}. Assuming that this can be done, the next
problem for more 
thorough studies of $\S$-Gowdy FAdS solutions
is to construct more general or at least different classes 
of initial data than those derived in
\Sectionref{ch:settingupCauchyProblems}.  In particular it will be interesting
to construct a family of $\S$-Gowdy initial data that contains the
polarized case to compare with the analytic results in the
$\lambda=0$-case. On the one hand, this can be considered as another code check
in the $\lambda=0$-case, and on the other hand, the outstanding
$\lambda>0$ case can be studied in this simpler setting first.
We have argued before that it is an outstanding problem 
for all topologies if  the presence of the cosmological constant
can change the 
phenomenology of the solutions. Already in the
much simpler case of  
$\T$-topology there are interesting expectations, in particular
regarding outstanding issues of certain theorems, see
\Sectionref{sec:ExpectT3GowdyLambda}. Indeed, with the current version
of the $\T$-commutator field equations code systematic
studies of these issues should
be straight forward and new insights can be expected.
Again regarding the initial data issue, eventually one would like to
have initial data on $\scrip$ which can be 
considered in some sense as generic within, say, the $\S$-Gowdy class,
i.e.\ one should be able to give convincing 
arguments that the corresponding solutions include all relevant
phenomena of generic solutions.

As we already said, the Gowdy case with spatial $\S$-topology is
outstanding from the numerical but also from the analytical point of
view. Indeed, there are, besides the analytical results for the
polarized case in \cite{Isenberg89}, 
problems in the analytical investigations in \cite{Stahl02}
in the two cases $\SoXSt$ and $\S$. It would be
enlightening to find out, possibly with our numerical method, if
these problems are really caused by unexpected phenomenology or rather by
the imperfectness of the analytical methods.  Note again that being
able to 
perform calculations in the $\S$-Gowdy case, we are also able to do
calculations in the $\SoXSt$-case as argued in
\Chapterref{ch:underlyingquestions}.

\paragraph{Studies of solutions with Cauchy horizons}
In the discussion in
\Sectionref{sec:S3singularGowdy} we speculated that the ``small
inhomogeneity'' run shown there develops a Cauchy horizon in the 
past. However, to make conclusive statements, further investigations
are necessary. If this turns out to be true then one has to
find out if and how this spacetime fits into Moncrief's class of generalized
Taub-NUT spacetimes \cite{moncrief84}. If, after all,
this solution turns out rather to have a curvature singularity in the
past, e.g.\ 
because the Kretschmann scalar blows up later, it
could still be enlightening to study the transition from the curvature
singularity of the perturbed Taub-NUT spacetime to the Cauchy horizon
of the unperturbed case by systematically decreasing the inhomogeneity
parameter. In the opposite direction, the general belief is that
Cauchy horizons decay in some way 
to BKL-singularities under generic perturbations so that a real
observer can never cross it. However, first,
it is not clear how this transition takes place and second there are
situations known with different behavior, for instance {weak null
  singularities} (\Sectionref{sec:bklconjecture}).
In any case, we should point out here that studies related to Cauchy horizons
are usually very subtle. It is  in general unsolved how to actually
determine, in 
particular numerically, if a
solution develops a Cauchy horizon. A first non-sufficient indication
for a Cauchy horizon is that the trace of the $2$nd fundamental form
blows up, because the slicing approaches a null hypersurface, while  
curvature invariants are bounded.
But, how do we distinguish
a Cauchy horizon from
a simple break down of the coordinate gauge?  How do we want to
conclude something about extendibility when the unknowns in the equations blow
up numerically? This is indeed a problem because variables blowing up
numerically without any analytical information on the type of
singularity can be expected to involve large error components.

In situation for which one knows analytically that Cauchy
horizons exist, there is an alternative way to study the character of
the horizon and the associated spacetime. Namely, we can try to
reconstruct such a solution
numerically in a gauge built upon timelike congruences which, at
least partly, are able to cross the horizon. Then all issues related to
Cauchy horizons like instability with respect to perturbations, in
particular those induced by numerical
errors, local non-uniqueness and possibly the existence of closed
causal curves could have an impact, and it could be interesting to see
how these show up in the numerical computation.
Dependent on the situation,
such investigations might already be possible in the general conformal
Gauß gauge 
since this gauge is motivated geometrically. The generalization of our current
implementation which is restricted to the Levi-Civita conformal Gauß gauge
(\Sectionref{sec:LCCGG}) to general conformal Gauß gauges is not
expected to be problematic in principle. 
However, when we want to compute solutions as just
suggested with the general conformal field equations, we have to
solve or at least weaken the problem of constraint growth. Recall that
our runs in  
\Sectionref{sec:runsGCFE} showed strong constraint growth and it
seemed that this issue was the main limitation for the precision of
the calculations.
It might be crucial to find another reduction of the Bianchi system
such that the 
constraint propagation is optimized. 

\newpage

\paragraph{The set dSSR and properties of corresponding solutions}
We have already mentioned that there is currently not much
understanding of the properties of the stability neighborhood of the
de-Sitter solution, e.g.\  boundedness, connectedness etc.\ It would
be interesting to obtain further insights on this and such could be obtained
by means of our numerical code.
This could be done by studying the initial data space systematically,
for example by investigating non-linear perturbations of the
$\lambda$-Taub-NUT family, because this family interpolates some of
the relevant regions in this space.

Related to this is the following. We have studied both singular
and regular $\S$-Gowdy spacetimes in \Sectionref{sec:comparison} and
\ref{sec:runsGCFE} and speculated about the non-linear stability (of
the continuum equations!)
of this class within the class of $\U$-symmetric spacetimes. It would
be interesting to continue such studies because such a stability might
give us indications about the
physical significance of Gowdy spacetimes.
For the regular class, in particular those $\S$-Gowdy solution
whose $\scrip$-initial data is in dSSR, the
current implementation of my code
seems to be well adapted to such stability studies because,
apart from the constraint 
growth probably associated with the formulation of choice of the conformal
field equations, we were able to reach high accuracy. In the singular
case, i.e.\ when we leave the stability region, many of the technical
problems above have to be considered before reliable conclusions about
these stability issues can be
obtained. 

A very challenging class of problems is related to the construction of 
solutions whose initial data is at (or maybe even close to) the
boundary of dSSR with partial 
$\scrim$. In this case 
one can expect that one has to find a gauge such that the time slices do not
approach the singular parts of $\scrim$ too closely while
simultaneously covering 
the regular parts. Maybe it is possible to realize such a gauge 
within the class of conformal Gauß gauges. 
If we were indeed successful to construct such solutions, then we could
maybe draw connections to the field of
cosmological black hole physics as we have argued in
\Sectionref{sec:situation_FADS}. Namely, it might be possible that
such solutions with partial $\scrim$ in the $\S$-case fail to be
asymptotically  simple in a non-trivial manner. Thus they could have
both collapsing and expanding regions, and this might give rise to this
interpretation.  One knows some families of explicit 
cosmological black hole solutions, for instance the
Schwarzschild de-Sitter solution (\Sectionref{sec:SSdS}); but this
does not correspond to our class because neither $\scrip$ nor \scrim
are compact. Further, so 
far, there are  
not many attempts to study the space of cosmological black hole
spacetime more generally. One of these attempts 
is  \cite{bicak1995} (see also
references therein) where the class of Robinson-Trautman spacetimes
with $\lambda>0$ is investigated. 

\paragraph{Other issues and applications}
There are further numerical problems whose
investigations can 
be interesting. It could be particularly
enlightening to get further insights into the problem of axis
instabilities  which were discussed in
\Sectionref{sec:comparisonGowdymethods}. One way of approaching this
problem is to
experiment with different time-marching schemes. 
Such studies on how
this influences both numerical dissipativity and efficiency, could be
of use even for completely different numerical applications.

Another numerical problem of more fundamental kind is the
following. During the analysis of my numerical
experiments (\Chapterref{ch:numexperiments}) I pointed out that
round-off errors can be an important error source and it can be
expected that this is true also in other applications.
To study systematically how important this error source really is, I
should, besides making statistical analyses as in \cite{Kaneko70},
repeat a few of my 
runs with ``quad'' precision. Up to now, I
used the standard ``double'' precision exclusively, meaning that the
numbers in the code are
represented with roughly $15$ decimal digits, internally 
with $64$ bits. I chose ``double'' precision because this
is the native internal precision of the processors. For 
``quad'' precision the numbers are represented with $128$
bits which gives an accuracy up to $33$ decimal digits. But note that,
quad precision numbers have to be emulated by the software, i.e.\ are not 
directly supported by the processor, and this slows down the code
significantly. More information on the binary representation of
floating point numbers used by the Intel compilers and processors can
be found in \cite{Intel}.
Nevertheless, it should be
sufficient to compute a few time steps in a few cases to draw further
conclusions.

There is another big class of applications which seems unrelated at
first sight, namely the \term{dS-CFT correspondence} which we have
completely ignored in this thesis so far. The basic idea is to formulate a
physical conformal field theory on the conformal boundary of a
spacetime with positive Ricci curvature, and, dependent on the
properties of the conformal boundary and the conformal field theory, to
make statements about the properties of the conformal
spacetime. The
underlying concept is the holo\-graphy principle
\cite{Susskind94}. The dS-CFT correspondence is the positive curvature
analogue of the famous 
AdS-CFT correspondence \cite{Maldacena97} which was first 
formulated to resolve certain issues in string theory. A mathematically
profound overview article is \cite{Anderson04b} and a brief
description and the status of the mathematical idea can be found in
the introduction of \cite{galloway2002}. Since
Friedrich's Cauchy problem is a well-posed formulation, at least in
$3+1$ dimensions, to study the outstanding issues of dS-CFT,
investigations of open issues in this correspondence are indeed
related to our discussions before.

\section{Summary}
\label{sec:summary}
Motivated by the observational evidence that our universe is in an
accelerated expanding phase presently, we have studied the class of
FAdS solutions of EFE in this thesis. 
Although these spacetimes are ``almost'' in accordance with the cosmic no-hair
picture in the future and hence ``simple'' in some
sense there, there is a difficult and so far not understood interrelation
of the past behavior and the properties of $\scrip$. 
If the topology of $\scrip$ is compact, then there are a number of
theorems, including singularity theorems, the Yamabe theorem and
stability theorems which draw a certain picture about this
issue. However, this picture is both incomplete and mysterious. First,
the theorems available are not able to cover all situations of
interest. Second, there are major outstanding issues in general
relativity which also show up here.
For example it is currently not
more than a hope that generic maximal globally hyperbolic past
incomplete solutions are  
$C^2$-inextendible as claimed by the strong cosmic
censorship conjecture. Further it is not clear in general under which
conditions 
curvature singularities are really of
BKL-type. In the class of FAdS solutions,
all these outstanding
questions and problems can be
expected to be influenced by a subtle interplay of the time evolution
and the geometrical
properties of $\scrip$.
FAdS solutions can be constructed by means of Friedrich's
Cauchy problem involving his conformal field equations. It is a
well-posed formulation of the problem which allows to
prescribe $\scrip$-data sets subject to relatively weak constraints;
in particular the topology and geometry of $\scrip$ can be given
almost freely. In this thesis, we decided to restrict $\scrip$ to the
topologies  
$\S$ and $\T$, since these provide the simplest paradigms to shed light
on these issues.

I decided to approach the outstanding issues in this class of
spacetimes numerically.
There are several techniques to treat spacetimes with spatial
topologies as $\S$ which are ``non-trivial'' from the numerical point of 
view. We started by discussing our motivation to develop a single patch
approach based on pseudospectral methods and elaborated on expected
advantages and disadvantages. After having motivated a choice of
coordinates on $\S$, we described the development of the
numerical method. Our implementation was done in a way such that spatial
$\T$- and $\S$-topologies can be treated with the same spectral
infrastructure; this was realized by studying a map
$\T\rightarrow\S$. Consider the 
pull-back of a smooth function on $\S$ via this map to $\T$. The
special properties of its Fourier series was
analyzed and
the formally singular terms in the equations at the coordinate
singularities were explicitly regularized in terms of Fourier expansions.
\Partref{part:treatment} is devoted to the explanation of these ideas
and of the implementation of the code. Afterwards in
\Partref{part:analysis}, applications and their analyses are discussed.

In this thesis we restricted to applications involving spacetimes with
Gowdy symmetry. The main motivation for this was to 
simplify the problem, however, the code is implemented such 
that it is not restricted to this symmetry. 
Gowdy symmetry was also
the motivation for our choice of coordinates on $\S$. Gowdy spacetimes can be
considered as one of 
the simplest inhomogeneous classes of solutions of EFE but which
nevertheless show high complexity. Due to several rigorous results on
this class of spacetimes one has a large catalog of ideas
for the outstanding cases, in particular for
$\S$-Gowdy solutions. With our code we were able to provide the first,
though so far unsystematic, studies in the published literature of
$\S$-Gowdy solutions.

In \Partref{part:analysis}, we started by investigating linearizations of the
field equations and by
comparing the numerical results to the known exact solutions in
presence of the coordinate 
singularity. Then I analyzed the errors
obtained with various variants of my spectral approach for
(non-linear) regular
$\lambda$-Gowdy spacetimes with $\S$-topology. In both cases I was
able to identify one method which is stable and highly accurate,
namely the so called down-to-up method.
Next, I applied this stable method to compute singular
$\lambda$-Gowdy spacetimes with spatial $\S$-topology, but also  with
$\T$-topology for
comparisons. It turns out that the Levi-Civita conformal Gauß gauge
used for both topologies is not suitable to approach the singularity
in a ``homogeneous'' way. Although the results are very accurate, this
did not allow us
to compare our results to those obtained by means of other methods.
Nevertheless, two main
preliminary observations were made besides the successful testing of the
implementation. The first of
these results 
suggests that the class of $\lambda$-Gowdy solutions with
$\S$-topology might be non-linearly stable within
the class of 
$\U$-symmetric solutions. The second of these results has to do with
the possible stability of the Cauchy horizon of a $\lambda$-Taub-NUT
spacetime under perturbations of Gowdy symmetry. However, further
discussions and 
analyses are necessary for reliable conclusions. 

The problems related to the Levi-Civita Gauß gauge, but also the lack of
understanding how a suitable gauge to approach the Gowdy singularity
can be formulated for the conformal field equations, motivated us to
consider the commutator field 
equations. These equations are so far restricted to Gowdy symmetry and
$\T$-slices, however, they are formulated in the highly adapted timelike
area gauge and do not require $\lambda>0$. Of course, the conformal
field equations
  can also be applied for any sign of the cosmological
  constant, however, the specific formulation
  \Eqsref{eq:gcfe_levi_cevita_evolution}, which I 
  implemented, requires $\lambda=3$.  
Our
numerical results obtained with the commutator field equations were
very accurate and 
promising, and we were able to reproduce the 
fine structure at the Gowdy singularity. Indeed, these were the
first published attempts to compute $\T$-Gowdy spacetimes with
$\lambda\not=0$ numerically.  
This case is
particularly interesting as we argued and, although we
have not been able to make
systematic studies due to lack of time, we 
elaborated on our expectations related to this outstanding issue at least. Note
however, that the solutions obtained with the commutator field
equations so far are not necessarily FAdS.

We further argued that if we were able to implement our
preliminary ideas to modify the commutator field equations for
$\S$-topology in a $1+1$-manner, we could reach sufficient resolutions in the
Gowdy $\S$-case to obtain at least as much accuracy as Garfinkle in
his work on Gowdy
$\SoXSt$-solutions, but with all advantages of a 
spectral method. 

In this thesis we have focused on two main aspects: first, the
numerical construction
of solutions of EFE under the presence of the \S coordinate
singularity, and second, the numerical 
approach to the Gowdy singularity for the spatial topologies $\S$ and
$\T$. Clearly, other interesting applications, outstanding problems
and questions had to be deferred. Some of those were summarized and
discussed in \Sectionref{sec:outluckfuture}. Although we obtained only a
limited amount of results on the outstanding issues of general relativity in
this thesis, the large number of possibilities for future research
suggests that our investigations here have contributed to open the
door for many interesting studies in the future.

In summary, the investigation of the class of FAdS spacetimes is a
promising challenge both from the fundamental and the 
numerical 
point of view. A rich phenomenology with the possibility
to obtain deeper insights on fundamental outstanding problems in general
relativity is expected. I believe that 
the mutual interaction of pure 
mathematics 
and numerical treatment can be very helpful to come up with new ideas
and new approaches for our research field by combining the power of
these two.
Maybe, by
making use of the forthcoming accurate observational results, even
new implications for the properties of our own universe can be
derived, for instance predictions about the distribution of the
temperature fluctuations in the cosmic microwave background and
primordial gravitational waves.

\bibliography{bibliography}

\phantomsection
\addcontentsline{toc}{chapter}{Acknowledgments} 
\chapter*{Acknowledgments
\markboth{Acknowledgments}{Acknowledgments}}
\parskip2ex
\parindent0cm

I would like to thank everybody who contributed to this
thesis in one or the other way. Of particular importance for this was
certainly Helmut Friedrich because he gave me this exciting thesis
topic, patiently answered all my questions
and advised me both scientifically and
personally so many times. I can surely say that I learned many
interesting things 
during my thesis time -- also things I had never even
heard about before.

Particular thanks also go to Alan Rendall who helped me with many
problems and misunderstandings.

During my time at the Albert-Einstein Institute in Potsdam, I had many
inspiring (not only) scientific discussions particularly with the
people from the 
mathematical group, including Andres Acena, Roger Bieli, Sergio Dain,
Mark Heinzle, Todd Oliynyk, Tilman Vogel and Anil Zenginoglu (sorry for
leaving out all the accent marks at this point).
I also found it very helpful that I could always contact Jörg
Frauendiener and David Garfinkle by
email.

My family and Dagmar Mrozik have always given me their
support and encouragement, although I have not often been very cooperative --
especially at the end of thesis writing. I am still astonished about
Dagmar's extraordinary 
sense for the English language. Unfortunately, I have ``used'' her
talents only very few times for writing this thesis, as the reader has
certainly  recognized. 

Finally, I would like to thank all the tea producers in the world. Without
your product this work would not have been possible. As a rough
estimate, there are about $200$ liters of black tea contained in this
thesis -- and this is only for writing up! 

\phantomsection
\addcontentsline{toc}{chapter}{Erklärung} 
\chapter*{Erklärung}
\vspace{1ex}
\vfill
\Large
\begin{flushleft}
Hiermit erkläre ich, dass ich diese Arbeit an keiner anderen
Hochschule eingereicht und dass ich sie selbstständig und
ausschließlich mit den angegebenen Mitteln angefertigt habe. Weiterhin ist
kein Teil der vorliegenden Arbeit bereits anderweitig veröffentlicht worden.
\end{flushleft}
\normalfont
\vspace{5cm}
Datum\hfill Unterschrift
\vfill

\end{document}